\documentclass[aps,prx,twocolumn,superscriptaddress,amsmath,amssymb]{revtex4}
\usepackage{xcolor}
\usepackage{tikz}
\newcommand\SRO{$\rm SrRE_2O_4$}
\newcommand\BRO{$\rm BaRE_2O_4$}
\newcommand\SEO{$\rm SrEr_2O_4$}
\newcommand\SHO{$\rm SrHo_2O_4$}
\newcommand\SDO{$\rm SrDy_2O_4$}
\newcommand\BDO{$\rm BaDy_2O_4$}

\newsavebox{\cldb}
\sbox{\cldb}{
\begin{tikzpicture}
\filldraw[color=blue!100, fill=white!100, thick](0,0) circle (0.065);
\end{tikzpicture}}

\newsavebox{\club}
\sbox{\club}{
\begin{tikzpicture}
\filldraw[color=blue!100, fill=black!100, thick](0,0) circle (0.065);
\end{tikzpicture}}

\newsavebox{\cldr}
\sbox{\cldr}{
\begin{tikzpicture}
\filldraw[color=red!100, fill=white!100, thick](0,0) circle (0.065);
\end{tikzpicture}}

\newsavebox{\clur}
\sbox{\clur}{
\begin{tikzpicture}
\filldraw[color=red!100, fill=black!100, thick](0,0) circle (0.065);
\end{tikzpicture}}

\newsavebox{\clgr}
\sbox{\clgr}{
\begin{tikzpicture}
\filldraw[color=red!100, fill=gray!50, thick](0,0) circle (0.08);
\end{tikzpicture}}

\newsavebox{\clgb}
\sbox{\clgb}{
\begin{tikzpicture}
\filldraw[color=blue!100, fill=gray!50, thick](0,0) circle (0.08);
\end{tikzpicture}}

\newsavebox{\cld}
\sbox{\cld}{
\begin{tikzpicture}
\filldraw[color=black!100, fill=white!100, thick](0,0) circle (0.065);
\end{tikzpicture}}

\newsavebox{\clu}
\sbox{\clu}{
\begin{tikzpicture}
\filldraw[color=black!100, fill=black!100, thick](0,0) circle (0.065);
\end{tikzpicture}}

\newsavebox{\clg}
\sbox{\clg}{
\begin{tikzpicture}
\filldraw[color=black!100, fill=gray!50, thick](0,0) circle (0.08);
\end{tikzpicture}}

\newsavebox{\tgr}
\sbox{\tgr}{
\unitlength=0.40mm%
\begin{picture}(9,5)
\put(0,-1.4){\usebox{\clg}}
\put(2.55,5.3){\usebox{\clg}}
\put(5.1,-1.4){\usebox{\clg}}
\end{picture}}

\newsavebox{\ddd}
\sbox{\ddd}{
\unitlength=0.40mm%
\begin{picture}(9,5)
\put(0,-1.4){\usebox{\cld}}
\put(2.55,5.3){\usebox{\cld}}
\put(5.1,-1.4){\usebox{\cld}}
\end{picture}}

\newsavebox{\ddu}
\sbox{\ddu}{
\unitlength=0.40mm%
\begin{picture}(9,5)
\put(0,-1.4){\usebox{\cld}}
\put(2.55,5.3){\usebox{\cld}}
\put(5.1,-1.4){\usebox{\clu}}
\end{picture}}

\newsavebox{\dud}
\sbox{\dud}{
\unitlength=0.40mm%
\begin{picture}(9,5)
\put(0,-1.4){\usebox{\cld}}
\put(2.55,5.3){\usebox{\clu}}
\put(5.1,-1.4){\usebox{\cld}}
\end{picture}}

\newsavebox{\uuu}
\sbox{\uuu}{
\unitlength=0.40mm%
\begin{picture}(9,5)
\put(0,-1.4){\usebox{\clu}}
\put(2.55,5.3){\usebox{\clu}}
\put(5.1,-1.4){\usebox{\clu}}
\end{picture}}

\newsavebox{\uud}
\sbox{\uud}{
\unitlength=0.40mm%
\begin{picture}(9,5)
\put(0,-1.4){\usebox{\clu}}
\put(2.55,5.3){\usebox{\cld}}
\put(5.1,-1.4){\usebox{\clu}}
\end{picture}}

\newsavebox{\duu}
\sbox{\duu}{
\unitlength=0.40mm%
\begin{picture}(9,5)
\put(0,-1.4){\usebox{\cld}}
\put(2.55,5.3){\usebox{\clu}}
\put(5.1,-1.4){\usebox{\clu}}
\end{picture}}

\newsavebox{\tgrr}
\sbox{\tgrr}{
\unitlength=0.40mm%
\begin{picture}(9,5)
\put(0,-1.4){\usebox{\clgr}}
\put(2.55,5.3){\usebox{\clgr}}
\put(5.1,-1.4){\usebox{\clgr}}
\end{picture}}

\newsavebox{\dddr}
\sbox{\dddr}{
\unitlength=0.40mm%
\begin{picture}(9,5)
\put(0,-1.4){\usebox{\cldr}}
\put(2.55,5.3){\usebox{\cldr}}
\put(5.1,-1.4){\usebox{\cldr}}
\end{picture}}

\newsavebox{\ddur}
\sbox{\ddur}{
\unitlength=0.40mm%
\begin{picture}(9,5)
\put(0,-1.4){\usebox{\cldr}}
\put(2.55,5.3){\usebox{\cldr}}
\put(5.1,-1.4){\usebox{\clur}}
\end{picture}}

\newsavebox{\dudr}
\sbox{\dudr}{
\unitlength=0.40mm%
\begin{picture}(9,5)
\put(0,-1.4){\usebox{\cldr}}
\put(2.55,5.3){\usebox{\clur}}
\put(5.1,-1.4){\usebox{\cldr}}
\end{picture}}

\newsavebox{\uuur}
\sbox{\uuur}{
\unitlength=0.40mm%
\begin{picture}(9,5)
\put(0,-1.4){\usebox{\clur}}
\put(2.55,5.3){\usebox{\clur}}
\put(5.1,-1.4){\usebox{\clur}}
\end{picture}}

\newsavebox{\uudr}
\sbox{\uudr}{
\unitlength=0.40mm%
\begin{picture}(9,5)
\put(0,-1.4){\usebox{\clur}}
\put(2.55,5.3){\usebox{\cldr}}
\put(5.1,-1.4){\usebox{\clur}}
\end{picture}}

\newsavebox{\duur}
\sbox{\duur}{
\unitlength=0.40mm%
\begin{picture}(9,5)
\put(0,-1.4){\usebox{\cldr}}
\put(2.55,5.3){\usebox{\clur}}
\put(5.1,-1.4){\usebox{\clur}}
\end{picture}}

\newsavebox{\tgrb}
\sbox{\tgrb}{
\unitlength=0.40mm%
\begin{picture}(9,5)
\put(0,-1.4){\usebox{\clgb}}
\put(2.55,5.3){\usebox{\clgb}}
\put(5.1,-1.4){\usebox{\clgb}}
\end{picture}}

\newsavebox{\dddb}
\sbox{\dddb}{
\unitlength=0.40mm%
\begin{picture}(9,5)
\put(0,-1.4){\usebox{\cldb}}
\put(2.55,5.3){\usebox{\cldb}}
\put(5.1,-1.4){\usebox{\cldb}}
\end{picture}}

\newsavebox{\ddub}
\sbox{\ddub}{
\unitlength=0.40mm%
\begin{picture}(9,5)
\put(0,-1.4){\usebox{\cldb}}
\put(2.55,5.3){\usebox{\cldb}}
\put(5.1,-1.4){\usebox{\club}}
\end{picture}}

\newsavebox{\dudb}
\sbox{\dudb}{
\unitlength=0.40mm%
\begin{picture}(9,5)
\put(0,-1.4){\usebox{\cldb}}
\put(2.55,5.3){\usebox{\club}}
\put(5.1,-1.4){\usebox{\cldb}}
\end{picture}}

\newsavebox{\uuub}
\sbox{\uuub}{
\unitlength=0.40mm%
\begin{picture}(9,5)
\put(0,-1.4){\usebox{\club}}
\put(2.55,5.3){\usebox{\club}}
\put(5.1,-1.4){\usebox{\club}}
\end{picture}}

\newsavebox{\uudb}
\sbox{\uudb}{
\unitlength=0.40mm%
\begin{picture}(9,5)
\put(0,-1.4){\usebox{\club}}
\put(2.55,5.3){\usebox{\cldb}}
\put(5.1,-1.4){\usebox{\club}}
\end{picture}}

\newsavebox{\duub}
\sbox{\duub}{
\unitlength=0.40mm%
\begin{picture}(9,5)
\put(0,-1.4){\usebox{\cldb}}
\put(2.55,5.3){\usebox{\club}}
\put(5.1,-1.4){\usebox{\club}}
\end{picture}}

\newsavebox{\dddrb}
\sbox{\dddrb}{
\unitlength=0.40mm%
\begin{picture}(9,5)
\put(0,-1.4){\usebox{\cldr}}
\put(2.55,5.3){\usebox{\cldb}}
\put(5.1,-1.4){\usebox{\cldr}}
\end{picture}}

\newsavebox{\ddurb}
\sbox{\ddurb}{
\unitlength=0.40mm%
\begin{picture}(9,5)
\put(0,-1.4){\usebox{\cldr}}
\put(2.55,5.3){\usebox{\cldb}}
\put(5.1,-1.4){\usebox{\clur}}
\end{picture}}

\newsavebox{\tgrrb}
\sbox{\tgrrb}{
\unitlength=0.40mm%
\begin{picture}(9,5)
\put(0,-1.4){\usebox{\clgr}}
\put(2.55,5.3){\usebox{\clgb}}
\put(5.1,-1.4){\usebox{\clgr}}
\end{picture}}

\newsavebox{\dudrb}
\sbox{\dudrb}{
\unitlength=0.40mm%
\begin{picture}(9,5)
\put(0,-1.4){\usebox{\cldr}}
\put(2.55,5.3){\usebox{\club}}
\put(5.1,-1.4){\usebox{\cldr}}
\end{picture}}

\newsavebox{\uuurb}
\sbox{\uuurb}{
\unitlength=0.40mm%
\begin{picture}(9,5)
\put(0,-1.4){\usebox{\clur}}
\put(2.55,5.3){\usebox{\club}}
\put(5.1,-1.4){\usebox{\clur}}
\end{picture}}

\newsavebox{\uudrb}
\sbox{\uudrb}{
\unitlength=0.40mm%
\begin{picture}(9,5)
\put(0,-1.4){\usebox{\clur}}
\put(2.55,5.3){\usebox{\cldb}}
\put(5.1,-1.4){\usebox{\clur}}
\end{picture}}

\newsavebox{\duurb}
\sbox{\duurb}{
\unitlength=0.40mm%
\begin{picture}(9,5)
\put(0,-1.4){\usebox{\cldr}}
\put(2.55,5.3){\usebox{\club}}
\put(5.1,-1.4){\usebox{\clur}}
\end{picture}}

\newsavebox{\tgrbr}
\sbox{\tgrbr}{
\unitlength=0.40mm%
\begin{picture}(9,5)
\put(0,-1.4){\usebox{\clgb}}
\put(2.55,5.3){\usebox{\clgr}}
\put(5.1,-1.4){\usebox{\clgb}}
\end{picture}}

\newsavebox{\dddbr}
\sbox{\dddbr}{
\unitlength=0.40mm%
\begin{picture}(9,5)
\put(0,-1.4){\usebox{\cldb}}
\put(2.55,5.3){\usebox{\cldr}}
\put(5.1,-1.4){\usebox{\cldb}}
\end{picture}}

\newsavebox{\ddubr}
\sbox{\ddubr}{
\unitlength=0.40mm%
\begin{picture}(9,5)
\put(0,-1.4){\usebox{\cldb}}
\put(2.55,5.3){\usebox{\cldr}}
\put(5.1,-1.4){\usebox{\club}}
\end{picture}}

\newsavebox{\dudbr}
\sbox{\dudbr}{
\unitlength=0.40mm%
\begin{picture}(9,5)
\put(0,-1.4){\usebox{\cldb}}
\put(2.55,5.3){\usebox{\clur}}
\put(5.1,-1.4){\usebox{\cldb}}
\end{picture}}

\newsavebox{\uuubr}
\sbox{\uuubr}{
\unitlength=0.40mm%
\begin{picture}(9,5)
\put(0,-1.4){\usebox{\club}}
\put(2.55,5.3){\usebox{\clur}}
\put(5.1,-1.4){\usebox{\club}}
\end{picture}}

\newsavebox{\uudbr}
\sbox{\uudbr}{
\unitlength=0.40mm%
\begin{picture}(9,5)
\put(0,-1.4){\usebox{\club}}
\put(2.55,5.3){\usebox{\cldr}}
\put(5.1,-1.4){\usebox{\club}}
\end{picture}}

\newsavebox{\duubr}
\sbox{\duubr}{
\unitlength=0.40mm%
\begin{picture}(9,5)
\put(0,-1.4){\usebox{\cldb}}
\put(2.55,5.3){\usebox{\clur}}
\put(5.1,-1.4){\usebox{\club}}
\end{picture}}

\newsavebox{\red}
\sbox{\red}{
\begin{tikzpicture}
\filldraw[color=red!100, fill=gray!50, thick](0,0) circle (0.065);
\end{tikzpicture}}

\newsavebox{\bl}
\sbox{\bl}{
\begin{tikzpicture}
\filldraw[color=blue!100, fill=gray!50, thick](0,0) circle (0.065);
\end{tikzpicture}}

\newsavebox{\rp}
\sbox{\rp}{
\unitlength=0.35mm%
\begin{picture}(9,9)
\put(0,-1.4){\usebox{\red}}
\put(2.55,5.3){\usebox{\red}}
\put(5.1,-1.4){\usebox{\red}}
\end{picture}}

\newsavebox{\bp}
\sbox{\bp}{
\unitlength=0.35mm%
\begin{picture}(9,9)
\put(0,-1.4){\usebox{\bl}}
\put(2.55,5.3){\usebox{\bl}}
\put(5.1,-1.4){\usebox{\bl}}
\end{picture}}

\newsavebox{\rbp}
\sbox{\rbp}{
\unitlength=0.35mm%
\begin{picture}(9,9)
\put(0,-1.4){\usebox{\red}}
\put(2.55,5.3){\usebox{\bl}}
\put(5.1,-1.4){\usebox{\red}}
\end{picture}}

\newsavebox{\brp}
\sbox{\brp}{
\unitlength=0.35mm%
\begin{picture}(9,9)
\put(0,-1.4){\usebox{\bl}}
\put(2.55,5.3){\usebox{\red}}
\put(5.1,-1.4){\usebox{\bl}}
\end{picture}}

\begin{document}

\title{An Ising model on a 3D honeycomb zigzag-ladder lattice: a solution to the ground-state problem and application to the SrRE$_2$O$_4$ and BaRE$_2$O$_4$ magnets}

\author{Yu.I. Dublenych}
\affiliation{Institute for Condensed Matter Physics, National Academy of Sciences of Ukraine, 1 Svientsitskii Street, 79011 Lviv, Ukraine}
\author{O.A. Petrenko}
\affiliation{Department of Physics, University of Warwick, Coventry CV4 7AL, United Kingdom}

\date{\today}

\begin{abstract}
An exact solution (incomplete) of the ground-state problem for an Ising model in an external field on a 3D honeycomb zigzag-ladder lattice with two types of sites is found.
It is shown that the geometrical frustration due to the presence of triangle elements leads to the emergence of a variety of magnetic phases.
The majority of these are partially disordered (highly degenerate).
The theoretical results are used to explain the sequence of experimentally observed phase transitions in the honeycomb zigzag-ladder magnets and to predict the appearance of new phases.
\end{abstract}

\maketitle

\section{Introduction}%%%%%%%%%%%%
%----------------------------------------------------------
Geometrically frustrated magnets, due to richness of their magnetic structures and behaviors, are the most intensively studied objects in physics of magnetism and magnetic materials during several last decades.
Their theoretical description is a rather difficult task, especially in the case, where quantum effects are essential.
However, among frustrated magnets, there are many compounds with large-moment magnetic atoms.
These magnets can be well described with classical Heisenberg spin models.
If, in addition, there are easy axes of magnetization, then Ising-type models could be applied.

In the present paper, we study geometrically frustrated magnets with magnetic atoms carrying large spins.
These are 3D honeycomb zigzag-ladder magnets such as \SRO\cite{Karunadasa_2005,Petrenko_2014} and \BRO~\cite{Doi_2006,Besara_2014,Aczel_2015,Prevost_2018}, where RE is a rare earth atom.
These families of compounds exhibit a very rich magnetic behavior, especially in an external magnetic field~\cite{Hayes_2012,Cheffings_2013,Young_2014,Aczel_2014,Bidaud_2016,Petrenko_2017,Gauthier_2017_b,Young_2019,Khalyavin_2019}.
Rare earth magnetic atoms in these compounds occupy two crystallographically inequivalent positions with substantially different values of magnetic moments (which can be considered as classical ones) and, very often, almost orthogonal axes of easy magnetization.

\begin{figure}[tb]
%\begin{center}
\includegraphics[scale = 1.0]{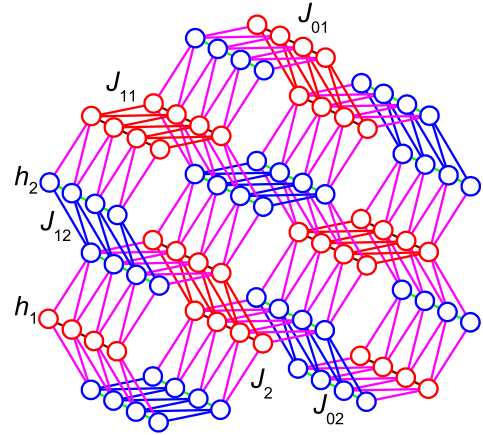}
\caption{A 3D honeycomb zigzag-ladder lattice.
There are two types of sites (spins): red and blue.
The coupling between two neighboring spins along ladder legs (runs) is $J_{01}$ for red sites and $J_{02}$ for blue ones ($J_{11}$ and $J_{12}$), respectively.
The coupling between ``red'' and ``blue'' neighboring spins is $J_2$.
External field parameters are $h_1$ and $h_2$ for the ``red'' and ``blue'' spins, respectively.}
\label{fig1}
%\end{center}
\end{figure}

One of the most important experimental and theoretical challenges is to determine magnetic structures of these magnets in an external magnetic field at low temperature.
If an appropriate Hamiltonian is established, then one can try to solve the ground-state problem for it.
This is difficult even for rather simple classical Hamiltonians and, although several methods have been developed, no general algorithm exists~\cite{Luttinger_1946,Luttinger_1951,Danielian_1961,Lyons_1964,Kanamori_1966,Morita_1974,Brandt_1986,KENNEDY_1994,Kaplan_2007,Hayden_2010}.

To give an appropriate description of honeycomb zigzag-ladder magnets, we consider an Ising-like Hamiltonian with seven parameters.
Since we deal with two types of spins, and the direction of the external field is arbitrary, two, rather than one, external field parameter should be introduced.
Although we refer to spins throughout the text, one has to keep in mind that the orbital contribution to the magnetic moments is significant in almost all \SRO\ and \BRO\ compounds.

The ground-state problem for Ising-like Hamiltonians can be solved by using the method of basic rays and basic sets of cluster configurations~\cite{Dublenych_2011_1,Dublenych_2011_2,Dublenych_2012}.
In the previous studies by one of the authors, this method was used to rigorously prove the completeness of solutions found in some other ways.
Here, we use the method as a tool for finding a solution to the ground-state problem for an Ising model on a honeycomb zigzag-ladder lattice with two types of sites.
We consider the smallest clusters, that is, triangular plaquettes.
However, there are four types of such plaquettes with six configurations for each type, therefore, the problem is rather complex.
We have found 22 basic rays (edges of ground-state regions in the parameter space), but it is not a complete set -- to determine all the basic rays, bigger clusters should be considered.
In principle, it should be possible to establish a complete solution for this ground-state problem using a specially developed computer software, but it is not elaborated yet and, for the purposes of this paper, we restrict ourselves to an incomplete solution.

However, even the incomplete solution of the ground-state problem sheds extra light on the magnetic properties of honeycomb zigzag-ladder magnets, \SRO\ and \BRO.

The paper is organized as follows. Subsection~A of Section~\ref{Sec_II} gives the description of the model under consideration and the cluster method used.
In Subsection~B, triangular plaquettes and their spin configurations are introduced, the Hamiltonian is presented as a sum of energies of all the plaquettes of the lattice.
In Subsection~C, all the basic rays (vectors) which can be found using the triangular plaquettes are listed.
Fully dimensional (that is, seven-dimensional) ground-state regions and corresponding ground-state structures found on the base of these basic rays are described in Subsection~D.
In Subsection~E, the disorder (degeneracy) of fully dimensional phases is analyzed.
Subsection~F addresses the completeness of the sets of basic rays for each phase.
In Subsection G, ``nontriangular'' fully dimensional structures neighboring ``triangular'' ones are constructed and analyzed and, in Section~H, six examples of ground-state phase diagrams are presented.
In Section~\ref{Sec_III}, the relation between the experimental and the theoretical results is discussed and, in Section~\ref{Sec_IV}, the conclusions are drawn.

\section{``Triangular'' ground-state structures}%%%%%%%%%%%%
\label{Sec_II}
%------------------------------------------------------------------------------------------------------------------------------------------------
\subsection{Model and method}%%%%%%%%%%%%
%------------------------------------------------------------------------------------------------------------------------------------------------
The magnetic lattice of 3D honeycomb zigzag-ladder magnets is shown in Fig.~\ref{fig1}.
The structure is composed of three types of zigzag ladders.
There are two species of nonequivalent sites, these are depicted with two colors: blue and red.
The value of spin at each site is equal to $-1$ or $+1$.
The coupling between two neighboring spins along ladder legs (runs) is $J_{01}$ for red sites and $J_{02}$ for blue ones ($J_{11}$ and $J_{12}$).
The coupling between spins at neighboring sites of different colors (along ladder runs) is $J_2$.
There are also two external field parameters, $h_1$ and $h_2$ for red and blue sites, respectively; these parameters depend on the components of an external field along the  two easy magnetization axes and the values of the magnetic moments at red and blue sites.
We therefore consider an Ising-type model with seven parameters and the Hamiltonian of the model reads
\begin{widetext}
\begin{eqnarray}
&&H = \sum_{\left\langle {{\color{brown}\textbf{brown}} \atop \text{bonds}} \right\rangle} J_{01}\sigma_i \sigma_j + \sum_{\left\langle {{\color{green}\textbf{green}} \atop \text{bonds}} \right\rangle} J_{02}\sigma_i \sigma_j
+ \sum_{\left\langle {{\color{red}\textbf{red}} \atop \text{bonds}} \right\rangle} J_{11}\sigma_i \sigma_j + \sum_{\left\langle {{\color{blue}\textbf{blue}} \atop \text{bonds}} \right\rangle} J_{12}\sigma_i \sigma_j + \sum_{\left\langle {{\color{magenta}\textbf{magenta}} \atop \text{bonds}} \right\rangle} J_2\sigma_i \sigma_j\nonumber\\
&&~~~~~ - \sum_{\left\langle {{\color{red}\textbf{red}} \atop \text{sites}} \right\rangle} h_1 \sigma_i - \sum_{\left\langle {{\color{blue}\textbf{blue}} \atop \text{sites}} \right\rangle} h_2 \sigma_i.
\label{eq1}
\end{eqnarray}
\end{widetext}

To find the ground states of such a model, we use a cluster method developed by one of the authors in the previous papers, the so-called method of basic rays and basic sets of cluster configurations.
Let us briefly elaborate on the main aspects of the method used.

The ground-state phase diagram for any Ising-type model is a set of convex polyhedral cones in the parameter space.
A polyhedral cone is the linear hull, that is, all linear combinations with nonnegative coefficients -- the so-called conic hull -- of a set of vectors.
It is fully determined by its edges or vectors along them.
The most important are fully dimensional polyhedral cones (seven-dimensional for the model considered).
These cones fill the parameter space without gaps and overlaps.
We refer to a structure, which is a ground-state structure in a fully dimensional polyhedral cone, as fully dimensional and to the corresponding edges (vectors) as basic rays (vectors)~\cite{Dublenych_2011_1,Dublenych_2011_2,Dublenych_2012}.
A ground-state problem can be considered as resolved if all the edges (basic rays or basic vectors) of all the fully dimensional polyhedral cones are determined as well as all the ground states at these edges.
The ground states in basic rays (the same along entire ray) are constructed with the lowest energy configurations of a cluster (or clusters).
We refer to the sets of these configurations as ``basic sets of cluster configurations."

\subsection{Triangular plaquettes and their energies} %%%%%%%%%%%%
\begin{figure}[bt]
%\begin{center}
\includegraphics[scale = 0.52]{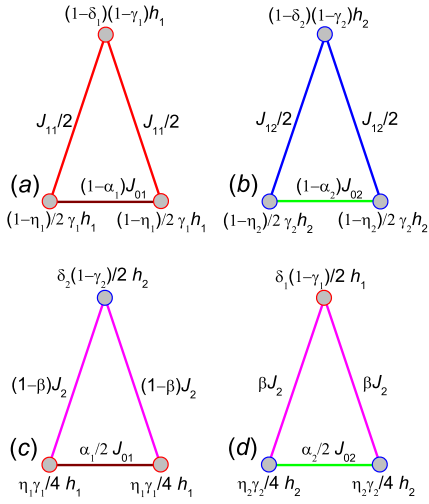}
\caption{Four types of triangular plaquettes and their energies (see Fig.~\ref{fig1}).
The arbitrariness in energy distribution between plaquettes of different types with common sites or bonds is taken into account by introducing a set of coefficients $\alpha_1$, $\alpha_2$, $\beta$, $\gamma_1$, $\gamma_2$, $\eta_1$, and $\eta_2$
which can take arbitrary values between zero and one and which we refer to as ``free'' coefficients.}
\label{fig2}
%\end{center}
\end{figure}

%------------------------------------------------------------------------------------------------------------------------------------------------
Let us consider the simplest plaquettes of the lattice shown in Fig.~\ref{fig1} -- triangular ones (Fig.~\ref{fig2}).
There are four types of triangular plaquettes, the total energy can be distributed between them in different ways, as every plaquette has vortexes and sides shared with the neighboring plaquettes.
The arbitrariness in energy distribution can be taken into account by introducing a set of coefficients $\alpha_1$, $\alpha_2$, $\beta$, $\gamma_1$, $\gamma_2$, $\eta_1$, and $\eta_2$ which can take arbitrary values between zero and one and which we refer to as ``free'' coefficients.
The four types of the triangular plaquettes and energy distribution between them are shown in Fig.~\ref{fig2}.
The Hamiltonian (\ref{eq1}) can be presented as a sum of energies for all the plaquettes,
\begin{widetext}
\begin{eqnarray}
&&H = \sum_{\usebox{\rp}\,_{_i}} \left[(1-\alpha_1)J_{01}\sigma_{i1} \sigma_{i2} + \frac{J_{11}}{2}(\sigma_{i2}\sigma_{i3} + \sigma_{i3}\sigma_{i1})
-\frac{1-\eta_1}{2}\gamma_1 h_1(\sigma_{i1} + \sigma_{i2}) - (1-\delta_1)(1-\gamma_1)h_1\sigma_{i3} \right]\nonumber\\
&&~~~+\sum_{\usebox{\bp}\,_{_i}} \left[(1-\alpha_2)J_{02}\sigma_{i1}\sigma_{i2} + \frac{J_{12}}{2}(\sigma_{i2}\sigma_{i3} + \sigma_{i3}\sigma_{i1})
-\frac{1-\eta_2}{2}\gamma_2 h_2(\sigma_{i1} + \sigma_{i2}) - (1-\delta_2)(1-\gamma_2)h_2\sigma_{i3} \right]\nonumber\\
&&~~~+\sum_{\usebox{\rbp}\,_{_i}} \left[\frac{\alpha_1}{2}J_{01}\sigma_{i1}\sigma_{i2} + (1-\beta)\frac{J_2}{2}(\sigma_{i2}\sigma_{i3} + \sigma_{i3}\sigma_{i1})
-\eta_1\frac{\gamma_1}{4}h_1(\sigma_{i1} + \sigma_{i2}) - \delta_2\frac{1-\gamma_2}{2}h_2\sigma_{i3} \right] \nonumber\\
&&~~~+\sum_{\usebox{\brp}\,_{_i}} \left[\frac{\alpha_2}{2}J_{02}\sigma_{i1}\sigma_{i2} + (1-\beta)\frac{J_2}{2}(\sigma_{i2}\sigma_{i3} + \sigma_{i3}\sigma_{i1})
-\eta_2\frac{\gamma_2}{4}h_2(\sigma_{i1} + \sigma_{i2}) - \delta_1\frac{1-\gamma_1}{2}h_1\sigma_{i3} \right],
\label{eq2}
\end{eqnarray}
\end{widetext}
where the first, second, third, and fourth summations go over all the plaquettes of the type \emph{a}, \emph{b}, \emph{c}, and \emph{d}, respectively.

Let us show, for instance, that the energy of every red site is taken into account only once in the sum of energies of all the plaquettes on the lattice, that is, in the Hamiltonian~(\ref{eq2}).
Every red site belongs to three plaquettes of type {\it a} (to one in upper position and to two in lower position), to four plaquettes of type {\it c}, and to two plaquettes of type {\it d}.
Therefore the one-site energy is
\begin{eqnarray}
&&e = -\sigma h_1\left((1 - \delta_1)(1 - \gamma_1)+2\frac{(1-\eta_1)\gamma_1}{2}\right.\nonumber\\
&&~~~\left.+4\frac{\eta_1\gamma_1}{4}+
2\frac{\delta_1(1-\gamma_1)}{2}\right) = -\sigma h_1,
\label{eq3}
\end{eqnarray}
where $\sigma$ is the value of spin at the red site.
It should be noted that the Hamiltonian~(\ref{eq2}) does not depend on free coefficients despite the fact that the four plaquette Hamiltonians do depend on them.

There are six configurations of each plaquette, $\usebox{\ddd}$, $\usebox{\ddu}$, $\usebox{\dud}$, $\usebox{\uud}$, $\usebox{\duu}$, and $\usebox{\uuu}$, where open and solid circles denote spins $\sigma = -1$ and $\sigma = +1$, respectively.
The energies of these configurations for all the four types of plaquettes are given in the Appendix.

\subsection{Basic rays (vectors) and basic sets of triangular plaquettes configurations}%%%%%%%%%%%%
%------------------------------------------------------------------------------------------------------------------------------------------------
Using the expressions for these energies (see Appendix), one can find 22 basic rays.
They are given in Table~\ref{table1}. In the first column of the table, the basic rays [7-vectors $(J_{01},J_{02},J_{11},J_{12},J_2,h_1,h_2)$] are listed.
Symbols ${}^\star$, $\widetilde{}~$, and ${}^{\overline{~~}}$ denote the following transformations: sublattices swap (red sublattice becomes blue and vice versa), spin flip on the blue sublattice, and spin flip on both sublattices.
In the second column, the ground-state configurations of the four types of plaquettes for the corresponding basic ray are given.
The symbol $\|$ separates configurations for four different types of triangular plaquettes.
The symbol \usebox{\tgr} denotes the set of all the six configurations.
In the last column, the ``free'' coefficients values that minimize the energies of the corresponding configurations in the basic ray are presented.

\begin{table*}[tb]
\caption{Basic rays and basic sets of configurations for the Ising model on a honeycomb zigzag-ladder lattice.}
\begin{ruledtabular}
\begin{tabular}{cccc}
\multicolumn{2}{c}{Basic ray}&\multicolumn{1}{c}{Basic set}&``Free''\\
\multicolumn{2}{c}{$(J_{01},J_{02},J_{11},J_{12},J_2,h_1,h_2)$}&\multicolumn{1}{c}{of configurations $\mathbf{R}_i$}&coefficients\\
\hline\\[-2mm]
$\mathbf{r}_{1}^{}$&$(-1,0,0,0,0,0,0)$&\usebox{\dddr} \usebox{\dudr}
\usebox{\uudr} \usebox{\uuur} $\|$
\usebox{\tgrb} $\|$
\usebox{\dddrb} \usebox{\dudrb} \usebox{\uudrb} \usebox{\uuurb} $\|$
\usebox{\tgrbr}&$\alpha_1 = 0$\\[1mm]
$\mathbf{r}_{1}^{\star}$&$(0,-1,0,0,0,0,0)$&
\usebox{\tgrr} $\|$
\usebox{\dddb} \usebox{\dudb} \usebox{\uudb} \usebox{\uuub} $\|$
\usebox{\tgrrb} $\|$
\usebox{\dddbr} \usebox{\dudbr} \usebox{\uudbr} \usebox{\uuubr}&$\alpha_2 = 0$\\[2mm]
$\mathbf{r}_2^{}$&$(1,0,-2,0,0,0,0)$&\usebox{\dddr} \usebox{\ddur}
\usebox{\duur} \usebox{\uuur} $\|$
\usebox{\tgrb} $\|$ \usebox{\tgrrb} $\|$
\usebox{\tgrbr}&$\alpha_1 = 0$\\[1mm]
$\mathbf{r}_2^{\star}$&$(0,1,0,-2,0,0,0)$&\usebox{\tgrr} $\|$
\usebox{\dddb} \usebox{\ddub} \usebox{\duub} \usebox{\uuub} $\|$ \usebox{\tgrrb} $\|$
\usebox{\tgrbr}&$\alpha_2 = 0$\\[1mm]
$\mathbf{r}_3^{}$&$(1,0,2,0,0,0,0)$&\usebox{\ddur} \usebox{\dudr}
\usebox{\uudr} \usebox{\duur} $\|$
\usebox{\tgrb} $\|$ \usebox{\tgrrb} $\|$ \usebox{\tgrbr} &$\alpha_1 = 0$\\[1mm]
$\mathbf{r}_3^{\star}$&$(0,1,0,2,0,0,0)$&\usebox{\tgrr} $\|$
\usebox{\ddub} \usebox{\dudb} \usebox{\uudb} \usebox{\duub} $\|$ \usebox{\tgrrb} $\|$
\usebox{\tgrbr} &$\alpha_2 = 0$\\[1mm]
$\mathbf{r}_{4}$&$(2,0,0,0,1,0,0)$&\usebox{\tgrr} $\|$ \usebox{\tgrb}
$\|$
\usebox{\ddurb} \usebox{\dudrb} \usebox{\uudrb} \usebox{\duurb} $\|$
\usebox{\tgrbr} &$\alpha_1 = 1$, $\beta = 0$\\[1mm]
$\mathbf{r}_{4}^{\thicksim}$&$(2,0,0,0,-1,0,0)$&\usebox{\tgrr} $\|$
\usebox{\tgrb} $\|$
\usebox{\dddrb} \usebox{\ddurb} \usebox{\duurb} \usebox{\uuurb} $\|$ \usebox{\tgrbr}&$\alpha_1 = 1$, $\beta = 0$\\[1mm]
$\mathbf{r}_{4}^{\star}$&$(0,2,0,0,1,0,0)$&\usebox{\tgrr} $\|$
\usebox{\tgrb} $\|$ \usebox{\tgrrb} $\|$ \usebox{\ddubr} \usebox{\dudbr} \usebox{\uudbr} \usebox{\duubr}
&$\alpha_2 = 1$, $\beta = 1$\\[1mm]
$\mathbf{r}_{4}^{\thicksim\star}$&$(0,2,0,0,-1,0,0)$&\usebox{\tgrr}
$\|$ \usebox{\tgrb} $\|$ \usebox{\tgrrb} $\|$ \usebox{\dddbr}
\usebox{\ddubr} \usebox{\duubr} \usebox{\uuubr}
&$\alpha_2 = 1$, $\beta = 1$\\[1mm]
$\mathbf{r}_{5}^{}$&$(1,0,0,0,0,2,0)$&
\usebox{\ddur} \usebox{\uudr} \usebox{\duur} \usebox{\uuur} $\|$ \usebox{\tgrb} $\|$
\usebox{\ddurb} \usebox{\uudrb} \usebox{\duurb} \usebox{\uuurb} $\|$
\usebox{\tgrbr} &$\alpha_1 = 0$, $\gamma_1 = 1$, $\eta_1 = 0$\\[1mm]
$\mathbf{r}_{5}^{-}$&$(1,0,0,0,0,-2,0)$&
\usebox{\dddr} \usebox{\ddur} \usebox{\dudr} \usebox{\duur} $\|$
\usebox{\tgrb} $\|$
\usebox{\dddrb} \usebox{\ddurb} \usebox{\dudrb} \usebox{\duurb} $\|$
\usebox{\tgrbr} &$\alpha_1 = 0$, $\gamma_1 = 1$, $\eta_1 = 0$\\[1mm]
$\mathbf{r}_{5}^{\star}$&$(0,1,0,0,0,0,2)$&\usebox{\tgrr} $\|$
\usebox{\ddub} \usebox{\uudb} \usebox{\duub} \usebox{\uuub} $\|$
\usebox{\tgrrb} $\|$
\usebox{\ddubr} \usebox{\uudbr} \usebox{\duubr} \usebox{\uuubr}
&$\alpha_2 = 0$, $\gamma_2 = 1$, $\eta_2 = 0$\\[1mm]
$\mathbf{r}_{5}^{\star{-}}$&$(0,1,0,0,0,0,-2)$&\usebox{\tgrr} $\|$
\usebox{\dddb} \usebox{\ddub} \usebox{\dudb} \usebox{\duub} $\|$
\usebox{\tgrrb} $\|$
\usebox{\dddbr} \usebox{\ddubr} \usebox{\dudbr} \usebox{\duubr}
&$\alpha_2 = 0$, $\gamma_2 = 1$, $\eta_2 = 0$\\[1mm]
$\mathbf{r}_{6}^{}$&$(0,0,1,0,0,2,0)$&\usebox{\dudr} \usebox{\uudr}
\usebox{\duur} \usebox{\uuur} $\|$
\usebox{\tgrb} $\|$
\usebox{\tgrrb} $\|$
\usebox{\tgrbr} &$\gamma_1 = \frac12$, $\delta_1 = 0$, $\eta_1 = 0$\\[1mm]
$\mathbf{r}_{6}^{-}$&$(0,0,1,0,0,-2,0)$&\usebox{\dddr} \usebox{\ddur}
\usebox{\dudr} \usebox{\uudr} $\|$
\usebox{\tgrb} $\|$
\usebox{\tgrrb} $\|$
\usebox{\tgrbr}
&$\gamma_1 = \frac12$, $\delta_1 = 0$, $\eta_1 = 0$\\[1mm]
$\mathbf{r}_{6}^{\star}$&$(0,0,0,1,0,0,2)$&\usebox{\tgrr} $\|$
\usebox{\dudb} \usebox{\uudb} \usebox{\duub} \usebox{\uuub} $\|$
\usebox{\tgrrb} $\|$
\usebox{\tgrbr}
&$\gamma_2 = \frac12$, $\delta_2 = 0$, $\eta_2 = 0$\\[1mm]
$\mathbf{r}_{6}^{\star{-}}$&$(0,0,0,1,0,0,-2)$&\usebox{\tgrr} $\|$
\usebox{\dddb} \usebox{\ddub} \usebox{\dudb} \usebox{\uudb} $\|$
\usebox{\tgrrb} $\|$
\usebox{\tgrbr}
&$\gamma_2 = \frac12$, $\delta_2 = 0$, $\eta_2 = 0$\\[1mm]
$\mathbf{r}_7^{}$&$(0,0,0,0,1,4,4)$&\usebox{\tgrr} $\|$
\usebox{\tgrb} $\|$
\usebox{\dudrb} \usebox{\uudrb} \usebox{\duurb} \usebox{\uuurb} $\|$
\usebox{\dudbr} \usebox{\uudbr} \usebox{\duubr} \usebox{\uuubr}
&$\beta = 0$, $\gamma_1 = 1$, $\gamma_2 = 0$,\\
&&&$\delta_2 = 1$, $\eta_1 = 1$\\[1mm]
$\mathbf{r}_7^{-}$&$(0,0,0,0,1,-4,-4)$&\usebox{\tgrr} $\|$
\usebox{\tgrb} $\|$
\usebox{\dddrb} \usebox{\ddurb} \usebox{\dudrb} \usebox{\uudrb}  $\|$
\usebox{\dddbr} \usebox{\ddubr} \usebox{\dudbr} \usebox{\uudbr}
&$\beta = 0$, $\gamma_1 = 1$, $\gamma_2 = 0$,\\
&&&$\delta_2 = 1$, $\eta_1 = 1$\\[1mm]
$\mathbf{r}_7^{\thicksim}$&$(0,0,0,0,-1,4,-4)$&
\usebox{\tgrr} $\|$
\usebox{\tgrb} $\|$
\usebox{\dddrb} \usebox{\ddurb} \usebox{\uudrb} \usebox{\uuurb} $\|$
\usebox{\dddbr} \usebox{\dudbr} \usebox{\duubr} \usebox{\uuubr}
&$\beta = 1$, $\gamma_1 = 1$, $\gamma_2 = 0$,\\
&&&$\delta_1 = 1$, $\eta_2 = 1$\\[1mm]
$\mathbf{r}_7^{\thicksim{-}}$&$(0,0,0,0,-1,-4,4)$&
\usebox{\tgrr} $\|$
\usebox{\tgrb} $\|$
\usebox{\dddrb} \usebox{\dudrb} \usebox{\duurb} \usebox{\uuurb} $\|$
\usebox{\dddbr} \usebox{\ddubr} \usebox{\uudbr} \usebox{\uuubr}
&$\beta = 1$, $\gamma_1 = 1$, $\gamma_2 = 0$,\\
&&&$\delta_1 = 1$, $\eta_2 = 1$\\[-3mm]
\label{table1}
\end{tabular}
\end{ruledtabular}
\end{table*}

Let us consider an example.
If $\alpha_1 = 0$, then, in ray $\mathbf{r}_{1}^{}$ for which $J_{01} < 0$ (ferromagnetic coupling) and all the other parameters are equal to zero, the following triangular plaquette configurations have the minimal energies: \usebox{\ddd},
\usebox{\dud}, \usebox{\uud}, and \usebox{\uuu} for the types {\it a} and {\it c} of triangular plaquettes (configurations \usebox{\ddur} and \usebox{\duur} have higher energies while \usebox{\ddurb} and \usebox{\duurb} configurations, despite lower energies, are incompatible with the remaining configurations) and all the possible configurations for the types {\it b} and {\it d} of triangular plaquettes.
Any global configuration, constructed with these local ones, is a ground-state configuration in this ray, that is, any global configuration, where local configurations \usebox{\ddur}, \usebox{\duur}, \usebox{\ddurb}, and \usebox{\duurb} are excluded, is a ground-state in this ray (and vice versa).
It is clear that for these ground states all the red chains are ferromagnetic while the blue chains could be arbitrary.
We refer to a structure constructed with a set of triangular plaquette configurations in such a way, that is, without any additional condition, as a ``triangular'' structure.

It should be noted that, at this stage, it is not yet proven that the $\mathbf{r}_{1}^{}$ ray is a basic one.
As it will become apparent below, the 22 rays listed in Table~\ref{table1} are indeed basic rays but they do not form a complete set.

\subsection{Fully dimensional ``triangular'' phases}%%%%%%%%%%%%
%------------------------------------------------------------------------------------------------------------------------------------------------
Although the set of basic rays is incomplete, many fully dimensional global ground-state configurations can be found using these basic rays, they are given in Table~\ref{table2} and Figs.~3-7.

\begin{figure*}[!]
\includegraphics[scale = 0.55]{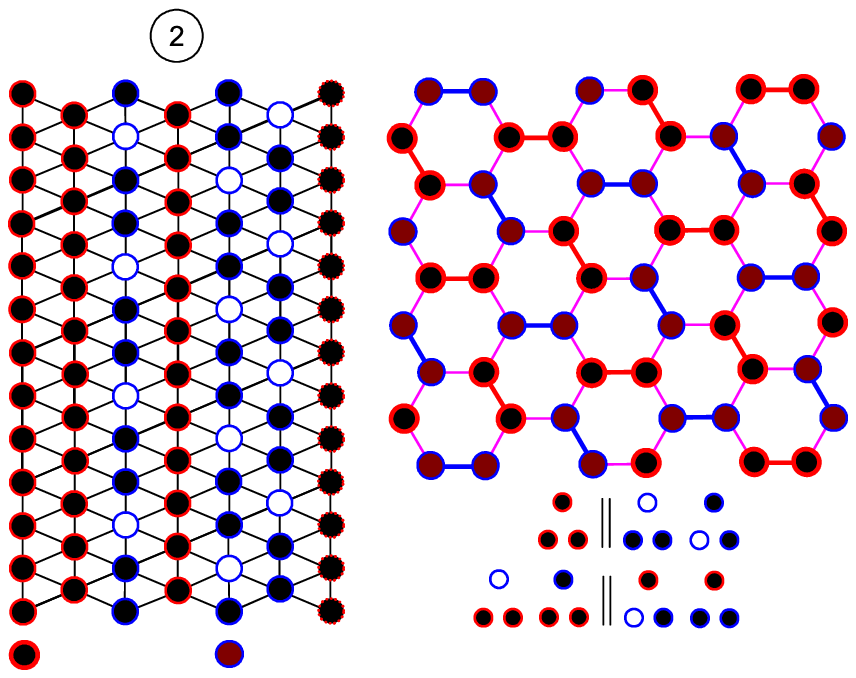}
\hspace{0.1cm}
\includegraphics[scale = 0.55]{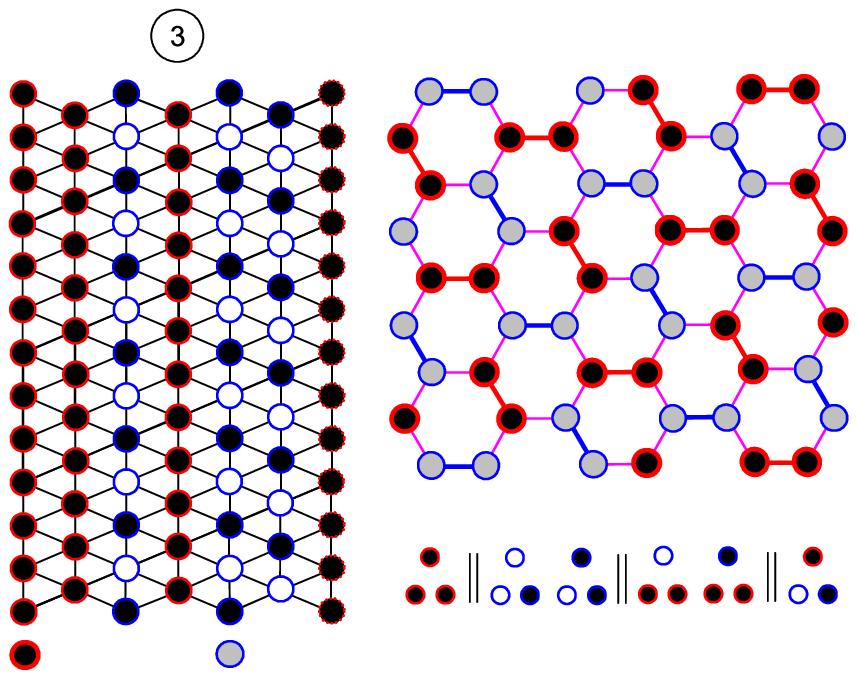}
\hspace{0.1cm}
\includegraphics[scale = 0.55]{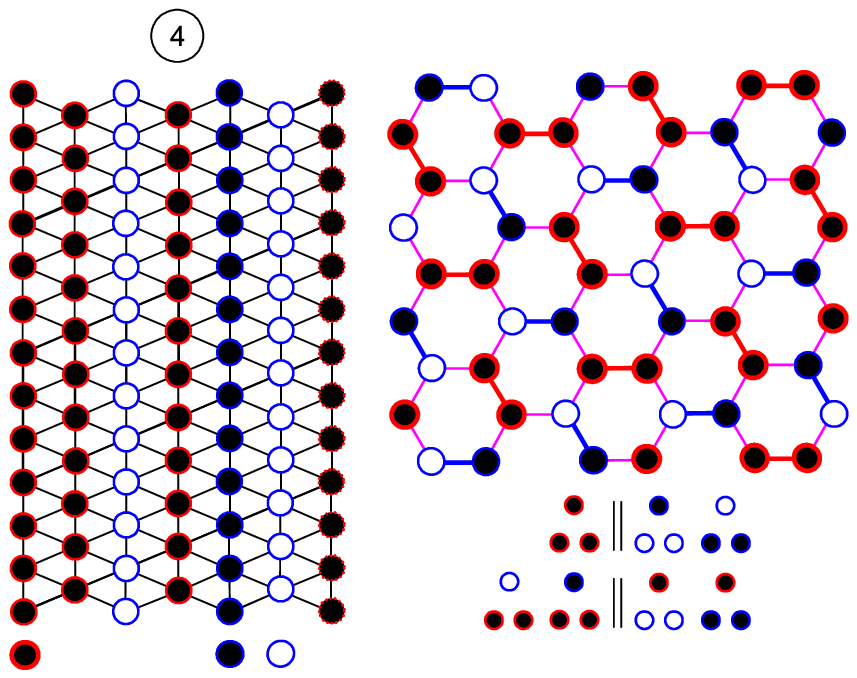}

\vspace{0.2cm}

\includegraphics[scale = 0.55]{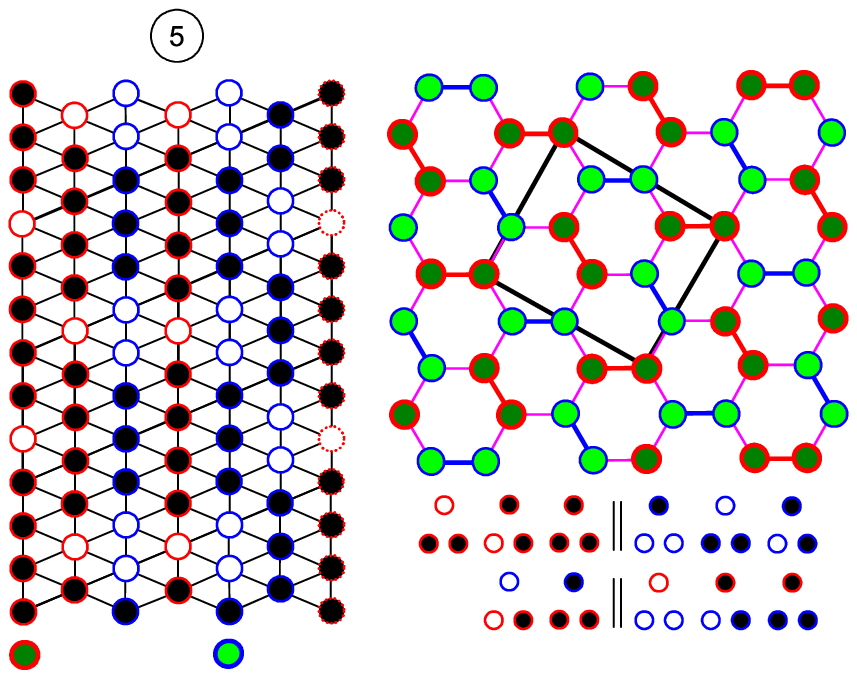}
\hspace{0.1cm}
\includegraphics[scale = 0.55]{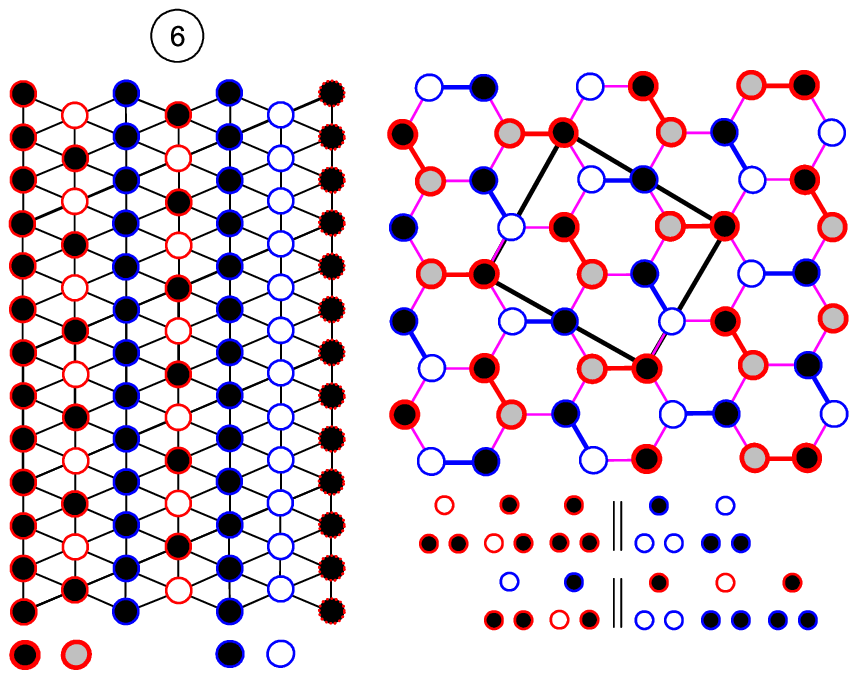}
\hspace{0.1cm}
\includegraphics[scale = 0.55]{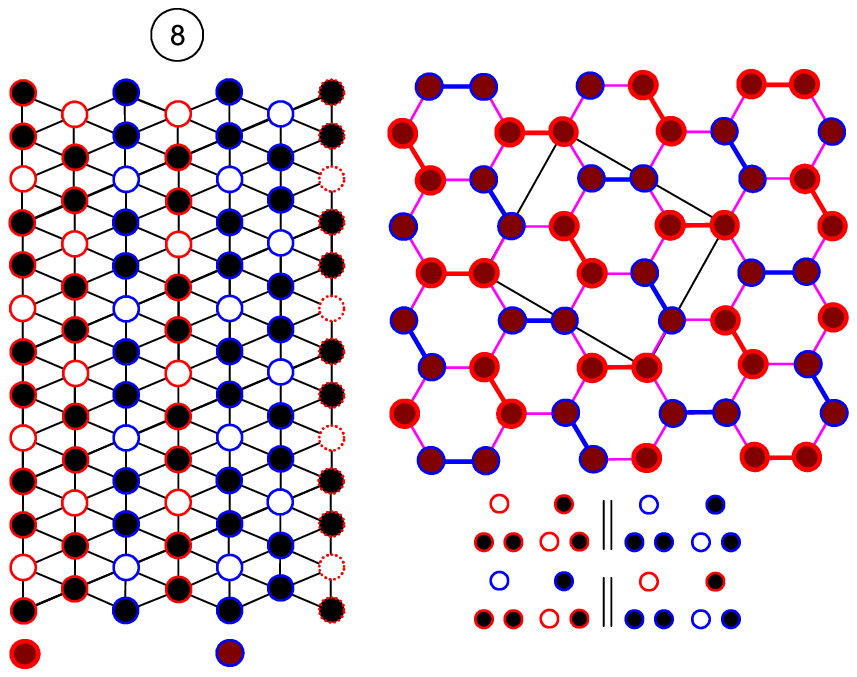}

\vspace{0.2cm}

\includegraphics[scale = 0.55]{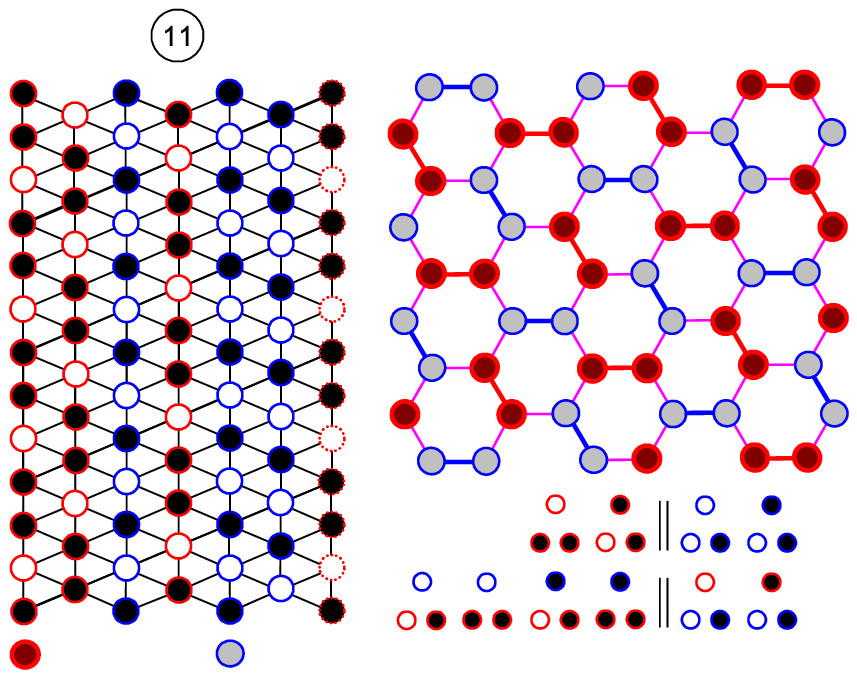}
\hspace{0.1cm}
\includegraphics[scale = 0.55]{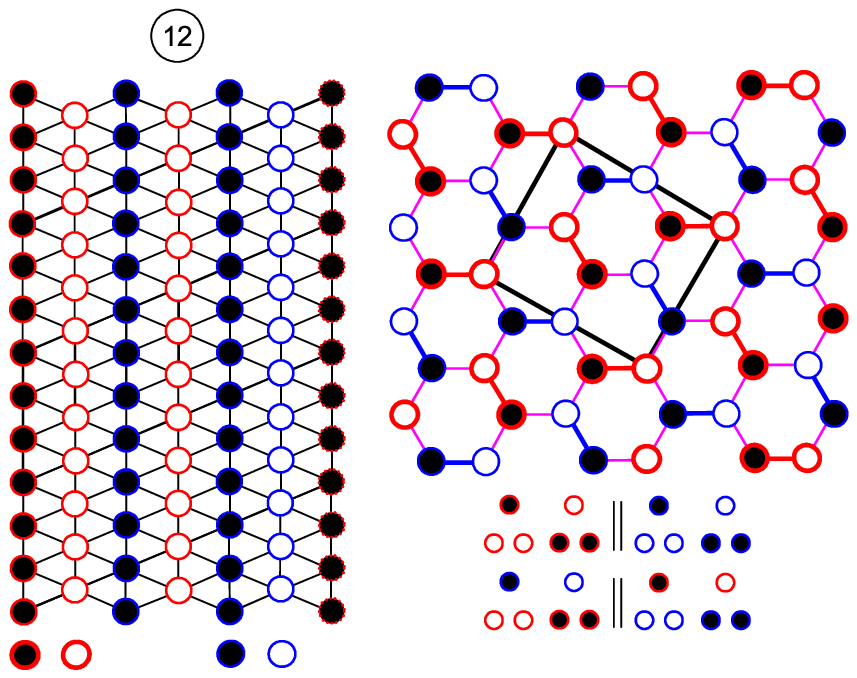}
\hspace{0.1cm}
\includegraphics[scale = 0.55]{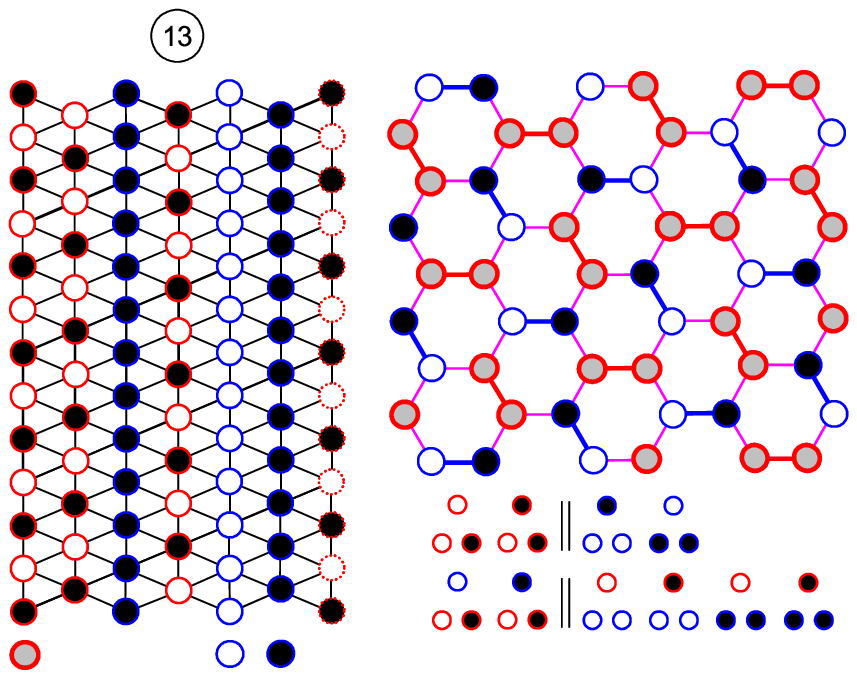}

\vspace{0.2cm}

\includegraphics[scale = 0.55]{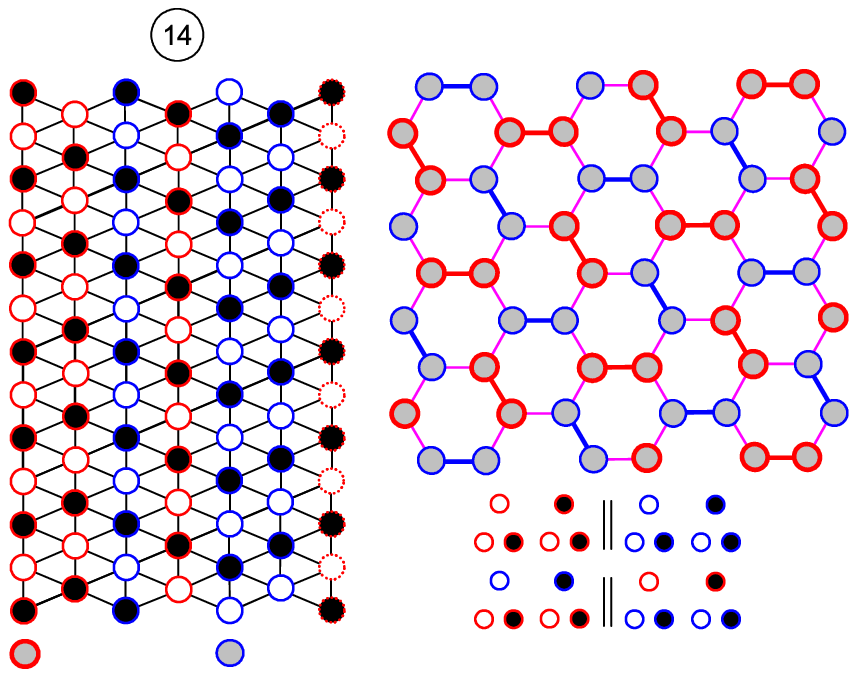}
\hspace{0.1cm}
\includegraphics[scale = 0.55]{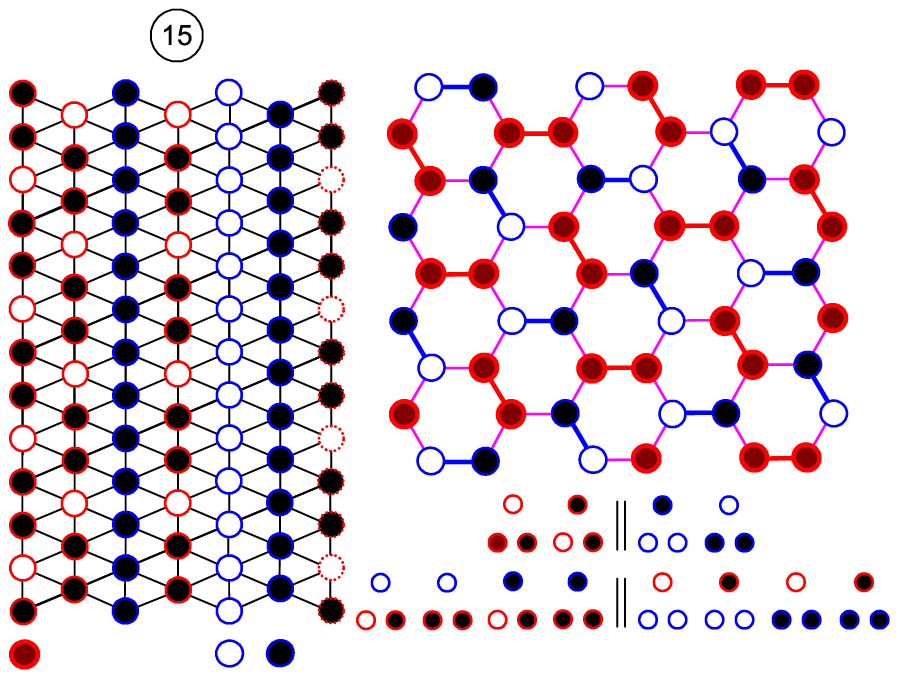}
\caption{Phases 2, 3, 4, 5, 6, 8, 11, 12, 13, 14, and 15.
For each phase, the left hand panel shows the configuration of the spins within each chain of a hexagonal well.
The larger colored circles at the bottom of some chains are a key that indicate how the spins along a chain are distributed in each of the arrangements shown in the right hand panels.
For each phase, the right hand panel shows the setting of the spins viewed down the chains from above.
The configurations of triangular plaquettes are also given.
Phases 5, 8, and 12 are ordered, while phases 2, 3, 4, 6, 11, 13, 14, and 15 are disordered (the disorder is two-dimensional).}
\label{fig3}
\end{figure*}

\begin{figure}[tb]
%\begin{center}
\includegraphics[scale = 0.55]{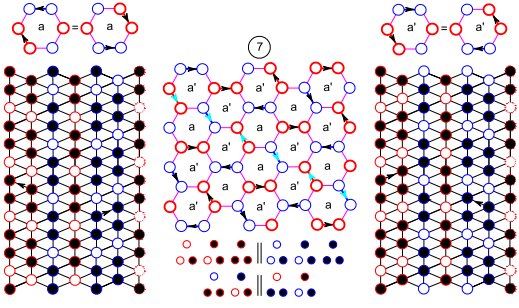}
\caption{Phase 7. The structures of this phase are constructed with two configurations of hexagonal well: $a$ and $a'$.
The arrows show the directions of shift for chains.
The global arrow configuration (constructed with two hexagon configurations) is fully determined by a line of arrows (depicted in cyan).
Therefore, the disorder is one-dimensional.}
\label{fig4}
%\end{center}
\end{figure}

\begin{table*}
\caption{Fully dimensional regions and ``triangular'' ground-state structures of the Ising model on the honeycomb zigzag-ladder lattice.
The structures (from first to 14th) are numbered in order of decreasing magnetization of the ``red'' sublattice.}
\begin{ruledtabular}
\begin{tabular}{ccccl}
Struc-&\multicolumn{1}{c}{Generating}&\multicolumn{1}{c}{Characteristics of the structure(s)}&Magneti-&\multicolumn{1}{c}{}\\
ture(s)&\multicolumn{1}{c}{configurations}&\multicolumn{1}{c}{(energy per six plaquettes)}&zation&\multicolumn{1}{c}{Basic rays}\\
\hline\\[-2mm]
$1$&~\usebox{\uuur} $\|$ \usebox{\uuub} $\|$ \usebox{\uuurb} $\|$
\usebox{\uuubr}
&$J_{01} + J_{02} + J_{11} + J_{12} + 4J_2 - h_1 - h_2$&1, 1&$\mathbf{r}_1^{}, \mathbf{r}_1^\star, \mathbf{r}_2^{}, \mathbf{r}_2^\star,$\\
&&$\left[~1~\|~1~\|~2~\|~2~\right]$,~ {\mbox order}&&$\mathbf{r}_4^\thicksim, \mathbf{r}_4^{\thicksim\star}, \mathbf{r}_5^{}, \mathbf{r}_5^\star,$\\
&&&&$\mathbf{r}_6^{}, \mathbf{r}_6^\star, \mathbf{r}_7^{}, \mathbf{r}_7^\thicksim, \mathbf{r}_7^{\thicksim-}$\\[1mm]

$2$&~\usebox{\uuur} $\|$ \usebox{\uudb} \usebox{\duub} $\|$
\usebox{\uudrb} \usebox{\uuurb} $\|$ \usebox{\duubr} \usebox{\uuubr}&
$\frac13(3J_{01} - J_{02} + 3J_{11} - J_{12} + 4J_2 - 3h_1 - h_2)$&1, 1/3&$\mathbf{r}_1^{}, \mathbf{r}_2^{}, \mathbf{r}_3^\star, \mathbf{r}_4^{\thicksim\star}, \mathbf{r}_5{}, $\\
&&$\left[~3~\|~1, 2~\|~2, 4~\|~4, 2~\right]$,~ {\mbox 2D disorder}&&$\mathbf{r}_5^\star, \mathbf{r}_6^{}, \mathbf{r}_6^\star, \mathbf{r}_7^{}, \mathbf{r}_7^\thicksim$\\[1mm]

$3$&~\usebox{\uuur} $\|$ \usebox{\ddub} \usebox{\duub} $\|$
\usebox{\uudrb} \usebox{\uuurb} $\|$ \usebox{\duubr}&
$J_{01} - J_{02} + J_{11} - h_1$&1, 0&$\mathbf{r}_1^{}, \mathbf{r}_2^{}, \mathbf{r}_2^\star, \mathbf{r}_3^\star,$\\
&&$\left[~2~\|~1, 1~\|~2, 2~\|~4~\right]$,~ {\mbox 2D disorder}&&$\mathbf{r}_4^\star, \mathbf{r}_4^{\thicksim\star}, \mathbf{r}_5^{}, \mathbf{r}_5^\star,$\\
&&&&$\mathbf{r}_5^{\star-}, \mathbf{r}_6^{}, \mathbf{r}_7^{}, \mathbf{r}_7^\thicksim$\\[1mm]

$4$&~\usebox{\uuur} $\|$ \usebox{\dudb}  \usebox{\uudb} $\|$
\usebox{\uudrb} \usebox{\uuurb} $\|$ \usebox{\dudbr} \usebox{\uuubr}&
$J_{01} + J_{02} + J_{11} - J_{12} - h_1$&1, 0&$\mathbf{r}_1^{}, \mathbf{r}_1^\star, \mathbf{r}_2^{}, \mathbf{r}_3^\star,$\\
&&$\left[~2~\|~1, 1~\|~2, 2~\|~2, 2~\right]$,~ {\mbox 2D disorder}&&$\mathbf{r}_5^{}, \mathbf{r}_6^{}, \mathbf{r}_6^\star, \mathbf{r}_6^{\star-},$\\
&&&&$\mathbf{r}_7^{}, \mathbf{r}_7^\thicksim$\\[1mm]

$5$&~\usebox{\uudr} \usebox{\duur} \usebox{\uuur} $\|$
\usebox{\dudb} \usebox{\uudb} \usebox{\duub} $\|$
\usebox{\ddurb} \usebox{\uuurb} $\|$ \usebox{\dddbr} \usebox{\duubr} \usebox{\uuubr}&
$\frac15(J_{01} + J_{02} + J_{11} - 3J_{12} + 12J_2 - 3h_1 - h_2)$&$3/5$, 1/5&$\mathbf{r}_3^\star, \mathbf{r}_4^\thicksim, \mathbf{r}_4^{\thicksim\star}, \mathbf{r}_5^{},$\\
&&$\left[~1, 2, 2~\|~1, 2, 2~\|~4, 6~\|~2, 4, 4~\right]$,~ {\mbox order}&&$\mathbf{r}_6^{}, \mathbf{r}_6^\star, \mathbf{r}_7^\thicksim$\\[1mm]%

$6$&~\usebox{\uudr} \usebox{\duur} \usebox{\uuur} $\|$
\usebox{\dudb} \usebox{\uudb} $\|$ \usebox{\uudrb} \usebox{\duurb} $\|$
\usebox{\dudbr} \usebox{\uudbr} \usebox{\uuubr}&
$\frac12(2J_{02} - 2J_{12} - 4J_2 - h_1)$&1/2, 0&$\mathbf{r}_1^\star, \mathbf{r}_3^\star, \mathbf{r}_4^{}, \mathbf{r}_5^{},$\\
&&$\left[~1, 2, 1~\|~2, 2~\|~4, 4~\|~4, 2, 2~\right]$,~ {\mbox 2D disorder}&&$\mathbf{r}_6^{}, \mathbf{r}_6^\star, \mathbf{r}_6^{\star-}, \mathbf{r}_7^{}$\\[1mm]

$7$&~\usebox{\ddur} \usebox{\duur} \usebox{\uuur} $\|$ \usebox{\ddub}
\usebox{\duub} \usebox{\uuub} $\|$ \usebox{\uudrb} \usebox{\duurb} $\|$
\usebox{\uudbr} \usebox{\duubr}&
$\frac13(-J_{01} - J_{02} + J_{11} + J_{12} - 4J_2 - h_1 - h_2)$&1/3, 1/3&$\mathbf{r}_2^{}, \mathbf{r}_2^\star, \mathbf{r}_4^{}, \mathbf{r}_4^\star,$\\
&&$\left[~1, 1, 1~\|~1, 1, 1~\|~2, 4~\|~2, 4~\right]$,~ {\mbox 1D disorder}&&$\mathbf{r}_5^{}, \mathbf{r}_5^\star, \mathbf{r}_7^{}$\\[1mm]

$8$&~\usebox{\uudr} \usebox{\duur} $\|$ \usebox{\uudb} \usebox{\duub}
$\|$ \usebox{\uudrb} \usebox{\duurb} $\|$ \usebox{\uudbr} \usebox{\duubr}
&$\frac13(-J_{01} - J_{02} - J_{11} - J_{12} - 4J_2 - h_1 - h_2)$&1/3, 1/3&$\mathbf{r}_3^{}, \mathbf{r}_3^\star, \mathbf{r}_4^{}, \mathbf{r}_4^\star,$\\
&&$\left[~1, 2 ~\|~ 1, 2~\|~2, 4~\|~2, 4~\right]$,~ {\mbox order}&&$\mathbf{r}_5^{}, \mathbf{r}_5^\star, \mathbf{r}_6^{}, \mathbf{r}_6^\star, \mathbf{r}_7^{}$\\[1mm]

$9$&~\usebox{\uudr} \usebox{\duur} $\|$ \usebox{\uudb} \usebox{\duub}
$\|$ \usebox{\ddurb} \usebox{\duurb} \usebox{\uuurb} $\|$ \usebox{\ddur}
\usebox{\duubr} \usebox{\uuubr}&
$\frac13(-J_{01} - J_{02} - J_{11} - J_{12} + 4J_2 - h_1 - h_2)$&1/3, 1/3&$\mathbf{r}_3^{}, \mathbf{r}_3^\star, \mathbf{r}_4^\thicksim, \mathbf{r}_4^{\thicksim\star},$\\
&&$\left[~1, 2~\|~1, 2~\|~2, 2, 2~\|~2, 2, 2~\right]$,~ {\mbox 2D disorder}&&$\mathbf{r}_5^{}, \mathbf{r}_5^\star, \mathbf{r}_6^{}, \mathbf{r}_6^\star$\\[1mm]

$10$&~\usebox{\uudr} \usebox{\duur} $\|$ \usebox{\ddub}
\usebox{\duub} \usebox{\uuub} $\|$ \usebox{\ddurb}
\usebox{\duurb} \usebox{\uuurb} $\|$ \usebox{\ddubr} \usebox{\duubr} \usebox{\uuubr}&
$\frac13(-J_{01} - J_{02} - J_{11} + J_{12} + 4J_2 - h_1 - h_2)$&1/3, 1/3&$\mathbf{r}_2^\star, \mathbf{r}_3^{}, \mathbf{r}_4^\thicksim, \mathbf{r}_4^{\thicksim\star},$\\
&&$\left[~1, 2~\|~1, 1, 1~\|~2, 2, 2~\|~2, 2, 2~\right]$,~ {\mbox 3D disorder}&&$\mathbf{r}_5^{}, \mathbf{r}_5^\star, \mathbf{r}_6^{}$\\[1mm]

$11$&~ \usebox{\uudr} \usebox{\duur} $\|$ \usebox{\ddub} \usebox{\duub} $\|$
\usebox{\ddurb} \usebox{\uudrb} \usebox{\duurb} \usebox{\uuurb}  $\|$ \usebox{\ddubr} \usebox{\duubr}&
$\frac13(- J_{01} -3J_{02} - J_{11} - h_1)$&1/3, 0&$\mathbf{r}_2^\star, \mathbf{r}_3^{}, \mathbf{r}_3^\star, \mathbf{r}_4^\star, \mathbf{r}_4^{\thicksim\star},$\\
&&$\left[~2, 4~\|~3, 3~\|~4, 2, 4, 2~\|~4, 8~\right]$,~ {\mbox 2D disorder}&&$\mathbf{r}_5^{}, \mathbf{r}_5^\star, \mathbf{r}_5^{\star-}, \mathbf{r}_6^{}$\\[1mm]

$12$&~\usebox{\dudr} \usebox{\uudr} $\|$ \usebox{\dudb} \usebox{\uudb}
$\|$ \usebox{\dudrb} \usebox{\uudrb} $\|$ \usebox{\dudbr} \usebox{\uudbr}
&$J_{01} + J_{02} - J_{11} - J_{12} - 4J_2$&0, 0&$\mathbf{r}_1^{}, \mathbf{r}_1^\star, \mathbf{r}_3^{}, \mathbf{r}_3^\star,$\\
&&$\left[~1, 1 ~\|~ 1, 1~\|~2, 2~\|~2, 2~\right]$,~ {\mbox order}&&$\mathbf{r}_4^{}, \mathbf{r}_4^\star, \mathbf{r}_6^{}, \mathbf{r}_6^{-},$\\
&&&&$\mathbf{r}_6^\star, \mathbf{r}_6^{\star-}, \mathbf{r}_7^{}, \mathbf{r}_7^{-}$\\[1mm]

$13$&~\usebox{\ddur} \usebox{\duur} $\|$ \usebox{\dudb} \usebox{\uudb}
$\|$ \usebox{\ddurb} \usebox{\duurb} $\|$ \usebox{\dddbr} \usebox{\dudbr}
\usebox{\uudbr} \usebox{\uuubr}&
$-J_{01} + J_{02} - J_{12}$&0, 0&$\mathbf{r}_1^\star, \mathbf{r}_2^{}, \mathbf{r}_3^{}, \mathbf{r}_3^\star,$\\
&&$\left[~1, 1~\|~1, 1~\|~2, 2~\|~1, 1, 1, 1~\right]$,~ {\mbox 2D disorder}&&$\mathbf{r}_4^{}, \mathbf{r}_4^\thicksim, \mathbf{r}_5^{}, \mathbf{r}_5^{-},$\\
&&&&$\mathbf{r}_6^\star, \mathbf{r}_6^{\star-}$\\[1mm]

$14$&~\usebox{\ddur} \usebox{\duur} $\|$ \usebox{\ddub} \usebox{\duub}
$\|$ \usebox{\ddurb} \usebox{\duurb} $\|$ \usebox{\ddubr} \usebox{\duubr}
&$-J_{01} - J_{02}$&0, 0&$\mathbf{r}_2^{}, \mathbf{r}_2^\star, \mathbf{r}_3^{}, \mathbf{r}_3^\star,$\\
&&$\left[~1, 1 ~\|~ 1, 1~\|~2, 2~\|~2, 2~\right]$,~ {\mbox 2D disorder}&&$\mathbf{r}_4^{}, \mathbf{r}_{4}^\thicksim, \mathbf{r}_4^\star, \mathbf{r}_4^{\thicksim \star}$\\
&&&&$\mathbf{r}_5^{}, \mathbf{r}_5^{-}, \mathbf{r}_5^\star, \mathbf{r}_5^{\star-}$\\[1mm]

\hline\\

$15$&~\usebox{\uudr} \usebox{\duur} $\|$ \usebox{\dudb} \usebox{\uudb} $\|$ &
$\frac13(-J_{01} + 3J_{02} - J_{11} - 3J_{12} - h_1)$&$1/3$, 0&$\mathbf{r}_1^\star, \mathbf{r}_3, \mathbf{r}_3^\star, \mathbf{r}_5^{}, \mathbf{r}_6^{},$\\[0.7mm]
&~ \usebox{\ddurb} \usebox{\uudrb} \usebox{\duurb} \usebox{\uuurb} $\|$ \usebox{\dddbr} \usebox{\dudbr} \usebox{\uudbr} \usebox{\uuubr} &$\left[~2, 4~\|~3, 3~\|~4, 2, 4, 2~\|~2, 4, 2, 4~\right]$,~ {\mbox disorder}&&$\mathbf{r}_6^\star, \mathbf{r}_6^{\star -}$\\[-3mm]

\label{table2}
\end{tabular}
\end{ruledtabular}
\end{table*}

The first column of Table~\ref{table2} labels the regions in the parameter space.
In the second column, the triangular plaquette configurations that generate all the ground-state structures in this region are shown.
The third column lists some characteristics of the structure(s), such as energy (per six plaquettes), relative number of each plaquette configuration in the structure(s), and dimensionality of disorder.
In the penultimate column, the magnetization of two sublattices per site (red and blue) is given.
The last column shows basic vectors (not all in most cases) for the region considered, among these there are necessarily seven linearly independent ones, except for region 15.
For this region seven basic vectors are determined but only six of them are linearly independent.
However, one can prove that phase 15 is fully dimensional.
To obtain all the basic vectors for this phase, larger clusters should be considered.

Let us show an example.
Structures 3 (see Fig.~\ref{fig3}), composed of red $uu$ chains and blue $ud$ chains, are generated with triangular plaquette configurations \usebox{\uuur}, \usebox{\ddub}, \usebox{\duub}, \usebox{\uudrb}, \usebox{\uuurb}, and \usebox{\duubr}, that is, every triangular plaquette in this structures should be one of these.
This set of triangular plaquette configurations is a subset of basic sets $\mathbf{R}_i$ for basic rays $\mathbf{r}_1^{}, \mathbf{r}_2^{}, \mathbf{r}_2^\star, \mathbf{r}_3^\star, \mathbf{r}_4^\star, \mathbf{r}_4^{\thicksim\star}, \mathbf{r}_5^{}, \mathbf{r}_5^\star, \mathbf{r}_5^{\star-}, \mathbf{r}_6^{}, \mathbf{r}_7^{}$, and $\mathbf{r}_7^\thicksim$ (see Table~\ref{table1}).
It means that, in the conic hull of this set of vectors, structures 3 are ground-state ones.
To calculate the energy of structures 3, it is sufficient to determine relative number of each plaquette configuration in these structures (see Appendix).
These numbers are 2, 1, 1, 2, 2, and 4, respectively.

In Table~\ref{table2}, we give only one representative per class of structures.
Other structures of the class can be obtained from given one by applying three transformations (${}^\star$, $\widetilde{}~$, and ${}^{\overline{~~}}$).
For instance, the class of structure 5 contains eight structures (see Table~\ref{table3} and Fig.~\ref{fig8}): 5, $5^\star$, $\widetilde 5$, $\widetilde 5^\star$, $\overline 5$, $\overline {5^\star}$, $\overline {\widetilde 5}$, and $\overline {\widetilde 5^\star}$.
This is the maximum number of structures in one class. It should be noted here that $\overline{\widetilde{n^\ast}} = ({\tilde n})^\ast$.
The class of structure 14 contains only structure 14 because this structure is symmetric with respect to all the three transformations.
There are 69 ``triangular'' phases in total.

\subsection{Disorder of the phases}%%%%%%%%%%%%
%------------------------------------------------------------------------------------------------------------------------------------------------
\begin{figure*}[tb]
%\begin{center}
\includegraphics[scale = 0.55]{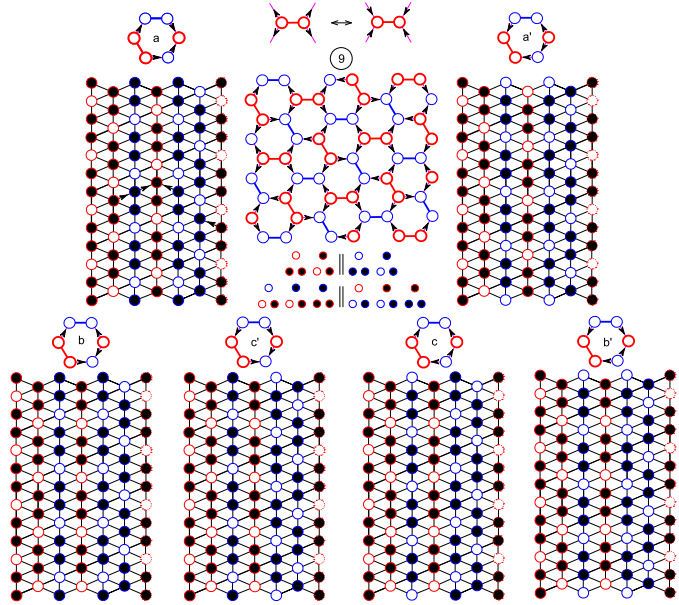}
\caption{Structure(s) 9 is composed of six hexagonal well configurations corresponding to six arrow configurations of hexagons ($a$, $b$, $c$, $a'$, $b'$, and $c'$) in which two arrows are pointing clockwise and two others anticlockwise or vice versa.
The global arrow configuration generated by local configuration $b$ is depicted.
The substitution shown in the upper part of the figure can be made locally without violating the ground state rules.
Therefore, the disorder is two-dimensional.}
\label{fig5}
%\end{center}
\end{figure*}

Now, let us analyze the disorder of phases.
Phases 1, 5, 8, and 12 (and the phases obtained from them by using the transformations described above) are ordered.
All the other phases are disordered, that is, there are an infinite number of structures with the same energy.
A disorder can be characterized by its dimensionality.
For instance, the disorder of phase~7 is one-dimensional, as shown in Fig.~\ref{fig4}.
The structures of this phase are constructed with two hexagonal well configurations that can be depicted as hexagons with two arrows showing the shift of the corresponding chains.
The global lattice configuration is mapped on two-dimensional arrow configurations.
It is easy to see, such an arrow configuration is determined by an arbitrary one-dimensional sequence of arrows.
So, the disorder is one-dimensional.

\begin{figure*}[tb]
%\begin{center}
\includegraphics[scale = 0.55]{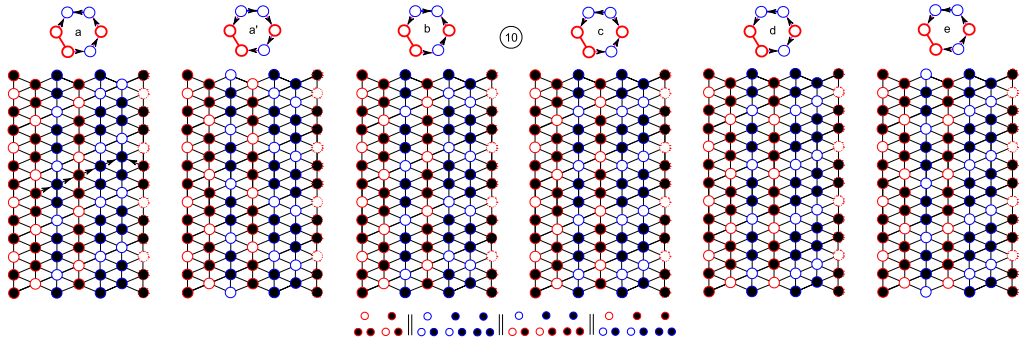}
\caption{Structures 10 are composed of ten hexagonal well configurations corresponding to ten arrow configurations of hexagons, $a$, $b$, $c$, $d$, $e$, $a'$, $b'$, $c'$, $d'$, and $e'$ (the last four are not shown), in which one arrow is pointing clockwise and four others anticlockwise or vice versa.}
\label{fig6}
%\end{center}
\end{figure*}

The disorder of phase 2 is two-dimensional, since all the chains are ordered but every up-up-down ($uud$) chain can be in three different positions.
That is, there is a perfect order along the $c$~direction (along the chains) but a disorder in the $ab$-plane.
The disorder of phase 4 is also two-dimensional, since every ``blue" ladder can be in two different positions.
It is shown in Fig.~\ref{fig5} that the disorder of phase 9 is two-dimensional as well, because all the chains are ordered and, in the structure generated by arrow configuration $b$, the local arrow configurations depicted in upper part of the figure are interchangeable.

\begin{figure*}[htb]
\begin{center}
\includegraphics[scale = 0.27]{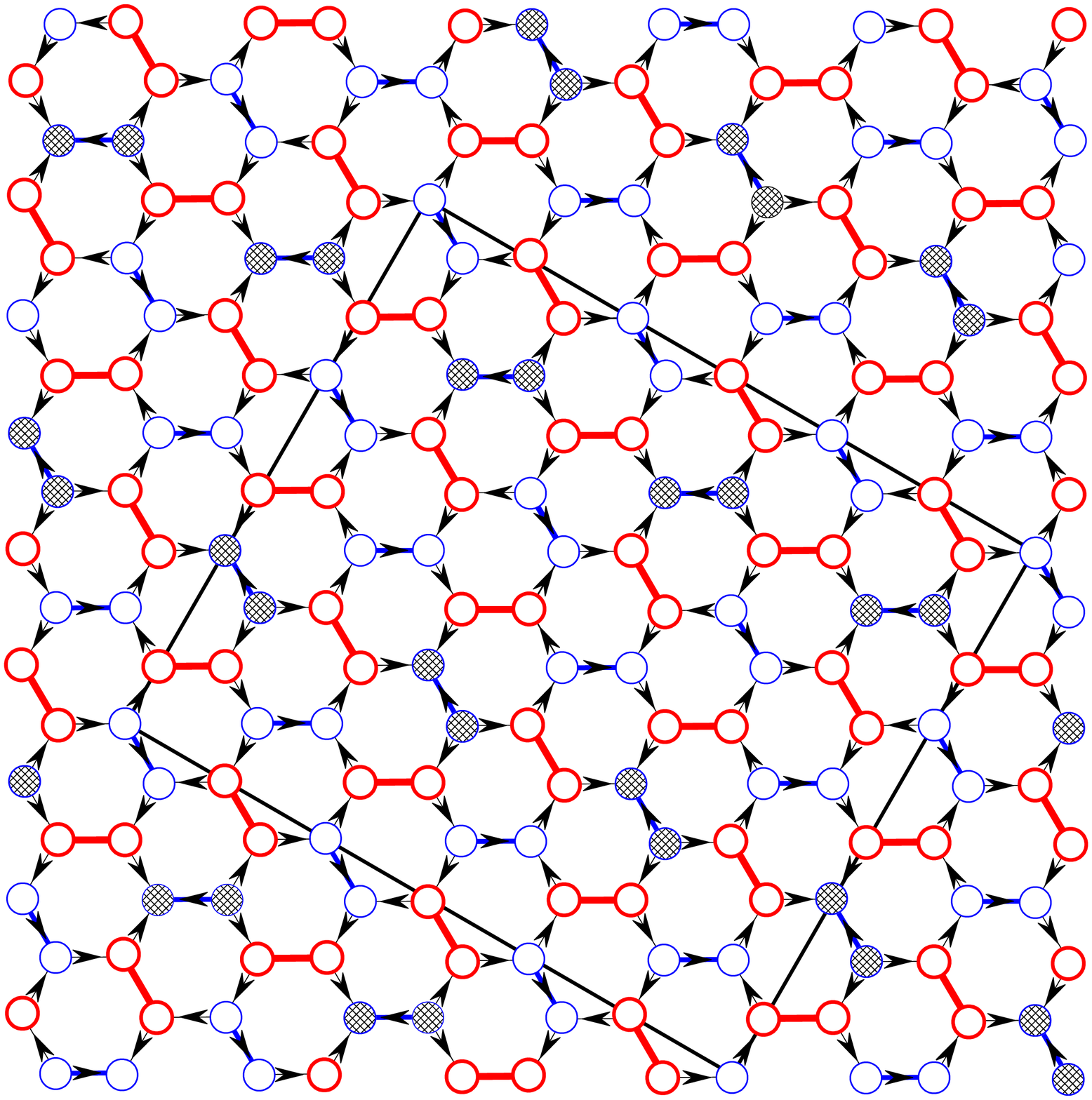}
\includegraphics[scale = 0.26]{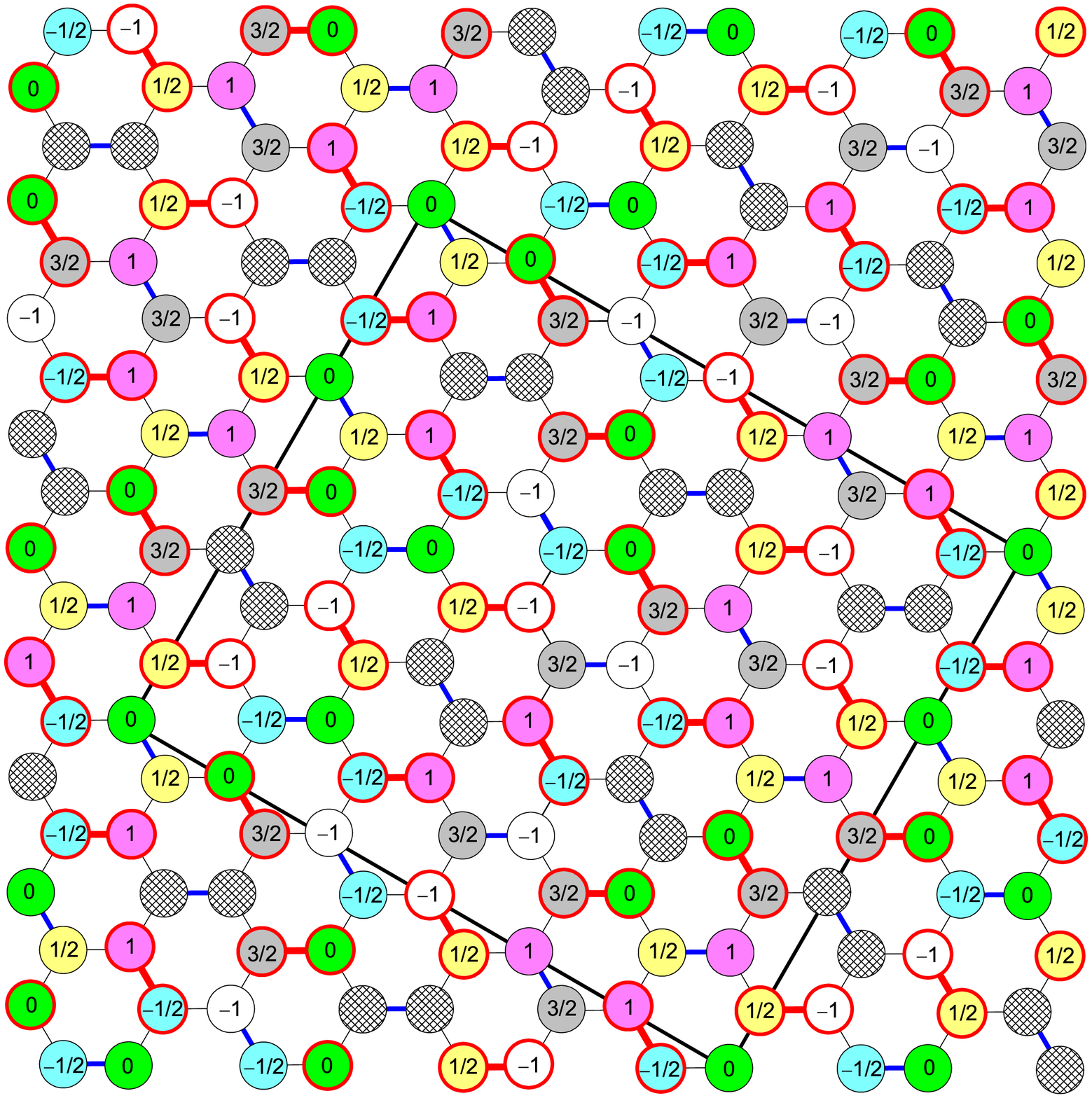}
\includegraphics[scale = 1.0]{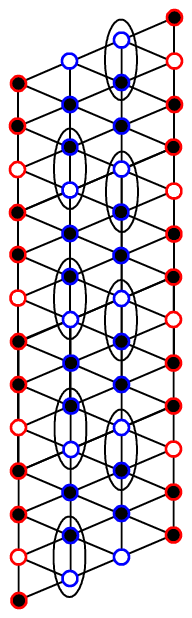}
\caption{One of possible structures of phase 10 is shown in two ways, (left panel) with the help of arrow configurations and (middle panel) by indicating the shift of each chain, in the units of the in-chain spin distance.
The unit cell is also indicated.
For the ``shaded'' chains the shifts of all the three neighboring chains are equal.
In such blue chains (right panel), one of two spins in each oval can be arbitrary, the other being of opposite direction.
The disorder in phase 10 is three-dimensional.}
\label{fig7}
\end{center}
\end{figure*}

The disorder of phase 10 is three-dimensional, that is, the degeneration is macroscopic and therefore there exists a residual entropy in this phase.
Let us prove this.
The structures of phase 10 are constructed with ten hexagonal well configurations, shown in Fig.~6.
These well configurations are composed of identical chains, $uud$.
The shift of a chain configuration when passing to a neighboring one can be indicated by an arrow.
So, we have ten arrow configurations of hexagons.
Notation $x^\prime$ means that all the arrows in the hexagon are opposite to those in hexagon configuration $x$.
In Fig.~\ref{fig7}, an example of arrow configurations and another representation of the same structures, by explicit indication of chain positions, are given.
For the ``shaded'' chains the shifts of all the three neighboring chains are equal.
In such blue chains, one of two spins in each oval (Fig.~\ref{fig7}, right panel) can point in an arbitrary direction with other being opposite.
So, the disorder in phase 10 is three-dimensional.

\begin{table*}[htb]
\caption{Structure 5 and seven other structures obtained from it using transformations ${}^\star$, $\widetilde{}~$, and
${}^{\overline{~~}}$.}
\begin{ruledtabular}
\begin{tabular}{ccccl}
Struc-&\multicolumn{1}{c}{Generating}&\multicolumn{1}{c}{Characteristics of the structure}&Magneti-&\multicolumn{1}{c}{Basic}\\
ture&\multicolumn{1}{c}{configurations}&\multicolumn{1}{c}{(energy per six plaquettes)}&zation&\multicolumn{1}{c}{rays}\\
\hline\\[-2mm]
$5$&~\usebox{\uudr} \usebox{\duur} \usebox{\uuur} $\|$ \usebox{\dudb} \usebox{\uudb} \usebox{\duub} $\|$
\usebox{\ddurb} \usebox{\uuurb} $\|$ \usebox{\dddbr} \usebox{\duubr} \usebox{\uuubr}&
$\frac15(J_{01} + J_{02} + J_{11} - 3J_{12} + 12J_2 - 3h_1 - h_2)$&$3/5$, 1/5&$\mathbf{r}_3^\star, \mathbf{r}_4^\thicksim, \mathbf{r}_4^{\thicksim\star}, \mathbf{r}_5^{},$\\
&&$\left[~1, 2, 2~\|~1, 2, 2~\|~4, 6~\|~2, 4, 4~\right]$,~ {\mbox order}&&$\mathbf{r}_6^{}, \mathbf{r}_6^\star, \mathbf{r}_7^\thicksim$\\[1mm]%

$5^\star$&~\usebox{\dudr} \usebox{\uudr} \usebox{\duur} $\|$
\usebox{\uudb} \usebox{\duub} \usebox{\uuub} $\|$
\usebox{\dddrb} \usebox{\duurb} \usebox{\uuurb} $\|$ \usebox{\ddubr} \usebox{\uuubr}&
$\frac15(J_{01} + J_{02} - 3J_{11} + J_{12} + 12J_2 - h_1 - 3h_2)$&1/5, 3/5&$\mathbf{r}_3^{}, \mathbf{r}_4^\thicksim, \mathbf{r}_4^{\thicksim\star}, \mathbf{r}_5^\star,$\\
&&$\left[~1, 2, 2~\|~1, 2, 2~\|~2, 4, 4~\|~4, 6~\right]$,~ {\mbox order}&&$\mathbf{r}_6^{}, \mathbf{r}_6^\star, \mathbf{r}_7^{\thicksim-}$\\[1mm]%

$\widetilde{5}$&~\usebox{\uudr} \usebox{\duur} \usebox{\uuur} $\|$
\usebox{\ddub} \usebox{\dudb} \usebox{\uudb} $\|$
\usebox{\uudrb} \usebox{\duurb} $\|$ \usebox{\dudbr} \usebox{\uudbr} \usebox{\duubr}&
$\frac15(J_{01} + J_{02} + J_{11} - 3J_{12} - 12J_2 - 3h_1 + h_2)$&$3/5$, $-1/5$&$\mathbf{r}_3^\star, \mathbf{r}_4, \mathbf{r}_4^\star, \mathbf{r}_5^{},$\\
&&$\left[~1, 2, 2~\|~2, 2, 1~\|~6, 4~\|~4, 2, 4~\right]$,~ {\mbox order}&&$\mathbf{r}_6^{}, \mathbf{r}_6^{\star-}, \mathbf{r}_7^{}$\\[1mm]

$\widetilde{5}^\star$&~\usebox{\ddur} \usebox{\dudr} \usebox{\uudr}
$\|$ \usebox{\uudb} \usebox{\duub} \usebox{\uuub} $\|$
\usebox{\dudrb} \usebox{\uudrb} \usebox{\duurb} $\|$ \usebox{\uudbr} \usebox{\duubr}&
$\frac15(J_{01} + J_{02} - 3J_{11} + J_{12} - 12J_2 + h_1 - 3h_2)$&$-1/5$, 3/5&$\mathbf{r}_3^{}, \mathbf{r}_4^{}, \mathbf{r}_4^\star, \mathbf{r}_5^\star,$\\
&&$\left[~2, 2, 1~\|~1, 2, 2~\|~4, 2, 4~\|~6, 4~\right]$,~ {\mbox order}&&$\mathbf{r}_6^{-}, \mathbf{r}_6^\star, \mathbf{r}_7^{}$\\[1mm]

$\overline 5$&~\usebox{\dddr} \usebox{\ddur} \usebox{\dudr} $\|$
\usebox{\ddub} \usebox{\dudb} \usebox{\uudb} $\|$
\usebox{\dddrb} \usebox{\duurb} $\|$ \usebox{\dddbr} \usebox{\ddubr} \usebox{\uuubr}&
$\frac15(J_{01} + J_{02} + J_{11} - 3J_{12} + 12J_2 + 3h_1 + h_2)$&$-3/5$, $-1/5$&$\mathbf{r}_3^\star, \mathbf{r}_4^\thicksim, \mathbf{r}_4^{\thicksim\star}, \mathbf{r}_5^{-},$\\
&&$\left[~2, 2, 1~\|~2, 2, 1~\|~6, 4~\|~4, 4, 2~\right]$,~ {\mbox order}&&$\mathbf{r}_6^{-}, \mathbf{r}_{6}^{\star{-}}, \mathbf{r}_7^{\thicksim -}$\\[1mm]%

$\overline{5^\star}$&~\usebox{\ddur} \usebox{\dudr} \usebox{\uudr} $\|$
\usebox{\dddb} \usebox{\ddub} \usebox{\dudb} $\|$
\usebox{\dddrb} \usebox{\ddurb} \usebox{\uuurb} $\|$ \usebox{\dddbr} \usebox{\duubr}&
$\frac15(J_{01} + J_{02} - 3J_{11} + J_{12} + 12J_2 + h_1 + 3h_2)$&$-1/5$, $-3/5$&$\mathbf{r}_3^{}, \mathbf{r}_4^\thicksim, \mathbf{r}_4^{\thicksim\star}, \mathbf{r}_5^{\star -},$\\
&&$\left[~2, 2, 1~\|~2, 2, 1~\|~4, 4, 2~\|~6, 4~\right]$,~ {\mbox order}&&$\mathbf{r}_6^{-}, \mathbf{r}_{6}^{\star{-}}, \mathbf{r}_7^{\thicksim}$\\[1mm]%

$\overline{\widetilde{5}}$&~\usebox{\dddr} \usebox{\ddur} \usebox{\dudr} $\|$
\usebox{\dudb} \usebox{\uudb} \usebox{\duub} $\|$
\usebox{\ddurb} \usebox{\dudrb} $\|$ \usebox{\ddubr} \usebox{\dudbr} \usebox{\uudbr}&
$\frac15(J_{01} + J_{02} + J_{11} - 3J_{12} - 12J_2 + 3h_1 - h_2)$&$-3/5$, 1/5&$\mathbf{r}_3^\star, \mathbf{r}_4, \mathbf{r}_4^\star, \mathbf{r}_5^{-},$\\
&&$\left[~2, 2, 1~\|~1, 2, 2~\|~4, 6~\|~4, 2, 4~\right]$,~ {\mbox order}&&$\mathbf{r}_6^{-}, \mathbf{r}_6^{\star}, \mathbf{r}_7^{-}$\\[1mm]

$\overline{\widetilde{5}^\star}$&~\usebox{\dudr} \usebox{\uudr} \usebox{\duur}
$\|$ \usebox{\dddb} \usebox{\ddub} \usebox{\dudb} $\|$
\usebox{\ddurb} \usebox{\dudrb} \usebox{\uudrb} $\|$ \usebox{\ddubr} \usebox{\dudbr}&
$\frac15(J_{01} + J_{02} - 3J_{11} + J_{12} - 12J_2 - h_1 + 3h_2)$&$1/5$,$-3/5$&$\mathbf{r}_3^{}, \mathbf{r}_4^{}, \mathbf{r}_4^\star, \mathbf{r}_5^{\star -},$\\
&&$\left[~1, 2, 2~\|~2, 2, 1~\|~4, 2, 4~\|~4, 6~\right]$,~ {\mbox order}&&$\mathbf{r}_6^{}, \mathbf{r}_6^{\star -}, \mathbf{r}_7^{-}$\\[-3mm]
\label{table3}
\end{tabular}
\end{ruledtabular}
\end{table*}

\begin{figure*}[htb]
\begin{center}
\includegraphics[scale = 0.7]{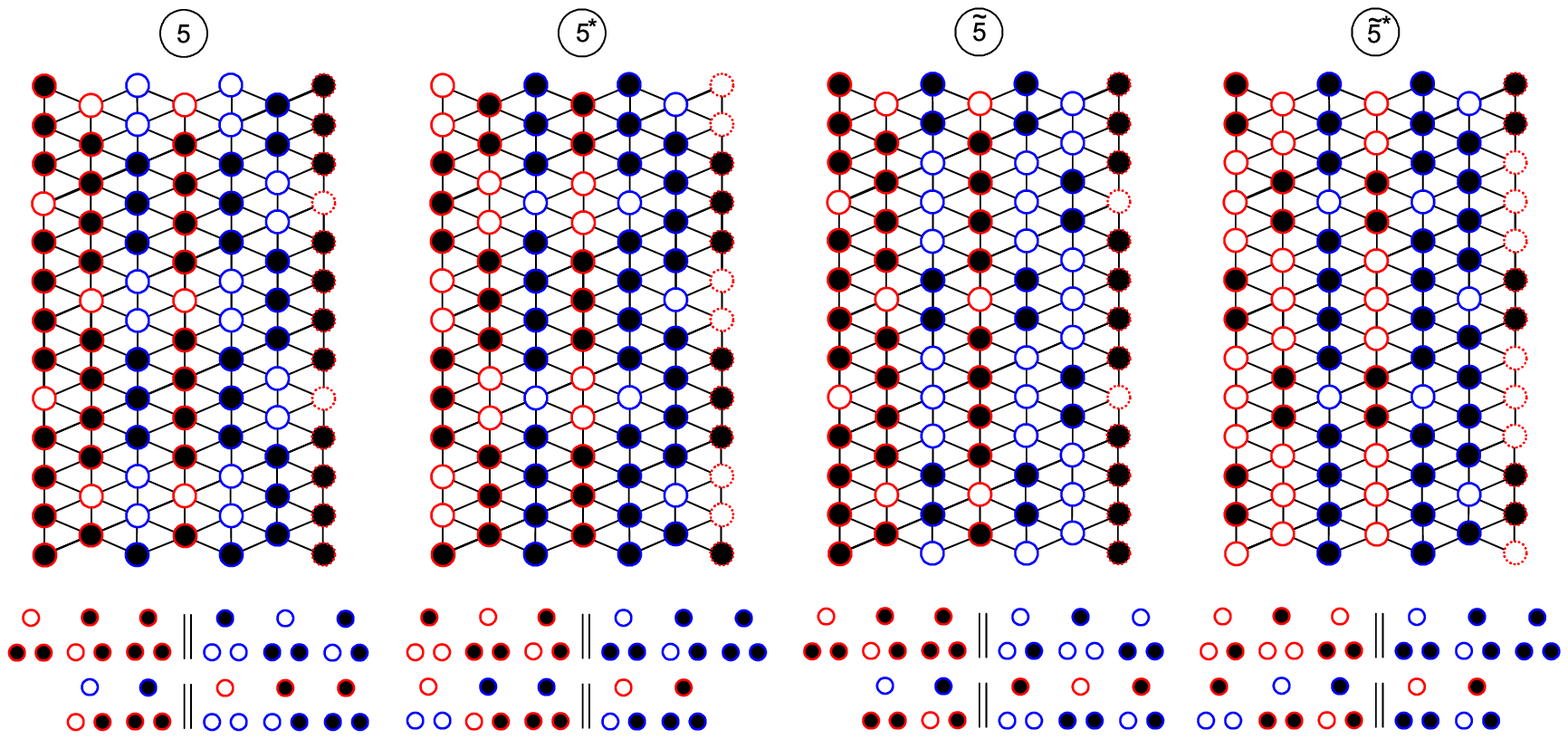}
\caption{Structure 5 and three related structures obtained from it by using transformations ${}^\star$, and~ $\widetilde{}~$.
Transformation ${}^{\overline{~~}}$ (spin flip on both sublattices) gives additional four structures.
Only one hexagonal well is shown for each structure.}
\label{fig8}
\end{center}
\end{figure*}

\subsection{Completeness of sets of basic rays for each phase}%%%%%%%%%%%%
%------------------------------------------------------------------------------------------------------------------------------------------------
Considering a set of basic rays (vectors) for a fully dimensional phase (see Table~\ref{table2}), we can check whether this set is complete.
First of all we should find all the six-dimensional faces (6-faces) of corresponding seven-dimensional polyhedral cone and the configurations of triangular plaquettes for these 6-faces.
As an example, let us consider phase 13.
The set of basic vectors for this phase is $\{\mathbf{r}_1^\star, \mathbf{r}_2^{}, \mathbf{r}_3^{}, \mathbf{r}_3^\star, \mathbf{r}_4^{}, \mathbf{r}_4^\thicksim, \mathbf{r}_5^{}, \mathbf{r}_5^{-}, \mathbf{r}_6^\star, \mathbf{r}_6^{\star-}\}$.\
The sets of basic vectors for its 6-faces and corresponding sets of plaquette configurations are given below. For each 6-face the corresponding neighboring phase is indicated in parentheses.\\[-3mm]

(1) $\{\mathbf{r}_1^\star, \mathbf{r}_2^{}, \mathbf{r}_3^{}, \mathbf{r}_4^{}, \mathbf{r}_4^\thicksim, \mathbf{r}_5^{}, \mathbf{r}_5^{-}, \mathbf{r}_6^\star\}$,~~ $(3^\star, 13)$\\[-3mm]

$\usebox{\ddur}~ \usebox{\duur}$ $\|$ $\usebox{\dudb}~ \usebox{\uudb}~ \usebox{\uuub}$ $\|$
$\usebox{\ddurb}~ \usebox{\duurb}$ $\|$ $\usebox{\dddbr}~ \usebox{\dudbr}~ \usebox{\uudbr}~ \usebox{\uuubr}$\\[-3mm]

(2) $\{\mathbf{r}_1^\star, \mathbf{r}_2^{}, \mathbf{r}_3^{}, \mathbf{r}_4^{}, \mathbf{r}_4^\thicksim, \mathbf{r}_5^{}, \mathbf{r}_5^{-}, \mathbf{r}_6^{\star-}\}$,~~ $(\overline{3}^\star, 13)$\\[-2mm]

$\usebox{\ddur}~ \usebox{\duur}$ $\|$ $\usebox{\dddb}~ \usebox{\dudb}~ \usebox{\uudb}$ $\|$
$\usebox{\ddurb}~ \usebox{\duurb}$ $\|$ $\usebox{\dddbr}~ \usebox{\dudbr}~ \usebox{\uudbr}~ \usebox{\uuubr}$\\[-3mm]

(3) $\{\mathbf{r}_2^{}, \mathbf{r}_3^{}, \mathbf{r}_3^\star, \mathbf{r}_4^{}, \mathbf{r}_4^\thicksim, \mathbf{r}_5^{}, \mathbf{r}_5^{-}, \mathbf{r}_6^\star\}$,~~ $(11^\star, 13)$\\[-3mm]

$\usebox{\ddur}~ \usebox{\duur}$ $\|$ $\usebox{\dudb}~ \usebox{\uudb}~ \usebox{\duub}$ $\|$
$\usebox{\ddurb}~ \usebox{\duurb}$ $\|$ $\usebox{\dddbr}~ \usebox{\ddubr}~ \usebox{\dudbr}~ \usebox{\uudbr}~ \usebox{\duubr}~ \usebox{\uuubr}$\\[-3mm]

(4) $\{\mathbf{r}_2^{}, \mathbf{r}_3^{}, \mathbf{r}_3^\star, \mathbf{r}_4^{}, \mathbf{r}_4^\thicksim, \mathbf{r}_5^{}, \mathbf{r}_5^{-}, \mathbf{r}_6^{\star-}\}$,~~ $(\overline{11}^\star, 13)$\\[-3mm]

$\usebox{\ddur}~ \usebox{\duur}$ $\|$ $\usebox{\ddub}~ \usebox{\dudb}~ \usebox{\uudb}$ $\|$
$\usebox{\ddurb}~ \usebox{\duurb}$ $\|$ $\usebox{\dddbr}~ \usebox{\ddubr}~ \usebox{\dudbr}~ \usebox{\uudbr}~ \usebox{\duubr}~ \usebox{\uuubr}$\\[-3mm]

(5) $\{\mathbf{r}_1^\star, \mathbf{r}_2^{}, \mathbf{r}_3^\star, \mathbf{r}_4^\thicksim, \mathbf{r}_5^{}, \mathbf{r}_6^\star, \mathbf{r}_6^{\star-}\}$,~~ $(22, 13)$\\[-3mm]

$\usebox{\ddur}~ \usebox{\duur}~ \usebox{\uuur}$ $\|$ $\usebox{\dudb}~ \usebox{\uudb}$ $\|$
$\usebox{\ddurb}~ \usebox{\duurb}~ \usebox{\uuurb}$ $\|$ $\usebox{\dddbr}~ \usebox{\dudbr}~ \usebox{\uudbr}~ \usebox{\uuubr}$\\[-3mm]

(6) $\{\mathbf{r}_1^\star, \mathbf{r}_2^{}, \mathbf{r}_3^\star, \mathbf{r}_4^\thicksim, \mathbf{r}_5^{-}, \mathbf{r}_6^\star, \mathbf{r}_6^{\star-}\}$,~~ $(\overline{22}, 13)$\\[-3mm]

$\usebox{\dddr}~ \usebox{\ddur}~ \usebox{\duur}$ $\|$ $\usebox{\dudb}~ \usebox{\duub}$ $\|$
$\usebox{\dddrb}~ \usebox{\ddurb}~ \usebox{\duurb}$ $\|$ $\usebox{\dddbr}~ \usebox{\dudbr}~ \usebox{\uudbr}~ \usebox{\uuubr}$\\[-3mm]

(7) $\{\mathbf{r}_1^\star, \mathbf{r}_2^{}, \mathbf{r}_3^\star, \mathbf{r}_4^{}, \mathbf{r}_5^{}, \mathbf{r}_6^\star, \mathbf{r}_6^{\star-}\}$,~~ $(\widetilde{22}, 13)$\\[-3mm]

$\usebox{\ddur}~ \usebox{\duur}~ \usebox{\uuur}$ $\|$ $\usebox{\dudb}~ \usebox{\uudb}$ $\|$
$\usebox{\ddurb}~ \usebox{\uudrb}~ \usebox{\duurb}$ $\|$ $\usebox{\dddbr}~ \usebox{\dudbr}~ \usebox{\uudbr}~ \usebox{\uuubr}$\\[-3mm]

(8) $\{\mathbf{r}_1^\star, \mathbf{r}_2^{}, \mathbf{r}_3^\star, \mathbf{r}_4^{}, \mathbf{r}_5^{-}, \mathbf{r}_6^\star, \mathbf{r}_6^{\star-}\}$,~~ $(\overline{\widetilde{22}}, 13)$\\[-3mm]

$\usebox{\dddr}~ \usebox{\ddur}~ \usebox{\duur}$ $\|$ $\usebox{\dudb}~ \usebox{\uudb}$ $\|$
$\usebox{\ddurb}~ \usebox{\dudrb}~ \usebox{\duurb}$ $\|$ $\usebox{\dddbr}~ \usebox{\dudbr}~ \usebox{\uudbr}~ \usebox{\uuubr}$\\[-3mm]

(9) $\{\mathbf{r}_1^\star, \mathbf{r}_3^{}, \mathbf{r}_3^\star, \mathbf{r}_4^\thicksim, \mathbf{r}_5^{}, \mathbf{r}_6^\star, \mathbf{r}_6^{\star-}\}$,~~ $(23, 13)$\\[-3mm]

$\usebox{\ddur}~ \usebox{\uudr}~ \usebox{\duur}$ $\|$ $\usebox{\dudb}~ \usebox{\uudb}$ $\|$
$\usebox{\ddurb}~ \usebox{\duurb}~ \usebox{\uuurb}$ $\|$ $\usebox{\dddbr}~ \usebox{\dudbr}~ \usebox{\uudbr}~ \usebox{\uuubr}$\\[-3mm]

(10) $\{\mathbf{r}_1^\star, \mathbf{r}_3^{}, \mathbf{r}_3^\star, \mathbf{r}_4^\thicksim, \mathbf{r}_5^{-}, \mathbf{r}_6^\star, \mathbf{r}_6^{\star-}\}$,~~ $(\overline{23}, 13)$\\[-3mm]

$\usebox{\ddur}~ \usebox{\dudr}~ \usebox{\duur}$ $\|$ $\usebox{\dudb}~ \usebox{\uudb}$ $\|$
$\usebox{\dddrb}~ \usebox{\ddurb}~ \usebox{\duurb}$ $\|$ $\usebox{\dddbr}~ \usebox{\dudbr}~ \usebox{\uudbr}~ \usebox{\uuubr}$\\[-3mm]

(11) $\{\mathbf{r}_1^\star, \mathbf{r}_3^{}, \mathbf{r}_3^\star, \mathbf{r}_4^{}, \mathbf{r}_5^{}, \mathbf{r}_6^\star, \mathbf{r}_6^{\star-}\}$,~~ $(\widetilde {23}, 13)$\\[-3mm]

$\usebox{\ddur}~ \usebox{\uudr}~ \usebox{\duur}$ $\|$ $\usebox{\dudb}~ \usebox{\uudb}$ $\|$
$\usebox{\ddurb}~ \usebox{\uudrb}~ \usebox{\duurb}$ $\|$ $\usebox{\dddbr}~ \usebox{\dudbr}~ \usebox{\uudbr}~ \usebox{\uuubr}$\\[-3mm]

(12) $\{\mathbf{r}_1^\star, \mathbf{r}_3^{}, \mathbf{r}_3^\star, \mathbf{r}_4^{}, \mathbf{r}_5^{-}, \mathbf{r}_6^\star, \mathbf{r}_6^{\star-}\}$,~~ $(\overline{\widetilde {23}}, 13)$\\[-3mm]

$\usebox{\ddur}~ \usebox{\dudr}~ \usebox{\duur}$ $\|$ $\usebox{\dudb}~ \usebox{\uudb}$ $\|$
$\usebox{\ddurb}~ \usebox{\dudrb}~ \usebox{\duurb}$ $\|$ $\usebox{\dddbr}~ \usebox{\dudbr}~ \usebox{\uudbr}~ \usebox{\uuubr}$\\[-2mm]

In the first four faces, in addition to the structures 13, there are other structures that we have previously identified.
This means that these faces indeed bound the region for phase 13.
The remaining eight faces also bound this region, since, in addition to structures 13, some new structures can be constructed for these faces.
The structures (among these new ones) containing the most of plaquette configurations which are absent in structures 13 (configuration $\usebox{\uuur}$ and $\usebox{\uuurb}$ for the fifth face; configurations $\usebox{\uudr}$ and $\usebox{\uuurb}$ for the ninth face) are the structures of fully dimensional phases whose regions have common boundaries with region 13.
These structures are shown in Fig.~\ref{fig9} and we therefore have found the complete set of basic vectors for phase 13.

The complete sets of basic vectors are also found for phases 4
($\{\mathbf{r}_1^{}, \mathbf{r}_1^\star, \mathbf{r}_2^{}, \mathbf{r}_3^\star, \mathbf{r}_5^{}, \mathbf{r}_6^{},
\mathbf{r}_6^\star,  \mathbf{r}_6^{\star-}, \mathbf{r}_7^{},
\mathbf{r}_7^{\thicksim}\}$), 11 ($\{\mathbf{r}_2^\star,
\mathbf{r}_3^{}, \mathbf{r}_3^\star, \mathbf{r}_4^\star,
\mathbf{r}_4^{\thicksim\star}, \mathbf{r}_5^{}, \mathbf{r}_5^\star,
\mathbf{r}_5^{\star-}, \mathbf{r}_6^{}\}$), and 14
($\{\mathbf{r}_2^{}, \mathbf{r}_2^\star, \mathbf{r}_3^{},
\mathbf{r}_3^\star, \mathbf{r}_4^{}, \mathbf{r}_{4}^\thicksim,
\mathbf{r}_4^\star, \mathbf{r}_4^{\thicksim \star}, \mathbf{r}_5^{},
\mathbf{r}_5^{-}, \mathbf{r}_5^\star, \mathbf{r}_5^{\star-}\}$).
For the rest of the phases the sets of basic vectors are incomplete (see Supplement~\cite{Note1}).
Consider, for instance, phase 10. Its set of basic vectors, $\{\mathbf{r}_2^\star, \mathbf{r}_3^{}, \mathbf{r}_4^\thicksim, \mathbf{r}_4^{\thicksim\star}, \mathbf{r}_5^{}, \mathbf{r}_5^\star, \mathbf{r}_6^{}\}$, is incomplete.
The sets of basic vectors for six-dimensional faces of region 10 and corresponding sets of plaquette configurations are as follows\\[-3mm]

(1) $\{\mathbf{r}_2^\star, \mathbf{r}_4^\thicksim, \mathbf{r}_4^{\thicksim\star}, \mathbf{r}_5^{}, \mathbf{r}_5^\star, \mathbf{r}_6^{}\}$,~~ $(1, 10)$\\[-3mm]

$\usebox{\uudr}~ \usebox{\duur}~ \usebox{\uuur}$ $\|$ $\usebox{\ddub}~ \usebox{\duub}~ \usebox{\uuub}$ $\|$
$\usebox{\ddurb}~ \usebox{\duurb}~ \usebox{\uuurb}$ $\|$ $\usebox{\ddubr}~ \usebox{\duubr}~ \usebox{\uuubr}$\\[-3mm]

(2) $\{\mathbf{r}_2^\star, \mathbf{r}_3^{}, \mathbf{r}_4^\thicksim, \mathbf{r}_5^{}, \mathbf{r}_5^\star, \mathbf{r}_6^{}\}$,~~ $(2^\star, 10)$\\[-3mm]

$\usebox{\uudr}~ \usebox{\duur}$ $\|$ $\usebox{\ddub}~ \usebox{\duub}~ \usebox{\uuub}$ $\|$
$\usebox{\ddurb}~ \usebox{\duurb}~ \usebox{\uuurb}$ $\|$ $\usebox{\ddur}~ \usebox{\uudbr}~ \usebox{\duubr}~ \usebox{\uuubr}$\\[-3mm]

(3) $\{\mathbf{r}_3^{}, \mathbf{r}_4^\thicksim, \mathbf{r}_4^{\thicksim\star}, \mathbf{r}_5^{}, \mathbf{r}_5^\star, \mathbf{r}_6^{}\}$,~~ $(9, 10)$\\[-3mm]

$\usebox{\uudr}~ \usebox{\duur}$ $\|$ $\usebox{\ddub}~ \usebox{\uudb}~ \usebox{\duub}~ \usebox{\uuub}$ $\|$
$\usebox{\ddurb}~ \usebox{\duurb}~ \usebox{\uuurb}$ $\|$ $\usebox{\ddubr}~ \usebox{\duubr}~ \usebox{\uuubr}$\\[-3mm]

(4) $\{\mathbf{r}_2^\star, \mathbf{r}_3^{}, \mathbf{r}_4^{\thicksim\star}, \mathbf{r}_5^{}, \mathbf{r}_5^\star, \mathbf{r}_6^{}\}$,~~ $(11, 10)$\\[-3mm]

$\usebox{\uudr}~ \usebox{\duur}$ $\|$ $\usebox{\ddub}~ \usebox{\duub}~ \usebox{\uuub}$ $\|$
$\usebox{\ddurb}~ \usebox{\uudrb}~ \usebox{\duurb}~ \usebox{\uuurb}$ $\|$ $\usebox{\ddubr}~ \usebox{\duubr}~ \usebox{\uuubr}$\\[-3mm]

(5) $\{\mathbf{r}_2^\star, \mathbf{r}_3^{}, \mathbf{r}_4^\thicksim, \mathbf{r}_4^{\thicksim\star}, \mathbf{r}_5^{}, \mathbf{r}_5^\star\}$,~~ $(14, 10)$\\[-3mm]

$\usebox{\ddur}~ \usebox{\uudr}~ \usebox{\duur}$ $\|$ $\usebox{\ddub}~ \usebox{\duub}~ \usebox{\uuub}$ $\|$
$\usebox{\ddurb}~ \usebox{\duurb}~ \usebox{\uuurb}$ $\|$ $\usebox{\ddubr}~ \usebox{\duubr}~ \usebox{\uuubr}$\\[-3mm]

(6) $\{\mathbf{r}_2^\star, \mathbf{r}_3^{}, \mathbf{r}_4^\thicksim, \mathbf{r}_4^{\thicksim\star}, \mathbf{r}_5^{}, \mathbf{r}_6^{}\}$,~~ $(20, 10)$\\[-3mm]

$\usebox{\uudr}~ \usebox{\duur}$ $\|$ $\usebox{\dddb}~ \usebox{\ddub}~ \usebox{\duub}~ \usebox{\uuub}$ $\|$
$\usebox{\ddurb}~ \usebox{\duurb}~ \usebox{\uuurb}$ $\|$ $\usebox{\dddbr}~ \usebox{\ddubr}~ \usebox{\duubr}~ \usebox{\uuubr}$\\[-3mm]

(7) $\{\mathbf{r}_2^\star, \mathbf{r}_3^{}, \mathbf{r}_4^\thicksim, \mathbf{r}_4^{\thicksim\star}, \mathbf{r}_5^\star, \mathbf{r}_6^{}\}$,~~ $(-, 10)$\\[-3mm]

$\usebox{\dudr}~ \usebox{\uudr}~ \usebox{\duur}$ $\|$ $\usebox{\ddub}~ \usebox{\duub}~ \usebox{\uuub}$ $\|$
$\usebox{\dddrb}~ \usebox{\ddurb}~ \usebox{\duurb}~ \usebox{\uuurb}$ $\|$ $\usebox{\ddubr}~ \usebox{\duubr}~ \usebox{\uuubr}$\\[-2mm]

Face 6 gives a new phase, phase 20 (Fig.~\ref{fig9}).
Face 7 is not a face between two fully dimensional phase regions because, with the corresponding set of triangular configurations, it is not possible to construct any structure different from structures 10 (although it is possible to construct a new configuration of a well).
Therefore the set of basic vectors for region 10 is incomplete.

\subsection{Fully dimensional ``nontriangular'' phases which have common boundaries with ``triangular'' phases}%%%%%%%%%%%%
\begin{figure*}[htb]
%\begin{center}
\includegraphics[scale = 0.55]{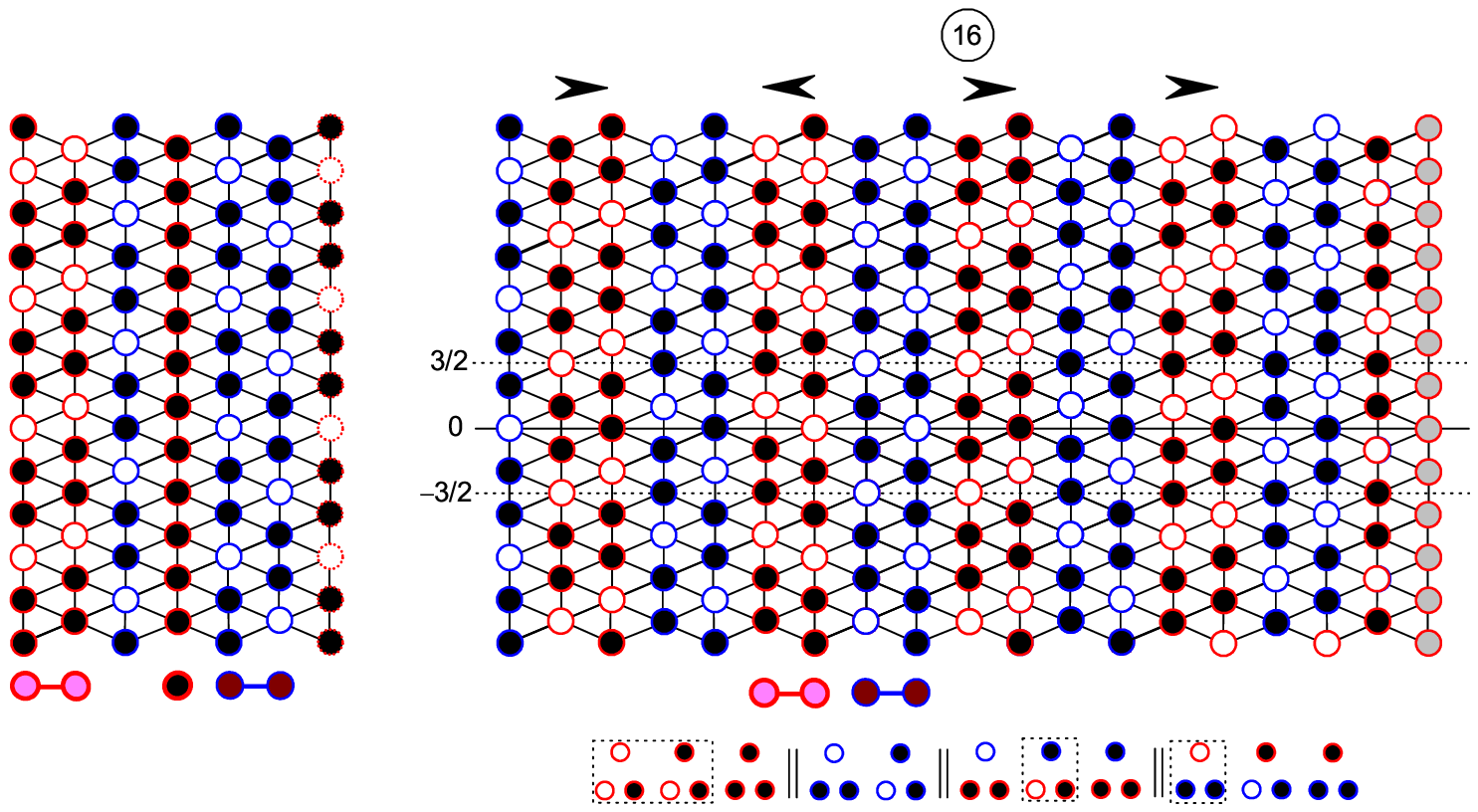}
\hspace{0.1cm}
\includegraphics[scale = 0.34]{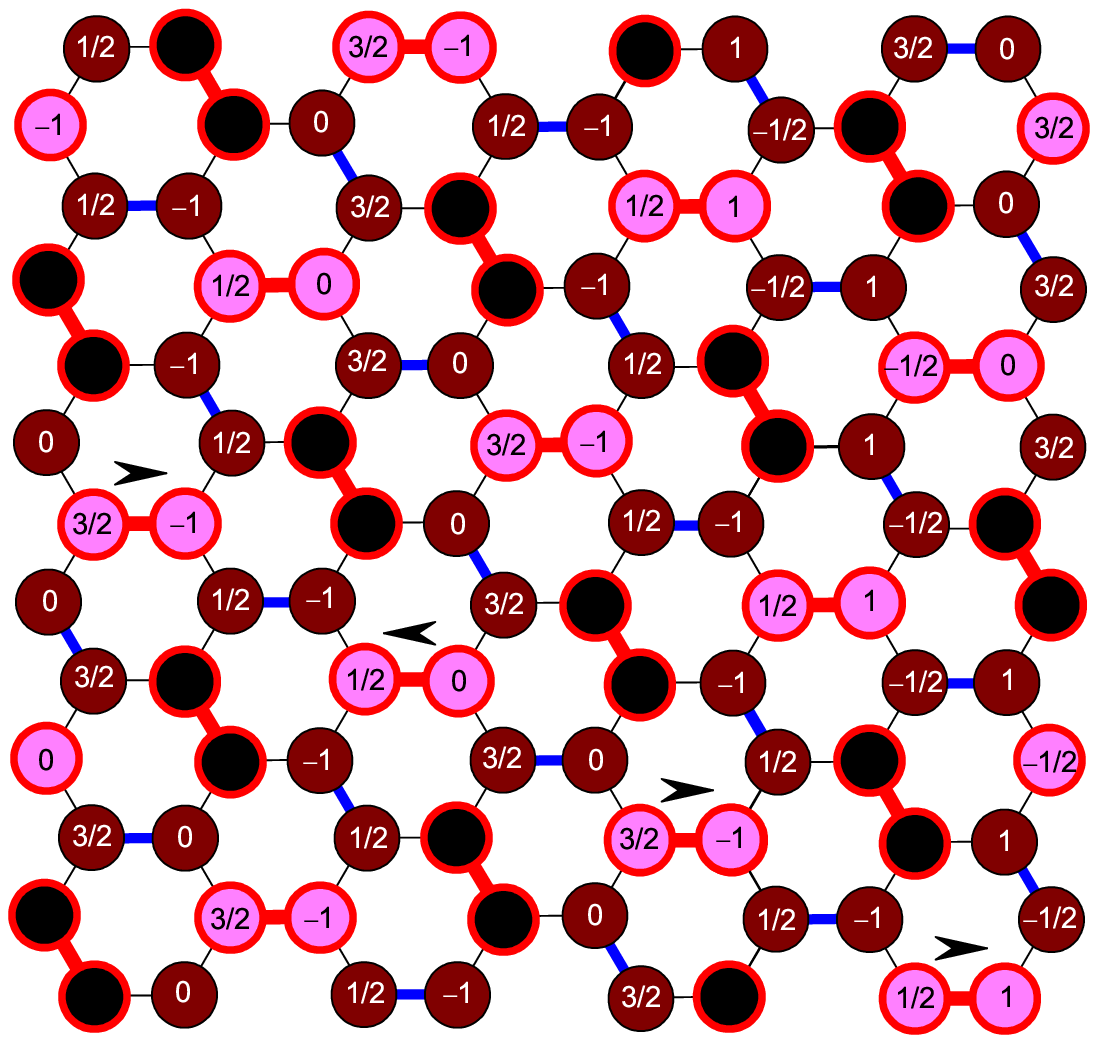}

\vspace{0.2cm}

\includegraphics[scale = 0.55]{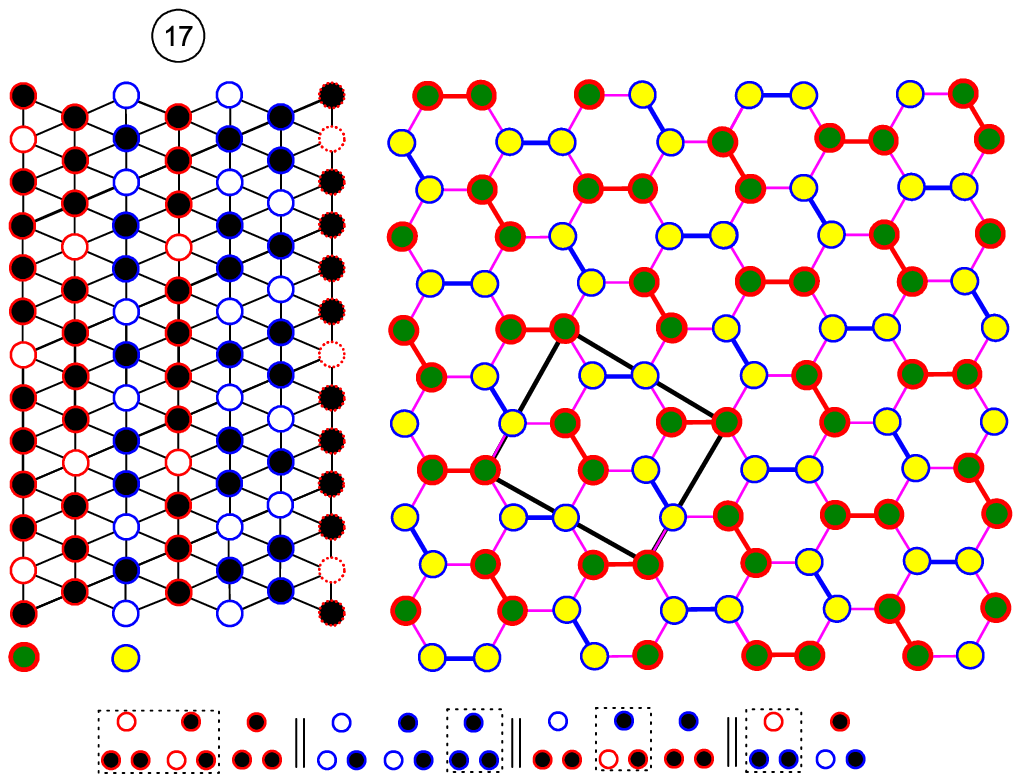}
\hspace{0.5cm}
\includegraphics[scale = 0.55]{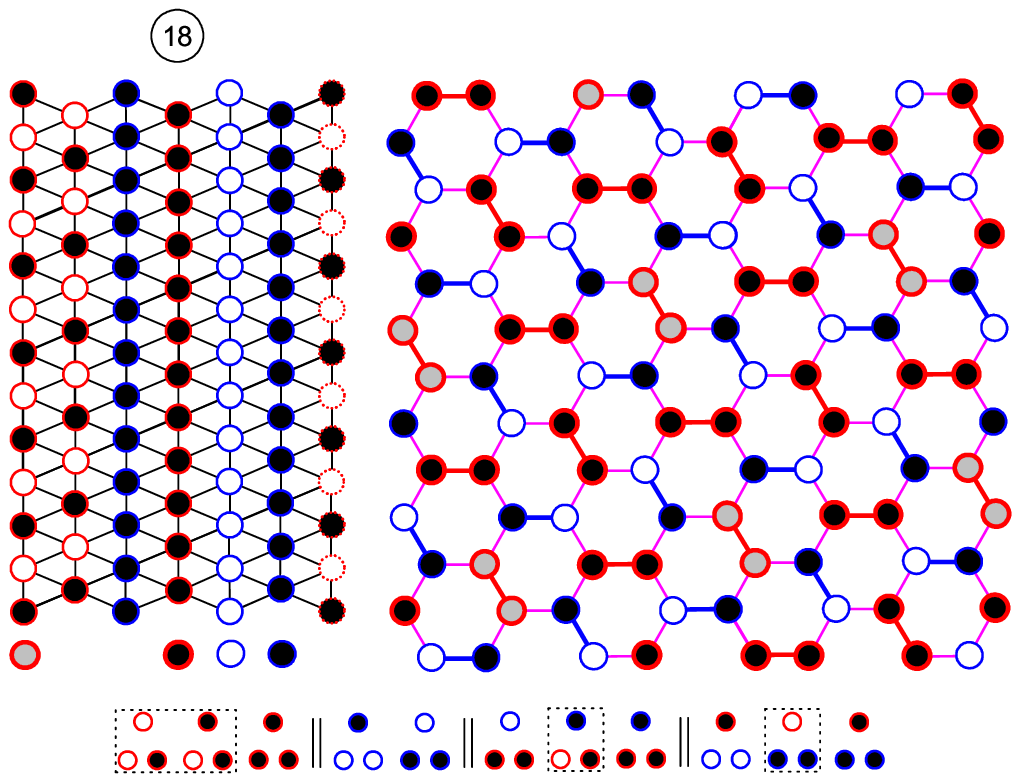}

\vspace{0.2cm}

\includegraphics[scale = 0.55]{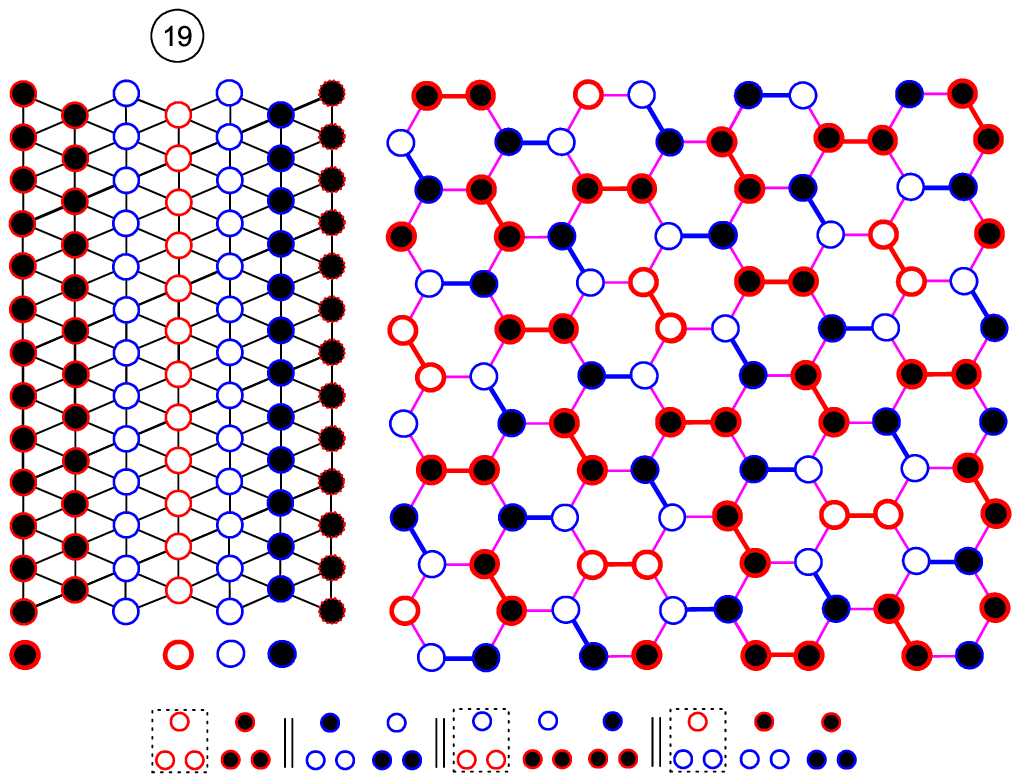}
\hspace{0.5cm}
\includegraphics[scale = 0.55]{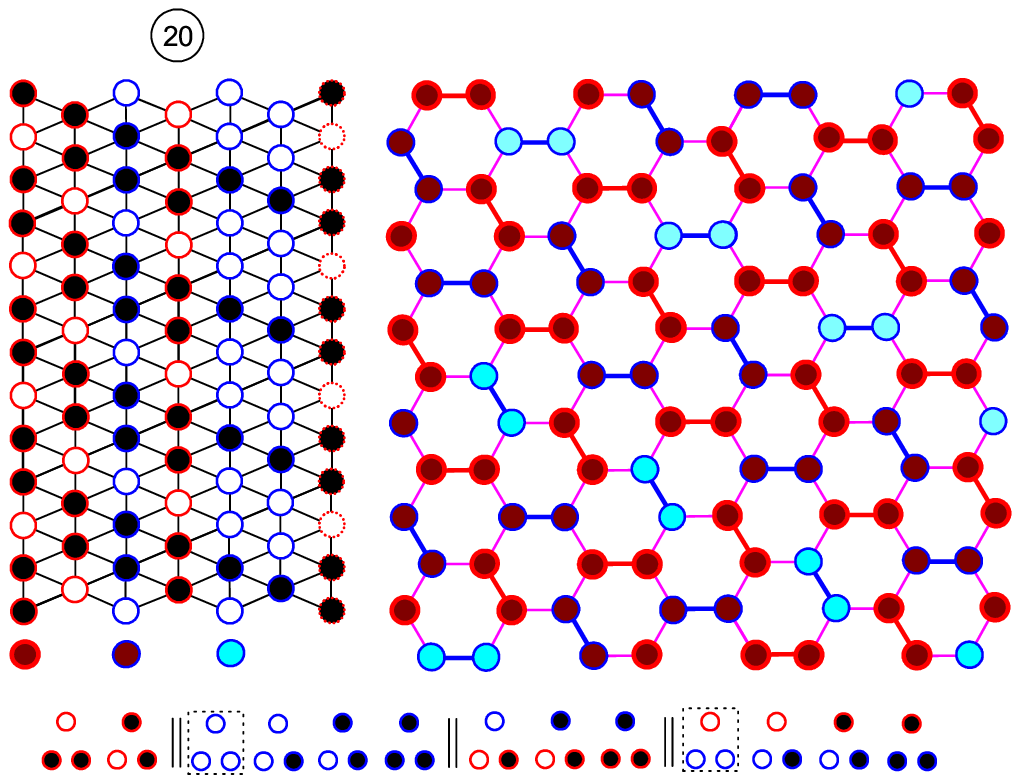}

\vspace{0.2cm}

\includegraphics[scale = 0.55]{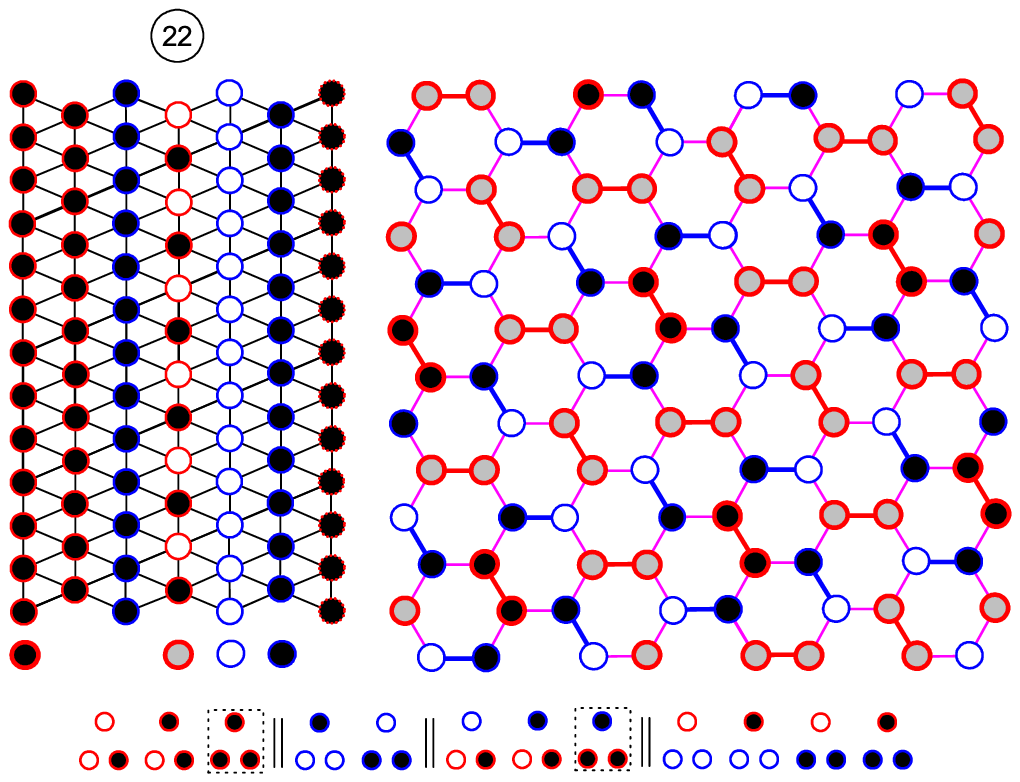}
\hspace{0.5cm}
\includegraphics[scale = 0.55]{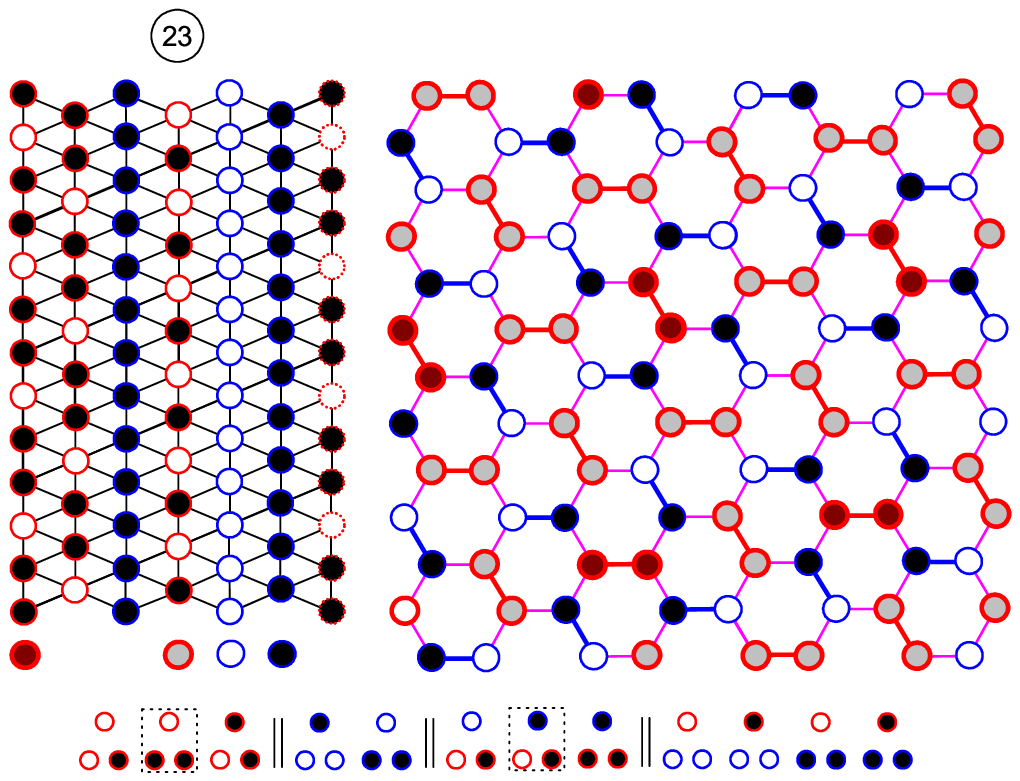}
\caption{Phases 16, 17, 18, 19, 20, 22, and 23. They appear at boundary of phases 2, 3, 4, 4, 10, 13, and 13, respectively.
The triangular configurations shown below the structures are the ground-state configurations at these boundaries.
New triangular configurations are surrounded by dotted squares.
The principe of these structures construction is to find at the given boundary the structures containing maximum number of such configurations.
To show chains, only one hexagonal well configuration is depicted for each phase.
A more detailed picture of phase 20 is shown in Fig.~\ref{fig7} (middle panel) with ``dashed'' $ddu$ chains.}
\label{fig9}
%\end{center}
\end{figure*}

\begin{figure}[tb]
%\begin{center}
\includegraphics[scale = 0.8]{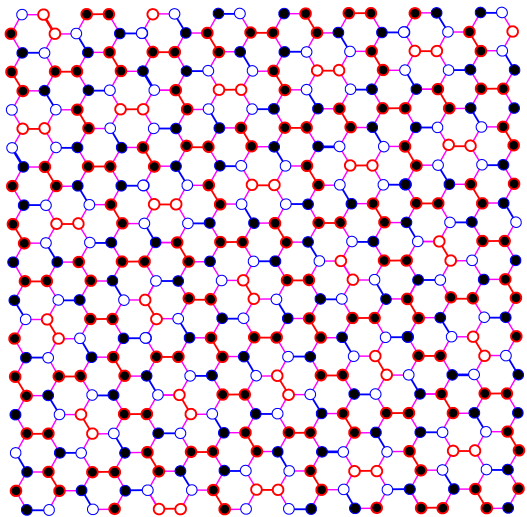}
\caption{Disorder of phase 19. Open and filled circles denote two types of ferromagnetic chains.
A similar disorder is present in phases 18, 22, and 23.}
\label{fig10}
%\end{center}
\end{figure}

\begin{figure}[tb]
%\begin{center}
\includegraphics[scale = 0.65]{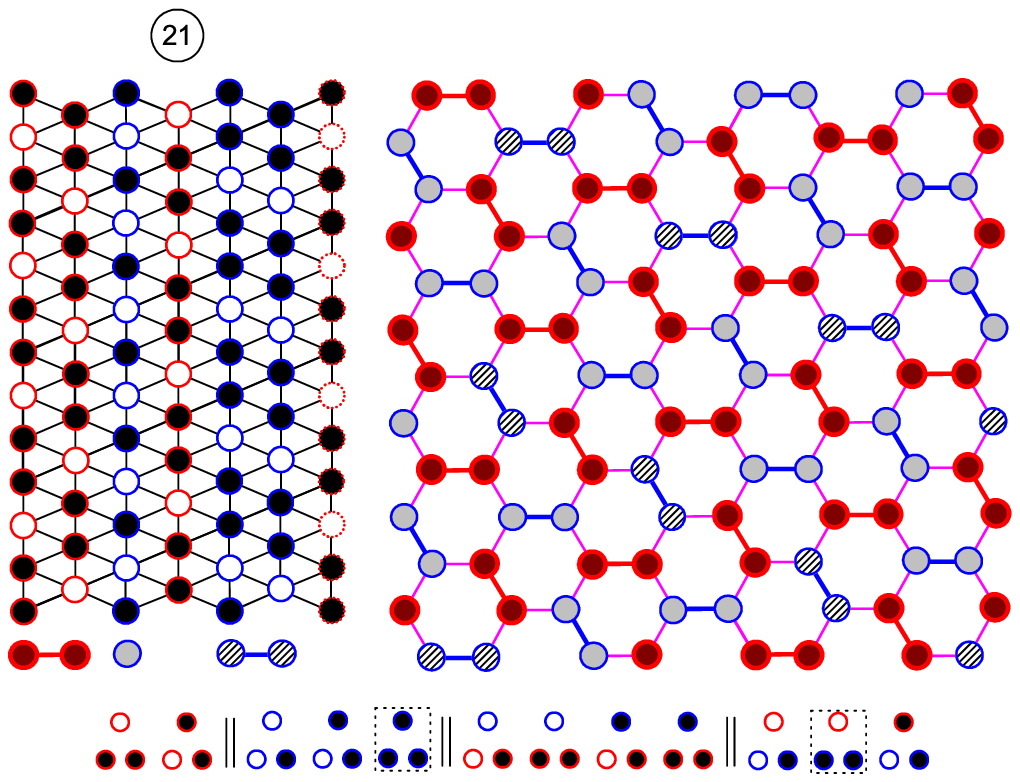}

\includegraphics[scale = 0.2]{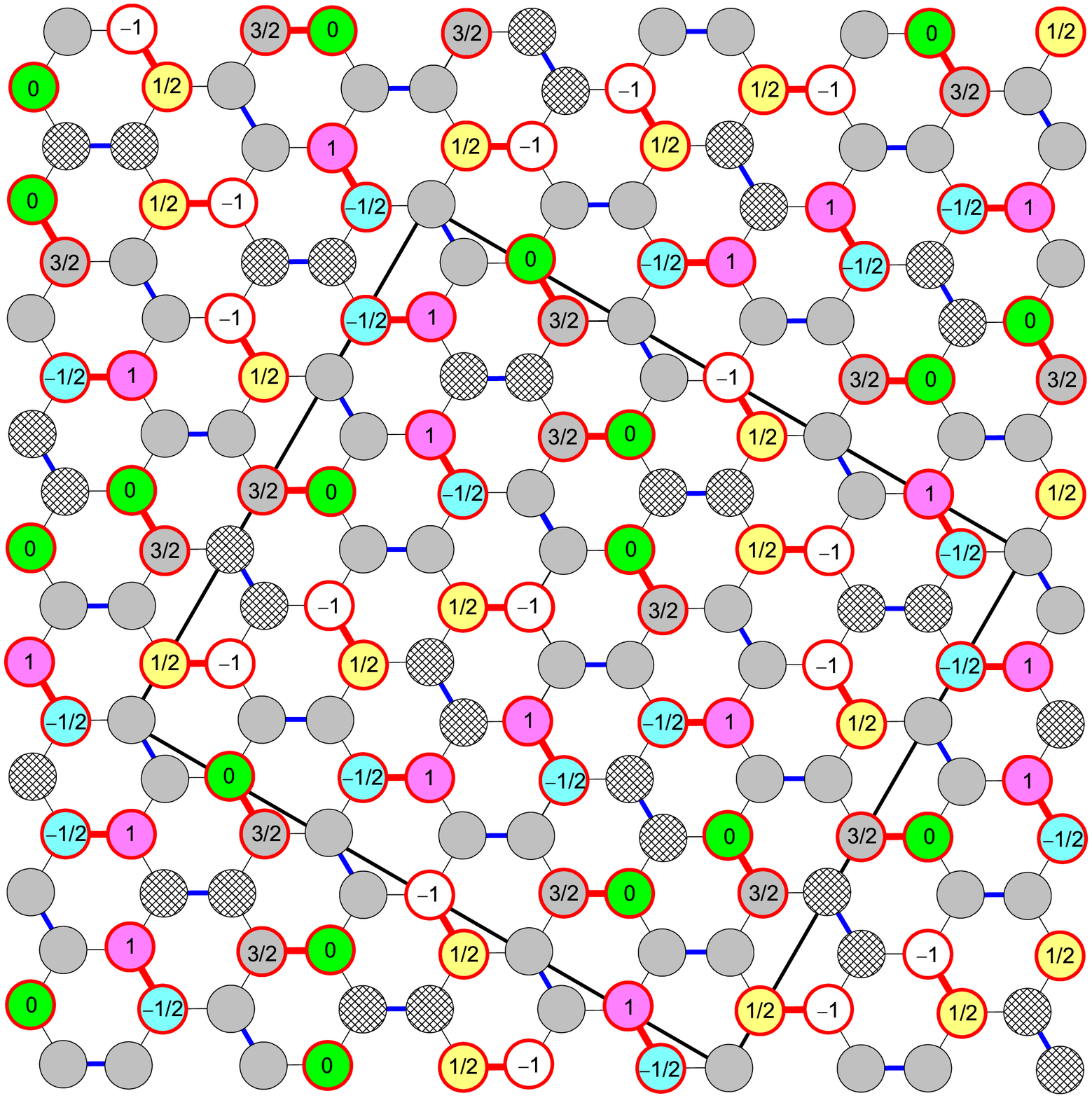}
\caption{Phase 21. There is a two-dimensional disorder in this phase
due to the presence of $ud$ chains. Lower panel gives more detailed
picture of the structure (compare with Fig.~\ref{fig7}, middle panel).}
\label{fig11}
%\end{center}
\end{figure}

%------------------------------------------------------------------------------------------------------------------------------------------------
As discussed above, the majority of 6-faces found with an incomplete set of basic vectors for a ``triangular'' phase region are real 6-faces of this region and, even if the neighboring phase is not a ``triangular'' one, it is possible to determine the
ground-state structure(s) for this phase.
Such structures should have a maximum number of new triangular configurations which are absent in structures of the neighboring phase already known but present in ground-state structures at the common boundary (6-face).
We found nine phases (more exactly, nine classes) of this type.
The list of these phases is given in Table~\ref{table4} (one representative per class) and the corresponding structures are depicted in Figs.~9-14.
In these figures, new triangular configurations are framed by dotted squares.
It should be noted that the sets of triangular configurations in Table~\ref{table4} and in the figures are the sets of ground-state triangular configurations for six-dimensional boundaries.
As one can see from Figs.~\ref{fig9}-\ref{fig11}, some ``nontriangular'' structures, in contrast to ``triangular'' ones, are composed of two different types of red ladder (phases 16, 18, 19, 22, and 23) or blue ladder (phases 20 and 21) configurations.
These phases are due to the interaction between red and blue ladders, they are therefore excluded from considerations in the one-dimensional models, such as 1D ANNNI model used in Ref.~\onlinecite{Wen_2015}.

It is worthwhile to study the disorder of these ``nontriangular'' phases.
Phases 17 and 20 are ordered.
As it is clear from Fig.~\ref{fig9} (upper panel), the disorder of phase 16 is one-dimensional, since the structure is completely determined by a sequence of arrows showing the shifts of neighboring red $uud$ chains.
The disorder of phases 18, 21, 22, and 23 is two-dimensional due to $ud$ chains.
A complex disorder is present in phase 24.
The structures of this phase can be mapped on two-dimensional arrow configurations composed of ten hexagon arrow configurations in which one arrow is aligned clockwise and the five others anticlockwise or vice versa, the arrow
between blue sites being aligned with the majority of the arrows.
Three arrows depicted in Fig.~\ref{fig13} (left hand panel) produce an infinite half-chain of hexagon arrow configurations.
At the first sight the local arrow configuration shown in Fig.~\ref{fig13} (middle panel) should produce a three-dimensional disorder.
However, the number of this arrow configurations is infinitesimal, since every configuration of this type generates at least two half-chains of hexagons.
So, the disorder is not three-dimensional but two- or, possibly, even one-dimensional.

\begin{figure*}[htb]
%\begin{center}
\includegraphics[scale = 0.7]{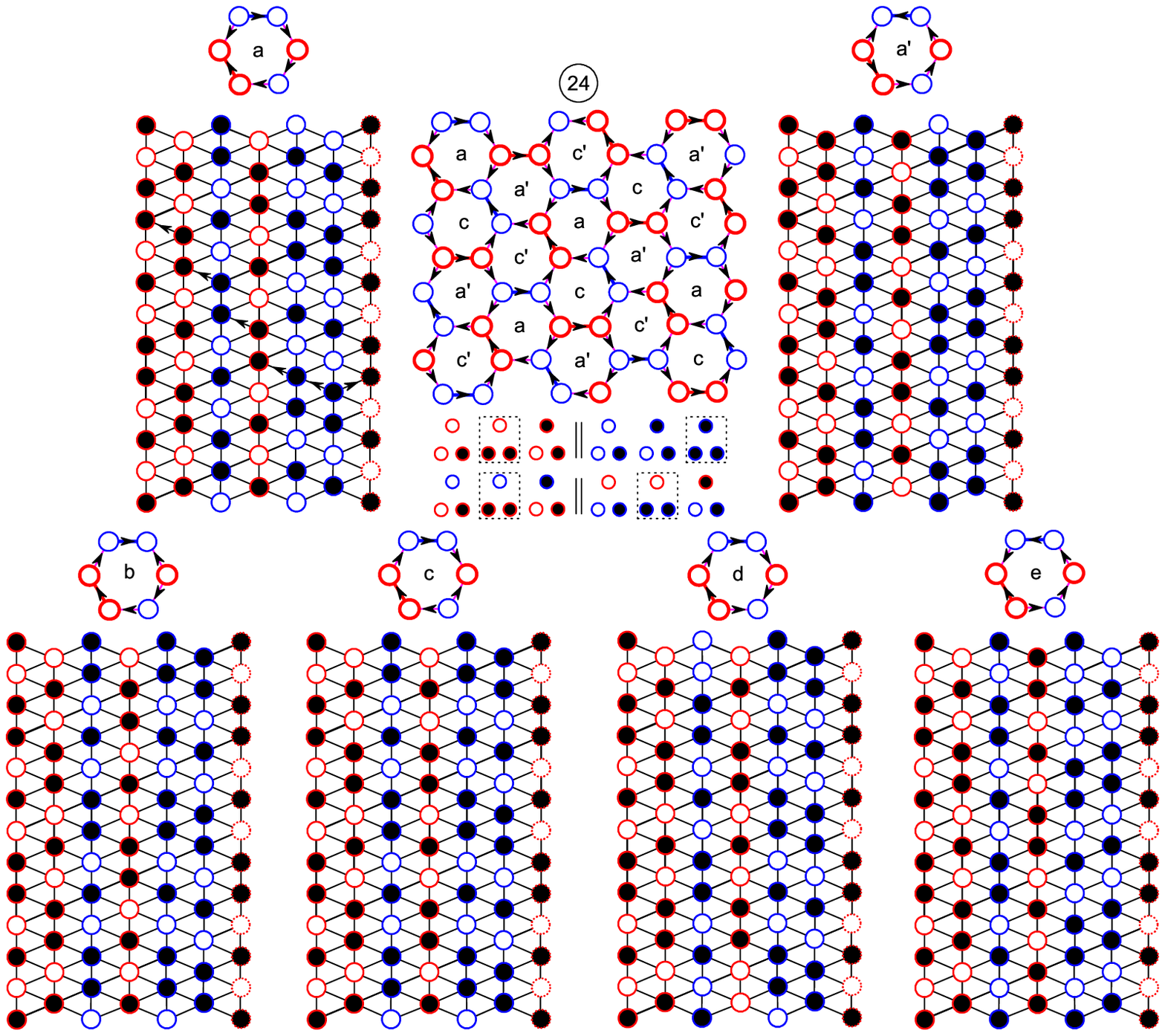}
\caption{Structures 24 are determined by ten arrow configurations of hexagon ($a$, $b$, $c$, $d$, $e$, $a'$, $b'$, $c'$, $d'$, and $e'$) in which one arrow is pointing clockwise and five others anticlockwise or vice versa.
The arrow between blue sites is aligned with the majority of the arrows.
An example of global arrow configuration is also shown.}
\label{fig12}
%\end{center}
\end{figure*}

\begin{figure*}[htb]
%\begin{center}
\includegraphics[scale = 0.25]{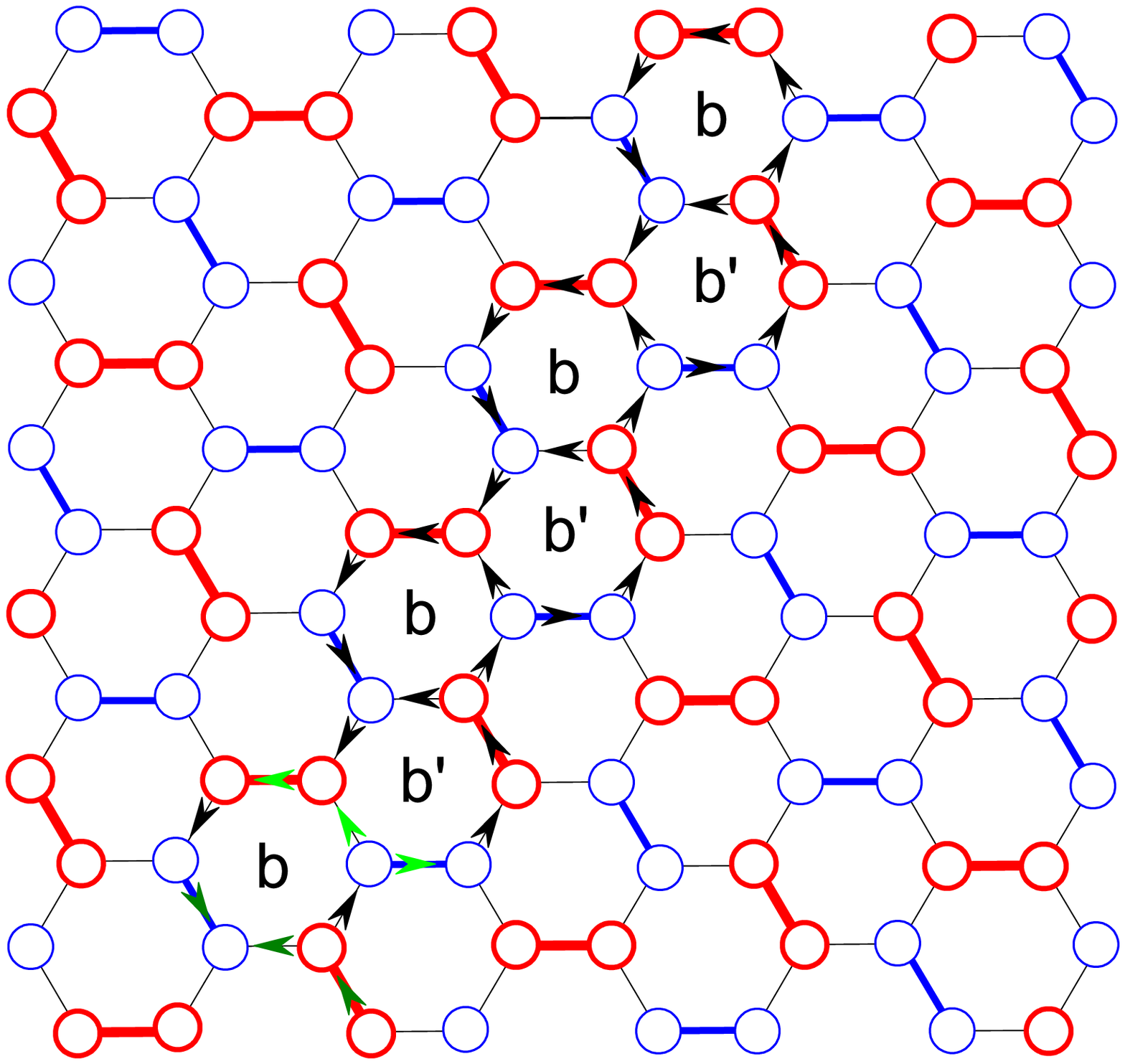}
\hspace{0.5cm}
\includegraphics[scale = 0.25]{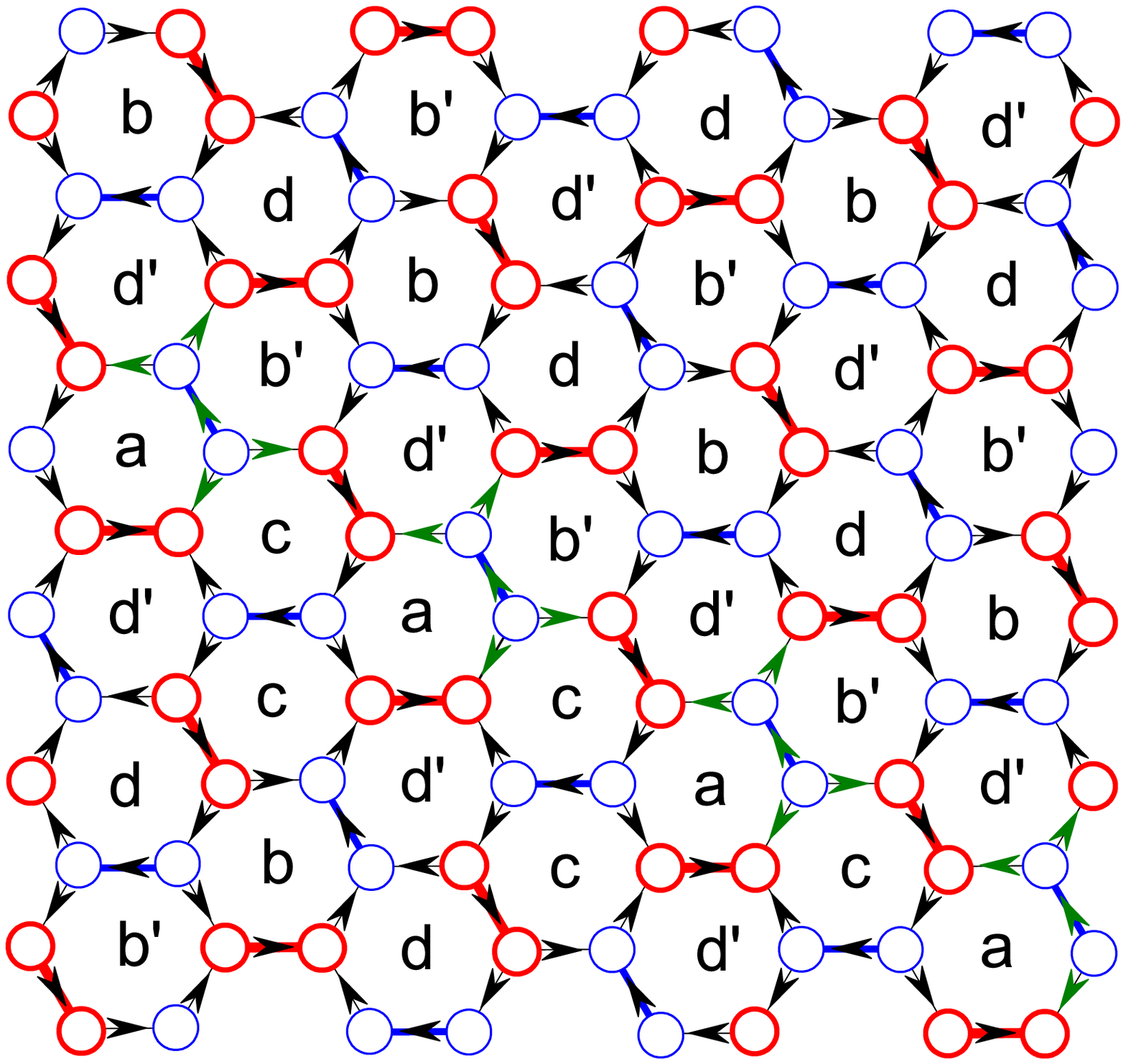}
\hspace{0.5cm}
\includegraphics[scale = 0.6]{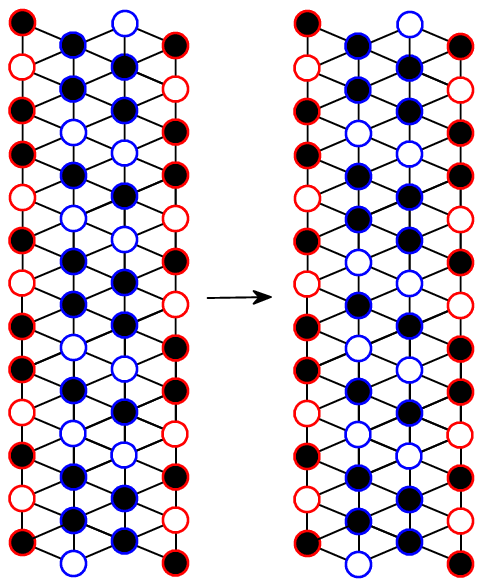}
\caption{Disorder of phase 24. Half-chain of hexagon arrow configurations for phase 24 (see Fig.~\ref{fig12}) is completely determined by the three arrow depicted in olive or in green (left hand panel).
Local arrow configuration shown in olive (middle panel) could lead to a three-dimensional disorder because a rearrangement of spins in blue chains of such configuration is possible (right hand panel).
However, the number of this arrow configurations is infinitely small, since every configuration of this type generates at least two half-chains of hexagons.}
\label{fig13}
%\end{center}
\end{figure*}

\begin{table*}[htb]
\caption{Fully dimensional regions and ``nontriangular'' ground-state structures of the Ising model on the honeycomb zigzag-ladder lattice.}
\begin{ruledtabular}
\begin{tabular}{ccccl}
Boun-&\multicolumn{1}{c}{Triangular}&\multicolumn{1}{c}{Characteristics}&Magneti-&\multicolumn{1}{c}{Basic}\\
dary&\multicolumn{1}{c}{configurations}&\multicolumn{1}{c}{of ``nontriangular'' structures}&zation&\multicolumn{1}{c}{rays}\\
\hline\\[-2mm]

$(2, 16)$&~\usebox{\ddur} \usebox{\duur} \usebox{\uuur} $\|$ \usebox{\uudb} \usebox{\duub} $\|$&
$\frac13(J_{01} - J_{02} + 2J_{11} - J_{12} - 2h_1 - h_2)$&2/3, 1/3&$\mathbf{r}_2^{}, \mathbf{r}_3^\star, \mathbf{r}_5^{},$\\[0.7mm]
&~\usebox{\uudrb} \usebox{\duurb} \usebox{\uuurb} $\|$ \usebox{\uudbr} \usebox{\duubr} \usebox{\uuubr}&$\left[~1, 1, 4~\|~2, 4~\|~4, 4, 4~\|~2, 8, 2~\right]$,~ {\mbox 1D disorder}&&$\mathbf{r}_5^\star, \mathbf{r}_6^\star,\mathbf{r}_7^{}$\\[2mm]

$(3, 17)$&~\usebox{\uudr} \usebox{\duur} \usebox{\uuur} $\|$ \usebox{\ddub} \usebox{\duub} \usebox{\uuub} $\|$&
$\frac15(J_{01} - 3J_{02} + J_{11} + J_{12} - 4J_2 - 3h_1 - h_2)$&3/5, 1/5&$\mathbf{r}_2^\star, \mathbf{r}_4^\star, \mathbf{r}_5^{},$\\[0.7mm]
&~\usebox{\uudrb} \usebox{\duurb} \usebox{\uuurb} $\|$ \usebox{\uudbr} \usebox{\duubr}&$\left[~1, 2, 2~\|~2, 2, 1~\|~4, 4, 2~\|~2, 8~\right]$,~ {\mbox order}&&$\mathbf{r}_5^\star, \mathbf{r}_6^{}, \mathbf{r}_7^{}$\\[2mm]

$(4, 18)$&~\usebox{\ddur} \usebox{\duur} \usebox{\uuur} $\|$ \usebox{\dudb} \usebox{\uudb} $\|$&
$\frac14(2J_{01} + 4J_{02} + 3J_{11} - 4J_{12} - 4J_2 - 3h_1)$&3/4, 0&$\mathbf{r}_1^\star, \mathbf{r}_2^{}, \mathbf{r}_3^\star, \mathbf{r}_5^{},$\\[0.7mm]
&~\usebox{\uudrb} \usebox{\duurb} \usebox{\uuurb} $\|$ \usebox{\dudbr} \usebox{\uudbr} \usebox{\uuubr}&$\left[~1, 1, 6~\|~4, 4~\|~8, 4, 4~\|~8, 2, 6~\right]$,~ {\mbox 2D disorder}&&$ \mathbf{r}_6^\star, \mathbf{r}_6^{\star-}, \mathbf{r}_7^{}$\\[2mm]

$(4, 19)$&~\usebox{\dddr} \usebox{\uuur} $\|$ \usebox{\dudb} \usebox{\uudb} $\|$&
$\frac12(2J_{01} + 2J_{02} + 2J_{11} - 2J_{12} + 4J_2 - h_1)$&1/2, 0&$\mathbf{r}_1^{}, \mathbf{r}_1^\star, \mathbf{r}_2^{}, \mathbf{r}_3^\star,$\\[0.7mm]
&~\usebox{\dddrb} \usebox{\uudrb} \usebox{\uuurb}$\|$ \usebox{\dddbr} \usebox{\dudbr} \usebox{\uuubr}&$\left[~1, 3~\|~2, 2~\|~2, 2, 4~\|~2, 2, 4~\right]$,~ {\mbox 2D disorder}&&$\mathbf{r}_6^\star, \mathbf{r}_6^{\star-}, \mathbf{r}_7^{\thicksim}$\\[2mm]

$(10, 20)$&~\usebox{\uudr} \usebox{\duur} $\|$ \usebox{\dddb} \usebox{\ddub} \usebox{\duub} \usebox{\uuub} $\|$&
$\frac19(-3J_{01} - 3J_{02} - 3J_{11} + 3J_{12} + 12J_2 - 3h_1 - h_2)$&1/3, 1/9&$\mathbf{r}_2^\star, \mathbf{r}_3^{}, \mathbf{r}_4^\thicksim,$\\[0.7mm]
&~\usebox{\ddurb} \usebox{\duurb} \usebox{\uuurb} $\|$ \usebox{\dddbr} \usebox{\ddubr} \usebox{\duubr} \usebox{\uuubr}&$\left[~3, 6~\|~1, 3, 3, 2~\|~8, 4, 6~\|~2, 4, 8, 4~\right]$,~ {\mbox order}&&$\mathbf{r}_4^{\thicksim\star}, \mathbf{r}_5^{}, \mathbf{r}_6^{}$\\[2mm]

$(11, 21)$&~\usebox{\uudr} \usebox{\duur} $\|$ \usebox{\ddub} \usebox{\duub} \usebox{\uuub} $\|$&
$\frac19(-3J_{01} - 7J_{02} - 3J_{11} + J_{12} - 4J_2 - 3h_1 - h_2)$&1/3, 1/9&$\mathbf{r}_2^\star, \mathbf{r}_3^{}, \mathbf{r}_4^\star, \mathbf{r}_5^{},$\\[0.7mm]
&~\usebox{\ddurb} \usebox{\uudrb} \usebox{\duurb} \usebox{\uuurb} $\|$ \usebox{\ddubr} \usebox{\uudbr} \usebox{\duubr}&$\left[~3, 6~\|~4, 4, 1~\|~4, 4, 8, 2~\|~4, 2, 12~\right]$,~ {\mbox 2D disorder}&&$\mathbf{r}_5^\star, \mathbf{r}_6^{}$\\[2mm]

$(13, 22)$&~\usebox{\ddur} \usebox{\duur} \usebox{\uuur} $\|$ \usebox{\dudb} \usebox{\uudb} $\|$&
$\frac14(-2J_{01} + 4J_{02} + J_{11}- 4J_{12} + 4J_2 - h_1)$&1/4, 0&$\mathbf{r}_1^\star, \mathbf{r}_2^{}, \mathbf{r}_3^\star, \mathbf{r}_4^{\thicksim},$\\[0.7mm]
&~\usebox{\ddurb} \usebox{\duurb} \usebox{\uuurb} $\|$ \usebox{\dddbr} \usebox{\dudbr} \usebox{\uudbr} \usebox{\uuubr}&$\left[~3, 3, 2~\|~4, 4~\|~8, 4, 4~\|~4, 4, 2, 6~\right]$,~ {\mbox 2D disorder}&&$\mathbf{r}_5^{}, \mathbf{r}_6^\star, \mathbf{r}_6^{\star-}$\\[2mm]

$(13, 23)$&~\usebox{\ddur} \usebox{\uudr} \usebox{\duur} $\|$ \usebox{\dudb} \usebox{\uudb} $\|$&
$\frac{1}{12}(-10J_{01} + 12J_{02} - J_{11} - 12J_{12} + 4J_2 - h_1)$&1/12, 0&$\mathbf{r}_1^\star, \mathbf{r}_3^{}, \mathbf{r}_3^\star, \mathbf{r}_4^\thicksim,$\\[0.7mm]
&~\usebox{\ddurb} \usebox{\duurb} \usebox{\uuurb} $\|$ \usebox{\dddbr} \usebox{\dudbr} \usebox{\uudbr} \usebox{\uuubr}&$\left[~9, 2, 13~\|~12, 12~\|~24, 20, 4~\|~12, 12, 10, 14~\right]$,&&$\mathbf{r}_5^{}, \mathbf{r}_6^\star, \mathbf{r}_6^{\star-}$\\
&&{\mbox 2D disorder}&&\\

$(14, 24)$&~\usebox{\ddur} \usebox{\uudr} \usebox{\duur} $\|$ \usebox{\ddub} \usebox{\duub} \usebox{\uuub} $\|$&
$\frac15(-3J_{01} - 3J_{02} - J_{11} + J_{12} - 4J_2 - h_1 - h_2)$&$1/5$, 1/5&$\mathbf{r}_2^\star, \mathbf{r}_3^{}, \mathbf{r}_4^{}, \mathbf{r}_4^\star,$\\[0.7mm]
&~\usebox{\ddurb} \usebox{\uudrb} \usebox{\duurb} $\|$ \usebox{\ddubr} \usebox{\uudbr} \usebox{\duubr}&$\left[~1, 1, 3~\|~2, 2, 1~\|~2, 2, 6~\|~2, 2, 6~\right]$,~ {\mbox disorder}&&$\mathbf{r}_5^{}, \mathbf{r}_5^\star$\\[-3mm]

\label{table4}
\end{tabular}
\end{ruledtabular}
\end{table*}

\subsection{Ground-state phase diagrams in the $(h_1, h_2)$-plane}%%%%%%%%%%%%%%%%%%%%%%%%%%
%------------------------------------------------------------------------------------------------------------------------------------------------
\begin{figure*}[htb]
%\begin{center}
\includegraphics[scale = 0.6]{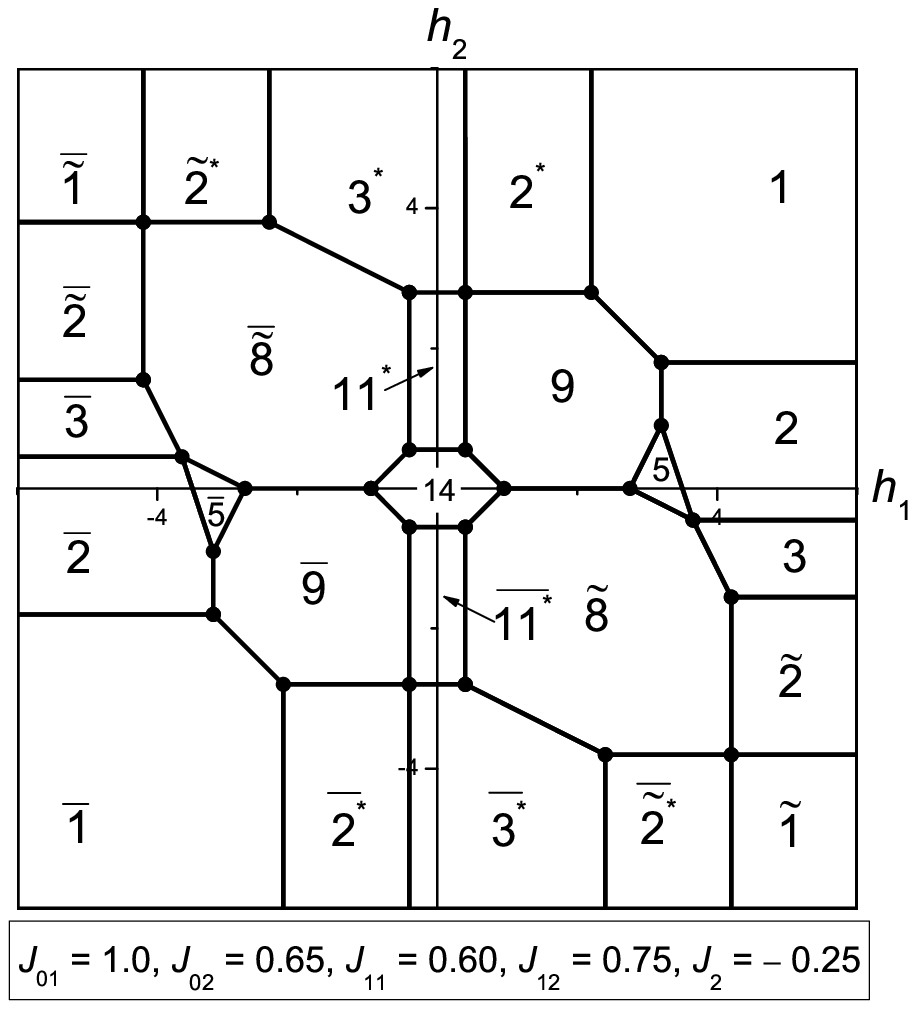}
\includegraphics[scale = 0.6]{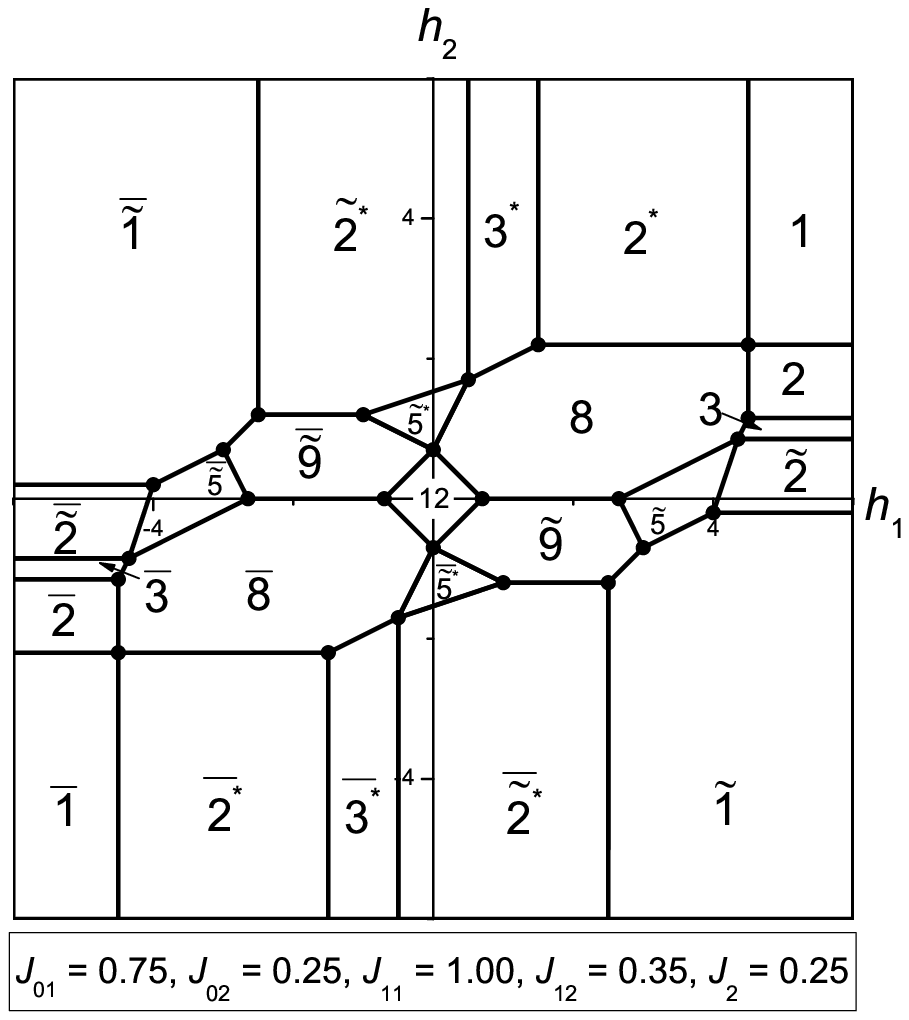}
\includegraphics[scale = 0.6]{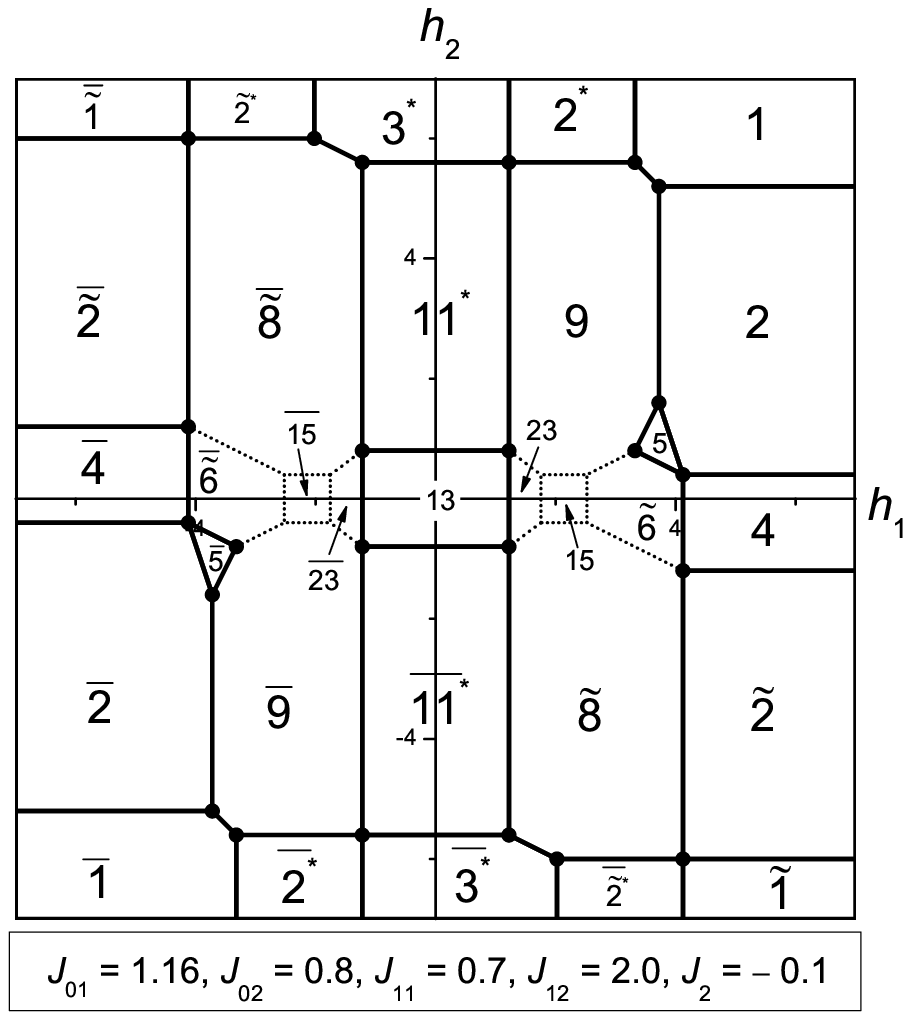}

\includegraphics[scale = 0.6]{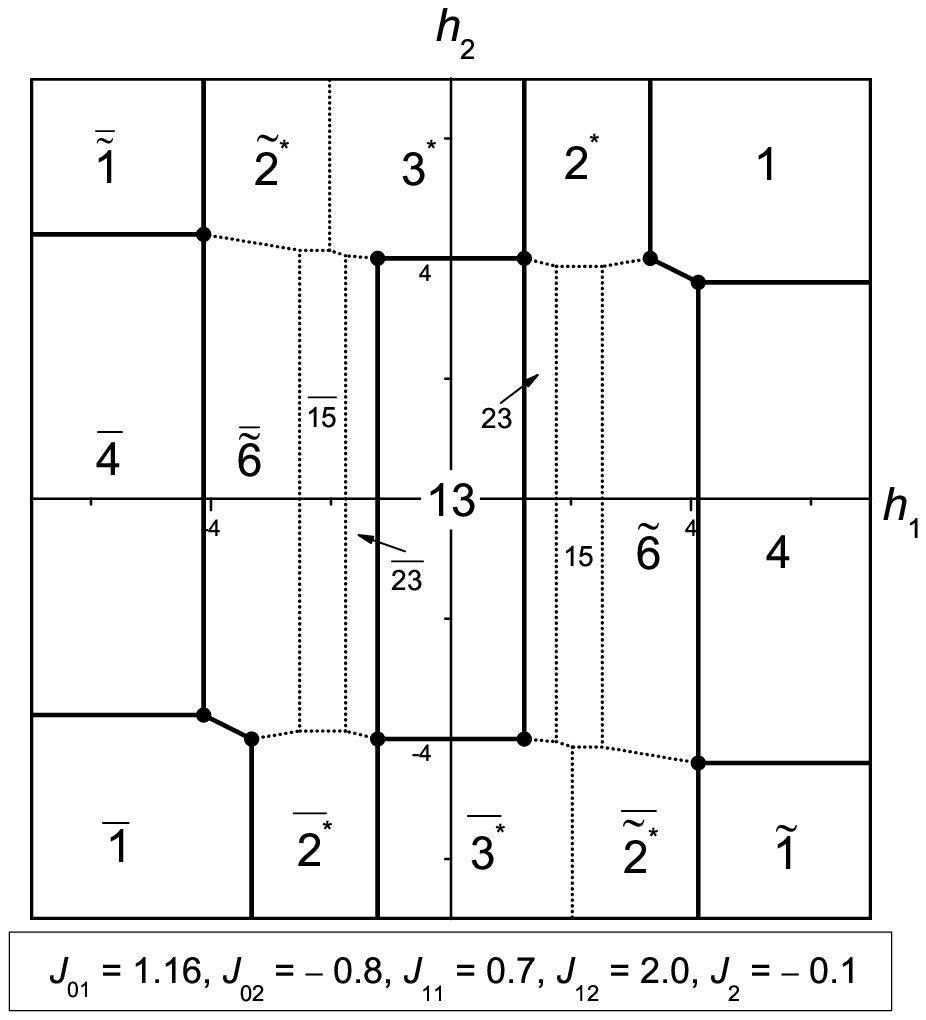}
\includegraphics[scale = 0.6]{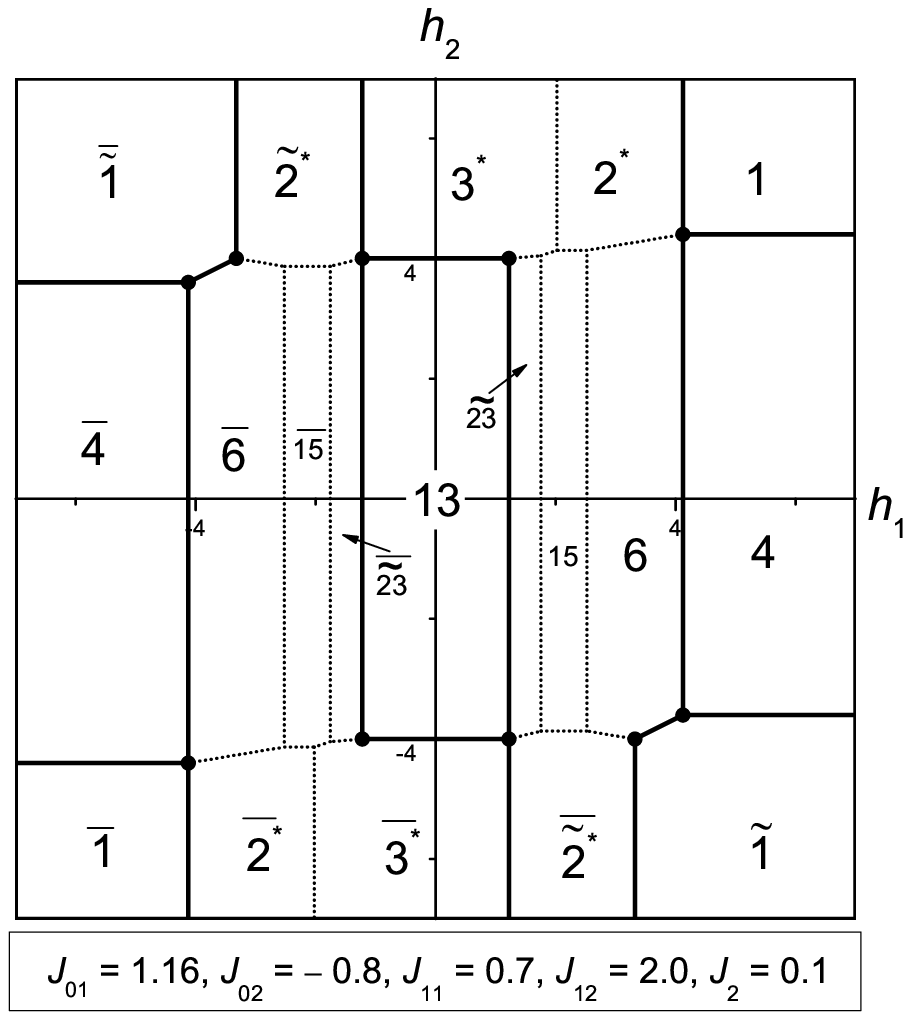}
\includegraphics[scale = 0.6]{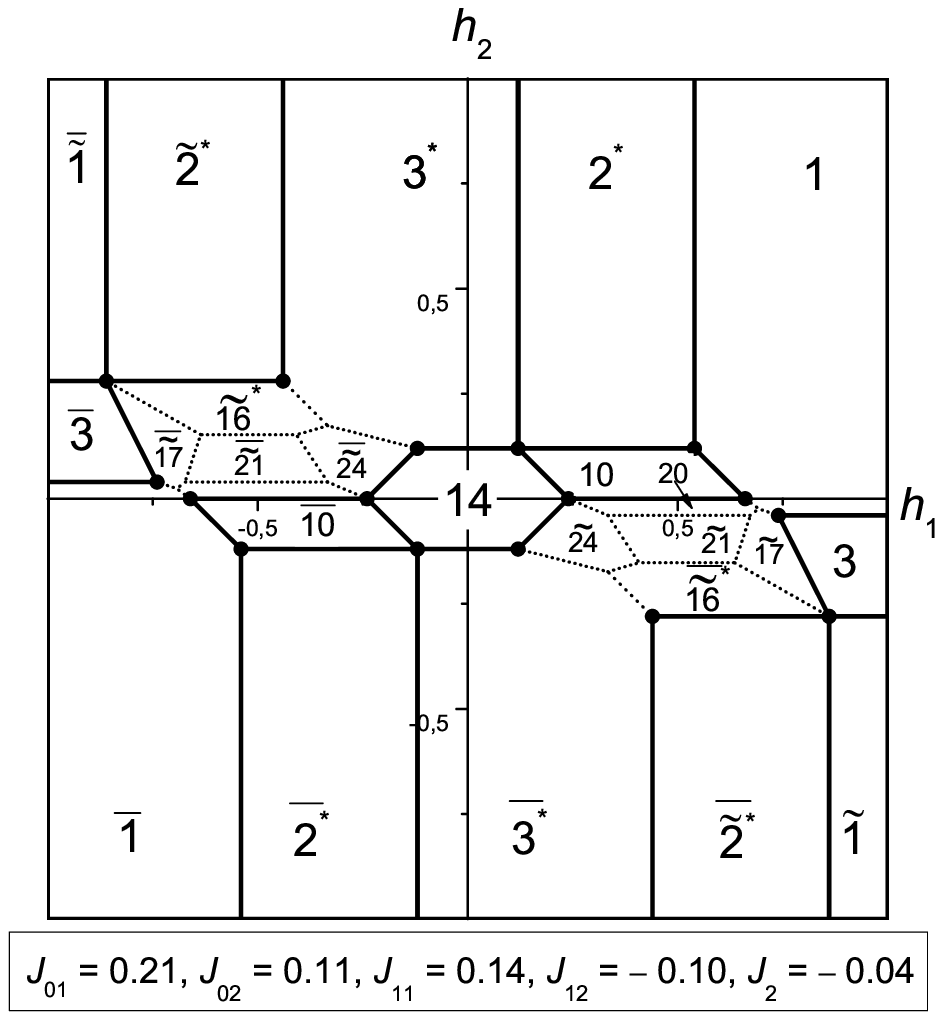}
\caption{Examples of ground-state phase diagrams in the $(h_1, h_2)$-plane (the fields and couplings are shown in arbitrary units).
Some diagrams are not completely proven, particularly the transitions depicted by the doted lines.}
\label{fig14}
%\end{center}
\end{figure*}

Consider the ground-state phase diagrams in the $(h_1, h_2)$-plane.
Although the solution of the ground-state problem is incomplete, at some particular values of the parameters $J_{01}$, $J_{02}$, $J_{11}$,$J_{12}$, and $J_2$, it is possible to construct exact and complete phase diagrams.
Six examples of such diagrams are given in Fig.~\ref{fig14}.
The boundaries shown with dotted lines are not strictly proven.

Let us show how to prove that the point where three phases, for instance, 2, 5, and 9, meet, exists in a ground-state phase diagram.
This point is determined by the following set of vectors (common for all the three phases), \{$\mathbf{r}_3^\star, \mathbf{r}_4^{\thicksim\star}, \mathbf{r}_5^{}, \mathbf{r}_6^{}, \mathbf{r}_6^{\star}$\}.
At fixed $J_{01}$, $J_{02}$, $J_{11}$ $J_{12}$, and $J_2$ the solution of the equation
\begin{eqnarray}
&&a_3^\star \mathbf{r}_3^\star +
a_4^{\thicksim\star}\mathbf{r}_4^{\thicksim\star} +
a_5^{}\mathbf{r}_5^{} + a_6^{}\mathbf{r}_6^{} +
a_6^\star\mathbf{r}_6^\star \nonumber \\
&&= (J_{01}, J_{02}, J_{11}, J_{12}, J_2, h_1, h_2)
\label{eq4}
\end{eqnarray}
is
\begin{eqnarray}
&&a_3^\star = J_{02} + 2J_2, \nonumber \\
&&a_4^{\thicksim\star} = -J_2, \nonumber \\
&&a_5^{} = J_{01}, \nonumber \\
&&a_6^{} = J_{11}, \nonumber \\
&&a_6^\star =  -2J_{02} + J_{12} - 4J_2, \nonumber \\
&&h_1 = 2J_{01} + 2J_{11}, \nonumber \\
&&h_2 = - 4J_{02} + 2J_{12} - 8J_2.
\label{eq5}
\end{eqnarray}
For $J_{01}=1.0$, $J_{02}=0.65$, $J_{11}=0.60$, $J_{12}=0.75$, and $J_2=-0.25$ all the five coefficients are nonnegative, so, the linear combination in the left side of Eq.~\ref{eq4} belongs to the conical hull of the set of vectors, and, therefore, for these values of parameters, the point where the phases 2, 5, and 9 meet exists in the ground-state phase diagram for these values of parameters.
It is the point $h_1=3.2$, $h_2=0.9$.

In a similar way one can, for instance, find conditions for the existence of region 13 -- region 23 boundary in the $(h_1, h_2)$-plane,
\begin{eqnarray}
&&J_{11} > 0,~ J_{12} > 0,~ J_2 < 0, \nonumber \\
&&2J_{01} - J_{11} + 4J_2 > 0, \nonumber \\
&&J_{12} - 2J_{02} > 0.
\label{eq6}
\end{eqnarray}

Then, for this boundary we have
\begin{eqnarray}
&&h_1 = 2J_{01} - J_{11} + 4J_2, \nonumber \\
&&4J_{02} - 2J_{12} < h_2 < 2J_{12} - 4J_{02}~\text{if}~J_{02} > 0,\nonumber \\
&&~~~~~~-2J_{12} < h_2 < 2J_{12}~\text{if}~J_{02} < 0.
\label{eq7}
\end{eqnarray}

\section{Application to \SRO\ and \BRO\ compounds}%%%%%%%%%%%%%%%%%%%%%%%%%
\label{Sec_III}
%------------------------------------------------------------------------------------------------------------------------------------------------
In this section, we consider an application of the theoretical approach discussed above to the magnetic properties of the two families of rare-earth compounds, \SRO\ and \BRO.
We start by briefly summarizing what is experimentally known about the ground state configurations of these zigzag-ladder magnets, particularly focusing on the in-field behaviour of \SEO, \SHO, \SDO\ and \BDO.

The crystal structure of these compounds is very close to the one depicted in Fig.~1, with two RE ions in different positions forming a set of triangular ladders running along the $c$ axis~\cite{Karunadasa_2005}.
The ladders are arranged in a honeycomb-like lattice in the $a-b$ plane, however, the honeycombs are significantly distorted so that the distances between the ions are not identical, which results in the need to introduce different exchange couplings, $J_{11} \neq J_{12} \neq J_2$ in our model.

One important question to address here is to what degree the \SRO\ and \BRO\ compounds could be characterized as Ising-type magnets.
The answer to this question should come most naturally from considering the effects of crystal fields (CFs), however, the task of establishing the sets of relevant CF parameters for the two ions in crystallographically inequivalent positions is far from trivial.
Because of the low overall symmetry and the large number of atoms in a unit cell, interpretation of inelastic neutron scattering data does not necessarily return a unique set of CF parameters unless supplemented by optical and electron paramagnetic resonance measurements, and so far this has only been done for \SEO~\cite{Malkin_2015}.

In zero field, the Er ions positioned in \SEO\ on different sites participate in the formation of two different magnetic systems acting almost independently of each other~\cite{Petrenko_2008,Hayes_2011}.
Er1 sites form a long-range antiferromagnetic order with the magnetic moments aligned parallel to the $c$~direction.
For this site, each ladder is made of the two ferromagnetic chains aligned antiparallel to each other.
Er2 sites participate in the formation of a short-range one-dimensional order, where the spins lay in the $a-b$ plane, and demonstrate very strong antiferromagnetic in-chain correlations (along the $c$~axis) with much weaker correlations between the chains (that is in the direction normal to the $c$~axis).
In the absence of an external field, phase~13 is realized in \SEO\ (without degeneracy of the Er1 subsystem).

In \SHO, the zero-field ground state is similar to that of \SEO, however, for the Ho1 sites, the magnetic order remains limited even at the lowest experimentally achievable temperature~\cite{Young_2013}.
This can be explained by the degeneracy (disorder) of both Ho1 and Ho2 subsystems in phase 13.

In \SDO, there are no long-range correlations between the magnetic moments in zero field, but they can be induced by applying a relatively weak magnetic field along the $b$~axis~\cite{Petrenko_2017}.
In fact the magnetization process in all the three compounds demonstrate similar features, as revealed by the low-$T$ single-crystal magnetization $M(H)$ measurements~\cite{Hayes_2012}.
For certain directions of an applied field, the process is characterized by the appearance of a magnetization plateau, albeit not very pronounced but still clearly visible on the $dM(H)/dH$ curves.
To stabilize the plateaus, the field should be applied along the $a$~axis in \SEO\ and along the $b$~axis in \SDO\ and \SHO.
The value of magnetization on the plateaux is approximately a third of the magnetization observed in higher fields~\cite{Hayes_2012}.
The 1/3 magnetization plateaux are, of course, a common feature of many triangular antiferromagnets, they correspond to the states with the two spins on each triangle pointing along the field and the third spin pointing in the opposite direction (so called up-up-down, $uud$, structure)~\cite{Korshunov_1986,Chubukov_1991}.
Overall magnetization data are consistent with the Ising behavior in these three \SRO\ compounds.
The two magnetic sites have their magnetization easy-axes aligned along (or very near) the two crystallographic axes, while when the field is applied along the third crystallographic axis, the measured magnetization is significantly lower (particularly for Ho and Er compounds~\cite{Hayes_2012}) suggesting that it is a hard magnetization axis for both sites.

Apart from the magnetization data, the evidence for the field-induced $uud$ structure comes from the results of neutron diffraction for \SDO~\cite{Petrenko_2017,Gauthier_2017_b}, \SHO~\cite{Young_2019}, \SEO~\cite{Riberolles_2020} and \BDO~\cite{Khalyavin_2019}.
The $uud$ structures are characterized by the appearance of the sharp, almost resolution-limited magnetic peaks at non-integer positions.
In \SHO, the observed peaks are at the $(h 0 \frac{1}{3})$, $(h 0 \frac{2}{3})$ and symmetry related positions~\cite{Young_2019}, in \SDO, they are indexed by the propagation vector ${\bf k}^\prime \! = \! [0~\frac{1}{3}~\frac{1}{3}]$~\cite{Petrenko_2017,Gauthier_2017_b} and in \BDO, the propagation vector is ${\bf k}^\prime \! = \! [0~0~\frac{1}{3}]$~\cite{Khalyavin_2019}.

\begin{figure*}[htb]
%\begin{center}
\includegraphics[scale = 0.55]{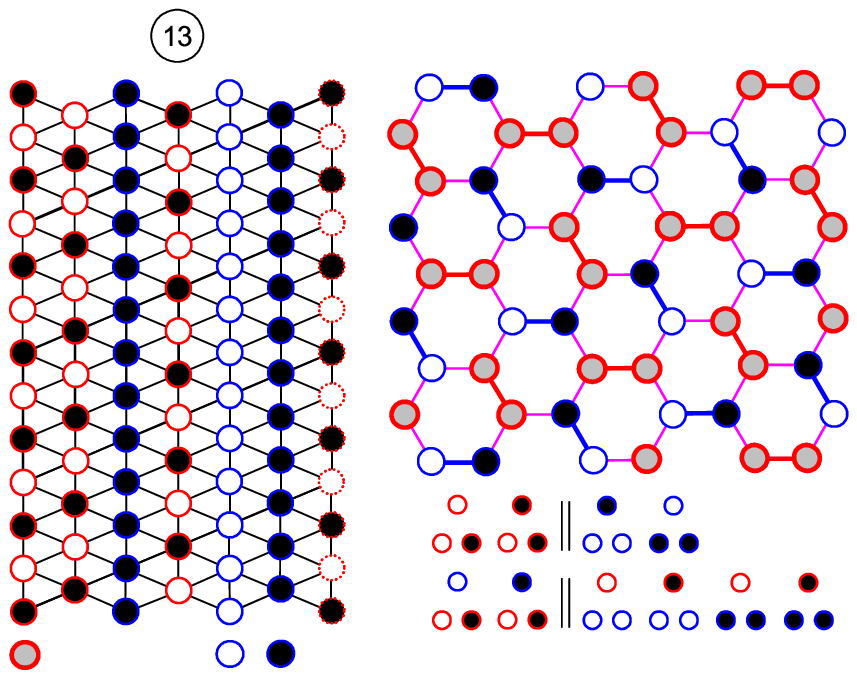}
\hspace{0.25cm}
\includegraphics[scale = 0.55]{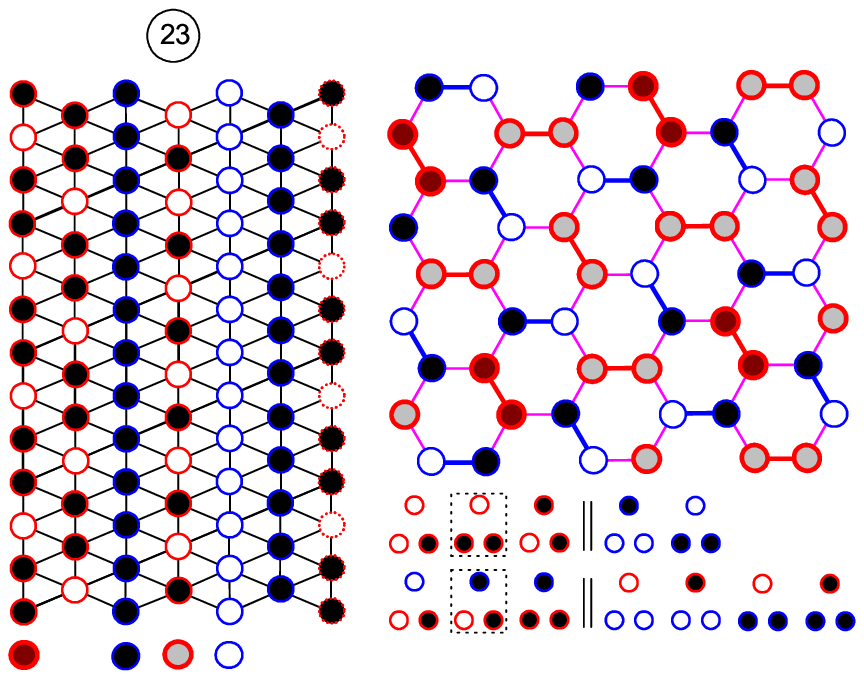}
\hspace{0.25cm}
\includegraphics[scale = 0.55]{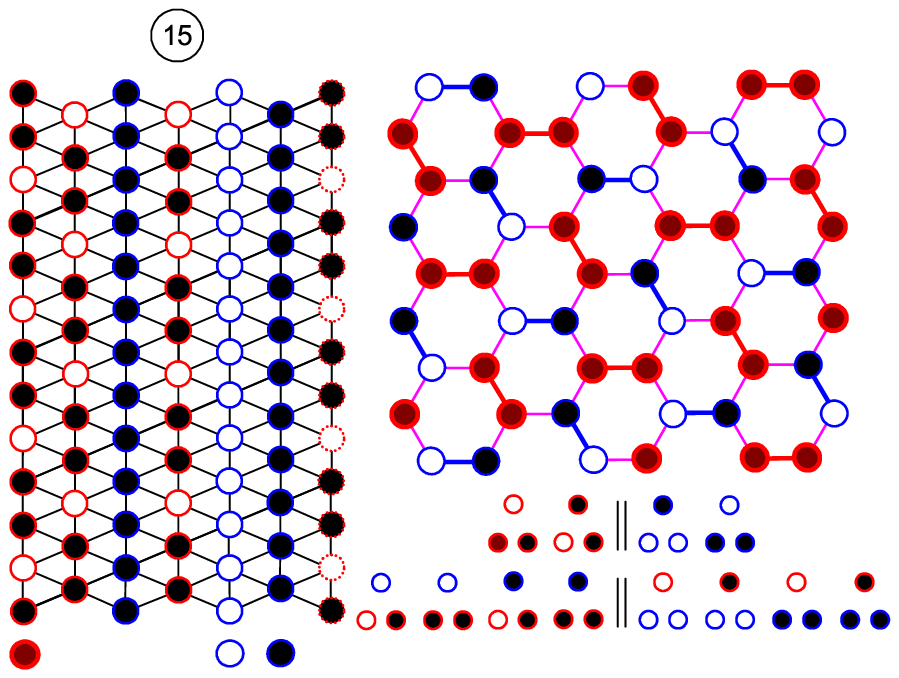}

\includegraphics[scale = 0.55]{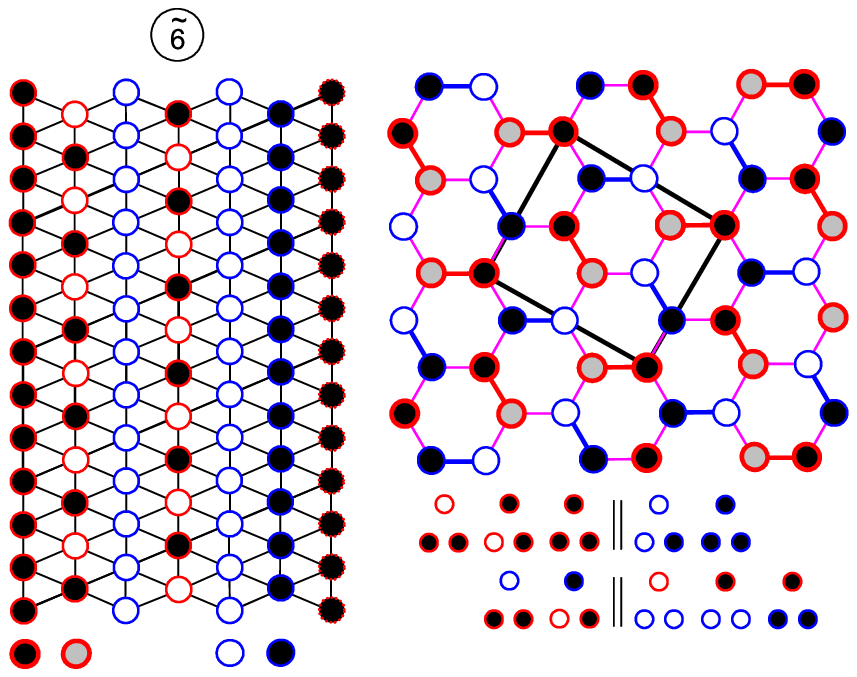}
\hspace{0.2cm}
\includegraphics[scale = 0.55]{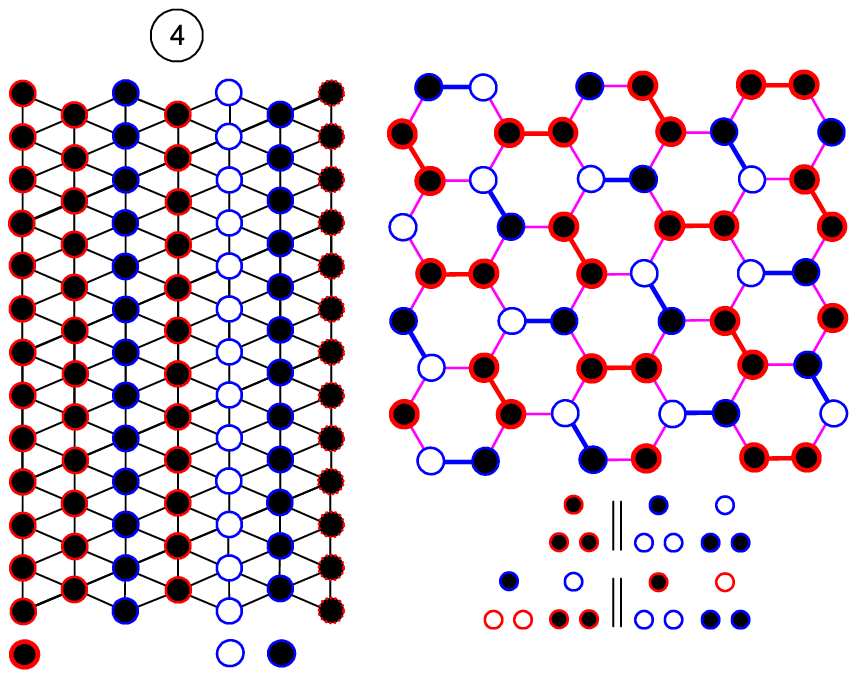}
\caption{Sequence of the proposed phase transitions in \SHO\ and \SEO\ for a magnetic field applied along the easy-magnetization direction for the RE2 site, $a$~axis in \SEO\ and $b$~axis in \SHO, for $J_2 < 0$, $J_{11} > 0$.
For $J_2 > 0$ ($J_{11} > 0$), the phase $\widetilde {23}$ should appear instead of the phase 23 and the phase 6 should appear instead of the phase $\tilde 6$.
The corresponding sequence of magnetization values for the red sublattice is 0, 1/12, 1/3, 1/2, and 1 (per one site of red sublattice).
Transition field values are $h_{1,13-23} = 2J_{01} - J_{11} + 4J_2$, $h_{1,23-15} = 2J_{01} - J_{11}- \frac43 J_2$, $h_{1,15-\tilde 6} = 2J_{01} + 2J_{11} + 12J_2$, $h_{1,{\tilde 6}-4} = 2J_{01} + 2J_{11} - 4J_2$.
The width of the region $\widetilde {6}$ is three times the width of the region 23 in $(h_1, h_2)$-plane (see Fig.~\ref{fig14}).}
\label{fig15}
%\end{center}
\end{figure*}

Let us consider the case of a field induced 1/3 magnetization plateau --- the field applied along the easy-magnetization direction for the RE2 site, $a$~axis in \SEO\ and $b$~axis in \SHO\ (see Fig.~\ref{fig15}).
The high-field phase with all the spins on the RE2 sites polarized along the field direction is phase 4.
The experimentally determined $uud$ structure is phase 15 (see Table~\ref{table2} and Fig.~\ref{fig3}).
However, it follows from our study that regions 13 and 15 as well as regions 4 and 15 have no common 6-face.
Therefore some intermediate phases should exist between them.
These are probably phases 23 and $\tilde 6$ ($\widetilde {23}$ and 6 if $J_2 > 0$) (see Fig.~\ref{fig15}).
We hope that in future the additional low temperature measurements will be able to verify the presence of these theoretically predicted phases in the \SRO\ magnets.

For $H\parallel b$ in \SDO, the situation should be somewhat similar, but the low-field transition is from a disordered state and therefore difficult to describe within the framework of our theory.
The transition from the field-induced $uud$ structure into a fully polarized state should, however also involve an intermediate phase. For $H\parallel b$ in \SDO, the zero-field phase is the disordered phase 14.
There are several possibilities for the field-induced $uud$ phase from phase 14 (see Supplement~\cite{Note1} and Fig.~\ref{fig14}).
The transition from the $uud$ structure into a fully polarized state should also involve an intermediate phase.

A very interesting case is found in \BDO.
Its low temperature zero-field structure is characterized by two half-integer propagation vectors, ${\bf k}_1=[\frac{1}{2}~0~\frac{1}{2}]$ and ${\bf k}_2=[\frac{1}{2}~\frac{1}{2}~\frac{1}{2}]$~\cite{Prevost_2018}.
In an applied field, the $uud$ structure is inferred from powder neutron diffraction measurements and from a pronounced plateau in the magnetization curve~\cite{Khalyavin_2019}.
The $uud$ structure appears to be much more stable than the zero field states, but finding any further intermediate magnetic states in \BDO\ will be experimentally challenging in the absence of a large size single crystal samples of this compound.
The zero-field phase for \BDO~ is phase 14 as for \SDO. For $\rm BaHo_2O_4$ it is phase 13.

\begin{figure*}[htb]
%\begin{center}
\includegraphics[scale = 0.55]{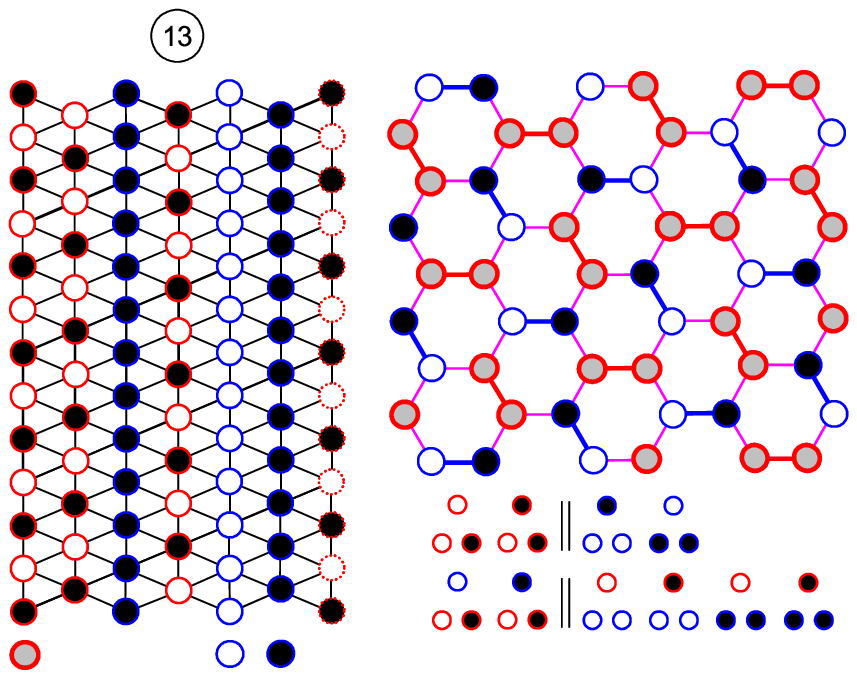}
\hspace{0.25cm}
\includegraphics[scale = 0.55]{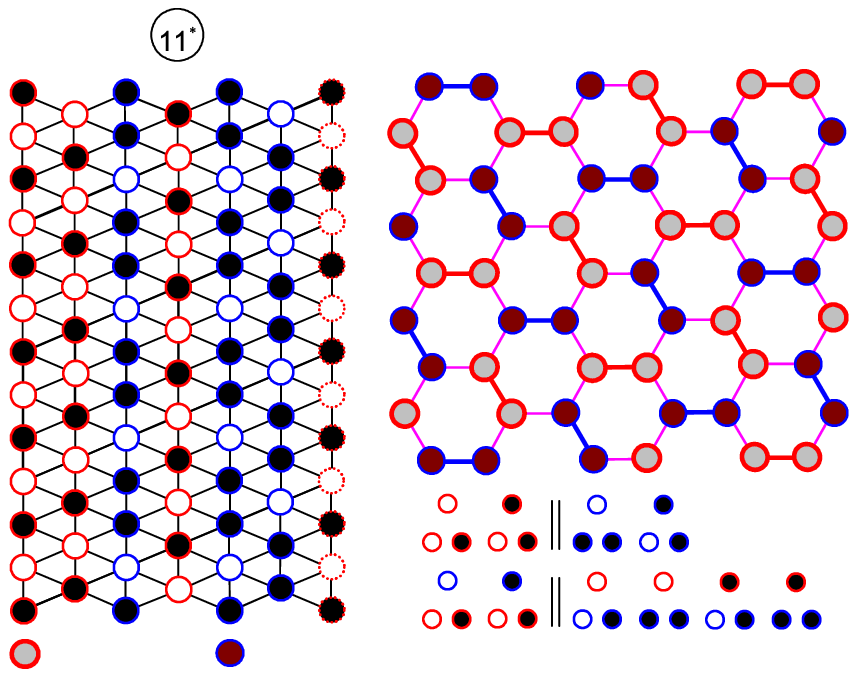}
\hspace{0.25cm}
\includegraphics[scale = 0.55]{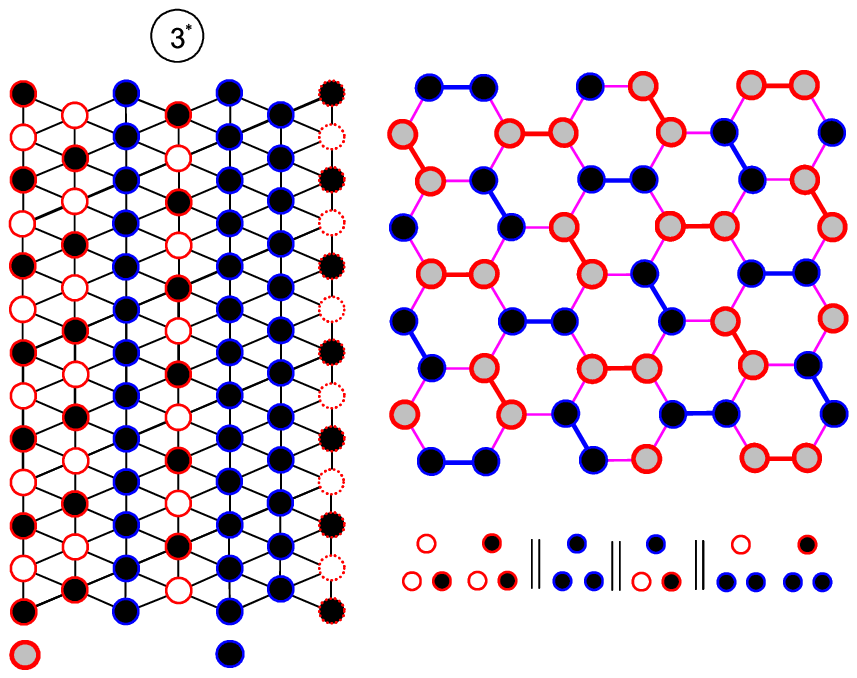}
\caption{A sequence of phase transitions proposed for an increasing field applied along the $c$~axis in \SHO\ and \SEO.
The corresponding numbers for magnetization of the blue sublattice are 0, 1/3, and 1. $J_{02} > 0$.
If $J_{02} < 0$, then there is a direct transition from phase 13 to phase 3$^*$.}
\label{fig16}
%\end{center}
\end{figure*}

Let us also consider the case of a field applied along the $c$~axis (direction of the chains of the magnetic atoms).
For this geometry, magnetization data for \SHO\ and \SEO\ suggest a single phase transition to a state with a full polarization of a site for which the easy-magnetization direction coincides with the $c$~axis.
In the language of this paper, the transition is from a zero-field phase 13 to phase 3$^*$ where all spins on one of the magnetic sites are parallel to the field while the other site remains the same as in zero-field.
The proposal is that with increasing field, structure 11$^*$ is stabilized between phases 13 and 3$^*$ (see Fig.~\ref{fig16}), although a direct transition between these phases is also possible (see Supplement~\cite{Note1}).

\section{Conclusions}%%%%%%%%%%%%%%%%%%%%%
\label{Sec_IV}
%------------------------------------------------------------------------------------------------------------------------------------------------
We present a solution to the ground-state problem for an Ising model in an external field for a honeycomb zigzag-ladder lattice with two different types of magnetic sites.
Although the solution is incomplete, the presence of a variety of ground-states is proved and, for several phases, the corresponding regions in seven-dimensional parameter space are completely determined.
Some of these phases are ordered but the majority are disordered with the disorder being one, two, or even three-dimensional.

The solution is used to explain the existence of experimentally detected spin arrangements in honeycomb zigzag-ladder magnets \SRO\ (and potentially \BRO) in an applied magnetic field.
New phases, yet to be detected experimentally, are predicted, especially those with two different types of magnetic configurations on the same ladder.

Since the set of basic rays that we found here is incomplete, we hope that the paper will inspire further efforts to find the remaining basic rays and to establish a complete set for this very interesting and complex ground-state problem.

\section{Appendix}

Here we present the energies for all the six configurations, $\usebox{\ddd}$, $\usebox{\ddu}$, $\usebox{\dud}$, $\usebox{\uud}$, $\usebox{\duu}$, and $\usebox{\uuu}$ (open and solid circles denote spins $\sigma = -1$ and $\sigma = +1$, respectively), of the four types of plaquettes (see Fig.~\ref{fig2} and Eq.~(\ref{eq2})).

\begin{eqnarray}
&&e_{11} = (1 - \alpha_1)J_{01} + J_{11} \nonumber\\
&&~~~~+ [(1 - \eta_1)\gamma_1 + (1 - \delta_1)(1 - \gamma_1)]h_1, \nonumber\\
&&e_{12} = -(1 - \alpha_1)J_{01} + (1 - \delta_1)(1 - \gamma_1)h_1, \nonumber\\
&&e_{13} = (1 - \alpha_1)J_{01} - J_{11} \nonumber\\
&&~~~~+ [(1 - \eta_1)\gamma_1 - (1 - \delta_1)(1 - \gamma_1)]h_1, \nonumber\\
&&e_{14} = (1- \alpha_1)J_{01} - J_{11} \nonumber\\
&&~~~~- [(1 - \eta_1)\gamma_1 - (1 - \delta_1)(1 - \gamma_1)]h_1, \nonumber\\
&&e_{15} = -(1 - \alpha_1)J_{01} - (1 - \delta_1)(1 - \gamma_1)h_1, \nonumber\\
&&e_{16} = (1 - \alpha_1)J_{01} + J_{11} \nonumber\\
&&~~~~- [(1 - \eta_1)\gamma_1 + (1 - \delta_1)(1 - \gamma_1)]h_1;
\label{eq6}
\end{eqnarray}

\begin{eqnarray}
&&e_{21} = (1 - \alpha_2)J_{02} + J_{12} \nonumber\\
&&~~~~+[(1 - \eta_2)\gamma_2 + (1 - \delta_2)(1 - \gamma_2)]h_2, \nonumber\\
&&e_{22} = -(1 - \alpha_2)J_{02} + (1 - \delta_2)(1 - \gamma_2)h_2, \nonumber\\
&&e_{23} = (1 - \alpha_2)J_{02} - J_{12} \nonumber\\
&&~~~~+ [(1 - \eta_2) \gamma_2 - (1 - \delta_2)(1 - \gamma_2)]h_2, \nonumber\\
&&e_{24} = (1 - \alpha_2)J_{02} - J_{12} \nonumber\\
&&~~~~- [(1 - \eta_2)\gamma_2 - (1 - \delta_2)(1 - \gamma_2)]h_2, \nonumber\\
&&e_{25} = -(1 - \alpha_2)J_{02} - (1 - \delta_2)(1 - \gamma_2)h_2, \nonumber\\
&&e_{26} = (1 - \alpha_2)J_{02} + J_{12} \nonumber\\
&&~~~~- [(1 - \eta_2)\gamma_2+(1 - \delta_2)(1 - \gamma_2)]h_2;
\label{eq7}
\end{eqnarray}

\begin{eqnarray}
&&e_{31} = \frac{\alpha_1}{2}J_{01} + 2(1 - \beta)J_2 + \eta_1 \frac{\gamma_1}{2}h_1 + \delta_2 \frac{(1 - \gamma_2)}{2}h_2, \nonumber\\
&&e_{32} = -\frac{\alpha_1}{2}J_{01} + \delta_2\frac{(1 - \gamma_2)}{2}h_2, \nonumber\\
&&e_{33} = \frac{\alpha_1}{2}J_{01} - 2(1 - \beta) J_2 + \eta_1 \frac{\gamma_1}{2}h_1 - \delta_2\frac{(1 - \gamma_2)}{2}h_2, \nonumber\\
&&e_{34} = \frac{\alpha_1}{2}J_{01} - 2(1 - \beta) J_2 - \eta_1\frac{\gamma_1}{2}h_1 + \delta_2\frac{(1 - \gamma_2)}{2}h_2, \nonumber\\
&&e_{35} = -\frac{\alpha_1}{2}J_{01} - \delta_2\frac{(1 - \gamma_2)}{2} h_2, \nonumber\\
&&e_{36} = \frac{\alpha_1}{2}J_{01} + 2(1 - \beta)J_2 - \eta_1\frac{\gamma_1}{2}h_1 - \delta_2\frac{(1 - \gamma_2)}{2}h_2; \nonumber\\
\label{eq8}
\end{eqnarray}

\begin{eqnarray}
&&e_{41} = \frac{\alpha_2}{2}J_{02} + 2\beta J_2 + \delta_1 \frac{(1 - \gamma_1)}{2} h_1 + \eta_2 \frac{\gamma_2}{2} h_2, \nonumber\\
&&e_{42} = -\frac{\alpha_2}{2}J_{02} + \delta_1\frac{(1 - \gamma_1)}{2}h_1, \nonumber\\
&&e_{43} = \frac{\alpha_2}{2}J_{02} - 2\beta J_2 - \delta_1\frac{(1 - \gamma_1)}{2}h_1 + \eta_2\frac{\gamma_2}{2} h_2, \nonumber\\
&&e_{44} = \frac{\alpha_2}{2}J_{02} - 2\beta J_2 + \delta_1 \frac{(1 - \gamma_1)}{2}h_1 - \eta_2 \frac{\gamma_2}{2} h_2, \nonumber\\
&&e_{45} = -\frac{\alpha_2}{2}J_{02} - \delta_1\frac{(1 - \gamma_1)}{2} h_1, \nonumber\\
&&e_{46} = \frac{\alpha_2}{2}J_{02} + 2\beta J_2 - \delta_1\frac{(1 - \gamma_1)}{2}h_1 - \eta_2\frac{\gamma_2}{2} h_2.
\label{eq9}
\end{eqnarray}

To calculate the energy of a structure (or structures in the case of degeneration), it is sufficient to know the relative numbers of plaquette configurations which generate this structure.
For instance, structures 2 are generated with seven configurations \usebox{\uuur}, \usebox{\uudb}, \usebox{\duub}, \usebox{\uudrb}, \usebox{\uuurb}, \usebox{\duubr}, and \usebox{\uuubr} (see Table~\ref{table2}), relative numbers of which in these structures are 3, 1, 2, 2, 4, 4, and 2, respectively.
Hence, the energy (per six plaquettes) of structures 2 is
\begin{eqnarray}
&&e_2 = \frac13(3e_{16} + e_{24} + 2e_{25} + 2e_{34} + 4e_{36} + 4e_{45} + 2e_{46})\nonumber\\
&&= \frac13(3J_{01} - J_{02} + 3J_{11} - J_{12} + 4J_2 - 3h_1 - h_2).
\label{eq10}
\end{eqnarray}
It should be noted that this energy does not depend on free coefficients although $e_{ij}$ do depend on these.
The magnetization of the red sublattice (per one red site) is equal to $3 \cdot 1/3 = 1$, for the blue sublattice, it is equal to 1/3.

Let us show how to find conditions for the existence of a region in the $(h_1, h_2)$-plane, for instance, region 4. This region is determined with the set of basic rays
$\{\mathbf{r}_1^{}, \mathbf{r}_1^\star, \mathbf{r}_2^{}, \mathbf{r}_3^\star, \mathbf{r}_5^{},
\mathbf{r}_6^{}, \mathbf{r}_6^\star,  \mathbf{r}_6^{\star-}, \mathbf{r}_7^{}, \mathbf{r}_7^{\thicksim}\}$.

From the equation
\begin{eqnarray}
&&a_1^{}\mathbf{r}_1^{} +  a_1^\star\mathbf{r}_1^\star + a_2^{}\mathbf{r}_2^{} + a_3^\star\mathbf{r}_3^\star + a_5^{}\mathbf{r}_5^{} \nonumber \\
&&+ a_6^{}\mathbf{r}_6^{} + a_6^\star\mathbf{r}_6^\star + a_6^{\star-}\mathbf{r}_6^{\star-} + a_7^{}\mathbf{r}_7^{} + a_7^{\thicksim}\mathbf{r}_7^{\thicksim}\nonumber \\
&&= (J_{01}, J_{02}, J_{11}, J_{12}, J_2, h_1, h_2)
\label{eq}
\end{eqnarray}
we have
\begin{eqnarray}
&&a_1^{} = a_2^{} + a_5^{} - J_{01}, \nonumber \\
&&a_1^\star = a_3^\star - J_{02}, \nonumber \\
&&a_6^{} = 2a_2^{} + J_{11}, \nonumber \\
&&a_6^{\star-} = -2a_3^\star - a_6^\star + J_{12}, \nonumber \\
&&a_7^{} = a_7^{\thicksim} + J_2, \nonumber \\
&&h_1 = 4a_2^{} + 2a_5^{} + 8a_7^{\thicksim} + 2J_{11} + 4J_2, \nonumber \\
&&h_2 =  4a_3^\star + 4a_6^\star - 2J_{12} + 4J_2.
\label{eq}
\end{eqnarray}
All the coefficients $a$ should be nonnegative, therefore, we obtain
\begin{equation}
J_{12} > 0,~ -2J_{02} + J_{12} > 0.
\end{equation}
These inequalities are the conditions for the existence of region 4 in $h_1, h_2$-plain at fixed $J_{01}$, $J_{02}$, $J_{11}$ $J_{12}$, and $J_2$.
For $h_1$ and $h_2$ of region 4 we have,
\begin{eqnarray}
&&h_1 > 2J_{11} + 4|J_2|~ \text{if}~ J_{01} < 0, \nonumber \\
&&h_1 > 2J_{01} + J_{11} + 4|J_2|~ \text{if}~ J_{01} > 0,~ J_{11} < 0,\nonumber \\
&&h_1 > 2J_{01} + 2J_{11} + 4|J_2|~ \text{if}~ J_{01} > 0,~ J_{11} > 0,\nonumber \\
&&-2J_{12} + 4J_2 < h_2 < 2J_{12} + 4J_2~ \text{if}~ J_{02} < 0, \nonumber \\
&&4J_{02} - 2J_{12} + 4J_2 < h_2 < -4J_{02} + 2J_{12} + 4J_2 \nonumber \\
&& \text{if}~ J_{02} > 0.
\label{eq}
\end{eqnarray}

\bibliography{Collection_2019}
\end{document}

\title{SUPPLEMENT for Yu. Dublenych and O. Petrenko, An Ising model on a 3D honeycomb zigzag-ladder lattice:
a solution to the ground-state problem and application to the SrRE$_2$O$_4$ and BaRE$_2$O$_4$ magnets}
\maketitle

For explanations see the main text, Section II, Subsection F.

Phase 1\\

$\{\mathbf{r}_1^{}, \mathbf{r}_1^\star, \mathbf{r}_2^{}, \mathbf{r}_2^\star, \mathbf{r}_4^\thicksim,
\mathbf{r}_4^{\thicksim\star}, \mathbf{r}_5^{}, \mathbf{r}_5^\star, \mathbf{r}_6^{}, \mathbf{r}_6^\star,
\mathbf{r}_7^{}, \mathbf{r}_7^\thicksim, \mathbf{r}_7^{\thicksim-}\}$\\

$\{\mathbf{r}_1^{}, \mathbf{r}_1^\star, \mathbf{r}_2^{}, \mathbf{r}_2^\star, \mathbf{r}_4^\thicksim, \mathbf{r}_4^{\thicksim\star}, \mathbf{r}_7^\thicksim, \mathbf{r}_7^{\thicksim-} \}$,~~ $(\overline{1}, 1)$\\[-3mm]

$\usebox{\dddr}~ \usebox{\uuur}$ $\|$ $\usebox{\dddb}~ \usebox{\uuub}$ $\|$
$\usebox{\dddrb}~ \usebox{\uuurb}$ $\|$ $\usebox{\dddbr}~ \usebox{\uuubr}$\\[-2mm]

$\{\mathbf{r}_1^{}, \mathbf{r}_1^\star, \mathbf{r}_2^{}, \mathbf{r}_2^\star, \mathbf{r}_5^{}, \mathbf{r}_6^{}, \mathbf{r}_7^{}, \mathbf{r}_7^\thicksim \}$,~~ $(\widetilde 1, 1)$\\[-3mm]

$\usebox{\uuur}$ $\|$ $\usebox{\dddb}~ \usebox{\uuub}$ $\|$
$\usebox{\uudrb}~ \usebox{\uuurb}$ $\|$ $\usebox{\dudbr}~ \usebox{\uuubr}$\\[-2mm]

$\{\mathbf{r}_1^{}, \mathbf{r}_1^\star, \mathbf{r}_2^{}, \mathbf{r}_2^\star, \mathbf{r}_5^\star, \mathbf{r}_6^\star, \mathbf{r}_7^{}, \mathbf{r}_7^{\thicksim-}\}$,~~ $(\overline{\widetilde 1}, 1)$\\[-3mm]

$\usebox{\dddr}~ \usebox{\uuur}$ $\|$ $\usebox{\uuub}$ $\|$
$\usebox{\dudrb}~ \usebox{\uuurb}$ $\|$ $\usebox{\uudbr}~ \usebox{\uuubr}$\\[-2mm]

$\{\mathbf{r}_1^{}, \mathbf{r}_2^{}, \mathbf{r}_4^{\thicksim\star}, \mathbf{r}_5^{}, \mathbf{r}_5^\star, \mathbf{r}_6^{}, \mathbf{r}_6^\star, \mathbf{r}_7^{}, \mathbf{r}_7^\thicksim\}$,~~ $(2, 1)$\\[-3mm]

$\usebox{\uuur}$ $\|$ $\usebox{\uudb}~ \usebox{\duub}~ \usebox{\uuub}$ $\|$
$\usebox{\uudrb}~ \usebox{\uuurb}$ $\|$ $\usebox{\duubr}~ \usebox{\uuubr}$\\[-2mm]

$\{\mathbf{r}_1^\star, \mathbf{r}_2^\star, \mathbf{r}_4^\thicksim, \mathbf{r}_5^{}, \mathbf{r}_5^\star, \mathbf{r}_6^{}, \mathbf{r}_6^\star, \mathbf{r}_7^{}, \mathbf{r}_7^{\thicksim-} \}$,~~ $(2^\star, 1)$\\[-3mm]

$\usebox{\uudr}~ \usebox{\duur}~ \usebox{\uuur}$ $\|$ $\usebox{\uuub}$ $\|$
$\usebox{\duurb}~ \usebox{\uuurb}$ $\|$ $\usebox{\uudbr}~ \usebox{\uuubr}$\\[-2mm]

$\{\mathbf{r}_1^{}, \mathbf{r}_2^{}, \mathbf{r}_2^\star, \mathbf{r}_4^{\thicksim\star}, \mathbf{r}_5^{}, \mathbf{r}_5^\star, \mathbf{r}_6^{}, \mathbf{r}_7^{}, \mathbf{r}_7^\thicksim \}$,~~ $(3, 1)$\\[-3mm]

$\usebox{\uuur}$ $\|$ $\usebox{\ddub}~ \usebox{\duub}~ \usebox{\uuub}$ $\|$
$\usebox{\uudrb}~ \usebox{\uuurb}$ $\|$ $\usebox{\duubr}~ \usebox{\uuubr}$\\[-2mm]

$\{\mathbf{r}_1^{}, \mathbf{r}_2^{}, \mathbf{r}_2^\star, \mathbf{r}_4^{\thicksim\star}, \mathbf{r}_5^\star, \mathbf{r}_7^{\thicksim-} \}$,~~ $(\overline{3}, 1)$\\[-3mm]

$\usebox{\dddr}~ \usebox{\uuur}$ $\|$ $\usebox{\ddub}~ \usebox{\duub}~ \usebox{\uuub}$ $\|$
$\usebox{\dddrb}~ \usebox{\dudrb}~ \usebox{\uuurb}$ $\|$ $\usebox{\ddubr}~ \usebox{\uuubr}$\\[-2mm]

$\{\mathbf{r}_1^\star, \mathbf{r}_2^{}, \mathbf{r}_2^\star, \mathbf{r}_4^\thicksim, \mathbf{r}_5^{}, \mathbf{r}_5^\star, \mathbf{r}_6^\star, \mathbf{r}_7^{}, \mathbf{r}_7^{\thicksim-} \}$,~~ $(3^\star, 1)$\\[-3mm]

$\usebox{\ddur}~ \usebox{\duur}~ \usebox{\uuur}$ $\|$ $\usebox{\uuub}$ $\|$
$\usebox{\duurb}~ \usebox{\uuurb}$ $\|$ $\usebox{\uudbr}~ \usebox{\uuubr}$\\[-2mm]

$\{\mathbf{r}_1^\star, \mathbf{r}_2^{}, \mathbf{r}_2^\star, \mathbf{r}_4^\thicksim, \mathbf{r}_5^{}, \mathbf{r}_7^\thicksim \}$,~~ $(\overline{3}^\star, 1)$\\[-3mm]

$\usebox{\ddur}~ \usebox{\duur}~ \usebox{\uuur}$ $\|$ $\usebox{\dddb}~ \usebox{\uuub}$ $\|$
$\usebox{\ddurb}~ \usebox{\uuurb}$ $\|$ $\usebox{\dddbr}~ \usebox{\dudbr}~ \usebox{\uuubr}$\\[-2mm]

$\{\mathbf{r}_1^{}, \mathbf{r}_1^\star, \mathbf{r}_2^{}, \mathbf{r}_5^{}, \mathbf{r}_6^{}, \mathbf{r}_6^\star, \mathbf{r}_7^{}, \mathbf{r}_7^\thicksim\}$,~~ $(4, 1)$\\[-3mm]

$\usebox{\uuur}$ $\|$ $\usebox{\dudb}~ \usebox{\uudb}~ \usebox{\uuub}$ $\|$
$\usebox{\uudrb}~ \usebox{\uuurb}$ $\|$ $\usebox{\dudbr}~ \usebox{\uuubr}$\\[-2mm]

$\{\mathbf{r}_1^{}, \mathbf{r}_1^\star, \mathbf{r}_2^\star, \mathbf{r}_5^\star, \mathbf{r}_6^{}, \mathbf{r}_6^\star, \mathbf{r}_7^{}, \mathbf{r}_7^{\thicksim-} \}$,~~ $(4^\star, 1)$\\[-3mm]

$\usebox{\dudr}~ \usebox{\uudr}~ \usebox{\uuur}$ $\|$ $\usebox{\uuub}$ $\|$
$\usebox{\dudrb}~ \usebox{\uuurb}$ $\|$ $\usebox{\uudbr}~ \usebox{\uuubr}$\\[-2mm]

$\{\mathbf{r}_4^\thicksim, \mathbf{r}_4^{\thicksim\star}, \mathbf{r}_5^{}, \mathbf{r}_6^{}, \mathbf{r}_6^\star, \mathbf{r}_7^\thicksim \}$,~~ $(5, 1)$\\[-3mm]

$\usebox{\uudr}~ \usebox{\duur}~ \usebox{\uuur}$ $\|$ $\usebox{\dudb}~ \usebox{\uudb}~ \usebox{\duub}~ \usebox{\uuub}$ $\|$
$\usebox{\ddurb}~\usebox{\uuurb}$ $\|$ $\usebox{\dddbr}~ \usebox{\duubr}~ \usebox{\uuubr}$\\[-2mm]

$\{\mathbf{r}_4^\thicksim, \mathbf{r}_4^{\thicksim\star}, \mathbf{r}_5^\star, \mathbf{r}_6^{}, \mathbf{r}_6^\star, \mathbf{r}_7^{\thicksim-} \}$,~~ $(5^\star, 1)$\\[-3mm]

$\usebox{\dudr}~ \usebox{\uudr}~ \usebox{\duur}~ \usebox{\uuur}$ $\|$ $\usebox{\uudb}~ \usebox{\duub}~ \usebox{\uuub}$ $\|$
$\usebox{\dddrb}~ \usebox{\duurb}~ \usebox{\uuurb}$ $\|$ $\usebox{\ddubr}~ \usebox{\uuubr}$\\[-2mm]

$\{\mathbf{r}_1^\star, \mathbf{r}_4^\thicksim, \mathbf{r}_5^{}, \mathbf{r}_6^{}, \mathbf{r}_6^\star, \mathbf{r}_7^\thicksim \}$,~~ $(\widetilde 6, 1)$\\[-3mm]

$\usebox{\uudr}~ \usebox{\duur}~ \usebox{\uuur}$ $\|$ $\usebox{\dudb}~ \usebox{\uudb}~ \usebox{\uuub}$ $\|$
$\usebox{\ddurb}~ \usebox{\uuurb}$ $\|$ $\usebox{\dddbr}~ \usebox{\dudbr}~ \usebox{\uuubr}$\\[-2mm]

$\{\mathbf{r}_1^{}, \mathbf{r}_4^{\thicksim\star}, \mathbf{r}_5^\star, \mathbf{r}_6^{}, \mathbf{r}_6^\star, \mathbf{r}_7^{\thicksim-} \}$,~~ $(\widetilde{6}^\star, 1)$\\[-3mm]

$\usebox{\dudr}~ \usebox{\uudr}~ \usebox{\uuur}$ $\|$ $\usebox{\uudb}~ \usebox{\duub}~ \usebox{\uuub}$ $\|$
$\usebox{\dddrb}~ \usebox{\dudrb}~ \usebox{\uuurb}$ $\|$ $\usebox{\ddubr}~ \usebox{\uuubr}$\\[-2mm]

$\{\mathbf{r}_2^{}, \mathbf{r}_2^\star, \mathbf{r}_4^\thicksim, \mathbf{r}_4^{\thicksim\star}, \mathbf{r}_5^{}, \mathbf{r}_7^\thicksim \}$,~~ $(\widetilde 7, 1)$\\[-3mm]

$\usebox{\ddur}~ \usebox{\duur}~ \usebox{\uuur}$ $\|$ $\usebox{\dddb}~ \usebox{\ddub}~ \usebox{\duub}~ \usebox{\uuub}$ $\|$
$\usebox{\ddurb}~ \usebox{\uuurb}$ $\|$ $\usebox{\dddbr}~ \usebox{\duubr}~ \usebox{\uuubr}$\\[-2mm]

$\{\mathbf{r}_2^{}, \mathbf{r}_2^\star, \mathbf{r}_4^\thicksim, \mathbf{r}_4^{\thicksim\star}, \mathbf{r}_5^\star, \mathbf{r}_7^{\thicksim-} \}$,~~ $(\overline{\widetilde 7}, 1)$\\[-3mm]

$\usebox{\dddr}~ \usebox{\ddur}~ \usebox{\duur}~ \usebox{\uuur}$ $\|$ $\usebox{\ddub}~ \usebox{\duub}~ \usebox{\uuub}$ $\|$
$\usebox{\dddrb}~ \usebox{\duurb}~ \usebox{\uuurb}$ $\|$ $\usebox{\ddubr}~ \usebox{\uuubr}$\\[-2mm]

$\{\mathbf{r}_4^\thicksim, \mathbf{r}_4^{\thicksim\star}, \mathbf{r}_5^{}, \mathbf{r}_5^\star, \mathbf{r}_6^{}, \mathbf{r}_6^\star \}$,~~ $(9, 1)$\\[-3mm]

$\usebox{\uudr}~ \usebox{\duur}~ \usebox{\uuur}$ $\|$ $\usebox{\uudb}~ \usebox{\duub}~ \usebox{\uuub}$ $\|$
$\usebox{\ddurb}~ \usebox{\duurb}~ \usebox{\uuurb}$ $\|$ $\usebox{\ddubr}~ \usebox{\duubr}~ \usebox{\uuubr}$\\[-2mm]

$\{\mathbf{r}_2^\star, \mathbf{r}_4^\thicksim, \mathbf{r}_4^{\thicksim\star}, \mathbf{r}_5^{}, \mathbf{r}_5^\star, \mathbf{r}_6^{} \}$,~~ $(10, 1)$\\[-3mm]

$\usebox{\uudr}~ \usebox{\duur}~ \usebox{\uuur}$ $\|$ $\usebox{\ddub}~ \usebox{\duub}~ \usebox{\uuub}$ $\|$
$\usebox{\ddurb}~ \usebox{\duurb}~ \usebox{\uuurb}$ $\|$ $\usebox{\ddubr}~ \usebox{\duubr}~ \usebox{\uuubr}$\\[-2mm]

$\{\mathbf{r}_2^{}, \mathbf{r}_4^\thicksim, \mathbf{r}_4^{\thicksim\star}, \mathbf{r}_5^{}, \mathbf{r}_5^\star, \mathbf{r}_6^\star \}$,~~ $(10^\star, 1)$\\[-3mm]

$\usebox{\ddur}~ \usebox{\duur}~ \usebox{\uuur}$ $\|$ $\usebox{\uudb}~ \usebox{\duub}~ \usebox{\uuub}$ $\|$
$\usebox{\ddurb}~ \usebox{\duurb}~ \usebox{\uuurb}$ $\|$ $\usebox{\ddubr}~ \usebox{\duubr}~ \usebox{\uuubr}$\\[-2mm]

$\{\mathbf{r}_1^{}, \mathbf{r}_1^\star, \mathbf{r}_4^\thicksim, \mathbf{r}_4^{\thicksim\star}, \mathbf{r}_6^{}, \mathbf{r}_6^\star, \mathbf{r}_7^\thicksim, \mathbf{r}_7^{\thicksim-} \}$,~~ $(\widetilde{12}, 1)$\\[-3mm]

$\usebox{\dudr}~ \usebox{\uudr}~ \usebox{\uuur}$ $\|$ $\usebox{\dudb}~ \usebox{\uudb}~ \usebox{\uuub}$ $\|$
$\usebox{\dddrb}~ \usebox{\uuurb}$ $\|$ $\usebox{\dddbr}~ \usebox{\uuubr}$\\[-2mm]

$\{\mathbf{r}_2^{}, \mathbf{r}_2^\star, \mathbf{r}_4^\thicksim, \mathbf{r}_4^{\thicksim\star}, \mathbf{r}_5^{}, \mathbf{r}_5^\star \}$,~~ $(14, 1)$\\[-3mm]

$\usebox{\ddur}~ \usebox{\duur}~ \usebox{\uuur}$ $\|$ $\usebox{\ddub}~ \usebox{\duub}~ \usebox{\uuub}$ $\|$
$\usebox{\ddurb}~ \usebox{\duurb}~ \usebox{\uuurb}$ $\|$ $\usebox{\ddubr}~ \usebox{\duubr}~ \usebox{\uuubr}$\\

$\{\mathbf{r}_1^\star, \mathbf{r}_2^{}, \mathbf{r}_4^\thicksim, \mathbf{r}_5^{}, \mathbf{r}_6^\star, \mathbf{r}_7^\thicksim \}$,~~ $(-, 1)$\\[-3mm]

$\usebox{\ddur}~ \usebox{\duur}~ \usebox{\uuur}$ $\|$ $\usebox{\dudb}~ \usebox{\uudb}~ \usebox{\uuub}$ $\|$
$\usebox{\ddurb}~ \usebox{\uuurb}$ $\|$ $\usebox{\dddbr}~ \usebox{\dudbr}~ \usebox{\uuubr}$\\[-2mm]

$\{\mathbf{r}_1^{}, \mathbf{r}_2^\star, \mathbf{r}_4^{\thicksim\star}, \mathbf{r}_5^\star, \mathbf{r}_6^{}, \mathbf{r}_7^{\thicksim-} \}$,~~ $(-, 1)$\\[-3mm]

$\usebox{\dudr}~ \usebox{\uudr}~ \usebox{\uuur}$ $\|$ $\usebox{\ddub}~ \usebox{\duub}~ \usebox{\uuub}$ $\|$
$\usebox{\dddrb}~ \usebox{\dudrb}~ \usebox{\uuurb}$ $\|$ $\usebox{\ddubr}~ \usebox{\uuubr}$\\[-2mm]

$\{\mathbf{r}_2^{}, \mathbf{r}_4^\thicksim, \mathbf{r}_4^{\thicksim\star}, \mathbf{r}_5^{}, \mathbf{r}_6^\star, \mathbf{r}_7^\thicksim \}$,~~ $(-, 1)$\\[-3mm]

$\usebox{\ddur}~ \usebox{\duur}~ \usebox{\uuur}$ $\|$ $\usebox{\dudb}~ \usebox{\uudb}~ \usebox{\duub}~ \usebox{\uuub}$ $\|$
$\usebox{\ddurb}~ \usebox{\uuurb}$ $\|$ $\usebox{\dddbr}~ \usebox{\duubr}~ \usebox{\uuubr}$\\[-2mm]

$\{\mathbf{r}_2^\star, \mathbf{r}_4^\thicksim, \mathbf{r}_4^{\thicksim\star}, \mathbf{r}_5^\star, \mathbf{r}_6^{}, \mathbf{r}_7^{\thicksim-} \}$,~~ $(-, 1)$\\[-3mm]

$\usebox{\dudr}~ \usebox{\uudr}~ \usebox{\duur}~ \usebox{\uuur}$ $\|$ $\usebox{\ddub}~ \usebox{\duub}~ \usebox{\uuub}$ $\|$
$\usebox{\dddrb}~ \usebox{\duurb}~ \usebox{\uuurb}$ $\|$ $\usebox{\ddubr}~ \usebox{\uuubr}$\\[-2mm]

$\{\mathbf{r}_2^{}, \mathbf{r}_4^\thicksim, \mathbf{r}_4^{\thicksim\star}, \mathbf{r}_5^\star, \mathbf{r}_6^\star, \mathbf{r}_7^{\thicksim-} \}$,~~ $(-, 1)$\\[-3mm]

$\usebox{\dddr}~ \usebox{\ddur}~ \usebox{\duur}~ \usebox{\uuur}$ $\|$ $\usebox{\uudb}~ \usebox{\duub}~ \usebox{\uuub}$ $\|$
$\usebox{\dddrb}~ \usebox{\duurb}~ \usebox{\uuurb}$ $\|$ $\usebox{\ddubr}~ \usebox{\uuubr}$\\[-2mm]

$\{\mathbf{r}_2^\star, \mathbf{r}_4^\thicksim, \mathbf{r}_4^{\thicksim\star}, \mathbf{r}_5^{}, \mathbf{r}_6^{}, \mathbf{r}_7^\thicksim \}$,~~ $(-, 1)$\\[-3mm]

$\usebox{\uudr}~ \usebox{\duur}~ \usebox{\uuur}$ $\|$ $\usebox{\dddb}~ \usebox{\ddub}~ \usebox{\duub}~ \usebox{\uuub}$ $\|$
$\usebox{\ddurb}~ \usebox{\uuurb}$ $\|$ $\usebox{\dddbr}~ \usebox{\duubr}~ \usebox{\uuubr}$\\[-2mm]

$\{\mathbf{r}_1^{}, \mathbf{r}_2^{}, \mathbf{r}_4^{\thicksim\star}, \mathbf{r}_5^\star, \mathbf{r}_6^\star, \mathbf{r}_7^{\thicksim-} \}$,~~ $(-, 1)$\\[-3mm]

$\usebox{\dddr}~ \usebox{\uuur}$ $\|$ $\usebox{\uudb}~ \usebox{\duub}~ \usebox{\uuub}$ $\|$
$\usebox{\dddrb}~ \usebox{\dudrb}~ \usebox{\uuurb}$ $\|$ $\usebox{\ddubr}~ \usebox{\uuubr}$\\[-2mm]

$\{\mathbf{r}_1^\star, \mathbf{r}_2^\star, \mathbf{r}_4^\thicksim, \mathbf{r}_5^{}, \mathbf{r}_6^{}, \mathbf{r}_7^\thicksim \}$,~~ $(-, 1)$\\[-3mm]

$\usebox{\uudr}~ \usebox{\duur}~ \usebox{\uuur}$ $\|$ $\usebox{\dddb}~ \usebox{\uuub}$ $\|$
$\usebox{\ddurb}~ \usebox{\uuurb}$ $\|$ $\usebox{\dddbr}~ \usebox{\dudbr}~ \usebox{\uuubr}$\\[-2mm]

$\{\mathbf{r}_1^{}, \mathbf{r}_1^\star, \mathbf{r}_2^{}, \mathbf{r}_4^\thicksim, \mathbf{r}_4^{\thicksim\star}, \mathbf{r}_6^\star, \mathbf{r}_7^\thicksim, \mathbf{r}_7^{\thicksim-} \}$,~~ $(-, 1)$\\[-3mm]

$\usebox{\dddr}~ \usebox{\uuur}$ $\|$ $\usebox{\dudb}~ \usebox{\uudb}~ \usebox{\uuub}$ $\|$
$\usebox{\dddrb}~ \usebox{\uuurb}$ $\|$ $\usebox{\dddbr}~ \usebox{\uuubr}$\\[-2mm]

$\{\mathbf{r}_1^{}, \mathbf{r}_1^\star, \mathbf{r}_2^\star, \mathbf{r}_4^\thicksim, \mathbf{r}_4^{\thicksim\star}, \mathbf{r}_6^{}, \mathbf{r}_7^\thicksim, \mathbf{r}_7^{\thicksim-} \}$,~~ $(-, 1)$\\[-3mm]

$\usebox{\dudr}~ \usebox{\uudr}~ \usebox{\dddr}$ $\|$ $\usebox{\dddb}~ \usebox{\uuub}$ $\|$
$\usebox{\dddrb}~ \usebox{\uuurb}$ $\|$ $\usebox{\dddbr}~ \usebox{\uuubr}$\\

\clearpage
\newpage

Phase 2\\[-2mm]

$\{\mathbf{r}_1^{}, \mathbf{r}_2^{}, \mathbf{r}_3^\star, \mathbf{r}_4^{\thicksim\star}, \mathbf{r}_5^{},
\mathbf{r}_5^\star, \mathbf{r}_6^{}, \mathbf{r}_6^\star, \mathbf{r}_7^{}, \mathbf{r}_7^\thicksim\}$\\[-2mm]

$\{\mathbf{r}_1^{}, \mathbf{r}_2^{}, \mathbf{r}_4^{\thicksim\star}, \mathbf{r}_5^{}, \mathbf{r}_5^\star,
\mathbf{r}_6^{}, \mathbf{r}_6^\star, \mathbf{r}_7^{}, \mathbf{r}_7^\thicksim\}$,~~ $(1, 2)$\\[-3mm]

$\usebox{\uuur}$ $\|$ $\usebox{\uudb}~ \usebox{\duub}~ \usebox{\uuub}$ $\|$
$\usebox{\uudrb}~ \usebox{\uuurb}$ $\|$ $\usebox{\duubr}~ \usebox{\uuubr}$\\[-2mm]

$\{\mathbf{r}_1^{}, \mathbf{r}_2^{}, \mathbf{r}_3^\star, \mathbf{r}_4^{\thicksim\star}, \mathbf{r}_5{}, \mathbf{r}_5^\star,
\mathbf{r}_6^{}, \mathbf{r}_7^{}, \mathbf{r}_7^\thicksim\}$,~~ $(3, 2)$\\[-3mm]

$\usebox{\uuur}$ $\|$ $\usebox{\ddub}~ \usebox{\uudb}~ \usebox{\duub}$ $\|$
$\usebox{\uudrb}~ \usebox{\uuurb}$ $\|$ $\usebox{\duubr}~ \usebox{\uuubr}$\\[-2mm]

$\{\mathbf{r}_1^{}, \mathbf{r}_2^{}, \mathbf{r}_3^\star,  \mathbf{r}_5{}, \mathbf{r}_6^{}, \mathbf{r}_6^\star,
\mathbf{r}_7^{}, \mathbf{r}_7^\thicksim\}$,~~ $(4, 2)$ \\[-3mm]%(1-st order)

$\usebox{\uuur}$ $\|$ $\usebox{\dudb}~ \usebox{\uudb}~ \usebox{\duub}$ $\|$
$\usebox{\uudrb}~ \usebox{\uuurb}$ $\|$ $\usebox{\dudbr}~ \usebox{\duubr}~ \usebox{\uuubr}$\\[-2mm]

$\{\mathbf{r}_3^\star, \mathbf{r}_4^{\thicksim\star}, \mathbf{r}_5^{}, \mathbf{r}_6^{}, \mathbf{r}_6^\star,
\mathbf{r}_7^\thicksim\}$,~~ $(5, 2)$\\[-3mm]

$\usebox{\uudr}~ \usebox{\duur}~ \usebox{\uuur}$ $\|$ $\usebox{\ddub}~ \usebox{\uudb}~ \usebox{\duub}$ $\|$
$\usebox{\ddurb}~ \usebox{\uudrb}~ \usebox{\uuurb}$ $\|$ $\usebox{\dddbr}~ \usebox{\duubr}~ \usebox{\uuubr}$\\[-2mm]

$\{\mathbf{r}_3^\star, \mathbf{r}_5{}, \mathbf{r}_5^\star, \mathbf{r}_6^{}, \mathbf{r}_6^\star, \mathbf{r}_7^{}\}$,~~ $(8, 2)$\\[-3mm]

$\usebox{\uudr}~ \usebox{\duur}~ \usebox{\uuur}$ $\|$ $\usebox{\uudb}~ \usebox{\duub}$ $\|$
$\usebox{\uudrb}~ \usebox{\duurb}~ \usebox{\uuurb}$ $\|$ $\usebox{\uudbr}~ \usebox{\duubr}~ \usebox{\uuubr}$\\[-2mm]

$\{\mathbf{r}_3^\star, \mathbf{r}_4^{\thicksim\star}, \mathbf{r}_5^{}, \mathbf{r}_5^\star, \mathbf{r}_6^{},
\mathbf{r}_6^\star\}$,~~ $(9, 2)$\\[-3mm]

$\usebox{\uudr}~ \usebox{\duur}~ \usebox{\uuur}$ $\|$ $\usebox{\uudb}~ \usebox{\duub}$ $\|$
$\usebox{\ddurb}~ \usebox{\uudrb}~ \usebox{\duurb}~ \usebox{\uuurb}$ $\|$ $\usebox{\ddubr}~ \usebox{\duubr}~ \usebox{\uuubr}$\\[-2mm]

$\{\mathbf{r}_2^{}, \mathbf{r}_3^\star, \mathbf{r}_4^{\thicksim\star}, \mathbf{r}_5{}, \mathbf{r}_5^\star,
\mathbf{r}_6^\star\}$,~~ $(10^\star, 2)$\\[-3mm]

$\usebox{\ddur}~ \usebox{\duur}~ \usebox{\uuur}$ $\|$ $\usebox{\uudb}~ \usebox{\duub}$ $\|$
$\usebox{\ddurb}~ \usebox{\uudrb}~ \usebox{\duurb}~ \usebox{\uuurb}$ $\|$ $\usebox{\ddubr}~ \usebox{\duubr}~ \usebox{\uuubr}$\\[-2mm]

$\{\mathbf{r}_1^{}, \mathbf{r}_3^\star, \mathbf{r}_4^{\thicksim\star}, \mathbf{r}_6^{}, \mathbf{r}_6^\star,
\mathbf{r}_7^\thicksim\}$,~~ $(\widetilde {12}, 2)$\\[-3mm]

$\usebox{\dudr}~ \usebox{\uudr}~ \usebox{\uuur}$ $\|$ $\usebox{\dudb}~ \usebox{\uudb}~ \usebox{\duub}$ $\|$
$\usebox{\dddrb}~ \usebox{\uudrb}~ \usebox{\uuurb}$ $\|$ $\usebox{\dddbr}~ \usebox{\duubr}~ \usebox{\uuubr}$\\[-2mm]

$\{\mathbf{r}_2^{}, \mathbf{r}_3^\star, \mathbf{r}_5^{}, \mathbf{r}_5^\star, \mathbf{r}_6^\star,\mathbf{r}_7^{}\}$,~~ $(16, 2)$\\[-3mm]

$\usebox{\ddur}~ \usebox{\duur}~ \usebox{\uuur}$ $\|$ $\usebox{\uudb}~ \usebox{\duub}$ $\|$
$\usebox{\uudrb}~ \usebox{\duurb}~ \usebox{\uuurb}$ $\|$ $\usebox{\uudbr}~ \usebox{\duubr}~ \usebox{\uuubr}$\\[-2mm]

$\{\mathbf{r}_2^{}, \mathbf{r}_3^\star, \mathbf{r}_4^{\thicksim\star}, \mathbf{r}_5^{}, \mathbf{r}_6^\star,
\mathbf{r}_7^\thicksim\}$,~~ $(-, 2)$\\[-3mm]

$\usebox{\ddur}~ \usebox{\duur}~ \usebox{\uuur}$ $\|$ $\usebox{\dudb}~ \usebox{\uudb}~ \usebox{\duub}$ $\|$
$\usebox{\ddurb}~ \usebox{\uudrb}~ \usebox{\uuurb}$ $\|$ $\usebox{\dddbr}~ \usebox{\duubr}~ \usebox{\uuubr}$\\

$\{\mathbf{r}_1^{}, \mathbf{r}_2^{}, \mathbf{r}_3^\star, \mathbf{r}_5^\star, \mathbf{r}_6^\star, \mathbf{r}_7^{}\}$,~~ $(-, 2)$\\[-3mm]

$\usebox{\dddr}~ \usebox{\uuur}$ $\|$ $\usebox{\uudb}~ \usebox{\duub}$ $\|$
$\usebox{\dudrb}~ \usebox{\uudrb}~ \usebox{\uuurb}$ $\|$ $\usebox{\uudbr}~ \usebox{\duubr}~ \usebox{\uuubr}$\\[-2mm]

$\{\mathbf{r}_1^{}, \mathbf{r}_3^\star, \mathbf{r}_4^{\thicksim\star}, \mathbf{r}_5^\star, \mathbf{r}_6^{},
\mathbf{r}_6^\star\}$,~~ $(-, 2)$\\[-3mm]

$\usebox{\dudr}~ \usebox{\uudr}~ \usebox{\uuur}$ $\|$ $\usebox{\uudb}~ \usebox{\duub}$ $\|$
$\usebox{\dddrb}~ \usebox{\dudrb}~ \usebox{\uudrb}~ \usebox{\uuurb}$ $\|$ $\usebox{\ddubr}~ \usebox{\duubr}~ \usebox{\uuubr}$\\[-2mm]

$\{\mathbf{r}_1^{}, \mathbf{r}_3^\star, \mathbf{r}_5^\star, \mathbf{r}_6^{},
\mathbf{r}_6^\star, \mathbf{r}_7^{}\}$,~~ $(-, 2)$\\[-3mm]

$\usebox{\dudr}~ \usebox{\uudr}~ \usebox{\uuur}$ $\|$ $\usebox{\uudb}~ \usebox{\duub}$ $\|$
$\usebox{\dudrb}~ \usebox{\uudrb}~ \usebox{\uuurb}$ $\|$ $\usebox{\uudbr}~ \usebox{\duubr}~ \usebox{\uuubr}$\\[-2mm]

$\{\mathbf{r}_1^{}, \mathbf{r}_2^{}, \mathbf{r}_3^\star, \mathbf{r}_4^{\thicksim\star},
\mathbf{r}_6^\star, \mathbf{r}_7^\thicksim\}$,~~ $(-, 2)$\\[-3mm]

$\usebox{\dddr}~ \usebox{\uuur}$ $\|$ $\usebox{\dudb}~ \usebox{\uudb}~ \usebox{\duub}$ $\|$
$\usebox{\dddrb}~ \usebox{\uudrb}~ \usebox{\uuurb}$ $\|$ $\usebox{\dddbr}~ \usebox{\duubr}~ \usebox{\uuubr}$\\[-2mm]

$\{\mathbf{r}_1^{}, \mathbf{r}_2^{}, \mathbf{r}_3^\star, \mathbf{r}_4^{\thicksim\star},
\mathbf{r}_5^\star, \mathbf{r}_6^\star\}$,~~ $(-, 2)$\\[-3mm]

$\usebox{\dddr}~ \usebox{\uuur}$ $\|$ $\usebox{\uudb}~ \usebox{\duub}$ $\|$
$\usebox{\dddrb}~ \usebox{\dudrb}~ \usebox{\uudrb}~ \usebox{\uuurb}$ $\|$ $\usebox{\ddubr}~ \usebox{\duubr}~ \usebox{\uuubr}$\\

Phase 3\\[-2mm]

$\{\mathbf{r}_1^{}, \mathbf{r}_2^{}, \mathbf{r}_2^\star, \mathbf{r}_3^\star, \mathbf{r}_4^\star,
\mathbf{r}_4^{\thicksim\star}, \mathbf{r}_5^{}, \mathbf{r}_5^\star, \mathbf{r}_5^{\star-},
\mathbf{r}_6^{}, \mathbf{r}_7^{}, \mathbf{r}_7^\thicksim\}$\\[-2mm]

$\{\mathbf{r}_1^{}, \mathbf{r}_2^{}, \mathbf{r}_2^\star, \mathbf{r}_4^{\thicksim\star}, \mathbf{r}_5^{},
\mathbf{r}_5^\star, \mathbf{r}_6^{}, \mathbf{r}_7^{}, \mathbf{r}_7^\thicksim\}$,~~ $(1, 3)$\\[-3mm]

$\usebox{\uuur}$ $\|$ $\usebox{\ddub}~ \usebox{\duub}~ \usebox{\uuub}$ $\|$
$\usebox{\uudrb}~ \usebox{\uuurb}$ $\|$ $\usebox{\duubr}~ \usebox{\uuubr}$\\[-2mm]

$\{\mathbf{r}_1^{}, \mathbf{r}_2^{}, \mathbf{r}_2^\star, \mathbf{r}_4^{\thicksim\star}, \mathbf{r}_5^{\star-},
\mathbf{r}_7^\thicksim\}$,~~ $(\overline{1}, 3)$\\[-3mm]

$\usebox{\dddr}~ \usebox{\uuur}$ $\|$ $\usebox{\dddb}~ \usebox{\ddub}~ \usebox{\duub}$ $\|$
$\usebox{\dddrb}~ \usebox{\dudrb}~ \usebox{\uuurb}$ $\|$ $\usebox{\dddbr}~ \usebox{\duubr}$\\[-2mm]

$\{\mathbf{r}_1^{}, \mathbf{r}_2^{}, \mathbf{r}_2^\star, \mathbf{r}_4^\star, \mathbf{r}_5^{}, \mathbf{r}_5^{\star-},
\mathbf{r}_6^{}, \mathbf{r}_7^{}, \mathbf{r}_7^\thicksim\}$,~~ $(\widetilde 1, 3)$\\[-3mm]

$\usebox{\uuur}$ $\|$ $\usebox{\dddb}~ \usebox{\ddub}~ \usebox{\duub}$ $\|$
$\usebox{\uudrb}~ \usebox{\uuurb}$ $\|$ $\usebox{\dudbr}~ \usebox{\duubr}$\\[-2mm]

$\{\mathbf{r}_1^{}, \mathbf{r}_2^{}, \mathbf{r}_2^\star, \mathbf{r}_4^\star, \mathbf{r}_5^\star,
\mathbf{r}_7^{}\}$,~~ $(\overline{\widetilde 1}, 3)$\\[-3mm]

$\usebox{\dddr}~ \usebox{\uuur}$ $\|$ $\usebox{\ddub}~ \usebox{\duub}~ \usebox{\dddb}$ $\|$
$\usebox{\dudrb}~ \usebox{\uudrb}~ \usebox{\uuurb}$ $\|$ $\usebox{\uudbr}~ \usebox{\duubr}$\\[-2mm]

$\{\mathbf{r}_1^{}, \mathbf{r}_2^{}, \mathbf{r}_3^\star, \mathbf{r}_4^{\thicksim\star}, \mathbf{r}_5^{},
\mathbf{r}_5^\star, \mathbf{r}_6^{}, \mathbf{r}_7^{},  \mathbf{r}_7^\thicksim\}$,~~ $(2, 3)$\\[-3mm]

$\usebox{\uuur}$ $\|$ $\usebox{\ddub}~ \usebox{\uudb}~ \usebox{\duub}$ $\|$
$\usebox{\uudrb}~ \usebox{\uuurb}$ $\|$ $\usebox{\duubr}~ \usebox{\uuubr}$\\[-2mm]

$\{\mathbf{r}_1^{}, \mathbf{r}_2^{}, \mathbf{r}_3^\star, \mathbf{r}_4^\star, \mathbf{r}_5^{},
\mathbf{r}_5^{\star-}, \mathbf{r}_6^{}, \mathbf{r}_7^{},  \mathbf{r}_7^\thicksim\}$,~~ $(\widetilde 2, 3)$\\[-3mm]

$\usebox{\uuur}$ $\|$ $\usebox{\ddub}~ \usebox{\dudb}~ \usebox{\duub}$ $\|$
$\usebox{\uudrb}~ \usebox{\uuurb}$ $\|$ $\usebox{\dudbr}~ \usebox{\duubr}$\\[-2mm]

$\{\mathbf{r}_1^{}, \mathbf{r}_2^{}, \mathbf{r}_2^\star, \mathbf{r}_3^\star, \mathbf{r}_4^\star,
\mathbf{r}_4^{\thicksim\star}, \mathbf{r}_5^\star, \mathbf{r}_5^{\star-}\}$,~~ $(\overline{3}, 3)$\\[-3mm]

$\usebox{\dddr}~ \usebox{\uuur}$ $\|$ $\usebox{\ddub}~ \usebox{\duub}$ $\|$
$\usebox{\dddrb}~ \usebox{\dudrb}~ \usebox{\uudrb}~ \usebox{\uuurb}$ $\|$ $\usebox{\ddubr}~ \usebox{\duubr}$\\[-2mm]

$\{\mathbf{r}_2^{}, \mathbf{r}_2^\star, \mathbf{r}_4^\star, \mathbf{r}_5^{}, \mathbf{r}_5^\star,
\mathbf{r}_7^{},\}$,~~ $(7, 3)$\\[-3mm]

$\usebox{\ddur}~ \usebox{\duur}~ \usebox{\uuur}$ $\|$ $\usebox{\ddub}~ \usebox{\duub}~ \usebox{\uuub}$ $\|$
$\usebox{\uudrb}~ \usebox{\duurb}~ \usebox{\uuurb}$ $\|$ $\usebox{\uudbr}~ \usebox{\duubr}$\\[-2mm]

$\{\mathbf{r}_2^{}, \mathbf{r}_2^\star, \mathbf{r}_4^{\thicksim\star}, \mathbf{r}_5^{}, \mathbf{r}_5^{\star-},
\mathbf{r}_7^\thicksim\}$,~~ $(\widetilde 7, 3)$\\[-3mm]

$\usebox{\ddur}~ \usebox{\duur}~ \usebox{\uuur}$ $\|$ $\usebox{\dddb}~ \usebox{\ddub}~ \usebox{\duub}$ $\|$
$\usebox{\ddurb}~ \usebox{\uudrb}~ \usebox{\uuurb}$ $\|$ $\usebox{\dddbr}~ \usebox{\duubr}$\\[-2mm]

$\{\mathbf{r}_3^\star, \mathbf{r}_4^\star, \mathbf{r}_5^{}, \mathbf{r}_5^\star,
\mathbf{r}_6^{}, \mathbf{r}_7^{},\}$,~~ $(8, 3)$\\[-3mm]

$\usebox{\uudr}~ \usebox{\duur}~ \usebox{\uuur}$ $\|$ $\usebox{\ddub}~ \usebox{\uudb}~ \usebox{\duub}$ $\|$
$\usebox{\uudrb}~ \usebox{\duurb}~ \usebox{\uuurb}$ $\|$ $\usebox{\uudbr}~ \usebox{\duubr}$\\[-2mm]

$\{\mathbf{r}_3^\star, \mathbf{r}_4^{\thicksim\star}, \mathbf{r}_5^{}, \mathbf{r}_5^{\star-},
\mathbf{r}_6^{}, \mathbf{r}_7^\thicksim\}$,~~ $(\widetilde 8, 3)$\\[-3mm]

$\usebox{\uudr}~ \usebox{\duur}~ \usebox{\uuur}$ $\|$ $\usebox{\ddub}~ \usebox{\dudb}~ \usebox{\duub}$ $\|$
$\usebox{\ddurb}~ \usebox{\uudrb}~ \usebox{\uuurb}$ $\|$ $\usebox{\dddbr}~ \usebox{\duubr}$\\[-2mm]

$\{\mathbf{r}_2^\star, \mathbf{r}_3^\star, \mathbf{r}_4^\star, \mathbf{r}_4^{\thicksim\star},
\mathbf{r}_5^{}, \mathbf{r}_5^\star, \mathbf{r}_5^{\star-}, \mathbf{r}_6^{}\}$,~~ $(11, 3)$\\[-3mm]

$\usebox{\uudr}~ \usebox{\duur}~ \usebox{\uuur}$ $\|$ $\usebox{\ddub}~ \usebox{\duub} $ $\|$
$\usebox{\ddurb}~ \usebox{\uudrb}~ \usebox{\duurb}~ \usebox{\uuurb}$ $\|$ $\usebox{\ddubr}~ \usebox{\duubr} $\\[-2mm]

$\{\mathbf{r}_1^{}, \mathbf{r}_2^\star, \mathbf{r}_3^\star, \mathbf{r}_4^\star,
\mathbf{r}_4^{\thicksim\star}, \mathbf{r}_5^\star, \mathbf{r}_5^{\star-}, \mathbf{r}_6^{}\}$,~~ $(13^\star, 3)$\\[-3mm]

$\usebox{\dudr}~ \usebox{\uudr}~ \usebox{\uuur}$ $\|$ $\usebox{\ddub}~ \usebox{\duub} $ $\|$
$\usebox{\dddrb}~ \usebox{\dudrb}~ \usebox{\uudrb}~ \usebox{\uuurb}$ $\|$ $\usebox{\ddubr}~ \usebox{\duubr}$\\[-2mm]

$\{\mathbf{r}_2^{}, \mathbf{r}_2^\star, \mathbf{r}_3^\star, \mathbf{r}_4^\star,  \mathbf{r}_4^{\thicksim\star},
\mathbf{r}_5^{}, \mathbf{r}_5^\star, \mathbf{r}_5^{\star-}\}$,~~ $(14, 3)$\\[-3mm]

$\usebox{\ddur}~ \usebox{\duur}~ \usebox{\uuur}$ $\|$ $\usebox{\ddub}~ \usebox{\duub}$ $\|$
$\usebox{\ddurb}~ \usebox{\uudrb}~ \usebox{\duurb}~ \usebox{\uuurb}$ $\|$ $\usebox{\ddubr}~ \usebox{\duubr}$\\[-2mm]

$\{\mathbf{r}_2^\star, \mathbf{r}_4^\star, \mathbf{r}_5^{}, \mathbf{r}_5^\star, \mathbf{r}_6^{}, \mathbf{r}_7^{}\}$,~~ $(17, 3)$\\[-3mm]

$\usebox{\uudr}~ \usebox{\duur}~ \usebox{\uuur}$ $\|$ $\usebox{\ddub}~ \usebox{\duub}~ \usebox{\uuub}$ $\|$
$\usebox{\uudrb}~ \usebox{\duurb}~ \usebox{\uuurb}$ $\|$ $\usebox{\uudbr}~ \usebox{\duubr}$\\[-2mm]

$\{\mathbf{r}_2^\star,  \mathbf{r}_4^{\thicksim\star}, \mathbf{r}_5^{}, \mathbf{r}_5^{\star-}, \mathbf{r}_6^{}, \mathbf{r}_7^\thicksim\}$,~~ $(\widetilde {17}, 3)$\\[-3mm]

$\usebox{\uudr}~ \usebox{\duur}~ \usebox{\uuur}$ $\|$ $\usebox{\dddb}~ \usebox{\ddub}~ \usebox{\duub}$ $\|$
$\usebox{\ddurb}~ \usebox{\uudrb}~ \usebox{\uuurb}$ $\|$ $\usebox{\dddbr}~ \usebox{\duubr}$\\

$\{\mathbf{r}_2^{}, \mathbf{r}_3^\star, \mathbf{r}_4^\star, \mathbf{r}_5^{}, \mathbf{r}_5^\star, \mathbf{r}_7^{}\}$,~~ $(-, 3)$\\[-3mm]

$\usebox{\ddur}~ \usebox{\duur}~ \usebox{\uuur}$ $\|$ $\usebox{\ddub}~ \usebox{\uudb}~ \usebox{\duub}$ $\|$
$\usebox{\uudrb}~ \usebox{\duurb}~ \usebox{\uuurb}$ $\|$ $\usebox{\uudbr}~ \usebox{\duubr}$\\[-2mm]

$\{\mathbf{r}_2^{}, \mathbf{r}_3^\star, \mathbf{r}_4^{\thicksim\star}, \mathbf{r}_5^{}, \mathbf{r}_5^{\star-}, \mathbf{r}_7^\thicksim\}$,~~ $(-, 3)$\\[-3mm]

$\usebox{\ddur}~ \usebox{\duur}~ \usebox{\uuur}$ $\|$ $\usebox{\ddub}~ \usebox{\dudb}~ \usebox{\duub} $ $\|$
$\usebox{\ddurb}~ \usebox{\uudrb}~ \usebox{\uuurb}$ $\|$ $\usebox{\dddbr}~ \usebox{\duubr}$\\[-2mm]

$\{\mathbf{r}_1^{}, \mathbf{r}_2^\star, \mathbf{r}_4^\star, \mathbf{r}_5^\star, \mathbf{r}_6^{}, \mathbf{r}_7^{}\}$,~~ $(-, 3)$\\[-3mm]

$\usebox{\dudr}~ \usebox{\uudr}~ \usebox{\uuur}$ $\|$ $\usebox{\ddub}~ \usebox{\duub}~ \usebox{\uuub}$ $\|$
$\usebox{\dudrb}~ \usebox{\uudrb}~ \usebox{\uuurb}$ $\|$ $\usebox{\uudbr}~ \usebox{\duubr}$\\[-2mm]

$\{\mathbf{r}_1^{}, \mathbf{r}_2^\star, \mathbf{r}_4^{\thicksim\star}, \mathbf{r}_5^{\star-}, \mathbf{r}_6^{}, \mathbf{r}_7^\thicksim\}$,~~ $(-, 3)$\\[-3mm]

$\usebox{\dudr}~ \usebox{\uudr}~ \usebox{\uuur}$ $\|$ $\usebox{\dddb}~ \usebox{\ddub}~ \usebox{\duub}$ $\|$
$\usebox{\dddrb}~ \usebox{\uudrb}~ \usebox{\uuurb}$ $\|$ $\usebox{\dddbr}~ \usebox{\duubr}$\\[-2mm]

$\{\mathbf{r}_1^{}, \mathbf{r}_2^{}, \mathbf{r}_3^\star, \mathbf{r}_4^\star, \mathbf{r}_5^\star, \mathbf{r}_7^{}\}$,~~ $(-, 3)$\\[-3mm]

$\usebox{\dddr}~ \usebox{\uuur}$ $\|$ $\usebox{\ddub}~ \usebox{\uudb}~ \usebox{\duub}$ $\|$
$\usebox{\dudrb}~ \usebox{\uudrb}~ \usebox{\uuurb}$ $\|$ $\usebox{\uudbr}~ \usebox{\duubr}$\\[-2mm]

$\{\mathbf{r}_1^{}, \mathbf{r}_2^{}, \mathbf{r}_3^\star, \mathbf{r}_4^{\thicksim\star}, \mathbf{r}_5^{\star-}, \mathbf{r}_7^\thicksim\}$,~~ $(-, 3)$\\[-3mm]

$\usebox{\dddr}~ \usebox{\uuur}$ $\|$ $\usebox{\ddub}~ \usebox{\dudb}~ \usebox{\duub}$ $\|$
$\usebox{\dddrb}~ \usebox{\uudrb}~ \usebox{\uuurb}$ $\|$ $\usebox{\dddbr}~ \usebox{\duubr}$\\[-2mm]

$\{\mathbf{r}_1^{}, \mathbf{r}_3^\star, \mathbf{r}_4^\star, \mathbf{r}_5^\star, \mathbf{r}_6^{}, \mathbf{r}_7^{}\}$,~~ $(-, 3)$\\[-3mm]

$\usebox{\dudr}~ \usebox{\uudr}~ \usebox{\uuur}$ $\|$ $\usebox{\ddub}~ \usebox{\uudb}~ \usebox{\duub}$ $\|$
$\usebox{\dudrb}~ \usebox{\uudrb}~ \usebox{\uuurb}$ $\|$ $\usebox{\uudbr}~ \usebox{\duubr}$\\[-2mm]

$\{\mathbf{r}_1^{}, \mathbf{r}_3^\star, \mathbf{r}_4^{\thicksim\star}, \mathbf{r}_5^{\star-}, \mathbf{r}_6^{}, \mathbf{r}_7^\thicksim\}$,~~ $(-, 3)$\\[-3mm]

$\usebox{\dudr}~ \usebox{\uudr}~ \usebox{\uuur}$ $\|$ $\usebox{\ddub}~ \usebox{\dudb}~ \usebox{\duub}$ $\|$
$\usebox{\dddrb}~ \usebox{\uudrb}~ \usebox{\uuurb}$ $\|$ $\usebox{\dddbr}~ \usebox{\duubr}$\\

\clearpage
\newpage

Phase 4\\

$\{\mathbf{r}_1^{}, \mathbf{r}_1^\star, \mathbf{r}_2^{}, \mathbf{r}_3^\star, \mathbf{r}_5^{},
\mathbf{r}_6^{}, \mathbf{r}_6^\star,  \mathbf{r}_6^{\star-}, \mathbf{r}_7^{}, \mathbf{r}_7^{\thicksim}\}$\\

$\{\mathbf{r}_1^{}, \mathbf{r}_1^\star, \mathbf{r}_2^{}, \mathbf{r}_5^{}, \mathbf{r}_6^{}, \mathbf{r}_6^\star,
\mathbf{r}_7^{}, \mathbf{r}_7^{\thicksim}\}$,~~ $(1, 4)$\\[-3mm] %(1-st order?)

$\usebox{\uuur}$ $\|$ $\usebox{\dudb}~ \usebox{\uudb}~ \usebox{\uuub}$ $\|$
$\usebox{\uudb}~ \usebox{\uuurb}$ $\|$ $\usebox{\dudbr}~ \usebox{\uuubr}$\\[-2mm]

$\{\mathbf{r}_1^{}, \mathbf{r}_1^\star, \mathbf{r}_2^{}, \mathbf{r}_5^{}, \mathbf{r}_6^{},
\mathbf{r}_6^{\star-}, \mathbf{r}_7^{}, \mathbf{r}_7^{\thicksim}\}$,~~ $(\widetilde 1, 4)$\\[-3mm]%(1-st order?)

$\usebox{\uuur}$ $\|$ $\usebox{\dddb}~ \usebox{\dudb}~ \usebox{\uudb}$ $\|$
$\usebox{\uudrb}~ \usebox{\uuurb}$ $\|$ $\usebox{\dudbr}~ \usebox{\uuubr}$\\[-2mm]

$\{\mathbf{r}_1^{}, \mathbf{r}_2^{}, \mathbf{r}_3^\star, \mathbf{r}_5^{}, \mathbf{r}_6^{}, \mathbf{r}_6^\star,
\mathbf{r}_7^{}, \mathbf{r}_7^{\thicksim}\}$,~~ $(2, 4)$\\[-3mm]

$\usebox{\uuur}$ $\|$ $\usebox{\dudb}~ \usebox{\uudb}~ \usebox{\duub}$ $\|$
$\usebox{\dudrb}~ \usebox{\uuurb}$ $\|$ $\usebox{\dudbr}~ \usebox{\duubr}~ \usebox{\uuubr}$\\[-2mm]

$\{\mathbf{r}_1^{}, \mathbf{r}_2^{}, \mathbf{r}_3^\star, \mathbf{r}_5^{}, \mathbf{r}_6^{},
\mathbf{r}_6^{\star-}, \mathbf{r}_7^{}, \mathbf{r}_7^{\thicksim}\}$,~~ $(\widetilde 2, 4)$\\[-3mm]

$\usebox{\uuur}$ $\|$ $\usebox{\ddub}~ \usebox{\dudb}~ \usebox{\uudb}$ $\|$
$\usebox{\uudrb}~ \usebox{\uuurb}$ $\|$ $\usebox{\dudbr}~ \usebox{\duubr}~ \usebox{\uuubr}$\\[-2mm]

$\{\mathbf{r}_1^\star, \mathbf{r}_3^\star, \mathbf{r}_5^{}, \mathbf{r}_6^{}, \mathbf{r}_6^\star,
\mathbf{r}_6^{\star-}, \mathbf{r}_7^{}\}$,~~ $(6, 4)$\\[-3mm]

$\usebox{\uudr}~ \usebox{\duur}~ \usebox{\uuur}$ $\|$ $\usebox{\dudb}~ \usebox{\uudb}$ $\|$
$\usebox{\uudrb}~ \usebox{\duurb}~ \usebox{\uuurb}$ $\|$ $\usebox{\dudbr}~ \usebox{\uudbr}~ \usebox{\uuubr}$\\[-2mm]

$\{\mathbf{r}_1^\star, \mathbf{r}_3^\star, \mathbf{r}_5^{}, \mathbf{r}_6^{}, \mathbf{r}_6^\star,
\mathbf{r}_6^{\star-}, \mathbf{r}_7^{\thicksim}\}$,~~ $(\widetilde 6, 4)$\\[-3mm]

$\usebox{\uudr}~ \usebox{\duur}~ \usebox{\uuur}$ $\|$ $\usebox{\dudb}~ \usebox{\uudb}$ $\|$
$\usebox{\ddurb}~ \usebox{\uudrb}~ \usebox{\uuurb}$ $\|$ $\usebox{\dddbr}~ \usebox{\dudbr}~ \usebox{\uuubr}$\\[-2mm]

$\{\mathbf{r}_1^{}, \mathbf{r}_1^\star, \mathbf{r}_3^\star, \mathbf{r}_6^{}, \mathbf{r}_6^\star, \mathbf{r}_6^{\star-},
\mathbf{r}_7^{}\}$,~~ $(12, 4)$\\[-3mm]%(1-st order?)

$\usebox{\dudr}~ \usebox{\uudr}~ \usebox{\uuur}$ $\|$ $\usebox{\dudb}~ \usebox{\uudb}$ $\|$
$\usebox{\dudrb}~ \usebox{\uudrb}~ \usebox{\uuurb}$ $\|$ $\usebox{\dudbr}~ \usebox{\uudbr}~ \usebox{\uuubr}$\\[-2mm]

$\{\mathbf{r}_1^{}, \mathbf{r}_1^\star, \mathbf{r}_3^\star, \mathbf{r}_6^{}, \mathbf{r}_6^\star,
\mathbf{r}_6^{\star-}, \mathbf{r}_7^{\thicksim}\}$,~~ $(\widetilde {12}, 4)$\\[-3mm]

$\usebox{\dudr}~ \usebox{\uudr}~ \usebox{\uuur}$ $\|$ $\usebox{\dudb}~ \usebox{\uudb}$ $\|$
$\usebox{\dddrb}~ \usebox{\uudrb}~ \usebox{\uuurb}$ $\|$ $\usebox{\dddbr}~ \usebox{\dudbr}~ \usebox{\uuubr}$\\[-2mm]

$\{\mathbf{r}_1^\star, \mathbf{r}_2^{}, \mathbf{r}_3^\star, \mathbf{r}_5^{}, \mathbf{r}_6^\star, \mathbf{r}_6^{\star-},
\mathbf{r}_7^{}\}$,~~ $(18, 4)$\\[-3mm]

$\usebox{\ddur}~ \usebox{\duur}~ \usebox{\uuur}$ $\|$ $\usebox{\dudb}~ \usebox{\uudb}$ $\|$
$\usebox{\uudrb}~ \usebox{\duurb}~ \usebox{\uuurb}$ $\|$ $\usebox{\dudbr}~ \usebox{\uudbr}~ \usebox{\uuubr}$\\[-2mm]

$\{\mathbf{r}_1^\star, \mathbf{r}_2^{}, \mathbf{r}_3^\star, \mathbf{r}_5^{}, \mathbf{r}_6^\star, \mathbf{r}_6^{\star-},
\mathbf{r}_7^{\thicksim}\}$,~~ $(\widetilde{18}, 4)$\\[-3mm]

$\usebox{\ddur}~ \usebox{\duur}~ \usebox{\uuur}$ $\|$ $\usebox{\dudb}~ \usebox{\uudb}$ $\|$
$\usebox{\ddurb}~ \usebox{\uudrb}~ \usebox{\uuurb}$ $\|$ $\usebox{\dddbr}~ \usebox{\dudbr}~ \usebox{\uuubr}$\\[-2mm]

$\{\mathbf{r}_1^{}, \mathbf{r}_1^\star, \mathbf{r}_2^{}, \mathbf{r}_3^\star, \mathbf{r}_6^\star, \mathbf{r}_6^{\star-}, \mathbf{r}_7^{\thicksim}\}$,~~ $(19, 4)$\\[-3mm]

$\usebox{\dddr}~ \usebox{\uuur}$ $\|$ $\usebox{\dudb}~ \usebox{\uudb}$ $\|$
$\usebox{\dddrb}~ \usebox{\uudrb}~ \usebox{\uuurb}$ $\|$ $\usebox{\dddbr}~ \usebox{\dudbr}~ \usebox{\uuubr}$\\[-2mm]

$\{\mathbf{r}_1^{}, \mathbf{r}_1^\star, \mathbf{r}_2^{}, \mathbf{r}_3^\star, \mathbf{r}_6^\star, \mathbf{r}_6^{\star-}, \mathbf{r}_7^{}\}$,~~ $(\widetilde {19}, 4)$\\[-3mm]

$\usebox{\dddr}~ \usebox{\uuur}$ $\|$ $\usebox{\dudb}~ \usebox{\uudb}$ $\|$
$\usebox{\dudrb}~ \usebox{\uudrb}~ \usebox{\uuurb}$ $\|$ $\usebox{\dudbr}~ \usebox{\uudbr}~ \usebox{\uuubr}$\\[0.25cm]

Phase 5\\

$\{\mathbf{r}_3^\star, \mathbf{r}_4^{\thicksim}, \mathbf{r}_4^{\thicksim\star}, \mathbf{r}_5^{}, \mathbf{r}_6^{}, \mathbf{r}_6^\star, \mathbf{r}_7^{\thicksim}\}$\\

$\{\mathbf{r}_4^{\thicksim}, \mathbf{r}_4^{\thicksim\star}, \mathbf{r}_5^{}, \mathbf{r}_6^{}, \mathbf{r}_6^\star, \mathbf{r}_7^{\thicksim}\}$,~~ $(1, 5)$\\[-3mm]

$\usebox{\uudr}~ \usebox{\duur}~ \usebox{\uuur}$ $\|$ $\usebox{\dudb}~ \usebox{\uudb}~ \usebox{\duub}~ \usebox{\uuub}$ $\|$
$\usebox{\ddurb}~ \usebox{\uuurb}$ $\|$ $\usebox{\dddbr}~ \usebox{\duubr}~ \usebox{\uuubr}$\\[-2mm]

$\{\mathbf{r}_3^\star, \mathbf{r}_4^{\thicksim\star}, \mathbf{r}_5^{}, \mathbf{r}_6^{}, \mathbf{r}_6^\star, \mathbf{r}_7^{\thicksim}\}$,~~ $(2, 5)$\\[-3mm]

$\usebox{\uudr}~ \usebox{\duur}~ \usebox{\uuur}$ $\|$ $\usebox{\dudb}~ \usebox{\uudb}~ \usebox{\duub}$ $\|$
$\usebox{\ddurb}~ \usebox{\uudrb}~ \usebox{\uuurb}$ $\|$ $\usebox{\dddbr}~ \usebox{\duubr}~ \usebox{\uuubr}$\\[-2mm]

$\{\mathbf{r}_3^\star, \mathbf{r}_4^{\thicksim}, \mathbf{r}_5^{}, \mathbf{r}_6^{}, \mathbf{r}_6^\star, \mathbf{r}_7^{\thicksim}\}$,~~ $(\widetilde 6, 5)$\\[-3mm]

$\usebox{\uudr}~ \usebox{\duur}~ \usebox{\uuur}$ $\|$ $\usebox{\dudb}~ \usebox{\uudb}~ \usebox{\duub}$ $\|$
$\usebox{\ddurb}~ \usebox{\uuurb}$ $\|$ $\usebox{\dddbr}~ \usebox{\dudbr}~ \usebox{\duubr}~ \usebox{\uuubr}$\\[-2mm]

$\{\mathbf{r}_3^\star, \mathbf{r}_4^{\thicksim}, \mathbf{r}_4^{\thicksim\star}, \mathbf{r}_5^{}, \mathbf{r}_6^{},  \mathbf{r}_7^{\thicksim}\}$,~~ $(\widetilde 8, 5)$\\[-3mm]

$\usebox{\uudr}~ \usebox{\duur}~ \usebox{\uuur}$ $\|$ $\usebox{\ddub}~ \usebox{\dudb}~ \usebox{\uudb}~ \usebox{\duub}$ $\|$
$\usebox{\ddurb}~ \usebox{\uuurb}$ $\|$ $\usebox{\dddrb}~ \usebox{\duubr}~ \usebox{\uuubr}$\\[-2mm]

$\{\mathbf{r}_3^\star, \mathbf{r}_4^{\thicksim}, \mathbf{r}_4^{\thicksim\star}, \mathbf{r}_5^{}, \mathbf{r}_6^{}, \mathbf{r}_6^\star\}$,~~ $(9, 5)$\\[-3mm]

$\usebox{\uudr}~ \usebox{\duur}~ \usebox{\uuur}$ $\|$ $\usebox{\dudb}~ \usebox{\uudb}~ \usebox{\duub}$ $\|$
$\usebox{\ddurb}~ \usebox{\duurb}~ \usebox{\uuurb}$ $\|$ $\usebox{\dddbr}~ \usebox{\ddubr}~ \usebox{\duubr}~ \usebox{\uuubr}$\\[-2mm]

\newpage

$\{\mathbf{r}_3^\star, \mathbf{r}_4^{\thicksim}, \mathbf{r}_4^{\thicksim\star},  \mathbf{r}_6^{}, \mathbf{r}_6^\star, \mathbf{r}_7^{\thicksim}\}$,~~ $(\widetilde {12}, 5)$\\[-3mm]

$\usebox{\dudr}~ \usebox{\uudr}~ \usebox{\duur}~ \usebox{\uuur}$ $\|$ $\usebox{\dudb}~ \usebox{\uudb}~ \usebox{\duub}$ $\|$
$\usebox{\dddrb}~ \usebox{\ddurb}~ \usebox{\uuurb}$ $\|$ $\usebox{\dddbr}~ \usebox{\duubr}~ \usebox{\uuubr}$\\

$\{\mathbf{r}_3^\star, \mathbf{r}_4^{\thicksim}, \mathbf{r}_4^{\thicksim\star}, \mathbf{r}_5^{}, \mathbf{r}_6^\star, \mathbf{r}_7^{\thicksim}\}$,~~ $(-, 5)$\\[-3mm]

$\usebox{\ddur}~ \usebox{\uudr}~ \usebox{\duur}~ \usebox{\uuur}$ $\|$ $\usebox{\dudb}~ \usebox{\uudb}~ \usebox{\duub}$ $\|$
$\usebox{\ddurb}~ \usebox{\uuurb}$ $\|$ $\usebox{\dddbr}~ \usebox{\duubr}~ \usebox{\uuubr}$\\[0.0cm]

Phase 6\\

$\{\mathbf{r}_1^\star, \mathbf{r}_3^\star, \mathbf{r}_4^{}, \mathbf{r}_5^{}, \mathbf{r}_6^{}, \mathbf{r}_6^\star, \mathbf{r}_6^{\star-}, \mathbf{r}_7^{}\}$\\[1mm]

$\{\mathbf{r}_1^\star, \mathbf{r}_4^{}, \mathbf{r}_5^{}, \mathbf{r}_6^{}, \mathbf{r}_6^{\star-}, \mathbf{r}_7^{}\}$,~~ $(\widetilde 1, 6)$\\[-3mm]

$\usebox{\uudr}~ \usebox{\duur}~ \usebox{\uuur}$ $\|$ $\usebox{\dddb}~ \usebox{\dudb}~ \usebox{\uudb}$ $\|$
$\usebox{\uudrb}~ \usebox{\duurb}$ $\|$ $\usebox{\dudbr}~ \usebox{\uudbr}~ \usebox{\uuubr}$\\[-2mm]

$\{\mathbf{r}_3^\star, \mathbf{r}_4^{}, \mathbf{r}_5^{}, \mathbf{r}_6^{}, \mathbf{r}_6^{\star-}, \mathbf{r}_7^{}\}$,~~ $(\widetilde 5, 6)$\\[-3mm]

$\usebox{\uudr}~ \usebox{\duur}~ \usebox{\uuur}$ $\|$ $\usebox{\ddub}~ \usebox{\dudb}~ \usebox{\uudb}$ $\|$
$\usebox{\uudrb}~ \usebox{\duurb}$ $\|$ $\usebox{\dudbr}~ \usebox{\uudbr}~ \usebox{\duubr}~ \usebox{\uuubr}$\\[-2mm]

$\{\mathbf{r}_3^\star, \mathbf{r}_4^{}, \mathbf{r}_5^{}, \mathbf{r}_6^{}, \mathbf{r}_6^\star, \mathbf{r}_7^{}\}$,~~ $(8, 6)$\\[-3mm]

$\usebox{\uudr}~ \usebox{\duur}~ \usebox{\uuur}$ $\|$ $\usebox{\dudb}~ \usebox{\uudb}~ \usebox{\duub}$ $\|$
$\usebox{\uudrb}~ \usebox{\duurb}$ $\|$ $\usebox{\dudbr}~ \usebox{\uudbr}~ \usebox{\duubr}~ \usebox{\uuubr}$\\[-2mm]

$\{\mathbf{r}_1^\star, \mathbf{r}_3^\star, \mathbf{r}_4^{}, \mathbf{r}_6^{}, \mathbf{r}_6^\star, \mathbf{r}_6^{\star-}, \mathbf{r}_7^{}\}$,~~ $(12, 6)$\\[-3mm]

$\usebox{\dudr}~ \usebox{\uudr}~ \usebox{\duur}~ \usebox{\uuur}$ $\|$ $\usebox{\dudb}~ \usebox{\uudb}$ $\|$
$\usebox{\dudrb}~ \usebox{\uudrb}~ \usebox{\duurb}$ $\|$ $\usebox{\dudbr}~ \usebox{\uudbr}~ \usebox{\uuubr}$\\[-2mm]

$\{\mathbf{r}_1^\star, \mathbf{r}_3^\star, \mathbf{r}_5^{}, \mathbf{r}_6^{}, \mathbf{r}_6^\star, \mathbf{r}_6^{\star-}, \mathbf{r}_7^{}\}$,~~ $(4, 6)$\\[-3mm]

$\usebox{\uudr}~ \usebox{\duur}~ \usebox{\uuur}$ $\|$ $\usebox{\dudb}~ \usebox{\uudb}$ $\|$
$\usebox{\uudrb}~ \usebox{\duurb}~ \usebox{\uuurb}$ $\|$ $\usebox{\dudbr}~ \usebox{\uudbr}~ \usebox{\uuubr}$\\

$\{\mathbf{r}_1^\star, \mathbf{r}_3^\star, \mathbf{r}_4^{}, \mathbf{r}_5^{}, \mathbf{r}_6^{}, \mathbf{r}_6^\star, \mathbf{r}_6^{\star-}\}$,~~ $(-, 6)$\\[-3mm]

$\usebox{\uudr}~ \usebox{\duur}~ \usebox{\uuur}$ $\|$ $\usebox{\dudb}~ \usebox{\uudb}$ $\|$
$\usebox{\ddurb}~ \usebox{\uudrb}~ \usebox{\duurb}$ $\|$ $\usebox{\dddbr}~ \usebox{\dudbr}~ \usebox{\uudbr}~ \usebox{\uuubr}$\\[-2mm]

$\{\mathbf{r}_1^\star, \mathbf{r}_3^\star, \mathbf{r}_4^{}, \mathbf{r}_5^{}, \mathbf{r}_6^\star, \mathbf{r}_6^{\star-}, \mathbf{r}_7^{}\}$,~~ $(-, 6)$\\[-3mm]

$\usebox{\ddur}~ \usebox{\uudr}~ \usebox{\duur}~ \usebox{\uuur}$ $\|$ $\usebox{\dudb}~ \usebox{\uudb}$ $\|$
$\usebox{\uudrb}~ \usebox{\duurb}$ $\|$ $\usebox{\dudbr}~ \usebox{\uudbr}~ \usebox{\uuubr}$\\[-2mm]

$\{\mathbf{r}_1^\star, \mathbf{r}_4^{}, \mathbf{r}_5^{}, \mathbf{r}_6^{}, \mathbf{r}_6^\star, \mathbf{r}_7^{}\}$,~~ $(-, 6)$\\[-3mm]

$\usebox{\uudr}~ \usebox{\duur}~ \usebox{\uuur}$ $\|$ $\usebox{\dudb}~ \usebox{\uudb}~ \usebox{\uuub}$ $\|$
$\usebox{\uudrb}~ \usebox{\duurb}$ $\|$ $\usebox{\dudbr}~ \usebox{\uudbr}~ \usebox{\uuubr}$\\[0.0cm]

Phase 7\\

$\{\mathbf{r}_2^{}, \mathbf{r}_2^\star, \mathbf{r}_4^{}, \mathbf{r}_4^\star, \mathbf{r}_5^{},
\mathbf{r}_5^\star, \mathbf{r}_7^{}\}$\\

$\{\mathbf{r}_2^{}, \mathbf{r}_2^\star, \mathbf{r}_4^{}, \mathbf{r}_4^\star, \mathbf{r}_5^{}, \mathbf{r}_7^{}\}$,~~ $(\widetilde 1, 7)$\\[-3mm]

$\usebox{\ddur}~ \usebox{\duur}~ \usebox{\uuur}$ $\|$ $\usebox{\dddb}~ \usebox{\ddub}~ \usebox{\duub}~ \usebox{\uuub}$ $\|$
$\usebox{\uudrb}~ \usebox{\duurb}$ $\|$ $\usebox{\dudbr}~ \usebox{\uudbr}~ \usebox{\duubr}$\\[-2mm]

$\{\mathbf{r}_2^{}, \mathbf{r}_2^\star, \mathbf{r}_4^{}, \mathbf{r}_4^\star, \mathbf{r}_5^\star, \mathbf{r}_7^{}\}$,~~ $(\overline{\widetilde 1}, 7)$\\[-3mm]

$\usebox{\dddr}~ \usebox{\ddur}~ \usebox{\duur}~ \usebox{\uuur}$ $\|$ $\usebox{\ddub}~ \usebox{\duub}~ \usebox{\uuub}$ $\|$
$\usebox{\dudrb}~ \usebox{\uudrb}~ \usebox{\duurb}$ $\|$ $\usebox{\uudbr}~ \usebox{\duubr}$\\[-2mm]

$\{\mathbf{r}_2^{}, \mathbf{r}_2^\star, \mathbf{r}_4^\star, \mathbf{r}_5^{}, \mathbf{r}_5^\star, \mathbf{r}_7^{}\}$,~~ $(3, 7)$\\[-3mm]

$\usebox{\ddur}~ \usebox{\duur}~ \usebox{\uuur}$ $\|$ $\usebox{\ddub}~ \usebox{\duub}~ \usebox{\uuub}$ $\|$
$\usebox{\uudrb}~ \usebox{\duurb}~ \usebox{\uuurb}$ $\|$ $\usebox{\uudbr}~ \usebox{\duubr}$\\[-2mm]

$\{\mathbf{r}_2^{}, \mathbf{r}_2^\star, \mathbf{r}_4^{}, \mathbf{r}_5^{}, \mathbf{r}_5^\star, \mathbf{r}_7^{}\}$,~~ $(3^\star, 7)$\\[-3mm]

$\usebox{\ddur}~ \usebox{\duur}~ \usebox{\uuur}$ $\|$ $\usebox{\ddub}~ \usebox{\duub}~ \usebox{\uuub}$ $\|$
$\usebox{\uudrb}~ \usebox{\duurb}$ $\|$ $\usebox{\uudbr}~ \usebox{\duubr}~ \usebox{\uuubr}$\\[-2mm]

$\{\mathbf{r}_2^{}, \mathbf{r}_2^\star, \mathbf{r}_4^{}, \mathbf{r}_4^\star, \mathbf{r}_5^{}, \mathbf{r}_5^\star\}$,~~ $(14, 7)$\\[-3mm]

$\usebox{\ddur}~ \usebox{\duur}~ \usebox{\uuur}$ $\|$ $\usebox{\ddub}~ \usebox{\duub}~ \usebox{\uuub}$ $\|$
$\usebox{\ddurb}~ \usebox{\uudrb}~ \usebox{\duurb}$ $\|$ $\usebox{\ddubr}~ \usebox{\uudbr}~ \usebox{\duubr}$\\

$\{\mathbf{r}_2^{}, \mathbf{r}_4^{}, \mathbf{r}_4^\star, \mathbf{r}_5^{}, \mathbf{r}_5^\star, \mathbf{r}_7^{}\}$,~~ $(-, 7)$\\[-3mm]

$\usebox{\ddur}~ \usebox{\duur}~ \usebox{\uuur}$ $\|$ $\usebox{\ddub}~ \usebox{\uudb}~ \usebox{\duub}~ \usebox{\uuub}$ $\|$
$\usebox{\uudrb}~ \usebox{\duurb}$ $\|$ $\usebox{\uudbr}~ \usebox{\duubr}$\\[-2mm]

$\{\mathbf{r}_2^\star, \mathbf{r}_4^{}, \mathbf{r}_4^\star, \mathbf{r}_5^{}, \mathbf{r}_5^\star, \mathbf{r}_7^{}\}$,~~ $(-, 7)$\\[-3mm]

$\usebox{\ddur}~ \usebox{\uudr}~ \usebox{\duur}~ \usebox{\uuur}$ $\|$ $\usebox{\ddub}~ \usebox{\duub}~ \usebox{\uuub}$ $\|$
$\usebox{\uudrb}~ \usebox{\duurb}$ $\|$ $\usebox{\uudbr}~ \usebox{\duubr}$\\

\clearpage
\newpage

Phase 8\\

$\{\mathbf{r}_3^{}, \mathbf{r}_3^\star, \mathbf{r}_4^{}, \mathbf{r}_4^\star,
\mathbf{r}_5^{}, \mathbf{r}_5^\star, \mathbf{r}_6^{}, \mathbf{r}_6^\star, \mathbf{r}_7^{}\}$\\

$\{\mathbf{r}_3^\star, \mathbf{r}_5^{}, \mathbf{r}_5^\star, \mathbf{r}_6^{}, \mathbf{r}_6^\star, \mathbf{r}_7^{}\}$,~~ $(2, 8)$\\[-3mm]

$\usebox{\uudr}~ \usebox{\duur}~ \usebox{\uuur}$ $\|$ $\usebox{\uudb}~ \usebox{\duub}$ $\|$
$\usebox{\uudrb}~ \usebox{\duurb}~ \usebox{\uuurb}$ $\|$ $\usebox{\uudbr}~ \usebox{\duubr}~ \usebox{\uuubr}$\\[-2mm]

$\{\mathbf{r}_3^{}, \mathbf{r}_5^{}, \mathbf{r}_5^\star, \mathbf{r}_6^{}, \mathbf{r}_6^\star, \mathbf{r}_7^{}\}$,~~ $(2^\star, 8)$\\[-3mm]

$\usebox{\uudr}~ \usebox{\duur}$ $\|$ $\usebox{\uudb}~ \usebox{\duub}~ \usebox{\uuub}$ $\|$
$\usebox{\uudrb}~ \usebox{\duurb}~ \usebox{\uuurb}$ $\|$ $\usebox{\uudbr}~ \usebox{\duubr}~ \usebox{\uuubr}$\\[-2mm]

$\{\mathbf{r}_3^\star, \mathbf{r}_4^\star, \mathbf{r}_5^{}, \mathbf{r}_5^\star, \mathbf{r}_6^{}, \mathbf{r}_7^{}\}$,~~ $(3, 8)$\\[-3mm]

$\usebox{\uudr}~ \usebox{\duur}~ \usebox{\uuur}$ $\|$ $\usebox{\ddub}~ \usebox{\uudb}~ \usebox{\duub}$ $\|$
$\usebox{\uudrb}~ \usebox{\duurb}~ \usebox{\uuurb}$ $\|$ $\usebox{\uudbr}~ \usebox{\duubr}$\\[-2mm]

$\{\mathbf{r}_3^{}, \mathbf{r}_4^{}, \mathbf{r}_5^{}, \mathbf{r}_5^\star, \mathbf{r}_6^\star, \mathbf{r}_7^{}\}$,~~ $(3^\star, 8)$\\[-3mm]

$\usebox{\ddur}~ \usebox{\uudr}~ \usebox{\duur}$ $\|$ $\usebox{\uudb}~ \usebox{\duub}~ \usebox{\uuub}$ $\|$
$\usebox{\uudrb}~ \usebox{\duurb}$ $\|$ $\usebox{\uudbr}~ \usebox{\duubr}~ \usebox{\uuubr}$\\[-2mm]

$\{\mathbf{r}_3^\star, \mathbf{r}_4^{}, \mathbf{r}_4^\star, \mathbf{r}_5^{}, \mathbf{r}_6^{}, \mathbf{r}_7^{}\}$,~~ $(\widetilde 5, 8)$\\[-3mm]

$\usebox{\uudr}~ \usebox{\duur}~ \usebox{\uuur}$ $\|$ $\usebox{\ddub}~ \usebox{\dudb}~ \usebox{\uudb}~ \usebox{\duub}$ $\|$
$\usebox{\uudrb}~ \usebox{\duurb}$ $\|$ $\usebox{\dudbr}~ \usebox{\uudbr}~ \usebox{\duubr}$\\[-2mm]

$\{\mathbf{r}_3^{}, \mathbf{r}_4^{}, \mathbf{r}_4^\star, \mathbf{r}_5^\star, \mathbf{r}_6^\star, \mathbf{r}_7^{}\}$,~~ $(\widetilde 5^\star, 8)$\\[-3mm]

$\usebox{\ddur}~ \usebox{\dudr}~ \usebox{\uudr}~ \usebox{\duur}$ $\|$ $\usebox{\uudb}~ \usebox{\duub}~ \usebox{\uuub}$ $\|$
$\usebox{\dudrb}~ \usebox{\uudrb}~ \usebox{\duurb}$ $\|$ $\usebox{\uudbr}~ \usebox{\duubr}$\\[-2mm]

$\{\mathbf{r}_3^\star, \mathbf{r}_4^{}, \mathbf{r}_5^{}, \mathbf{r}_6^{}, \mathbf{r}_6^\star, \mathbf{r}_7^{}\}$,~~ $(6, 8)$\\[-3mm]

$\usebox{\uudr}~ \usebox{\duur}~ \usebox{\uuur}$ $\|$ $\usebox{\dudb}~ \usebox{\uudb}~ \usebox{\duub}$ $\|$
$\usebox{\uudrb}~ \usebox{\duurb}$ $\|$ $\usebox{\dudbr}~ \usebox{\uudbr}~ \usebox{\duubr}~ \usebox{\uuubr}$\\[-2mm]

$\{\mathbf{r}_3^{}, \mathbf{r}_4^\star, \mathbf{r}_5^\star, \mathbf{r}_6^{}, \mathbf{r}_6^\star, \mathbf{r}_7^{}\}$,~~ $(6^\star, 8)$\\[-3mm]

$\usebox{\dudr}~ \usebox{\uudr}~ \usebox{\duur}$ $\|$ $\usebox{\uudb}~ \usebox{\duub}~ \usebox{\uuub}$ $\|$
$\usebox{\dudrb}~ \usebox{\uudrb}~ \usebox{\duurb}~ \usebox{\uuurb}$ $\|$ $\usebox{\uudbr}~ \usebox{\duubr}$\\[-2mm]

$\{\mathbf{r}_3^{}, \mathbf{r}_3^\star, \mathbf{r}_5^{}, \mathbf{r}_5^\star, \mathbf{r}_6^{}, \mathbf{r}_6^\star,\}$,~~ $(9, 8)$\\[-3mm]

$\usebox{\uudr}~ \usebox{\duur}$ $\|$ $\usebox{\uudb}~ \usebox{\duub}$ $\|$
$\usebox{\ddurb}~ \usebox{\uudrb}~ \usebox{\duurb}~ \usebox{\uuurb}$ $\|$ $\usebox{\ddubr}~ \usebox{\uudbr}~ \usebox{\duubr}~ \usebox{\uuubr}$\\[-2mm]

$\{\mathbf{r}_3^{}, \mathbf{r}_3^\star, \mathbf{r}_4^{}, \mathbf{r}_4^\star, \mathbf{r}_5^{}, \mathbf{r}_6^{}\}$,~~ $(\widetilde 9, 8)$\\[-3mm]

$\usebox{\uudr}~ \usebox{\duur}$ $\|$ $\usebox{\ddub}~ \usebox{\dudb}~ \usebox{\uudb}~ \usebox{\duub}$ $\|$
$\usebox{\ddurb}~ \usebox{\uudrb}~ \usebox{\duurb}$ $\|$ $\usebox{\ddubr}~ \usebox{\dudbr}~ \usebox{\uudbr}~ \usebox{\duubr}$\\[-2mm]

$\{\mathbf{r}_3^{}, \mathbf{r}_3^\star, \mathbf{r}_4^{}, \mathbf{r}_4^\star, \mathbf{r}_5^\star, \mathbf{r}_6^\star,\}$,~~ $(\overline{\widetilde 9}, 8)$\\[-3mm]

$\usebox{\ddur}~ \usebox{\dudr}~ \usebox{\uudr}~ \usebox{\duur}$ $\|$ $\usebox{\uudb}~ \usebox{\duub}$ $\|$
$\usebox{\ddurb}~ \usebox{\dudrb}~ \usebox{\uudrb}~ \usebox{\duurb}$ $\|$ $\usebox{\ddubr}~ \usebox{\uudbr}~ \usebox{\duubr}$\\[-2mm]

$\{\mathbf{r}_3^{}, \mathbf{r}_3^\star, \mathbf{r}_4^\star, \mathbf{r}_5^{}, \mathbf{r}_5^\star, \mathbf{r}_6^{}\}$,~~ $(11, 8)$\\[-3mm]

$\usebox{\uudr}~ \usebox{\duur}$ $\|$ $\usebox{\ddub}~ \usebox{\uudb}~ \usebox{\duub}$ $\|$
$\usebox{\ddurb}~ \usebox{\uudrb}~ \usebox{\duurb}~ \usebox{\uuurb}$ $\|$ $\usebox{\ddubr}~ \usebox{\uudbr}~ \usebox{\duubr}$\\[-2mm]

$\{\mathbf{r}_3^{}, \mathbf{r}_3^\star, \mathbf{r}_4^{}, \mathbf{r}_5^{}, \mathbf{r}_5^\star, \mathbf{r}_6^\star,\}$,~~ $(11^\star, 8)$\\[-3mm]

$\usebox{\ddur}~ \usebox{\uudr}~ \usebox{\duur}$ $\|$ $\usebox{\uudb}~ \usebox{\duub}$ $\|$
$\usebox{\ddurb}~ \usebox{\uudrb}~ \usebox{\duurb}$ $\|$ $\usebox{\ddubr}~ \usebox{\uudbr}~ \usebox{\duubr}~ \usebox{\uuubr}$\\[-2mm]

$\{\mathbf{r}_3^{}, \mathbf{r}_3^\star, \mathbf{r}_4^{}, \mathbf{r}_4^\star, \mathbf{r}_6^{}, \mathbf{r}_6^\star, \mathbf{r}_7^{}\}$,~~ $(12, 8)$\\[-3mm]

$\usebox{\dudr}~ \usebox{\uudr}~ \usebox{\duur}$ $\|$ $\usebox{\dudb}~ \usebox{\uudb}~ \usebox{\duub}$ $\|$
$\usebox{\dudrb}~ \usebox{\uudrb}~ \usebox{\duurb}$ $\|$ $\usebox{\dudbr}~ \usebox{\uudbr}~ \usebox{\duubr}$\\[-2mm]

$\{\mathbf{r}_3^{}, \mathbf{r}_3^\star, \mathbf{r}_4^{}, \mathbf{r}_4^\star, \mathbf{r}_5^{}, \mathbf{r}_5^\star,\}$,~~ $(14, 8)$\\[-3mm]

$\usebox{\ddur}~ \usebox{\uudr}~ \usebox{\duur}$ $\|$ $\usebox{\ddub}~ \usebox{\uudb}~ \usebox{\duub}$ $\|$
$\usebox{\ddurb}~ \usebox{\uudrb}~ \usebox{\duurb}$ $\|$ $\usebox{\ddubr}~ \usebox{\uudbr}~ \usebox{\duubr}$\\

$\{\mathbf{r}_3^{}, \mathbf{r}_4^{}, \mathbf{r}_4^\star, \mathbf{r}_5^{}, \mathbf{r}_5^\star, \mathbf{r}_6^{}, \mathbf{r}_7^{}\}$,~~ $(-, 8)$\\[-3mm]

$\usebox{\uudr}~ \usebox{\duur}$ $\|$ $\usebox{\ddub}~ \usebox{\uudb}~ \usebox{\duub}~ \usebox{\uuub}$ $\|$
$\usebox{\uudrb}~ \usebox{\duurb}$ $\|$ $\usebox{\uudbr}~ \usebox{\duubr}$\\[-2mm]

$\{\mathbf{r}_3^\star, \mathbf{r}_4^{}, \mathbf{r}_4^\star, \mathbf{r}_5^{}, \mathbf{r}_5^\star, \mathbf{r}_6^\star, \mathbf{r}_7^{}\}$,~~ $(-, 8)$\\[-3mm]

$\usebox{\ddur}~ \usebox{\uudr}~ \usebox{\duur}~ \usebox{\uuur}$ $\|$ $\usebox{\uudb}~ \usebox{\duub}$ $\|$
$\usebox{\uudrb}~ \usebox{\duurb}$ $\|$ $\usebox{\uudbr}~ \usebox{\duubr}$\\[-2mm]

$\{\mathbf{r}_3^{}, \mathbf{r}_3^\star, \mathbf{r}_4^{}, \mathbf{r}_5^{}, \mathbf{r}_6^{}, \mathbf{r}_6^\star\}$,~~ $(-, 8)$\\[-3mm]

$\usebox{\uudr}~ \usebox{\duur}$ $\|$ $\usebox{\dudb}~ \usebox{\uudb}~ \usebox{\duub}$ $\|$
$\usebox{\ddurb}~ \usebox{\uudrb}~ \usebox{\duurb}$ $\|$ $\usebox{\dddbr}~ \usebox{\ddubr}~ \usebox{\dudbr}~ \usebox{\uudbr}~ \usebox{\duubr}~ \usebox{\uuubr}$\\[-2mm]

$\{\mathbf{r}_3^{}, \mathbf{r}_3^\star, \mathbf{r}_4^\star, \mathbf{r}_5^\star, \mathbf{r}_6^{}, \mathbf{r}_6^\star\}$,~~ $(-, 8)$\\[-3mm]

$\usebox{\dudr}~ \usebox{\uudr}~ \usebox{\duur}$ $\|$ $\usebox{\uudb}~ \usebox{\duub}$ $\|$
$\usebox{\dddrb}~ \usebox{\ddurb}~ \usebox{\dudrb}~ \usebox{\uudrb}~ \usebox{\duurb}~ \usebox{\uuurb}$ $\|$ $\usebox{\ddubr}~ \usebox{\uudbr}~ \usebox{\duubr}$\\[-2mm]

\newpage

$\{\mathbf{r}_3^{}, \mathbf{r}_4^{}, \mathbf{r}_5^{}, \mathbf{r}_6^{}, \mathbf{r}_6^\star, \mathbf{r}_7^{}\}$,~~ $(-, 8)$\\[-3mm]

$\usebox{\uudr}~ \usebox{\duur}$ $\|$ $\usebox{\dudb}~ \usebox{\uudb}~ \usebox{\duub}~ \usebox{\uuub}$ $\|$
$\usebox{\uudrb}~ \usebox{\duurb}$ $\|$ $\usebox{\dudbr}~ \usebox{\uudbr}~ \usebox{\duubr}~ \usebox{\uuubr}$\\[-2mm]

$\{\mathbf{r}_3^\star, \mathbf{r}_4^\star, \mathbf{r}_5^\star, \mathbf{r}_6^{}, \mathbf{r}_6^\star, \mathbf{r}_7^{}\}$,~~ $(-, 8)$\\[-3mm]

$\usebox{\dudr}~ \usebox{\uudr}~ \usebox{\duur}~ \usebox{\uuur}$ $\|$ $\usebox{\uudb}~ \usebox{\duub}$ $\|$
$\usebox{\dudrb}~ \usebox{\uudrb}~ \usebox{\duurb}~ \usebox{\uuurb}$ $\|$ $\usebox{\uudbr}~ \usebox{\duubr}$\\[0.25cm]

Phase 9\\

$\{\mathbf{r}_3^{}, \mathbf{r}_3^\star, \mathbf{r}_4^\thicksim, \mathbf{r}_4^{\thicksim\star},
\mathbf{r}_5^{}, \mathbf{r}_5^\star, \mathbf{r}_6^{}, \mathbf{r}_6^\star\}$\\

$\{\mathbf{r}_4^\thicksim, \mathbf{r}_4^{\thicksim\star}, \mathbf{r}_5^{}, \mathbf{r}_5^\star, \mathbf{r}_6^{}, \mathbf{r}_6^\star\}$,~~ $(1, 9)$\\[-3mm]

$\usebox{\uudr}~ \usebox{\duur}~ \usebox{\uuur}$ $\|$ $\usebox{\uudb}~ \usebox{\duub}~ \usebox{\uuub}$ $\|$
$\usebox{\ddurb}~ \usebox{\duurb}~ \usebox{\uuurb}$ $\|$ $\usebox{\ddubr}~ \usebox{\duubr}~ \usebox{\uuubr}$\\[-2mm]

$\{\mathbf{r}_3^\star, \mathbf{r}_4^{\thicksim\star}, \mathbf{r}_5^{}, \mathbf{r}_5^\star, \mathbf{r}_6^{}, \mathbf{r}_6^\star\}$,~~ $(2, 9)$\\[-3mm]

$\usebox{\uudr}~ \usebox{\duur}~ \usebox{\uuur}$ $\|$ $\usebox{\uudb}~ \usebox{\duub}$ $\|$
$\usebox{\ddurb}~ \usebox{\uudrb}~\usebox{\duurb}~ \usebox{\uuurb}$ $\|$ $\usebox{\ddubr}~ \usebox{\duubr}~ \usebox{\uuubr}$\\[-2mm]

$\{\mathbf{r}_3^{}, \mathbf{r}_4^\thicksim, \mathbf{r}_5^{}, \mathbf{r}_5^\star, \mathbf{r}_6^{}, \mathbf{r}_6^\star\}$,~~ $(2^\star, 9)$\\[-3mm]

$\usebox{\uudr}~ \usebox{\duur}$ $\|$ $\usebox{\uudb}~ \usebox{\duub}~ \usebox{\uuub}$ $\|$
$\usebox{\ddurb}~ \usebox{\duurb}~ \usebox{\uuurb}$ $\|$ $\usebox{\ddubr}~ \usebox{\uudbr}~ \usebox{\duubr}~ \usebox{\uuubr}$\\[-2mm]

$\{\mathbf{r}_3^\star, \mathbf{r}_4^\thicksim, \mathbf{r}_4^{\thicksim\star}, \mathbf{r}_5^{}, \mathbf{r}_6^{}, \mathbf{r}_6^\star\}$,~~ $(5, 9)$\\[-3mm]

$\usebox{\uudr}~ \usebox{\duur}~ \usebox{\uuur}$ $\|$ $\usebox{\dudb}~ \usebox{\uudb}~ \usebox{\duub}$ $\|$
$\usebox{\ddurb}~ \usebox{\duurb}~ \usebox{\uuurb}$ $\|$ $\usebox{\dddbr}~ \usebox{\ddubr}~ \usebox{\duubr}~ \usebox{\uuubr}$\\[-2mm]

$\{\mathbf{r}_3^{}, \mathbf{r}_4^\thicksim, \mathbf{r}_4^{\thicksim\star}, \mathbf{r}_5^\star, \mathbf{r}_6^{}, \mathbf{r}_6^\star\}$,~~ $(5^\star, 9)$\\[-3mm]

$\usebox{\dudr}~ \usebox{\uudr}~ \usebox{\duur}$ $\|$ $\usebox{\uudb}~ \usebox{\duub}~ \usebox{\uuub}$ $\|$
$\usebox{\dddrb}~ \usebox{\ddurb}~ \usebox{\duurb}~ \usebox{\uuurb}$ $\|$ $\usebox{\ddubr}~ \usebox{\duubr}~ \usebox{\uuubr}$\\[-2mm]

$\{\mathbf{r}_3^{}, \mathbf{r}_3^\star, \mathbf{r}_5^{}, \mathbf{r}_5^\star, \mathbf{r}_6^{}, \mathbf{r}_6^\star\}$,~~ $(8, 9)$\\[-3mm]

$\usebox{\uudr}~ \usebox{\duur}$ $\|$ $\usebox{\uudb}~ \usebox{\duub}$ $\|$
$\usebox{\ddurb}~ \usebox{\uudrb}~ \usebox{\duurb}~ \usebox{\uuurb}$ $\|$ $\usebox{\ddubr}~ \usebox{\uudbr}~ \usebox{\duubr}~ \usebox{\uuubr}$\\[-2mm]

$\{\mathbf{r}_3^{}, \mathbf{r}_3^\star, \mathbf{r}_4^\thicksim, \mathbf{r}_4^{\thicksim\star}, \mathbf{r}_5^{}, \mathbf{r}_6^{}\}$,~~ $(\widetilde 8, 9)$\\[-3mm]

$\usebox{\uudr}~ \usebox{\duur}$ $\|$ $\usebox{\ddub}~ \usebox{\dudb}~ \usebox{\uudb}~ \usebox{\duub}$ $\|$
$\usebox{\ddurb}~ \usebox{\duurb}~ \usebox{\uuurb}$ $\|$ $\usebox{\dddbr}~ \usebox{\ddubr}~ \usebox{\duubr}~ \usebox{\uuubr}$\\[-2mm]

$\{\mathbf{r}_3^{}, \mathbf{r}_3^\star, \mathbf{r}_4^\thicksim, \mathbf{r}_4^{\thicksim\star}, \mathbf{r}_5^\star, \mathbf{r}_6^\star\}$,~~ $(\overline{\widetilde 8}, 9)$\\[-3mm]

$\usebox{\ddur}~ \usebox{\dudr}~ \usebox{\uudr}~ \usebox{\duur}$ $\|$ $\usebox{\uudb}~ \usebox{\duub}$ $\|$
$\usebox{\dddrb}~ \usebox{\ddurb}~ \usebox{\duurb}~ \usebox{\uuurb}$ $\|$ $\usebox{\ddubr}~ \usebox{\duubr}~ \usebox{\uuubr}$\\[-2mm]

$\{\mathbf{r}_3^{}, \mathbf{r}_4^\thicksim, \mathbf{r}_4^{\thicksim\star}, \mathbf{r}_5^{}, \mathbf{r}_5^\star,  \mathbf{r}_6^{}\}$,~~ $(10, 9)$\\[-3mm]

$\usebox{\uudr}~ \usebox{\duur}$ $\|$ $\usebox{\ddub}~ \usebox{\uudb}~ \usebox{\duub}~ \usebox{\uuub}$ $\|$
$\usebox{\ddurb}~ \usebox{\uudrb}~ \usebox{\uuurb}$ $\|$ $\usebox{\ddubr}~ \usebox{\uudbr}~ \usebox{\uuubr}$\\[-2mm]

$\{\mathbf{r}_3^\star, \mathbf{r}_4^\thicksim, \mathbf{r}_4^{\thicksim\star}, \mathbf{r}_5^{}, \mathbf{r}_5^\star, \mathbf{r}_6^\star\}$,~~ $(10^\star, 9)$\\[-3mm]

$\usebox{\ddur}~ \usebox{\uudr}~ \usebox{\duur}~ \usebox{\uuur}$ $\|$ $\usebox{\uudb}~ \usebox{\duub}$ $\|$
$\usebox{\ddurb}~ \usebox{\duurb}~ \usebox{\uuurb}$ $\|$ $\usebox{\ddubr}~ \usebox{\duubr}~ \usebox{\uuubr}$\\[-2mm]

$\{\mathbf{r}_3^{}, \mathbf{r}_3^\star, \mathbf{r}_4^{\thicksim\star}, \mathbf{r}_5^{}, \mathbf{r}_5^\star, \mathbf{r}_6^{}\}$,~~ $(11, 9)$\\[-3mm]

$\usebox{\uudr}~ \usebox{\duur}$ $\|$ $\usebox{\ddub}~ \usebox{\uudb}~ \usebox{\duub}$ $\|$
$\usebox{\ddurb}~ \usebox{\uudrb}~ \usebox{\duurb}~ \usebox{\uuurb}$ $\|$ $\usebox{\ddubr}~ \usebox{\duubr}~ \usebox{\uuubr}$\\[-2mm]

$\{\mathbf{r}_3^{}, \mathbf{r}_3^\star, \mathbf{r}_4^\thicksim, \mathbf{r}_5^{}, \mathbf{r}_5^\star, \mathbf{r}_6^\star\}$,~~ $(11^\star, 9)$\\[-3mm]

$\usebox{\ddur}~ \usebox{\uudr}~ \usebox{\duur}$ $\|$ $\usebox{\uudb}~ \usebox{\duub}$ $\|$
$\usebox{\ddurb}~ \usebox{\duurb}~ \usebox{\uuurb}$ $\|$ $\usebox{\ddubr}~ \usebox{\uudbr}~ \usebox{\duubr}~ \usebox{\uuubr}$\\[-2mm]

$\{\mathbf{r}_3^{}, \mathbf{r}_3^\star, \mathbf{r}_4^\thicksim, \mathbf{r}_4^{\thicksim\star}, \mathbf{r}_6^{}, \mathbf{r}_6^\star\}$,~~ $(\widetilde {12}, 9)$\\[-3mm]

$\usebox{\dudr}~ \usebox{\uudr}~ \usebox{\duur}$ $\|$ $\usebox{\dudb}~ \usebox{\uudb}~ \usebox{\duub}$ $\|$
$\usebox{\dddrb}~ \usebox{\ddurb}~ \usebox{\duurb}~ \usebox{\uuurb}$ $\|$ $\usebox{\dddbr}~ \usebox{\ddubr}~ \usebox{\duubr}~ \usebox{\uuubr}$\\[-2mm]

$\{\mathbf{r}_3^{}, \mathbf{r}_3^\star, \mathbf{r}_4^\thicksim, \mathbf{r}_4^{\thicksim\star}, \mathbf{r}_5^{}, \mathbf{r}_5^\star\}$,~~ $(14, 9)$\\[-3mm]

$\usebox{\ddur}~ \usebox{\uudr}~ \usebox{\duur}$ $\|$ $\usebox{\ddub}~ \usebox{\uudb}~ \usebox{\duub}$ $\|$
$\usebox{\ddurb}~ \usebox{\duurb}~ \usebox{\uuurb}$ $\|$ $\usebox{\ddubr}~ \usebox{\duubr}~ \usebox{\uuubr}$\\

$\{\mathbf{r}_3^{}, \mathbf{r}_3^\star, \mathbf{r}_4^\thicksim, \mathbf{r}_5^{}, \mathbf{r}_6^{}, \mathbf{r}_6^\star\}$,~~ $(-, 9)$\\[-3mm]

$\usebox{\uudr}~ \usebox{\duur}$ $\|$ $\usebox{\dudb}~ \usebox{\uudb}~ \usebox{\duub}$ $\|$
$\usebox{\ddurb}~ \usebox{\duurb}~ \usebox{\uuurb}$ $\|$ $\usebox{\tgrbr}$\\[-2mm]

$\{\mathbf{r}_3^{}, \mathbf{r}_3^\star, \mathbf{r}_4^{\thicksim\star}, \mathbf{r}_5^\star, \mathbf{r}_6^{}, \mathbf{r}_6^\star\}$,~~ $(-, 9)$\\[-3mm]

$\usebox{\dudr}~ \usebox{\uudr}~ \usebox{\duur}$ $\|$ $\usebox{\uudb}~ \usebox{\duub}$ $\|$
$\usebox{\tgrrb}$ $\|$ $\usebox{\ddubr}~ \usebox{\duubr}~ \usebox{\uuubr}$\\

\clearpage
\newpage

Phase 10\\

$\{\mathbf{r}_2^\star, \mathbf{r}_3^{}, \mathbf{r}_4^\thicksim,
\mathbf{r}_4^{\thicksim\star}, \mathbf{r}_5^{}, \mathbf{r}_5^\star, \mathbf{r}_6^{}\}$\\

$\{\mathbf{r}_2^\star, \mathbf{r}_4^\thicksim, \mathbf{r}_4^{\thicksim\star}, \mathbf{r}_5^{}, \mathbf{r}_5^\star, \mathbf{r}_6^{}\}$,~~ $(1, 10)$\\[-3mm]

$\usebox{\uudr}~ \usebox{\duur}~ \usebox{\uuur}$ $\|$ $\usebox{\ddub}~ \usebox{\duub}~ \usebox{\uuub}$ $\|$
$\usebox{\ddurb}~ \usebox{\duurb}~ \usebox{\uuurb}$ $\|$ $\usebox{\ddubr}~ \usebox{\duubr}~ \usebox{\uuubr}$\\[-2mm]

$\{\mathbf{r}_2^\star, \mathbf{r}_3^{}, \mathbf{r}_4^\thicksim, \mathbf{r}_5^{}, \mathbf{r}_5^\star, \mathbf{r}_6^{}\}$,~~ $(2^\star, 10)$\\[-3mm]

$\usebox{\uudr}~ \usebox{\duur}$ $\|$ $\usebox{\ddub}~ \usebox{\duub}~ \usebox{\uuub}$ $\|$
$\usebox{\ddurb}~ \usebox{\duurb}~ \usebox{\uuurb}$ $\|$ $\usebox{\ddubr}~ \usebox{\uudbr}~ \usebox{\duubr}~ \usebox{\uuubr}$\\[-2mm]

$\{\mathbf{r}_3^{}, \mathbf{r}_4^\thicksim, \mathbf{r}_4^{\thicksim\star}, \mathbf{r}_5^{}, \mathbf{r}_5^\star, \mathbf{r}_6^{}\}$,~~ $(9, 10)$\\[-3mm]

$\usebox{\uudr}~ \usebox{\duur}$ $\|$ $\usebox{\ddub}~ \usebox{\uudb}~ \usebox{\duub}~ \usebox{\uuub}$ $\|$
$\usebox{\ddurb}~ \usebox{\duurb}~ \usebox{\uuurb}$ $\|$ $\usebox{\ddubr}~ \usebox{\duubr}~ \usebox{\uuubr}$\\[-2mm]

$\{\mathbf{r}_2^\star, \mathbf{r}_3^{}, \mathbf{r}_4^{\thicksim\star}, \mathbf{r}_5^{}, \mathbf{r}_5^\star, \mathbf{r}_6^{}\}$,~~ $(11, 10)$\\[-3mm]

$\usebox{\uudr}~ \usebox{\duur}$ $\|$ $\usebox{\ddub}~ \usebox{\duub}~ \usebox{\uuub}$ $\|$
$\usebox{\ddurb}~ \usebox{\uudrb}~ \usebox{\duurb}~ \usebox{\uuurb}$ $\|$ $\usebox{\ddubr}~ \usebox{\duubr}~ \usebox{\uuubr}$\\[-2mm]

$\{\mathbf{r}_2^\star, \mathbf{r}_3^{}, \mathbf{r}_4^\thicksim, \mathbf{r}_4^{\thicksim\star}, \mathbf{r}_5^{}, \mathbf{r}_5^\star\}$,~~ $(14, 10)$\\[-3mm]

$\usebox{\ddur}~ \usebox{\uudr}~ \usebox{\duur}$ $\|$ $\usebox{\ddub}~ \usebox{\duub}~ \usebox{\uuub}$ $\|$
$\usebox{\ddurb}~ \usebox{\duurb}~ \usebox{\uuurb}$ $\|$ $\usebox{\ddubr}~ \usebox{\duubr}~ \usebox{\uuubr}$\\[-2mm]

$\{\mathbf{r}_2^\star, \mathbf{r}_3^{}, \mathbf{r}_4^\thicksim, \mathbf{r}_4^{\thicksim\star}, \mathbf{r}_5^{}, \mathbf{r}_6^{}\}$,~~ $(20, 10)$\\[-3mm]

$\usebox{\uudr}~ \usebox{\duur}$ $\|$ $\usebox{\dddb}~ \usebox{\ddub}~ \usebox{\duub}~ \usebox{\uuub}$ $\|$
$\usebox{\ddurb}~ \usebox{\duurb}~ \usebox{\uuurb}$ $\|$ $\usebox{\dddbr}~ \usebox{\ddubr}~ \usebox{\duubr}~ \usebox{\uuubr}$\\

$\{\mathbf{r}_2^\star, \mathbf{r}_3^{}, \mathbf{r}_4^\thicksim, \mathbf{r}_4^{\thicksim\star}, \mathbf{r}_5^\star, \mathbf{r}_6^{}\}$,~~ $(-, 10)$\\[-3mm]

$\usebox{\dudr}~ \usebox{\uudr}~ \usebox{\duur}$ $\|$ $\usebox{\ddub}~ \usebox{\duub}~ \usebox{\uuub}$ $\|$
$\usebox{\dddrb}~ \usebox{\ddurb}~ \usebox{\duurb}~ \usebox{\uuurb}$ $\|$ $\usebox{\ddubr}~ \usebox{\duubr}~ \usebox{\uuubr}$\\[0.25cm]

\newpage

Phase 11\\

$\{\mathbf{r}_2^\star, \mathbf{r}_3^{}, \mathbf{r}_3^\star, \mathbf{r}_4^\star, \mathbf{r}_4^{\thicksim\star},
\mathbf{r}_5^{}, \mathbf{r}_5^\star, \mathbf{r}_5^{\star-}, \mathbf{r}_6^{}\}$\\

$\{\mathbf{r}_2^\star, \mathbf{r}_3^\star, \mathbf{r}_4^\star, \mathbf{r}_4^{\thicksim\star}, \mathbf{r}_5^{},
\mathbf{r}_5^\star, \mathbf{r}_5^{\star-}, \mathbf{r}_6^{}\}$,~~ $(3, 11)$\\[-3mm]

$\usebox{\uudr}~ \usebox{\duur}~ \usebox{\uuur}$ $\|$ $\usebox{\ddub}~ \usebox{\duub}$ $\|$
$\usebox{\ddurb}~ \usebox{\uudrb}~ \usebox{\duurb}~ \usebox{\uuurb}$ $\|$ $\usebox{\ddubr}~ \usebox{\duubr}$\\[-2mm]

$\{\mathbf{r}_3^{}, \mathbf{r}_3^\star, \mathbf{r}_4^\star, \mathbf{r}_5^{}, \mathbf{r}_5^\star,
\mathbf{r}_6^{}\}$,~~ $(8, 11)$\\[-3mm]

$\usebox{\uudr}~ \usebox{\duur}$ $\|$ $\usebox{\ddub}~ \usebox{\uudb}~ \usebox{\duub}$ $\|$
$\usebox{\ddurb}~ \usebox{\uudrb}~ \usebox{\duurb}~ \usebox{\uuurb}$ $\|$ $\usebox{\ddubr}~ \usebox{\uudbr}~ \usebox{\duubr}$\\[-2mm]

$\{\mathbf{r}_3^{}, \mathbf{r}_3^\star, \mathbf{r}_4^{\thicksim\star}, \mathbf{r}_5^{}, \mathbf{r}_5^{\star-},
\mathbf{r}_6^{}\}$,~~ $(\widetilde 8, 11)$\\[-3mm]

$\usebox{\uudr}~ \usebox{\duur}$ $\|$ $\usebox{\ddub}~ \usebox{\dudb}~ \usebox{\duub}$ $\|$
$\usebox{\ddurb}~ \usebox{\uudrb}~ \usebox{\duurb}~ \usebox{\uuurb}$ $\|$ $\usebox{\dddbr}~ \usebox{\ddubr}~ \usebox{\duubr}$ \\[-2mm]

$\{\mathbf{r}_3^{}, \mathbf{r}_3^\star, \mathbf{r}_4^{\thicksim\star}, \mathbf{r}_5^{},\mathbf{r}_5^\star,
\mathbf{r}_6^{}\}$,~~ $(9, 11)$\\[-3mm]

$\usebox{\uudr}~ \usebox{\duur}$ $\|$ $\usebox{\ddub}~ \usebox{\uudb}~ \usebox{\duub}$ $\|$
$\usebox{\ddurb}~ \usebox{\uudrb}~ \usebox{\duurb}~ \usebox{\uuurb}$ $\|$ $\usebox{\ddubr}~ \usebox{\duubr}~ \usebox{\uuubr}$\\[-2mm]

$\{\mathbf{r}_3^{}, \mathbf{r}_3^\star, \mathbf{r}_4^\star, \mathbf{r}_5^{}, \mathbf{r}_5^{\star-},
\mathbf{r}_6^{}\}$,~~ $(\widetilde 9, 11)$\\[-3mm]

$\usebox{\uudr}~ \usebox{\duur}$ $\|$ $\usebox{\ddub}~ \usebox{\dudb}~ \usebox{\duub}$ $\|$
$\usebox{\ddurb}~ \usebox{\uudrb}~ \usebox{\duurb}~ \usebox{\uuurb}$ $\|$ $\usebox{\ddubr}~ \usebox{\dudbr}~ \usebox{\duubr}$\\[-2mm]

$\{\mathbf{r}_2^\star, \mathbf{r}_3^{}, \mathbf{r}_4^{\thicksim\star}, \mathbf{r}_5^{},
\mathbf{r}_5^\star,  \mathbf{r}_6^{}\}$,~~ $(10, 11)$\\[-3mm]

$\usebox{\uudr}~ \usebox{\duur}$ $\|$ $\usebox{\ddub}~ \usebox{\duub}~ \usebox{\uuub}$ $\|$
$\usebox{\ddurb}~ \usebox{\uudrb}~ \usebox{\duurb}~ \usebox{\uuurb}$ $\|$ $\usebox{\ddubr}~ \usebox{\duubr}~ \usebox{\uuubr}$\\[-2mm]

$\{\mathbf{r}_2^\star, \mathbf{r}_3^{}, \mathbf{r}_4^\star, \mathbf{r}_5^{}, \mathbf{r}_5^{\star-},
\mathbf{r}_6^{}\}$,~~ $(\widetilde {10}, 11)$\\[-3mm]

$\usebox{\uudr}~ \usebox{\duur}$ $\|$ $\usebox{\dddb}~ \usebox{\ddub}~ \usebox{\duub}$ $\|$
$\usebox{\ddurb}~ \usebox{\uudrb}~ \usebox{\duurb}~ \usebox{\uuurb}$ $\|$ $\usebox{\ddubr}~ \usebox{\dudbr}~ \usebox{\duubr}$\\[-2mm]

$\{\mathbf{r}_2^\star, \mathbf{r}_3^{}, \mathbf{r}_3^\star,  \mathbf{r}_4^\star, \mathbf{r}_4^{\thicksim\star}, \mathbf{r}_5^\star, \mathbf{r}_5^{\star-},
\mathbf{r}_6^{}\}$,~~ $(13^\star, 11)$\\[-3mm]

$\usebox{\dudr}~ \usebox{\uudr}~ \usebox{\duur}$ $\|$ $\usebox{\ddub}~ \usebox{\duub}$ $\|$
$\usebox{\dddrb}~ \usebox{\ddurb}~ \usebox{\dudrb}~ \usebox{\uudrb}~ \usebox{\duurb}~ \usebox{\uuurb}$ $\|$ $\usebox{\ddubr}~ \usebox{\duubr}$\\[-2mm]

$\{\mathbf{r}_2^\star, \mathbf{r}_3^{}, \mathbf{r}_3^\star, \mathbf{r}_4^\star, \mathbf{r}_4^{\thicksim\star}, \mathbf{r}_5^{},
\mathbf{r}_5^\star, \mathbf{r}_5^{\star-}\}$,~~ $(14, 11)$\\[-3mm]

$\usebox{\ddur}~ \usebox{\uudr}~ \usebox{\duur}$ $\|$ $\usebox{\ddub}~ \usebox{\duub}$ $\|$
$\usebox{\ddurb}~ \usebox{\uudrb}~ \usebox{\duurb}~ \usebox{\uuurb}$ $\|$ $\usebox{\ddubr}~ \usebox{\duubr}$\\[-2mm]

$\{\mathbf{r}_2^\star, \mathbf{r}_3^{}, \mathbf{r}_4^\star, \mathbf{r}_5^{}, \mathbf{r}_5^\star,
\mathbf{r}_6^{}\}$,~~ $(21, 11)$\\[-3mm]

$\usebox{\uudr}~ \usebox{\duur}$ $\|$ $\usebox{\ddub}~ \usebox{\duub}~ \usebox{\uuub}$ $\|$
$\usebox{\ddurb}~ \usebox{\uudrb}~ \usebox{\duurb}~ \usebox{\uuurb}$ $\|$ $\usebox{\ddubr}~ \usebox{\uudbr}~ \usebox{\duubr}$\\[-2mm]

$\{\mathbf{r}_2^\star, \mathbf{r}_3^{}, \mathbf{r}_4^{\thicksim\star}, \mathbf{r}_5^{}, \mathbf{r}_5^{\star-},
\mathbf{r}_6^{}\}$,~~ $(\widetilde{21}, 11)$\\[-3mm]

$\usebox{\uudr}~ \usebox{\duur}$ $\|$ $\usebox{\dddb}~ \usebox{\ddub}~ \usebox{\duub}$ $\|$
$\usebox{\ddurb}~ \usebox{\uudrb}~ \usebox{\duurb}~ \usebox{\uuurb}$ $\|$ $\usebox{\dddbr}~ \usebox{\ddubr}~ \usebox{\duubr}$\\

\clearpage
\newpage

Phase 12\\

$\{\mathbf{r}_1^{}, \mathbf{r}_1^\star, \mathbf{r}_3^{}, \mathbf{r}_3^\star, \mathbf{r}_4^{}, \mathbf{r}_4^\star,
\mathbf{r}_6^{}, \mathbf{r}_6^{-}, \mathbf{r}_6^\star, \mathbf{r}_6^{\star-}, \mathbf{r}_7^{}, \mathbf{r}_7^{-}\}$\\

$\{\mathbf{r}_1^{}, \mathbf{r}_1^\star, \mathbf{r}_4^{}, \mathbf{r}_4^\star, \mathbf{r}_6^{}, \mathbf{r}_6^{\star-}, \mathbf{r}_7^{}, \mathbf{r}_7^{-}\}$,~~ $(\widetilde 1, 12)$\\[-3mm]

$\usebox{\dudr}~ \usebox{\uudr}~ \usebox{\uuur}$ $\|$ $\usebox{\dddb}~ \usebox{\dudb}~ \usebox{\uudb}$ $\|$
$\usebox{\dudrb}~ \usebox{\uudrb}$ $\|$ $\usebox{\dudbr}~ \usebox{\uudbr}$\\[-2mm]

$\{\mathbf{r}_1^{}, \mathbf{r}_1^\star, \mathbf{r}_4^{}, \mathbf{r}_4^\star, \mathbf{r}_6^{-}, \mathbf{r}_6^{\star}, \mathbf{r}_7^{}, \mathbf{r}_7^{-}\}$,~~ $(\overline{\widetilde 1}, 12)$\\[-3mm]

$\usebox{\dddr}~ \usebox{\dudr}~ \usebox{\uudr}$ $\|$ $\usebox{\dudb}~ \usebox{\uudb}~ \usebox{\uuub}$ $\|$
$\usebox{\dudrb}~ \usebox{\uudrb}$ $\|$ $\usebox{\dudbr}~ \usebox{\uudbr}$\\[-2mm]

$\{\mathbf{r}_1^{}, \mathbf{r}_3^\star, \mathbf{r}_4^\star, \mathbf{r}_6^{}, \mathbf{r}_6^{\star-}, \mathbf{r}_7^{}\}$,~~ $(\widetilde 2, 12)$\\[-3mm]

$\usebox{\dudr}~ \usebox{\uudr}~ \usebox{\uuur}$ $\|$ $\usebox{\ddub}~ \usebox{\dudb}~ \usebox{\uudb}$ $\|$
$\usebox{\dudrb}~ \usebox{\uudrb}~ \usebox{\uuurb}$ $\|$ $\usebox{\dudbr}~ \usebox{\uudbr}~ \usebox{\duubr}$\\[-2mm]

$\{\mathbf{r}_1^{}, \mathbf{r}_3^\star, \mathbf{r}_4^\star, \mathbf{r}_6^\star, \mathbf{r}_6^{-}, \mathbf{r}_7^{-}\}$,~~ $(\overline{\widetilde 2}, 12)$\\[-3mm]

$\usebox{\dddr}~ \usebox{\dudr}~ \usebox{\uudr}$ $\|$ $\usebox{\dudb}~ \usebox{\uudb}~ \usebox{\duub}$ $\|$
$\usebox{\dddrb}~ \usebox{\dudrb}~ \usebox{\uudrb}$ $\|$ $\usebox{\ddubr}~ \usebox{\dudbr}~ \usebox{\uudbr}$\\[-2mm]

$\{\mathbf{r}_1^\star, \mathbf{r}_3^{}, \mathbf{r}_4^{}, \mathbf{r}_6^{-}, \mathbf{r}_6^\star, \mathbf{r}_7^{}\}$,~~ $(\widetilde 2^\star, 12)$\\[-3mm]

$\usebox{\ddur}~ \usebox{\dudr}~ \usebox{\uudr}$ $\|$ $\usebox{\dudb}~ \usebox{\uudb}~ \usebox{\uuub}$ $\|$
$\usebox{\dudrb}~ \usebox{\uudrb}~ \usebox{\duurb}$ $\|$ $\usebox{\dudbr}~ \usebox{\uudbr}~ \usebox{\uuubr}$\\[-2mm]

$\{\mathbf{r}_1^\star, \mathbf{r}_3^{}, \mathbf{r}_4^{}, \mathbf{r}_6^{}, \mathbf{r}_6^{\star-}, \mathbf{r}_7^{-}\}$,~~ $(\overline{\widetilde 2^\star}, 12)$\\[-3mm]

$\usebox{\dudr}~ \usebox{\uudr}~ \usebox{\duur}$ $\|$ $\usebox{\dddb}~ \usebox{\dudb}~ \usebox{\uudb}$ $\|$
$\usebox{\ddurb}~ \usebox{\dudrb}~ \usebox{\uudrb}$ $\|$ $\usebox{\dddbr}~ \usebox{\dudbr}~ \usebox{\uudbr}$\\[-2mm]

$\{\mathbf{r}_1^{}, \mathbf{r}_1^\star, \mathbf{r}_3^\star, \mathbf{r}_6^{}, \mathbf{r}_6^\star, \mathbf{r}_6^{\star-}, \mathbf{r}_7^{}\}$,~~ $(4, 12)$\\[-3mm]

$\usebox{\dudr}~ \usebox{\uudr}~ \usebox{\uuur}$ $\|$ $\usebox{\dudb}~ \usebox{\uudb}$ $\|$
$\usebox{\dudrb}~ \usebox{\uudrb}~ \usebox{\uuurb}$ $\|$ $\usebox{\dudbr}~ \usebox{\uudbr}~ \usebox{\uuubr}$\\[-2mm]

$\{\mathbf{r}_1^{}, \mathbf{r}_1^\star, \mathbf{r}_3^\star, \mathbf{r}_6^\star, \mathbf{r}_6^{-}, \mathbf{r}_6^{\star-}, \mathbf{r}_7^{-}\}$,~~ $(\overline{4}, 12)$\\[-3mm]

$\usebox{\dddr}~ \usebox{\dudr}~ \usebox{\uudr}$ $\|$ $\usebox{\dudb}~ \usebox{\uudb}$ $\|$
$\usebox{\dddrb}~ \usebox{\dudrb}~ \usebox{\uudrb}$ $\|$ $\usebox{\dddbr}~ \usebox{\dudbr}~ \usebox{\uudbr}$\\[-2mm]

$\{\mathbf{r}_1^{}, \mathbf{r}_1^\star, \mathbf{r}_3^{}, \mathbf{r}_6^{}, \mathbf{r}_6^\star, , \mathbf{r}_6^{-}, \mathbf{r}_7^{}\}$,~~ $(4^\star, 12)$\\[-3mm]

$\usebox{\dudr}~ \usebox{\uudr}$ $\|$ $\usebox{\dudb}~ \usebox{\uudb}~ \usebox{\uuub}$ $\|$
$\usebox{\dudrb}~ \usebox{\uudrb}~ \usebox{\uuurb}$ $\|$ $\usebox{\dudbr}~ \usebox{\uudbr}~ \usebox{\uuubr}$\\[-2mm]

$\{\mathbf{r}_1^{}, \mathbf{r}_1^\star, \mathbf{r}_3^{}, \mathbf{r}_6^{}, \mathbf{r}_6^{-}, \mathbf{r}_6^{\star-}, \mathbf{r}_7^{-}\}$,~~ $(\overline{4}^\star, 12)$\\[-3mm]

$\usebox{\dudr}~ \usebox{\uudr}$ $\|$ $\usebox{\dddb}~ \usebox{\dudb}~ \usebox{\uudb}$ $\|$
$\usebox{\dddrb}~ \usebox{\dudrb}~ \usebox{\uudrb}$ $\|$ $\usebox{\dddbr}~ \usebox{\dudbr}~ \usebox{\uudbr}$\\[-2mm]

$\{\mathbf{r}_3^\star, \mathbf{r}_4^{}, \mathbf{r}_4^\star, \mathbf{r}_6^{}, \mathbf{r}_6^{\star-}, \mathbf{r}_7^{}\}$,~~ $(\widetilde 5, 12)$\\[-3mm]

$\usebox{\dudr}~ \usebox{\uudr}~ \usebox{\duur}~ \usebox{\uuur}$ $\|$ $\usebox{\ddub}~ \usebox{\dudb}~ \usebox{\uudb}$ $\|$
$\usebox{\dudrb}~ \usebox{\uudrb}~ \usebox{\duurb}$ $\|$ $\usebox{\dudbr}~ \usebox{\uudbr}~ \usebox{\duubr}$\\[-2mm]

$\{\mathbf{r}_3^\star, \mathbf{r}_4^{}, \mathbf{r}_4^\star, \mathbf{r}_6^\star, \mathbf{r}_6^{-}, \mathbf{r}_7^{-}\}$,~~ $(\overline{\widetilde 5}, 12)$\\[-3mm]

$\usebox{\dddr}~ \usebox{\ddur}~ \usebox{\dudr}~ \usebox{\uudr}$ $\|$ $\usebox{\dudb}~ \usebox{\uudb}~ \usebox{\duub}$ $\|$
$\usebox{\ddurb}~ \usebox{\dudrb}~ \usebox{\uudrb}$ $\|$ $\usebox{\ddubr}~ \usebox{\dudbr}~ \usebox{\uudbr}$\\[-2mm]

$\{\mathbf{r}_3^{}, \mathbf{r}_4^{}, \mathbf{r}_4^\star, \mathbf{r}_6^{-}, \mathbf{r}_6^\star, \mathbf{r}_7^{}\}$,~~ $(\widetilde 5^\star, 12)$\\[-3mm]

$\usebox{\ddur}~ \usebox{\dudr}~ \usebox{\uudr}$ $\|$ $\usebox{\dudb}~ \usebox{\uudb}~ \usebox{\duub}~ \usebox{\uuub}$ $\|$
$\usebox{\dudrb}~ \usebox{\uudrb}~ \usebox{\duurb}$ $\|$ $\usebox{\dudbr}~ \usebox{\uudbr}~ \usebox{\duubr}$\\[-2mm]

$\{\mathbf{r}_3^{}, \mathbf{r}_4^{}, \mathbf{r}_4^\star, \mathbf{r}_6^{}, \mathbf{r}_6^{\star-}, \mathbf{r}_7^{-}\}$,~~ $(\overline{\widetilde 5^\star}, 12)$\\[-3mm]

$\usebox{\dudr}~ \usebox{\uudr}~ \usebox{\duur}$ $\|$ $\usebox{\dddb}~ \usebox{\ddub}~ \usebox{\dudb}~ \usebox{\uudb}$ $\|$
$\usebox{\ddurb}~ \usebox{\dudrb}~ \usebox{\uudrb}$ $\|$ $\usebox{\ddubr}~ \usebox{\dudbr}~ \usebox{\uudbr}$\\[-2mm]

$\{\mathbf{r}_1^\star, \mathbf{r}_3^\star, \mathbf{r}_4^{}, \mathbf{r}_6^{}, \mathbf{r}_6^\star, \mathbf{r}_6^{\star-}, \mathbf{r}_7^{}\}$,~~ $(6, 12)$\\[-3mm]

$\usebox{\dudr}~ \usebox{\uudr}~ \usebox{\duur}~ \usebox{\uuur}$ $\|$ $\usebox{\dudb}~ \usebox{\uudb}$ $\|$
$\usebox{\dudrb}~ \usebox{\uudrb}~ \usebox{\duurb}$ $\|$ $\usebox{\dudbr}~ \usebox{\uudbr}~ \usebox{\uuubr}$\\[-2mm]

$\{\mathbf{r}_1^\star, \mathbf{r}_3^\star, \mathbf{r}_4^{}, \mathbf{r}_6^\star, \mathbf{r}_6^{-}, \mathbf{r}_6^{\star-}, \mathbf{r}_7^{-}\}$,~~ $(\overline{6}, 12)$\\[-3mm]

$\usebox{\dddr}~ \usebox{\ddur}~ \usebox{\dudr}~ \usebox{\uudr}$ $\|$ $\usebox{\dudb}~ \usebox{\uudb}$ $\|$
$\usebox{\ddurb}~ \usebox{\dudrb}~ \usebox{\uudrb}$ $\|$ $\usebox{\dddbr}~ \usebox{\dudbr}~ \usebox{\uudbr}$\\[-2mm]

$\{\mathbf{r}_1^{}, \mathbf{r}_3^{}, \mathbf{r}_4^\star, \mathbf{r}_6^{}, \mathbf{r}_6^{-}, \mathbf{r}_6^\star, \mathbf{r}_7^{}\}$,~~ $(6^\star, 12)$\\[-3mm]

$\usebox{\dudr}~ \usebox{\uudr}$ $\|$ $\usebox{\dudb}~ \usebox{\uudb}~ \usebox{\duub}~ \usebox{\uuub}$ $\|$
$\usebox{\dudrb}~ \usebox{\uudrb}~ \usebox{\uuurb}$ $\|$ $\usebox{\dudbr}~ \usebox{\uudbr}~ \usebox{\duubr}$\\[-2mm]

$\{\mathbf{r}_1^{}, \mathbf{r}_3^{}, \mathbf{r}_4^\star, \mathbf{r}_6^{}, \mathbf{r}_6^{-}, \mathbf{r}_6^{\star-}, \mathbf{r}_7^{-}\}$,~~ $(\overline{6}^\star, 12)$\\[-3mm]

$\usebox{\dudr}~ \usebox{\uudr}$ $\|$ $\usebox{\dddb}~ \usebox{\ddub}~ \usebox{\dudb}~ \usebox{\uudb}$ $\|$
$\usebox{\dddrb}~ \usebox{\dudrb}~ \usebox{\uudrb}$ $\|$ $\usebox{\ddubr}~ \usebox{\dudbr}~ \usebox{\uudbr}$\\[-2mm]

\newpage

$\{\mathbf{r}_3^{}, \mathbf{r}_3^\star, \mathbf{r}_4^{}, \mathbf{r}_4^\star, \mathbf{r}_6^{}, \mathbf{r}_6^\star, \mathbf{r}_7^{}\}$,~~ $(8, 12)$\\[-3mm]

$\usebox{\dudr}~ \usebox{\uudr}~ \usebox{\duur} $ $\|$ $\usebox{\dudb}~ \usebox{\uudb}~ \usebox{\duub}$ $\|$
$\usebox{\dudrb}~ \usebox{\uudrb}~ \usebox{\duurb}$ $\|$ $\usebox{\dudbr}~ \usebox{\uudbr}~ \usebox{\duubr}$\\[-2mm]

$\{\mathbf{r}_3^{}, \mathbf{r}_3^\star, \mathbf{r}_4^{}, \mathbf{r}_4^\star, \mathbf{r}_6^{-}, \mathbf{r}_6^{\star-}, \mathbf{r}_7^{-}\}$,~~ $(\overline{8}, 12)$\\[-3mm]

$\usebox{\ddur}~ \usebox{\dudr}~ \usebox{\uudr}$ $\|$ $\usebox{\ddub}~ \usebox{\dudb}~ \usebox{\uudb}$ $\|$
$\usebox{\ddurb}~ \usebox{\dudrb}~ \usebox{\uudrb}$ $\|$ $\usebox{\ddubr}~ \usebox{\dudbr}~ \usebox{\uudbr}$\\[-2mm]

$\{\mathbf{r}_3^{}, \mathbf{r}_3^\star, \mathbf{r}_4^{}, \mathbf{r}_4^\star, \mathbf{r}_6^{}, \mathbf{r}_6^{\star-}\}$,~~ $(\widetilde 9, 12)$\\[-3mm]

$\usebox{\dudr}~ \usebox{\uudr}~ \usebox{\duur} $ $\|$ $\usebox{\ddub}~ \usebox{\dudb}~ \usebox{\uudb} $ $\|$
$\usebox{\ddurb}~ \usebox{\dudrb}~ \usebox{\uudrb}~ \usebox{\duurb}$ $\|$ $\usebox{\ddubr}~ \usebox{\dudbr}~ \usebox{\uudbr}~ \usebox{\duubr}$\\[-2mm]

$\{\mathbf{r}_3^{}, \mathbf{r}_3^\star, \mathbf{r}_4^{}, \mathbf{r}_4^\star, \mathbf{r}_6^\star, \mathbf{r}_6^{-}\}$,~~ $(\overline{\widetilde 9}, 12)$\\[-3mm]

$\usebox{\ddur}~ \usebox{\dudr}~ \usebox{\uudr}$ $\|$ $\usebox{\dudb}~ \usebox{\uudb}~ \usebox{\duub}$ $\|$
$\usebox{\ddurb}~ \usebox{\dudrb}~ \usebox{\uudrb}~ \usebox{\duurb}$ $\|$ $\usebox{\ddubr}~ \usebox{\dudbr}~ \usebox{\uudbr}~ \usebox{\duubr}$\\[-2mm]

$\{\mathbf{r}_1^{}, \mathbf{r}_1^\star, \mathbf{r}_3^{}, \mathbf{r}_3^\star, \mathbf{r}_6^{}, \mathbf{r}_6^\star, \mathbf{r}_6^{-}, \mathbf{r}_6^{\star-}\}$,~~ $(\widetilde {12}, 12)$\\[-3mm]

$\usebox{\dudr}~ \usebox{\uudr}$ $\|$ $\usebox{\dudb}~ \usebox{\uudb}$ $\|$
$\usebox{\dddrb}~ \usebox{\dudrb}~ \usebox{\uudrb}~ \usebox{\uuurb}$ $\|$ $\usebox{\dddbr}~ \usebox{\dudbr}~ \usebox{\uudbr}~ \usebox{\uuubr}$\\

$\{\mathbf{r}_1^{}, \mathbf{r}_3^\star, \mathbf{r}_4^\star, \mathbf{r}_6^{}, \mathbf{r}_6^\star, \mathbf{r}_7^{}\}$,~~ $(-, 12)$\\[-3mm]

$\usebox{\dudr}~ \usebox{\uudr}~ \usebox{\uuur}$ $\|$ $\usebox{\dudb}~ \usebox{\uudb}~ \usebox{\duub}$ $\|$
$\usebox{\dudrb}~ \usebox{\uudrb}~ \usebox{\uuurb}$ $\|$ $\usebox{\dudbr}~ \usebox{\uudbr}~ \usebox{\duubr}$\\[-2mm]

$\{\mathbf{r}_1^{}, \mathbf{r}_3^\star, \mathbf{r}_4^\star, \mathbf{r}_6^{-}, \mathbf{r}_6^{\star-}, \mathbf{r}_7^{-}\}$,~~ $(-, 12)$\\[-3mm]

$\usebox{\dddr}~ \usebox{\dudr}~ \usebox{\uudr}$ $\|$ $\usebox{\ddub}~ \usebox{\dudb}~ \usebox{\uudb}$ $\|$
$\usebox{\dddrb}~ \usebox{\dudrb}~ \usebox{\uudrb}$ $\|$ $\usebox{\ddubr}~ \usebox{\dudbr}~ \usebox{\uudbr}$\\[-2mm]

$\{\mathbf{r}_1^\star, \mathbf{r}_3^{}, \mathbf{r}_4^{}, \mathbf{r}_6^{}, \mathbf{r}_6^\star, \mathbf{r}_7^{}\}$,~~ $(-, 12)$\\[-3mm]

$\usebox{\dudr}~ \usebox{\uudr}~ \usebox{\duur}$ $\|$ $\usebox{\dudb}~ \usebox{\uudb}~ \usebox{\uuub}$ $\|$
$\usebox{\dudrb}~ \usebox{\uudrb}~ \usebox{\duurb}$ $\|$ $\usebox{\dudbr}~ \usebox{\uudbr}~ \usebox{\uuubr}$\\[-2mm]

$\{\mathbf{r}_1^\star, \mathbf{r}_3^{}, \mathbf{r}_4^{}, \mathbf{r}_6^{-}, \mathbf{r}_6^{\star-}, \mathbf{r}_7^{-}\}$,~~ $(-, 12)$\\[-3mm]

$\usebox{\ddur}~ \usebox{\dudr}~ \usebox{\uudr}$ $\|$ $\usebox{\dddb}~ \usebox{\dudb}~ \usebox{\uudb}$ $\|$
$\usebox{\ddurb}~ \usebox{\dudrb}~ \usebox{\uudrb}$ $\|$ $\usebox{\dddbr}~ \usebox{\dudbr}~ \usebox{\uudbr}$\\[-2mm]

$\{\mathbf{r}_1^{}, \mathbf{r}_3^{}, \mathbf{r}_3^\star, \mathbf{r}_4^\star, \mathbf{r}_6^{}, \mathbf{r}_6^\star, \mathbf{r}_6^{-}\}$,~~ $(-, 12)$\\[-3mm]

$\usebox{\dudr}~ \usebox{\uudr}$ $\|$ $\usebox{\dudb}~ \usebox{\uudb}~ \usebox{\duub}$ $\|$
$\usebox{\dddrb}~ \usebox{\dudrb}~ \usebox{\uudrb}~ \usebox{\uuurb}$ $\|$ $\usebox{\ddubr}~ \usebox{\dudbr}~ \usebox{\uudbr}~ \usebox{\duubr}$\\[-2mm]

$\{\mathbf{r}_1^{}, \mathbf{r}_3^{}, \mathbf{r}_3^\star, \mathbf{r}_4^\star, \mathbf{r}_6^{}, \mathbf{r}_6^{-}, \mathbf{r}_6^{\star-}\}$,~~ $(-, 12)$\\[-3mm]

$\usebox{\dudr}~ \usebox{\uudr}$ $\|$ $\usebox{\ddub}~ \usebox{\dudb}~ \usebox{\uudb}$ $\|$
$\usebox{\dddrb}~ \usebox{\dudrb}~ \usebox{\uudrb}~ \usebox{\uuurb}$ $\|$ $\usebox{\ddubr}~ \usebox{\dudbr}~ \usebox{\uudbr}~ \usebox{\duubr}$\\[-2mm]

$\{\mathbf{r}_1^\star, \mathbf{r}_3^{}, \mathbf{r}_3^\star, \mathbf{r}_4^{}, \mathbf{r}_6^{}, \mathbf{r}_6^\star, \mathbf{r}_6^{\star-}\}$,~~ $(-, 12)$\\[-3mm]

$\usebox{\dudr}~ \usebox{\uudr}~ \usebox{\duur}$ $\|$ $\usebox{\dudb}~ \usebox{\uudb}$ $\|$
$\usebox{\ddurb}~ \usebox{\dudrb}~ \usebox{\uudrb}~ \usebox{\duurb}$ $\|$ $\usebox{\dddbr}~ \usebox{\dudbr}~ \usebox{\uudbr}~ \usebox{\uuubr}$\\[-2mm]

$\{\mathbf{r}_1^\star, \mathbf{r}_3^{}, \mathbf{r}_3^\star, \mathbf{r}_4^{}, \mathbf{r}_6^\star, \mathbf{r}_6^{-}, \mathbf{r}_6^{\star-}\}$,~~ $(-, 12)$\\[-3mm]

$\usebox{\ddur}~ \usebox{\dudr}~ \usebox{\uudr}$ $\|$ $\usebox{\dudb}~ \usebox{\uudb}$ $\|$
$\usebox{\ddurb}~ \usebox{\dudrb}~ \usebox{\uudrb}~ \usebox{\duurb}$ $\|$ $\usebox{\dddbr}~ \usebox{\dudbr}~ \usebox{\uudbr}~ \usebox{\uuubr}$\\[-2mm]

$\{\mathbf{r}_1^{}, \mathbf{r}_1^\star, \mathbf{r}_3^{}, \mathbf{r}_4^{}, \mathbf{r}_4^\star, \mathbf{r}_6^{}, \mathbf{r}_7^{}, \mathbf{r}_6^{-}, \mathbf{r}_7^{-}\}$,~~ $(-, 12)$\\[-3mm]

$\usebox{\dudr}~ \usebox{\uudr}$ $\|$ $\usebox{\dddb}~ \usebox{\dudb}~ \usebox{\uudb}~ \usebox{\uuub}$ $\|$
$\usebox{\dudrb}~ \usebox{\uudrb}$ $\|$ $\usebox{\dudbr}~ \usebox{\uudbr}$\\[-2mm]

$\{\mathbf{r}_1^{}, \mathbf{r}_1^\star, \mathbf{r}_3^\star, \mathbf{r}_4^{}, \mathbf{r}_4^\star, \mathbf{r}_6^\star, \mathbf{r}_7^{}, \mathbf{r}_6^{\star-}, \mathbf{r}_7^{-}\}$,~~ $(-, 12)$\\[-3mm]

$\usebox{\dddr}~ \usebox{\dudr}~ \usebox{\uudr}~ \usebox{\uuur}$ $\|$ $\usebox{\dudb}~ \usebox{\uudb}$ $\|$
$\usebox{\dudrb}~ \usebox{\uudrb}$ $\|$ $\usebox{\dudbr}~ \usebox{\uudbr}$\\

\clearpage
\newpage

Phase 13\\

$\{\mathbf{r}_1^\star, \mathbf{r}_2^{}, \mathbf{r}_3^{}, \mathbf{r}_3^\star,
\mathbf{r}_4^{}, \mathbf{r}_4^\thicksim, \mathbf{r}_5^{}, \mathbf{r}_5^{-},
\mathbf{r}_6^\star, \mathbf{r}_6^{\star-}\}$\\

$\{\mathbf{r}_1^\star, \mathbf{r}_2^{}, \mathbf{r}_3^{}, \mathbf{r}_4^{}, \mathbf{r}_4^\thicksim, \mathbf{r}_5^{}, \mathbf{r}_5^{-}, \mathbf{r}_6^\star\}$,~~ $(3^\star, 13)$\\[-3mm]

$\usebox{\ddur}~ \usebox{\duur}$ $\|$ $\usebox{\dudb}~ \usebox{\uudb}~ \usebox{\uuub}$ $\|$
$\usebox{\ddurb}~ \usebox{\duurb}$ $\|$ $\usebox{\dddbr}~ \usebox{\dudbr}~ \usebox{\uudbr}~ \usebox{\uuubr}$\\[-2mm]

$\{\mathbf{r}_1^\star, \mathbf{r}_2^{}, \mathbf{r}_3^{}, \mathbf{r}_4^{}, \mathbf{r}_4^\thicksim, \mathbf{r}_5^{}, \mathbf{r}_5^{-}, \mathbf{r}_6^{\star-}\}$,~~ $(\overline{3}^\star, 13)$\\[-2mm]

$\usebox{\ddur}~ \usebox{\duur}$ $\|$ $\usebox{\dddb}~ \usebox{\dudb}~ \usebox{\uudb}$ $\|$
$\usebox{\ddurb}~ \usebox{\duurb}$ $\|$ $\usebox{\dddbr}~ \usebox{\dudbr}~ \usebox{\uudbr}~ \usebox{\uuubr}$\\[-2mm]

$\{\mathbf{r}_2^{}, \mathbf{r}_3^{}, \mathbf{r}_3^\star, \mathbf{r}_4^{}, \mathbf{r}_4^\thicksim, \mathbf{r}_5^{}, \mathbf{r}_5^{-}, \mathbf{r}_6^\star\}$,~~ $(11^\star, 13)$\\[-3mm]

$\usebox{\ddur}~ \usebox{\duur}$ $\|$ $\usebox{\dudb}~ \usebox{\uudb}~ \usebox{\duub}$ $\|$
$\usebox{\ddurb}~ \usebox{\duurb}$ $\|$ $\usebox{\dddbr}~ \usebox{\ddubr}~ \usebox{\dudbr}~ \usebox{\uudbr}~ \usebox{\duubr}~ \usebox{\uuubr}$\\[-2mm]

$\{\mathbf{r}_2^{}, \mathbf{r}_3^{}, \mathbf{r}_3^\star, \mathbf{r}_4^{}, \mathbf{r}_4^\thicksim, \mathbf{r}_5^{}, \mathbf{r}_5^{-}, \mathbf{r}_6^{\star-}\}$,~~ $(\overline{11}^\star, 13)$\\[-3mm]

$\usebox{\ddur}~ \usebox{\duur}$ $\|$ $\usebox{\ddub}~ \usebox{\dudb}~ \usebox{\uudb}$ $\|$
$\usebox{\ddurb}~ \usebox{\duurb}$ $\|$ $\usebox{\dddbr}~ \usebox{\ddubr}~ \usebox{\dudbr}~ \usebox{\uudbr}~ \usebox{\duubr}~ \usebox{\uuubr}$\\[-2mm]

$\{\mathbf{r}_1^\star, \mathbf{r}_2^{}, \mathbf{r}_3^\star, \mathbf{r}_4^\thicksim, \mathbf{r}_5^{}, \mathbf{r}_6^\star, \mathbf{r}_6^{\star-}\}$,~~ $(22, 13)$\\[-3mm]

$\usebox{\ddur}~ \usebox{\duur}~ \usebox{\uuur}$ $\|$ $\usebox{\dudb}~ \usebox{\uudb}$ $\|$
$\usebox{\ddurb}~ \usebox{\duurb}~ \usebox{\uuurb}$ $\|$ $\usebox{\dddbr}~ \usebox{\dudbr}~ \usebox{\uudbr}~ \usebox{\uuubr}$\\[-2mm]

$\{\mathbf{r}_1^\star, \mathbf{r}_2^{}, \mathbf{r}_3^\star, \mathbf{r}_4^\thicksim, \mathbf{r}_5^{-}, \mathbf{r}_6^\star, \mathbf{r}_6^{\star-}\}$,~~ $(\overline{22}, 13)$\\[-3mm]

$\usebox{\dddr}~ \usebox{\ddur}~ \usebox{\duur}$ $\|$ $\usebox{\dudb}~ \usebox{\duub}$ $\|$
$\usebox{\dddrb}~ \usebox{\ddurb}~ \usebox{\duurb}$ $\|$ $\usebox{\dddbr}~ \usebox{\dudbr}~ \usebox{\uudbr}~ \usebox{\uuubr}$\\[-2mm]

$\{\mathbf{r}_1^\star, \mathbf{r}_2^{}, \mathbf{r}_3^\star, \mathbf{r}_4^{}, \mathbf{r}_5^{}, \mathbf{r}_6^\star, \mathbf{r}_6^{\star-}\}$,~~ $(\widetilde{22}, 13)$\\[-3mm]

$\usebox{\ddur}~ \usebox{\duur}~ \usebox{\uuur}$ $\|$ $\usebox{\dudb}~ \usebox{\uudb}$ $\|$
$\usebox{\ddurb}~ \usebox{\uudrb}~ \usebox{\duurb}$ $\|$ $\usebox{\dddbr}~ \usebox{\dudbr}~ \usebox{\uudbr}~ \usebox{\uuubr}$\\[-2mm]

$\{\mathbf{r}_1^\star, \mathbf{r}_2^{}, \mathbf{r}_3^\star, \mathbf{r}_4^{}, \mathbf{r}_5^{-}, \mathbf{r}_6^\star, \mathbf{r}_6^{\star-}\}$,~~ $(\overline{\widetilde{22}}, 13)$\\[-3mm]

$\usebox{\dddr}~ \usebox{\ddur}~ \usebox{\duur}$ $\|$ $\usebox{\dudb}~ \usebox{\uudb}$ $\|$
$\usebox{\ddurb}~ \usebox{\dudrb}~ \usebox{\duurb}$ $\|$ $\usebox{\dddbr}~ \usebox{\dudbr}~ \usebox{\uudbr}~ \usebox{\uuubr}$\\[-2mm]

$\{\mathbf{r}_1^\star, \mathbf{r}_3^{}, \mathbf{r}_3^\star, \mathbf{r}_4^\thicksim, \mathbf{r}_5^{}, \mathbf{r}_6^\star, \mathbf{r}_6^{\star-}\}$,~~ $(23, 13)$\\[-3mm]

$\usebox{\ddur}~ \usebox{\uudr}~ \usebox{\duur}$ $\|$ $\usebox{\dudb}~ \usebox{\uudb}$ $\|$
$\usebox{\ddurb}~ \usebox{\duurb}~ \usebox{\uuurb}$ $\|$ $\usebox{\dddbr}~ \usebox{\dudbr}~ \usebox{\uudbr}~ \usebox{\uuubr}$\\[-2mm]

$\{\mathbf{r}_1^\star, \mathbf{r}_3^{}, \mathbf{r}_3^\star, \mathbf{r}_4^\thicksim, \mathbf{r}_5^{-}, \mathbf{r}_6^\star, \mathbf{r}_6^{\star-}\}$,~~ $(\overline{23}, 13)$\\[-3mm]

$\usebox{\ddur}~ \usebox{\dudr}~ \usebox{\duur}$ $\|$ $\usebox{\dudb}~ \usebox{\uudb}$ $\|$
$\usebox{\dddrb}~ \usebox{\ddurb}~ \usebox{\duurb}$ $\|$ $\usebox{\dddbr}~ \usebox{\dudbr}~ \usebox{\uudbr}~ \usebox{\uuubr}$\\[-2mm]

$\{\mathbf{r}_1^\star, \mathbf{r}_3^{}, \mathbf{r}_3^\star, \mathbf{r}_4^{}, \mathbf{r}_5^{}, \mathbf{r}_6^\star, \mathbf{r}_6^{\star-}\}$,~~ $(\widetilde {23}, 13)$\\[-3mm]

$\usebox{\ddur}~ \usebox{\uudr}~ \usebox{\duur}$ $\|$ $\usebox{\dudb}~ \usebox{\uudb}$ $\|$
$\usebox{\ddurb}~ \usebox{\uudrb}~ \usebox{\duurb}$ $\|$ $\usebox{\dddbr}~ \usebox{\dudbr}~ \usebox{\uudbr}~ \usebox{\uuubr}$\\[-2mm]

$\{\mathbf{r}_1^\star, \mathbf{r}_3^{}, \mathbf{r}_3^\star, \mathbf{r}_4^{}, \mathbf{r}_5^{-}, \mathbf{r}_6^\star, \mathbf{r}_6^{\star-}\}$,~~ $(\overline{\widetilde {23}}, 13)$\\[-3mm]

$\usebox{\ddur}~ \usebox{\dudr}~ \usebox{\duur}$ $\|$ $\usebox{\dudb}~ \usebox{\uudb}$ $\|$
$\usebox{\ddurb}~ \usebox{\dudrb}~ \usebox{\duurb}$ $\|$ $\usebox{\dddbr}~ \usebox{\dudbr}~ \usebox{\uudbr}~ \usebox{\uuubr}$\\

\clearpage
\newpage

Phase 14\\

$\{\mathbf{r}_2^{}, \mathbf{r}_2^\star, \mathbf{r}_3^{}, \mathbf{r}_3^\star,
\mathbf{r}_4^{}, \mathbf{r}_{4}^\thicksim, \mathbf{r}_4^\star, \mathbf{r}_4^{\thicksim \star},
\mathbf{r}_5^{}, \mathbf{r}_5^{-}, \mathbf{r}_5^\star, \mathbf{r}_5^{\star-}\}$\\

$\{\mathbf{r}_2^{}, \mathbf{r}_2^\star, \mathbf{r}_4^\thicksim, \mathbf{r}_4^{\thicksim\star}, \mathbf{r}_5^{}, \mathbf{r}_5^\star\}$,~~~~~~~~~~~~~~$(1, 14)$\\[-3mm]

$\usebox{\ddur}~ \usebox{\duur}~ \usebox{\uuur}$ $\|$ $\usebox{\ddub}~ \usebox{\duub}~ \usebox{\uuub}$ $\|$
$\usebox{\ddurb}~ \usebox{\duurb}~ \usebox{\uuurb}$ $\|$ $\usebox{\ddubr}~ \usebox{\duubr}~ \usebox{\uuubr}$\\[-2mm]

$\{\mathbf{r}_2^{}, \mathbf{r}_2^\star, \mathbf{r}_4^\thicksim, \mathbf{r}_4^{\thicksim\star}, \mathbf{r}_5^{-}, \mathbf{r}_5^{\star-}\}$,~~~~~~~~~~~$(\bar 1, 14)$\\[-3mm]

$\usebox{\dddr}~ \usebox{\ddur}~ \usebox{\duur}$ $\|$ $\usebox{\dddb}~ \usebox{\ddub}~ \usebox{\duub}$ $\|$
$\usebox{\dddrb}~ \usebox{\ddurb}~ \usebox{\duurb}$ $\|$ $\usebox{\dddbr}~ \usebox{\ddubr}~ \usebox{\duubr}$\\[-2mm]

$\{\mathbf{r}_2^{}, \mathbf{r}_2^\star, \mathbf{r}_4^{}, \mathbf{r}_4^\star, \mathbf{r}_5^{}, \mathbf{r}_5^{\star-}\}$,~~~~~~~~~~~~~~$(\widetilde{1}, 14)$\\[-3mm]

$\usebox{\ddur}~ \usebox{\duur}~ \usebox{\uuur}$ $\|$ $\usebox{\dddb}~ \usebox{\ddub}~ \usebox{\duub}$ $\|$
$\usebox{\ddurb}~ \usebox{\uudrb}~ \usebox{\duurb}$ $\|$ $\usebox{\ddubr}~ \usebox{\dudbr}~ \usebox{\duubr}$\\[-2mm]

$\{\mathbf{r}_2^{}, \mathbf{r}_2^\star, \mathbf{r}_4^{}, \mathbf{r}_4^\star, \mathbf{r}_5^{-}, \mathbf{r}_5^\star\}$,~~~~~~~~~~~~~~$(\overline{\widetilde{1}}, 14)$\\[-3mm]

$\usebox{\dddr}~ \usebox{\ddur}~ \usebox{\duur}$ $\|$ $\usebox{\ddub}~ \usebox{\duub}~ \usebox{\uuub}$ $\|$
$\usebox{\ddurb}~ \usebox{\dudrb}~ \usebox{\duurb}$ $\|$ $\usebox{\ddubr}~ \usebox{\uudbr}~ \usebox{\dddbr}$\\[-2mm]

$\{\mathbf{r}_2^{}, \mathbf{r}_2^\star, \mathbf{r}_3^\star, \mathbf{r}_4^\star, \mathbf{r}_4^{\thicksim\star}, \mathbf{r}_5^{}, \mathbf{r}_5^\star, \mathbf{r}_5^{\star-}\}$,~~~~~~~$(3, 14)$\\[-3mm]

$\usebox{\ddur}~ \usebox{\duur}~ \usebox{\uuur}$ $\|$ $\usebox{\ddub}~ \usebox{\duub}$ $\|$
$\usebox{\ddurb}~ \usebox{\uudrb}~ \usebox{\duurb}~ \usebox{\uuurb}$ $\|$ $\usebox{\ddubr}~ \usebox{\duubr}$\\[-2mm]

$\{\mathbf{r}_2^{}, \mathbf{r}_2^\star, \mathbf{r}_3^\star, \mathbf{r}_4^\star, \mathbf{r}_4^{\thicksim\star}, \mathbf{r}_5^{-}, \mathbf{r}_5^\star, \mathbf{r}_5^{\star-}\}$,~~~~~$(\overline{3}, 14)$\\[-3mm]

$\usebox{\dddr}~ \usebox{\ddur}~ \usebox{\duur}$ $\|$ $\usebox{\ddub}~ \usebox{\duub}$ $\|$
$\usebox{\dddrb}~ \usebox{\ddurb}~ \usebox{\dudrb}~ \usebox{\duurb}$ $\|$ $\usebox{\ddubr}~ \usebox{\duubr}$\\[-2mm]

$\{\mathbf{r}_2^{}, \mathbf{r}_2^\star, \mathbf{r}_3^{}, \mathbf{r}_4^{}, \mathbf{r}_4^\thicksim, \mathbf{r}_5^{}, \mathbf{r}_5^{-}, \mathbf{r}_5^\star\}$,~~~~~~~~$(3^\star, 14)$\\[-3mm]

$\usebox{\ddur}~ \usebox{\duur}$ $\|$ $\usebox{\ddub}~ \usebox{\duub}~ \usebox{\uuub}$ $\|$
$\usebox{\ddurb}~ \usebox{\duurb}$ $\|$ $\usebox{\ddubr}~ \usebox{\uudbr}~ \usebox{\duubr}~ \usebox{\uuubr}$\\[-2mm]

$\{\mathbf{r}_2^{}, \mathbf{r}_2^\star, \mathbf{r}_3^{}, \mathbf{r}_4^{}, \mathbf{r}_4^\thicksim, \mathbf{r}_5^{}, \mathbf{r}_5^{-}, \mathbf{r}_5^{\star-}\}$,~~~~~~~$(\overline{3}^\star, 14)$\\[-3mm]

$\usebox{\ddur}~ \usebox{\duur}$ $\|$ $\usebox{\dddb}~ \usebox{\ddub}~ \usebox{\duub}$ $\|$
$\usebox{\ddurb}~ \usebox{\duurb}$ $\|$ $\usebox{\dddbr}~ \usebox{\ddubr}~ \usebox{\dudbr}~ \usebox{\duubr}$\\[-2mm]

$\{\mathbf{r}_2^{}, \mathbf{r}_2^\star, \mathbf{r}_4^{}, \mathbf{r}_4^\star, \mathbf{r}_5^{}, \mathbf{r}_5^\star\}$,~~~~~~~~~~~~~~~$(7, 14)$\\[-3mm]

$\usebox{\ddur}~ \usebox{\duur}~ \usebox{\uuur}$ $\|$ $\usebox{\ddub}~ \usebox{\duub}~ \usebox{\uuub}$ $\|$
$\usebox{\ddurb}~ \usebox{\uudrb}~ \usebox{\duurb}$ $\|$ $\usebox{\ddubr}~ \usebox{\uudbr}~ \usebox{\duubr}$\\[-2mm]

$\{\mathbf{r}_2^{}, \mathbf{r}_2^\star, \mathbf{r}_4^{}, \mathbf{r}_4^\star, \mathbf{r}_5^{-}, \mathbf{r}_5^{\star-}\}$,~~~~~~~~~~~~$(\bar 7, 14)$\\[-3mm]

$\usebox{\dddr}~ \usebox{\ddur}~ \usebox{\duur}$ $\|$ $\usebox{\dddb}~ \usebox{\ddub}~ \usebox{\duub}$ $\|$
$\usebox{\ddurb}~ \usebox{\dudrb}~ \usebox{\duurb}$ $\|$ $\usebox{\ddubr}~ \usebox{\dudbr}~ \usebox{\duubr}$\\[-2mm]

$\{\mathbf{r}_2^{}, \mathbf{r}_2^\star, \mathbf{r}_4^\thicksim, \mathbf{r}_4^{\thicksim\star}, \mathbf{r}_5^{}, \mathbf{r}_5^{\star-}\}$,~~~~~~~~~~~~$(\widetilde{7}, 14)$\\[-3mm]

$\usebox{\ddur}~ \usebox{\duur}~ \usebox{\uuur}$ $\|$ $\usebox{\dddb}~ \usebox{\ddub}~ \usebox{\duub}$ $\|$
$\usebox{\ddurb}~ \usebox{\duurb}~ \usebox{\uuurb}$ $\|$ $\usebox{\dddbr}~ \usebox{\ddubr}~ \usebox{\duubr}$\\[-2mm]

$\{\mathbf{r}_2^{}, \mathbf{r}_2^\star, \mathbf{r}_4^\thicksim, \mathbf{r}_4^{\thicksim\star},\mathbf{r}_5^{-}, \mathbf{r}_5^\star\}$,~~~~~~~~~~~~$(\overline{\widetilde{7}}, 14)$\\[-3mm]

$\usebox{\dddr}~ \usebox{\ddur}~ \usebox{\duur}$ $\|$ $\usebox{\ddub}~ \usebox{\duub}~ \usebox{\uuub}$ $\|$
$\usebox{\dddrb}~ \usebox{\ddurb}~ \usebox{\duurb}$ $\|$ $\usebox{\ddubr}~ \usebox{\duubr}~ \usebox{\uuubr}$\\[-2mm]

$\{\mathbf{r}_3^{}, \mathbf{r}_3^\star, \mathbf{r}_4^{}, \mathbf{r}_4^\star, \mathbf{r}_5^{}, \mathbf{r}_5^\star\}$,~~~~~~~~~~~~~~~$(8, 14)$\\[-3mm]

$\usebox{\ddur}~ \usebox{\uudr}~ \usebox{\duur}$ $\|$ $\usebox{\ddub}~ \usebox{\uudb}~ \usebox{\duub}$ $\|$
$\usebox{\ddurb}~ \usebox{\uudrb}~ \usebox{\duurb}$ $\|$ $\usebox{\ddubr}~ \usebox{\uudbr}~ \usebox{\duubr}$\\[-2mm]

$\{\mathbf{r}_3^{}, \mathbf{r}_3^\star, \mathbf{r}_4^{}, \mathbf{r}_4^\star, \mathbf{r}_5^{-}, \mathbf{r}_5^{\star-}\}$,~~~~~~~~~~~~$(\bar 8, 14)$\\[-3mm]

$\usebox{\ddur}~ \usebox{\dudr}~ \usebox{\duur}$ $\|$ $\usebox{\ddub}~ \usebox{\dudb}~ \usebox{\duub}$ $\|$
$\usebox{\ddurb}~ \usebox{\dudrb}~ \usebox{\duurb}$ $\|$ $\usebox{\ddubr}~ \usebox{\dudbr}~ \usebox{\duubr}$\\[-2mm]

$\{\mathbf{r}_3^{}, \mathbf{r}_3^\star, \mathbf{r}_4^\thicksim, \mathbf{r}_4^{\thicksim\star}, \mathbf{r}_5^{}, \mathbf{r}_5^{\star-}\}$,~~~~~~~~~~~~~$(\widetilde{8}, 14)$\\[-3mm]

$\usebox{\ddur}~ \usebox{\uudr}~ \usebox{\duur}$ $\|$ $\usebox{\ddub}~ \usebox{\dudb}~ \usebox{\duub}$ $\|$
$\usebox{\ddurb}~ \usebox{\duurb}~ \usebox{\uuurb}$ $\|$ $\usebox{\dddbr}~ \usebox{\ddubr}~ \usebox{\duubr}$\\[-2mm]

$\{\mathbf{r}_3^{}, \mathbf{r}_3^\star, \mathbf{r}_4^\thicksim, \mathbf{r}_4^{\thicksim\star}, \mathbf{r}_5^{-}, \mathbf{r}_5^\star\}$,~~~~~~~~~~~~~$(\overline{\widetilde{8}}, 14)$\\[-3mm]

$\usebox{\ddur}~ \usebox{\dudr}~ \usebox{\duur}$ $\|$ $\usebox{\ddub}~ \usebox{\uudb}~ \usebox{\duub}$ $\|$
$\usebox{\dddrb}~ \usebox{\ddurb}~ \usebox{\duurb}$ $\|$ $\usebox{\ddubr}~ \usebox{\duubr}~ \usebox{\uuubr}$\\[-2mm]

$\{\mathbf{r}_3^{}, \mathbf{r}_3^\star, \mathbf{r}_4^\thicksim, \mathbf{r}_4^{\thicksim\star}, \mathbf{r}_5^{}, \mathbf{r}_5^\star\}$,~~~~~~~~~~~~~~$(9, 14)$\\[-3mm]

$\usebox{\ddur}~ \usebox{\uudr}~ \usebox{\duur}$ $\|$ $\usebox{\ddub}~ \usebox{\uudb}~ \usebox{\duub}$ $\|$
$\usebox{\ddurb}~ \usebox{\duurb}~ \usebox{\uuurb}$ $\|$ $\usebox{\ddubr}~ \usebox{\duubr}~ \usebox{\uuubr}$\\[-2mm]

$\{\mathbf{r}_3^{}, \mathbf{r}_3^\star, \mathbf{r}_4^\thicksim, \mathbf{r}_4^{\thicksim\star}, \mathbf{r}_5^{-}, \mathbf{r}_5^{\star-}\}$,~~~~~~~~~~~$(\bar 9, 14)$\\[-3mm]

$\usebox{\ddur}~ \usebox{\dudr}~ \usebox{\duur}$ $\|$ $\usebox{\ddub}~ \usebox{\dudb}~ \usebox{\duub}$ $\|$
$\usebox{\dddrb}~ \usebox{\ddurb}~ \usebox{\duurb}$ $\|$ $\usebox{\dddbr}~ \usebox{\ddubr}~ \usebox{\duubr}$\\[-2mm]

$\{\mathbf{r}_3^{}, \mathbf{r}_3^\star, \mathbf{r}_4^{}, \mathbf{r}_4^\star, \mathbf{r}_5^{}, \mathbf{r}_5^{\star-}\}$,~~~~~~~~~~~~~~$(\widetilde{9}, 14)$\\[-3mm]

$\usebox{\ddur}~ \usebox{\uudr}~ \usebox{\duur}$ $\|$ $\usebox{\ddub}~ \usebox{\dudb}~ \usebox{\duub}$ $\|$
$\usebox{\ddurb}~ \usebox{\uudrb}~ \usebox{\duurb}$ $\|$ $\usebox{\ddubr}~ \usebox{\dudbr}~ \usebox{\duubr}$\\[-2mm]

$\{\mathbf{r}_3^{}, \mathbf{r}_3^\star, \mathbf{r}_4^{}, \mathbf{r}_4^\star, \mathbf{r}_5^{-}, \mathbf{r}_5^\star\}$,~~~~~~~~~~~~~~$(\overline{\widetilde{9}}, 14)$\\[-3mm]

$\usebox{\ddur}~ \usebox{\dudr}~ \usebox{\duur}$ $\|$ $\usebox{\ddub}~ \usebox{\uudb}~ \usebox{\duub}$ $\|$
$\usebox{\ddurb}~ \usebox{\dudrb}~ \usebox{\duurb}$ $\|$ $\usebox{\ddubr}~ \usebox{\uudbr}~ \usebox{\duubr}$\\[-2mm]

$\{\mathbf{r}_2^\star, \mathbf{r}_3^{}, \mathbf{r}_4^\thicksim, \mathbf{r}_4^{\thicksim\star}, \mathbf{r}_5^{}, \mathbf{r}_5^\star\}$,~~~~~~~~~~~~~~$(10, 14)$\\[-3mm]

$\usebox{\ddur}~ \usebox{\uudr}~ \usebox{\duur}$ $\|$ $\usebox{\ddub}~ \usebox{\duub}~ \usebox{\uuub}$ $\|$
$\usebox{\ddurb}~ \usebox{\duurb}~ \usebox{\uuurb}$ $\|$ $\usebox{\ddubr}~ \usebox{\duubr}~ \usebox{\uuubr}$\\[-2mm]

$\{\mathbf{r}_2^\star, \mathbf{r}_3^{}, \mathbf{r}_4^\thicksim, \mathbf{r}_4^{\thicksim\star}, \mathbf{r}_5^{-}, \mathbf{r}_5^{\star-}\}$,~~~~~~~~~~~$(\overline{10}, 14)$\\[-3mm]

$\usebox{\ddur}~ \usebox{\dudr}~ \usebox{\duur}$ $\|$ $\usebox{\dddb}~ \usebox{\ddub}~ \usebox{\duub}$ $\|$
$\usebox{\dddrb}~ \usebox{\ddurb}~ \usebox{\duurb}$ $\|$ $\usebox{\dddbr}~ \usebox{\ddubr}~ \usebox{\duubr}$\\[-2mm]

$\{\mathbf{r}_2^{}, \mathbf{r}_3^\star, \mathbf{r}_4^\thicksim, \mathbf{r}_4^{\thicksim\star}, \mathbf{r}_5^{}, \mathbf{r}_5^\star\}$,~~~~~~~~~~~~~~$(10^\star, 14)$\\[-3mm]

$\usebox{\ddur}~ \usebox{\duur}~ \usebox{\uuur}$ $\|$ $\usebox{\ddub}~ \usebox{\uudb}~ \usebox{\duub}$ $\|$
$\usebox{\ddurb}~ \usebox{\duurb}~ \usebox{\uuurb}$ $\|$ $\usebox{\ddubr}~ \usebox{\duubr}~ \usebox{\uuubr}$\\[-2mm]

$\{\mathbf{r}_2^{}, \mathbf{r}_3^\star, \mathbf{r}_4^\thicksim, \mathbf{r}_4^{\thicksim\star}, \mathbf{r}_5^{-}, \mathbf{r}_5^{\star-}\}$,~~~~~~~~~~~$(\overline{10^\star}, 14)$\\[-3mm]

$\usebox{\dddr}~ \usebox{\ddur}~ \usebox{\duur}$ $\|$ $\usebox{\ddub}~ \usebox{\dudb}~ \usebox{\duub}$ $\|$
$\usebox{\dddrb}~ \usebox{\ddurb}~ \usebox{\duurb}$ $\|$ $\usebox{\dddbr}~ \usebox{\ddubr}~ \usebox{\duubr}$\\[-2mm]

$\{\mathbf{r}_2^\star, \mathbf{r}_3^{}, \mathbf{r}_4^{}, \mathbf{r}_4^\star, \mathbf{r}_5^{}, \mathbf{r}_5^{\star-}\}$,~~~~~~~~~~~~~~$(\widetilde{10}, 14)$\\[-3mm]

$\usebox{\ddur}~ \usebox{\uudr}~ \usebox{\duur}$ $\|$ $\usebox{\dddb}~ \usebox{\ddub}~ \usebox{\duub}$ $\|$
$\usebox{\ddurb}~ \usebox{\uudrb}~ \usebox{\duurb}$ $\|$ $\usebox{\ddubr}~ \usebox{\dudbr}~ \usebox{\duubr}$\\[-2mm]

$\{\mathbf{r}_2^\star, \mathbf{r}_3^{}, \mathbf{r}_4^{}, \mathbf{r}_4^\star, \mathbf{r}_5^{-}, \mathbf{r}_5^\star\}$,~~~~~~~~~~~~~~$(\overline{\widetilde{10}}, 14)$\\[-3mm]

$\usebox{\ddur}~ \usebox{\dudr}~ \usebox{\duur}$ $\|$ $\usebox{\ddub}~ \usebox{\duub}~ \usebox{\uuub}$ $\|$
$\usebox{\ddurb}~ \usebox{\dudrb}~ \usebox{\duurb}$ $\|$ $\usebox{\ddubr}~ \usebox{\uudbr}~ \usebox{\duubr}$\\[-2mm]

$\{\mathbf{r}_2^{}, \mathbf{r}_3^\star, \mathbf{r}_4^{}, \mathbf{r}_4^\star, \mathbf{r}_5^{}, \mathbf{r}_5^{\star-}\}$,~~~~~~~~~~~~~$(^\star\widetilde{10}, 14)$\\[-3mm]

$\usebox{\ddur}~ \usebox{\duur}~ \usebox{\uuur}$ $\|$ $\usebox{\ddub}~ \usebox{\dudb}~ \usebox{\duub}$ $\|$
$\usebox{\ddurb}~ \usebox{\uudrb}~ \usebox{\duurb}$ $\|$ $\usebox{\ddubr}~ \usebox{\dudbr}~ \usebox{\duubr}$\\[-2mm]

$\{\mathbf{r}_2^{}, \mathbf{r}_3^\star, \mathbf{r}_4^{}, \mathbf{r}_4^\star, \mathbf{r}_5^{-}, \mathbf{r}_5^\star\}$,~~~~~~~~~~~~~$(\overline{^\star\widetilde{10}}, 14)$\\[-3mm]

$\usebox{\dddr}~ \usebox{\ddur}~ \usebox{\duur}$ $\|$ $\usebox{\ddub}~ \usebox{\uudb}~ \usebox{\duub}$ $\|$
$\usebox{\ddurb}~ \usebox{\dudrb}~ \usebox{\duurb}$ $\|$ $\usebox{\ddubr}~ \usebox{\uudbr}~ \usebox{\duubr}$\\[-2mm]

$\{\mathbf{r}_2^\star, \mathbf{r}_3^{}, \mathbf{r}_3^\star, \mathbf{r}_4^\star, \mathbf{r}_4^{\thicksim\star}, \mathbf{r}_5^{}, \mathbf{r}_5^\star, \mathbf{r}_5^{\star-}\}$,~~~~~~~$(11, 14)$\\[-3mm]

$\usebox{\ddur}~ \usebox{\uudr}~ \usebox{\duur}$ $\|$ $\usebox{\ddub}~ \usebox{\duub}$ $\|$
$\usebox{\ddurb}~ \usebox{\uudrb}~ \usebox{\duurb}~ \usebox{\uuurb}$ $\|$ $\usebox{\ddubr}~ \usebox{\duubr}$\\[-2mm]

$\{\mathbf{r}_2^\star, \mathbf{r}_3^{}, \mathbf{r}_3^\star, \mathbf{r}_4^\star, \mathbf{r}_4^{\thicksim\star}, \mathbf{r}_5^{-}, \mathbf{r}_5^\star, \mathbf{r}_5^{\star-}\}$,~~~~~$(\overline{11}, 14)$\\[-3mm]

$\usebox{\ddur}~ \usebox{\dudr}~ \usebox{\duur}$ $\|$ $\usebox{\ddub}~ \usebox{\duub}$ $\|$
$\usebox{\dddrb}~ \usebox{\ddurb}~ \usebox{\dudrb}~ \usebox{\duurb}$ $\|$ $\usebox{\ddubr}~ \usebox{\duubr}$\\[-2mm]

$\{\mathbf{r}_2^{}, \mathbf{r}_3^{}, \mathbf{r}_3^\star, \mathbf{r}_4^{}, \mathbf{r}_4^\thicksim, \mathbf{r}_5^{}, \mathbf{r}_5^{-}, \mathbf{r}_5^\star\}$,~~~~~~~~$(11^\star, 14)$\\[-3mm]

$\usebox{\ddur}~ \usebox{\duur}$ $\|$ $\usebox{\ddub}~ \usebox{\uudb}~ \usebox{\duub}$ $\|$
$\usebox{\ddurb}~ \usebox{\duurb}$ $\|$ $\usebox{\ddubr}~ \usebox{\uudbr}~ \usebox{\duubr}~ \usebox{\uuubr}$\\[-2mm]

$\{\mathbf{r}_2^{}, \mathbf{r}_3^{}, \mathbf{r}_3^\star, \mathbf{r}_4^{}, \mathbf{r}_4^\thicksim, \mathbf{r}_5^{}, \mathbf{r}_5^{-}, \mathbf{r}_5^{\star-}\}$,~~~~~~~$(\overline{11}^\star, 14)$\\[-3mm]

$\usebox{\ddur}~ \usebox{\duur}$ $\|$ $\usebox{\ddub}~ \usebox{\dudb}~ \usebox{\duub}$ $\|$
$\usebox{\ddurb}~ \usebox{\duurb}$ $\|$ $\usebox{\dddbr}~ \usebox{\ddubr}~ \usebox{\dudbr}~ \usebox{\duubr}$\\[-2mm]

$\{\mathbf{r}_2^\star, \mathbf{r}_3^{}, \mathbf{r}_4^{}, \mathbf{r}_4^\star, \mathbf{r}_5^{}, \mathbf{r}_5^\star\}$,~~~~~~~~~~~~~~~$(24, 14)$\\[-3mm]

$\usebox{\ddur}~ \usebox{\uudr}~ \usebox{\duur}$ $\|$ $\usebox{\ddub}~ \usebox{\duub}~ \usebox{\uuub}$ $\|$
$\usebox{\ddurb}~ \usebox{\uudrb}~ \usebox{\duurb}$ $\|$ $\usebox{\ddubr}~ \usebox{\uudbr}~ \usebox{\duubr}$\\[-2mm]

$\{\mathbf{r}_2^\star, \mathbf{r}_3^{}, \mathbf{r}_4^{}, \mathbf{r}_4^\star, \mathbf{r}_5^{-}, \mathbf{r}_5^{\star-}\}$,~~~~~~~~~~~~$(\overline{24}, 14)$\\[-3mm]

$\usebox{\ddur}~ \usebox{\dudr}~ \usebox{\duur}$ $\|$ $\usebox{\dddb}~ \usebox{\ddub}~ \usebox{\duub}$ $\|$
$\usebox{\ddurb}~ \usebox{\dudrb}~ \usebox{\duurb}$ $\|$ $\usebox{\ddubr}~ \usebox{\dudbr}~ \usebox{\duubr}$\\[-2mm]

$\{\mathbf{r}_2^{}, \mathbf{r}_3^\star, \mathbf{r}_4^{}, \mathbf{r}_4^\star, \mathbf{r}_5^{}, \mathbf{r}_5^\star\}$,~~~~~~~~~~~~~~~$(24^\star, 14)$\\[-3mm]

$\usebox{\ddur}~ \usebox{\duur}~ \usebox{\uuur}$ $\|$ $\usebox{\ddub}~ \usebox{\uudb}~ \usebox{\duub}$ $\|$
$\usebox{\ddurb}~ \usebox{\uudrb}~ \usebox{\duurb}$ $\|$ $\usebox{\ddubr}~ \usebox{\uudbr}~ \usebox{\duubr}$\\[-2mm]

$\{\mathbf{r}_2^{}, \mathbf{r}_3^\star, \mathbf{r}_4^{}, \mathbf{r}_4^\star, \mathbf{r}_5^{-}, \mathbf{r}_5^{\star-}\}$,~~~~~~~~~~~~$(\overline{24}^\star, 14)$\\[-3mm]

$\usebox{\dddr}~ \usebox{\ddur}~ \usebox{\duur}$ $\|$ $\usebox{\ddub}~ \usebox{\dudb}~ \usebox{\duub}$ $\|$
$\usebox{\ddurb}~ \usebox{\dudrb}~ \usebox{\duurb}$ $\|$ $\usebox{\ddubr}~ \usebox{\dudbr}~ \usebox{\duubr}$\\[-2mm]

$\{\mathbf{r}_2^\star, \mathbf{r}_3^{}, \mathbf{r}_4^\thicksim, \mathbf{r}_4^{\thicksim\star}, \mathbf{r}_5^{}, \mathbf{r}_5^{\star-}\}$,~~~~~~~~~~~~$(\widetilde{24}, 14)$\\[-3mm]

$\usebox{\ddur}~ \usebox{\uudr}~ \usebox{\duur}$ $\|$ $\usebox{\dddb}~ \usebox{\ddub}~ \usebox{\duub}$ $\|$
$\usebox{\ddurb}~ \usebox{\duurb}~ \usebox{\uuurb}$ $\|$ $\usebox{\dddbr}~ \usebox{\ddubr}~ \usebox{\duubr}$\\[-2mm]

$\{\mathbf{r}_2^\star, \mathbf{r}_3^{}, \mathbf{r}_4^\thicksim, \mathbf{r}_4^{\thicksim\star}, \mathbf{r}_5^{-}, \mathbf{r}_5^\star\}$,~~~~~~~~~~~~$(\overline{\widetilde{24}}, 14)$\\[-3mm]

$\usebox{\ddur}~ \usebox{\dudr}~ \usebox{\duur}$ $\|$ $\usebox{\ddub}~ \usebox{\duub}~ \usebox{\uuub}$ $\|$
$\usebox{\dddrb}~ \usebox{\ddurb}~ \usebox{\duurb}$ $\|$ $\usebox{\ddubr}~ \usebox{\duubr}~ \usebox{\uuubr}$\\[-2mm]

$\{\mathbf{r}_2^{}, \mathbf{r}_3^\star, \mathbf{r}_4^\thicksim, \mathbf{r}_4^{\thicksim\star}, \mathbf{r}_5^{}, \mathbf{r}_5^{\star-}\}$,~~~~~~~~~~~~$(^\star\widetilde{24}, 14)$\\[-3mm]

$\usebox{\ddur}~ \usebox{\duur}~ \usebox{\uuur}$ $\|$ $\usebox{\ddub}~ \usebox{\dudb}~ \usebox{\duub}$ $\|$
$\usebox{\ddurb}~ \usebox{\duurb}~ \usebox{\uuurb}$ $\|$ $\usebox{\dddbr}~ \usebox{\ddubr}~ \usebox{\duubr}$\\[-2mm]

$\{\mathbf{r}_2^{}, \mathbf{r}_3^\star, \mathbf{r}_4^\thicksim, \mathbf{r}_4^{\thicksim\star}, \mathbf{r}_5^{-}, \mathbf{r}_5^\star\}$,~~~~~~~~~~~~$(\overline{^\star\widetilde{24}}, 14)$\\[-3mm]

$\usebox{\dddr}~ \usebox{\ddur}~ \usebox{\duur}$ $\|$ $\usebox{\ddub}~ \usebox{\uudb}~ \usebox{\duub}$ $\|$
$\usebox{\dddrb}~ \usebox{\ddurb}~ \usebox{\duurb}$ $\|$ $\usebox{\ddubr}~ \usebox{\duubr}~ \usebox{\uuubr}$\\